\documentclass[12pt]{report}
\usepackage{graphicx,amsmath,amssymb}
\usepackage{indentfirst}
\usepackage[usenames]{color}
\usepackage[colorlinks=true, urlcolor=navyblue, linkcolor=navyblue, citecolor=navyblue]{hyperref}
\usepackage{epstopdf}
\usepackage{enumerate}
\usepackage{appendix}
\usepackage{lineno}
\usepackage{setspace}

\definecolor{navyblue}{rgb}{0,0.08,0.45}
\definecolor{darkred}{rgb}{0.7,0.0,0.0}
\definecolor{darkgreen}{rgb}{0,0.6,0.2}
\definecolor{green}{rgb}{0.4,0.6,0}

\def\Dslash{\raise.15ex\hbox{/}\kern-.7em D}
\def\Pslash{\raise.15ex\hbox{/}\kern-.7em P}

\newcommand{\beq}{\begin{equation}}
\newcommand{\enq}{\end{equation}}
\newcommand{\beqa}{\begin{eqnarray}}
\newcommand{\beqast}{\begin{eqnarray*}}
\newcommand{\enqa}{\end{eqnarray}}
\newcommand{\enqast}{\end{eqnarray*}}
\newcommand{\nn}{\nonumber}
\newcommand{\req}[1]{(\ref{#1})}

\renewcommand{\arraystretch}{1.3}

\newcommand{\mbf}[1]{\mathbf{#1}}

\newcommand{\half}{{\frac{1}{2}}}
 \newcommand{\threehalf}{{\frac{3}{2}}}
 \newcommand{\fivehalf}{{\frac{5}{2}}}
 \newcommand{\sevenhalf}{{\frac{7}{2}}}
 \newcommand{\ninehalf}{{\frac{9}{2}}}
 \newcommand{\elevenhalf}{{\frac{11}{2}}}
 \newcommand{\thirteenhalf}{{\frac{13}{2}}}

\newcommand{\pa}{\partial}

\newcommand{\bec}{\begin{center}}
\newcommand{\enc}{\end{center}}
\newcommand{\beqo}{\begin{quote}}
\newcommand{\enqo}{\end{quote}}

\newcommand{\cL}{{\cal L}}

\newcommand{\al}{\alpha}
\newcommand{\be}{\beta}
\newcommand{\ga}{\gamma}
\newcommand{\de}{\delta}
\newcommand{\ep}{\epsilon}
\newcommand{\ze}{\zeta}
\newcommand{\et}{\eta}

\newcommand{\ka}{\kappa}
\newcommand{\la}{\lambda}
\newcommand{\rh}{\rho}
\newcommand{\si}{\sigma}

\newcommand{\om}{\omega}

\newcommand{\vp}{\varphi}

\newcommand{\Ga}{\Gamma}
\newcommand{\De}{\Delta}

\newcommand{\La}{\Lambda}

\newcommand{\Si}{\Sigma}

\newcommand{\Ph}{\Phi}

\newcommand{\Om}{\Omega}

\begin{document}

\begin{flushright}
{\small SLAC--PUB--15972 \\ \vspace{3pt}
}
\end{flushright}

\vspace{40pt}

\centerline{\huge Light-Front Holographic QCD}

\vspace{10pt}

\centerline{\huge {and Emerging Confinement}}

\vspace{50pt}

\centerline{Stanley J. Brodsky}

\vspace{5pt}

\centerline {\it SLAC National Accelerator Laboratory,
Stanford University, Stanford, CA 94309, USA}

\vspace{15pt}

\centerline{Guy F. de T\'eramond}

\vspace{5pt}

{\centerline {\it Universidad de Costa Rica, San Jos\'e, Costa Rica}

\vspace{15pt}

\centerline{Hans G\"unter Dosch}

\vspace{5pt}

\centerline {\it Institut f\"ur Theoretische Physik, Philosophenweg 16, D-6900 Heidelberg, Germany}

\vspace{15pt}

\centerline{Joshua Erlich}

\vspace{5pt}

\centerline {\it College of William and Mary, Williamsburg, VA 23187, USA}

\vspace{60pt}

{\small
\centerline{\href{mailto:sjbth@slac.stanford.edu}{\tt sjbth@slac.stanford.edu}, \, \href{mailto:gdt@asterix.crnet.cr}{\tt gdt@asterix.crnet.cr},}

\vspace{1pt}

\centerline{\href{mailto:dosch@thphys.uni-heidelberg.de}{\tt h.g.dosch@thphys.uni-heidelberg.de}, \, \href{mailto:jxerli@wm.edu}{\tt jxerli@wm.edu}}
}

 \vspace{40pt}

 \centerline{ \it (Invited review article to appear in Physics Reports)}

\newpage


\begin{abstract}

In this report we explore the remarkable connections between light-front dynamics, its holographic mapping to gravity in a higher-dimensional anti-de Sitter (AdS) space, and conformal quantum mechanics. This approach provides new insights into the origin of a fundamental mass scale and the physics underlying confinement dynamics in QCD in the limit of massless quarks.  The result is a relativistic light-front wave equation for arbitrary spin with an effective confinement potential derived from a conformal action and its embedding in AdS space. This equation allows for the computation of essential features of hadron spectra  in terms of a single  scale. The light-front holographic methods described here gives a  precise  interpretation of holographic variables and quantities in AdS space in terms of  light-front variables and quantum numbers.  This leads to a  relation between the AdS wave functions  and the boost-invariant light-front wave functions describing the internal structure of hadronic bound-states in physical space-time. The pion is massless in the chiral limit and the excitation spectra of relativistic light-quark meson and baryon bound states lie on linear Regge trajectories with identical slopes in the radial and orbital quantum numbers.  In the light-front holographic approach described here currents are expressed as an infinite sum of poles, and form factors as a product of poles.  At large $q^2$ the form factor incorporates the correct power-law fall-off for hard scattering independent of the specific dynamics and is dictated by the twist. At low $q^2$ the form factor leads to vector dominance. The approach is also extended to include small quark masses. We  briefly review in this report other holographic approaches to QCD, in particular top-down and bottom-up models based on chiral symmetry breaking.   We also include a discussion of open problems and future applications.

\end{abstract}

\tableofcontents

\chapter{Introduction \label{ch1}}

\section{Motivation}

Quantum Chromodynamics (QCD),  the $SU(3)$ color gauge field theory of quarks and gluons,  is the standard theory of strong interactions. High energy experiments, such as  the deep inelastic electron-proton scattering pioneered at SLAC~\cite{Breidenbach:1969kd},  which revealed the quark structure of the proton,  and continued  at DESY~\cite{Perez:2012um} to extremely short distances, have shown that the basic elementary interactions of quarks and gluons are remarkably well described by QCD~\cite{Marciano:1977su}.  Yet, because of its strong-coupling nature, it has been difficult to make precise predictions outside of its short-distance perturbative domain where it has been tested to high precision.  Unlike  Quantum Electrodynamics (QED), the fundamental theory of electrons and photons, the strong couplings  of quarks and gluons at large-distances  makes the calculation of hadronic properties, such as hadron masses, a very difficult problem to solve, notwithstanding that the fundamental QCD Lagrangian is well established. In particular, one has no analytical understanding of how quarks and gluons are  permanently confined and how hadrons  emerge as asymptotic states in this theory~\cite{Brambilla:2014aaa}.  In fact, in the limit of massless quarks no scale appears  in the QCD Lagrangian.  The classical  Lagrangian of QCD is thus  invariant under conformal transformations~\cite{Parisi:1972zy, Braun:2003rp}.  Nonetheless, the quantum theory built upon this conformal theory displays color confinement,  a mass gap, and asymptotic freedom.  One then confronts a fundamental question: how does the mass scale which determines the masses of the light-quark hadrons, the range of color confinement, as well as the running of the coupling appear in QCD?

Euclidean lattice methods~\cite{Wilson:1974sk} provide an important first-principle numerical simulation of nonperturbative QCD.   However, the excitation spectrum of hadrons represents a difficult challenge to lattice QCD due to the enormous computational complexity beyond ground-state configurations and the unavoidable presence of multi-hadron thresholds~\cite{Fodor:2012gf}. Furthermore, dynamical observables in Minkowski space-time are not  obtained directly from Euclidean space lattice computations.  Other methods, as for example the Dyson-Schwinger equations,  have also led to many important insights, such as the infrared fixed-point behavior of the  strong coupling constant and the pattern of dynamical quark mass generation~\cite{Cornwall:1981zr, Roberts:1994dr, Alkofer:2000wg, Fischer:2006ub}. In practice, however, these analyses have been limited to ladder approximation in Landau gauge.

A problem, common to all realistic relativistic quantum field theories, is especially flagrant in QCD: the only known analytically tractable treatment is perturbation theory, which obviously is not the most appropriate tool for solving a strongly interacting theory with permanently confined constituents. In fact, according to the Kinoshita-Lee-Nauenberg theorem, which applies to any order of perturbation theory, a  description of confinement using perturbative QCD is not possible in a simple way~\cite{Kinoshita:1962ur, Lee:1964is}. Thus, an important theoretical goal is  to find an initial approximation to QCD in its strongly coupled regime relevant at large distances,  which is both analytically tractable and can be systematically improved.  In fact,  even in weakly interacting theories, like QED, there is a need for semiclassical equations in order to treat bound states. The Schr\"odinger and Dirac equations play a central role in atomic physics, providing simple,  but effective, first approximations  of  the spectrum and wave functions of bound states which can  be systematically improved using the Bethe-Salpeter formalism~\cite{Salpeter:1951sz} and including corrections for quantum fluctuations, such as the Lamb shift and vacuum polarization. A long-sought goal in hadron physics is to find a simple analytic first approximation to QCD, analogous to the Schr\"odinger equation of atomic physics.  This  task is particularly challenging since the formalism must be fully relativistic,  give a good description of the hadron spectrum, and should also explain essential dynamical properties of hadrons.  There are several indications that such a goal might  well be within reach:

\begin{enumerate}[i)]

\item The quark model, based mainly on the Schr\"odinger equation with relativistic corrections is qualitatively very successful (See {\it e.g.}, \cite{PDG:2014}, Sect. 14).

\item There are striking regularities in the hadronic spectra, notably Regge trajectories~\cite{Regge:1959mz, Regge:1960zc}, which show a linear relation between the squared mass and the intrinsic angular momentum of hadrons (See {\it e.g.}, \cite{Collins:1977jy, Donnachie:2002en}).

\item There exists a convenient frame-independent Hamiltonian framework for treating bound-states in relativistic theories using light-front quantization. It is based on the front-form or relativistic dynamics~\cite{Dirac:1949cp}, where initial conditions are specified in the light-cone null-plane $x^0 + x^3 = 0$, not on the usual initial conditions at equal time, $x^0=0$.

\end{enumerate}

\noindent As an effective theory, we expect also that the resulting model incorporates underlying symmetries of the QCD Lagrangian.

\section{The AdS/CFT correspondence and holographic QCD}

The search for semiclassical equations in QCD obtained a strong advance some 15 years ago by the Maldacena Conjecture~\cite{Maldacena:1997re}. Roughly speaking, the conjecture states that a quantum gauge field theory in 4 dimensions corresponds to a classical gravitational theory in 5 dimensions.  In this type of correspondence the higher-dimensional gravitational theory is referred to as the holographic dual, or gravity dual, of the lower-dimensional quantum field theory. Holographic ideas in  physical theories have their origin in the seminal work of  Bekenstein and Hawking in the 1970s \cite{Bekenstein:1973ur, Hawking:1974sw}, which led to the surprising conclusion that black holes are thermodynamic systems which radiate at a temperature which depends on the size of the black hole.  The most unusual aspect of black-hole thermodynamics is that  the entropy of a black hole is proportional to the area of its horizon,  contrary to the typical situation in non-gravitational systems, in which entropy is an extensive quantity proportional to the volume of the system.  The maximal entropy of a system is a measure of the number of degrees of freedom in that system, so the distinction between gravitational and non-gravitational systems appears to limit the number of degrees of freedom of a gravitational system to that of a non-gravitational system in one fewer spatial dimension.  This idea was formalized  as the holographic principle, which postulates that a gravitational system may indeed be equivalent to a non-gravitational system in one fewer dimension~\cite{'tHooft:1993gx, Susskind:1994vu}.

The AdS/CFT correspondence between gravity on a higher-dimensional anti--de Sitter (AdS) space and conformal field theories (CFT) in a lower-dimensional space-time~\cite{Maldacena:1997re}, is an explicit realization of the holographic principle, and it remains a major focus of string theory research.   This correspondence has led to a  semiclassical gravity approximation for strongly-coupled quantum field theories, providing physical insights into its nonperturbative dynamics.  In practice, it provides an effective gravity description in a ($d+1$)-dimensional AdS, or other curved space-time, in terms of a flat $d$-dimensional conformally-invariant quantum field theory defined on the AdS asymptotic boundary, the boundary theory.  In the semiclassical approximation, the generating functional of the quantum field theory is given by the minimum of the classical action of the gravitational theory at the  4-dimensional asymptotic border of the 5-dimensional space~\cite{Gubser:1998bc, Witten:1998qj}. Thus, in principle, one can compute physical observables in a strongly coupled gauge theory  in terms of a weakly coupled classical gravity theory, which encodes information of the boundary theory.

In the prototypical example~\cite{Maldacena:1997re} of this duality, the gauge theory is ${\cal N}=4$ supersymmetric $SU(N_C)$ Yang-Mills theory (SYM), the maximally  supersymmetric gauge field theory in four-dimensional space-time.  The gravitational dual is Type IIB supergravity or string theory~\cite{Polchinski:1998rr} \footnote{A brief discussion of  holographic top-down duality with string theory is given in Chapter \ref{ch7}.}, depending on the gauge coupling and the number of colors $N_C$, in a direct product of five-dimensional AdS space-time and a five-sphere: AdS$_5\times S^5$. If $g$ is the gauge coupling of the Yang-Mills theory, then in the limit $N_C \to \infty$, with $g^2 N_C \gg 1$ but finite,  the limit of large 't Hooft coupling, $g^2 N_C$, ensures that the space-time geometry has curvature $\mathcal{R}$ much smaller than the string scale $1/l_s^2$ so that classical gravity is a good approximation.  A small curvature $\mathcal{R}$, thus implies a large AdS radius $R$, $\mathcal{R} \sim 1/R^2$, where $R=(4\pi g^2 N_C)^{1/4} l_s$~\cite{Maldacena:1997re}. Since the gauge coupling $g$ and string coupling $g_s$ are related by $g^2 = g_s$, the limit $N_C\rightarrow\infty$  ensures that the string coupling is small, so that stringy effects decouple~\footnote{A recent review of large $N_C$ gauge theories is given in Ref. \cite{Lucini:2012gg}.}.

Anti-de Sitter AdS$_{d+1}$-dimensional space-time is the maximally symmetric $d+1$ space with negative constant curvature and a $d$-dimensional flat space-time boundary.  In Poincar\'e  coordinates   $x^0, x^1, \cdots , x^d, z \equiv x^{d+1}$, where the asymptotic border to the physical four-dimensional space-time  is given by $z = 0$, the line element  is
\begin{equation} \label{AdSm}
ds^2 = \frac{R^2}{z^2} \left(\eta_{\mu \nu} dx^\mu dx^\nu - dz^2\right),
\end{equation}
where $\eta_{\mu \nu}$ is the usual  Minkowski metric in $d$ dimensions. The most general group of transformations that leave the AdS$_{d+1}$ differential line element invariant, the isometry group $SO(2,d)$ has dimension $(d+1)(d+2)/2$. In the AdS/CFT correspondence, the consequence of the $SO(2,4)$  isometry of AdS$_5$ is the conformal invariance of the  dual field theory.  Five-dimensional anti-de Sitter space AdS$_5$ has 15 isometries, which induce in the  Minkowski-space boundary theory the symmetry under the conformal group   ${\it Conf}\!\left(R^{1,3}\right)$ with 15 generators in four dimensions: 6 Lorentz transformations plus 4 space-time translations plus 4 special conformal transformations plus 1 dilatation~\cite{Mack:1969rr}.   This conformal symmetry implies that there can be no scale in the theory and therefore also no discrete spectrum. Indeed, ${\cal N}=4$ supersymmetric  $SU(N_C)$ Yang-Mills theory is a conformal field theory.

The AdS/CFT correspondence can  be extended to non-conformal and supersymmetric or non-supersymmetric quantum field theories, a duality also known as ``gauge/gravity'' or ``gauge/string'' duality, which expresses well the  generality of the conjectured duality.  In particular, it is important to note that the conformal invariance of the prototypical example, ${\cal N}$=4 supersymmetric $SU(N)$ Yang-Mills theory in 3+1 dimensions, is not required for the existence of a higher-dimensional gravity dual, and one can deform the original background geometry, giving rise to less symmetric gravity duals of confining theories with large 't Hooft coupling $g^2 N_C$~\cite{Polchinski:2000uf, Klebanov:2000hb}. For example Polchinski and Strassler considered a modification of ${\cal N}$=4 Yang-Mills theory which includes ${\cal N}$=1 supersymmetry-preserving masses for some of the fields (the ${\cal N}$=1 chiral multiplets), and they describe the gravity dual of this theory in a certain limit of scales and 't Hooft coupling~\cite{Polchinski:2000uf}. The nonvanishing masses break the conformal symmetry, and the resulting theory is confining at low energies.  Another way to arrive at a non-conformal theory is to consider  systems with nonvanishing temperature~\cite{Witten:1998zw, Rey:1998bq, Brandhuber:1998bs, Gross:1998gk},  where one coordinate is compactified. Yet another example is the Sakai-Sugimoto (SS) model \cite{Sakai:2004cn, Sakai:2005yt}, based on a specific brane construction in Type IIA string theory~\cite{Polchinski:1998rr}; however since it is similar to finite temperature models, it is neither conformal nor supersymmetric. The SS model is notable in that it is confining  and contains vector mesons and pions in its spectrum from the breaking of  $SU(N_f) \times SU(N_f)$ chiral symmetry. We will describe this model in Chapter \ref{ch7}.

The AdS/CFT duality provides a useful guide in the attempt to model QCD as a higher-dimensional gravitational theory, but in contrast with the ``top-down'' holographic approach described above, which is to a great extent  constrained by the symmetries,  no gravity theory dual to QCD  is known.    The boundary (four-dimensional) quantum field theory, defined at the asymptotic AdS boundary at $z = 0$, becomes the initial state of the higher-dimensional gravity theory (the bulk theory). However,  to construct a dual holographic theory starting from a given quantum field theory in physical flat space-time, one would require in addition to the boundary conditions -- the boundary theory, precise knowledge of the dynamical evolution in the bulk. Therefore, for phenomenological purposes it is more promising to follow a ``bottom-up" approach, that is to start from a realistic 4-dimensional quantum field theory and look for a corresponding higher dimensional classical gravitational theory which encodes basic aspects of the boundary theory.

QCD is fundamentally different from the supersymmetric Yang-Mills  theory occurring in the Maldacena correspondence. In contrast with QCD, where quarks transform under the fundamental representation of $SU(3)$, in SYM all quark fields transform under the adjoint representations of  $SU(N_C)$.  The conformal invariance of SYM theories implies that the  $\beta$-function vanishes and, therefore, the coupling is scale independent. On the AdS side, the conformal symmetry corresponds to the maximal symmetry of this space. The classical QCD Lagrangian with massless quarks is also conformally invariant in four dimensions where its coupling $g_s$ is  dimensionless.  A scale,  however,  is introduced by quantum effects, and therefore its conformal invariance is broken and its coupling depends on the energy scale $\mu$ at which it is measured. We may compute the scale at which $g_s^2(\mu)/4\pi$ becomes of order 1, as we follow the evolution of the coupling from high energy scales.  This roughly defines the scale $\Lambda_{\rm QCD}$ which signals the transition  from the perturbative region with quark and gluon degrees of freedom to the nonperturbative regime where hadrons should emerge.   This mechanism is know as `dimensional transmutation', whereby the conformal symmetry of the classical theory is anomalously broken by quantization,  thus introducing a dimensionfull parameter, the mass scale  $\Lambda_{\rm QCD}$.

QCD is asymptotically free~\cite{Gross:1973id, Politzer:1973fx}, so at high energies it resembles a rather simple scale invariant theory. This is in fact one important argument for the relevance of anti-de Sitter space in applications of the AdS/CFT correspondence to QCD.  For high energies or small distances the small coupling $g_s$ allows one to compute the corrections to scale invariance. This is certainly not the case in the infrared regime (IR), for distances comparable to the hadronic size, where perturbation theory breaks down. There is however evidence from lattice gauge theory~\cite{Furui:2006py}, Dyson Schwinger equations~\cite{vonSmekal:1997is, Binosi:2009qm},  and empirical effective charges~\cite{Deur:2005cf},  that the QCD $\beta$-function vanishes in the infrared.   In a confining theory where the gluons have an effective mass or maximal wavelength, all vacuum polarization corrections to the gluon self-energy should decouple at long wavelengths~\cite{Cornwall:1981zr}. Thus, from a physical perspective an infrared fixed point appears to be a natural consequence of confinement~\cite{Brodsky:2008be}.   In fact, the running of the QCD coupling  in the infrared region for $Q^2 <  4 \la$, where $\sqrt\la$ represents the hadronic mass scale, is expected to have the  form $\alpha_s(Q^2) \propto \exp{\left(-Q^2/4 \la\right)}$~\cite{Brodsky:2010ur}, which agrees with the shape of the effective charge defined from the Bjorken sum rule, displaying an infrared fixed point.  In the nonperturbative domain soft gluons are in effect sublimated into the effective confining potential. Above this region, hard-gluon exchange becomes important, leading to asymptotic freedom.  The scale $\Lambda$ entering the evolution of the  perturbative QCD running constant  in a given renormalization scheme, such as $\Lambda_{\overline{MS}}$,  can be determined in terms of the primary scheme-independent scale $\sqrt\la$~\cite{Deur:2014qfa}. This result is consistent with the hadronic flux-tube model~\cite{Isgur:1984bm} where soft gluons interact so strongly that they are sublimated into a color confinement potential for quarks. It is also consistent with the lack of empirical evidence confirming constituent gluons at small virtualities~\cite{Isgur:1985vy, Brodsky:2011pw}. At higher energy scales, $Q^2 > 4 \la$ we expect the usual perturbative QCD (PQCD) logarithmic dependence in $\al_s$ from the appearance of dynamical gluon degrees of freedom.

The relation between the dilatation symmetry and the symmetries in AdS$_5$ can be seen directly from the AdS metric. The line element (\ref{AdSm}) is invariant under a dilatation of all coordinates. Since a dilatation of the Minkowski coordinates $x^\mu  \to \rho x^\mu$  is compensated by a dilatation of the holographic variable $z \to \rho z$, it follows that the variable $z$ acts like a scaling variable in Minkowski space: different values of $z$ correspond to different energy scales at which a measurement is made.  As a result, short space- and time-like intervals map to the boundary in AdS space-time  near $z=0$~\footnote{As quark and gluons can only travel over short distances as compared to the confinement scale $\Lambda^{-1}_{\rm QCD}$, the space-time region for their propagation is adjacent to the light-cone~\cite{Bjorken:1989xw}.}. This corresponds to the ultraviolet (UV) region of AdS space.

A large  four-dimensional  interval of confinement dimensions $x_\mu x^\mu \sim 1/\Lambda_{\rm QCD}^2$ maps to the large infrared  region of AdS space $z \sim 1 / \Lambda_{\rm QCD}$.   In order to incorporate the mechanisms of confinement in the gravity dual the conformal invariance encoded in the isometries of AdS$_5$ must be broken. In bottom-up models the breaking of conformal symmetry is introduced  by modifying  the background AdS space-time at  an infrared region of the geometries which sets the scale of the strong interactions.  In this effective  approach, one considers the propagation of hadronic modes in a fixed effective gravitational background asymptotic to AdS space, thus encoding  prominent properties for QCD, such as the ultraviolet conformal limit at the AdS boundary at $z \to 0$, as well as modifications of the AdS background geometry in the large $z$ infrared region  to describe confinement.

On the other hand,  in models based on string theory  -- top-down models, the space-time geometry is dictated by the corresponding brane configuration and may be quite different from AdS$_5$  \cite{Sakai:2004cn, Sakai:2005yt, Kruczenski:2003uq}.  A comparison of the predictions of AdS/QCD models in various space-time backgrounds appears in Ref.~\cite{Becciolini:2009fu}.  The result of such a comparison is that, for a wide class of background space-time geometries, naive predictions based on five-dimensional AdS models (ignoring quantum corrections) are the most accurate.  One of the reasons for the phenomenological success of models based on the AdS geometry might be that they capture a conformal window in QCD at the hadronic scale~\cite{Brodsky:2008be}.

A simple way to obtain confinement and discrete normalizable modes is to truncate AdS space with the introduction of a sharp cut-off in the infrared region of AdS space, as in the ``hard-wall'' model~\cite{Polchinski:2001tt},  where one considers  a slice  of AdS space, $0 \leq z \leq z_0$, and imposes boundary conditions on the fields at the IR border $z_0 \sim 1/\Lambda_{\rm QCD}$.  As first shown by Polchinski and Strassler~\cite{Polchinski:2001tt}, the modified AdS space provides a derivation of dimensional counting rules~\cite{Brodsky:1973kr, Brodsky:1974vy, Matveev:ra} in QCD for the leading power-law fall-off of hard scattering beyond the perturbative regime. The modified theory generates the point-like hard behavior expected from QCD, instead of the soft behavior characteristic of extended objects~\cite{Polchinski:2001tt}.  On AdS space the physical  states are represented by normalizable modes $\Phi_P(x,z) = e^{iP \cdot x} \Phi(z)$, with plane waves along Minkowski coordinates $x^\mu$  to represent a physical free hadron with momentum $P^\mu$, and a wave function $\Phi(z)$ along the holographic coordinate $z$. The hadronic invariant mass $P_\mu P^\mu = M^2$  is found by solving the eigenvalue problem  for the AdS wave function $\Phi(z)$. 
This simple model fails however to reproduce the observed linear Regge behavior of hadronic excitations in $M^2$, a feature which is typical to many holographic models~\cite{Schreiber:2004ie, Shifman:2005zn}.

One can also introduce a ``dilaton" background in the holographic coordinate to produce a smooth cutoff at large distances  as  in the ``soft-wall'' model~\cite{Karch:2006pv} which explicitly breaks the maximal  AdS symmetry.   In this  bottom-up approach to AdS gravity, an effective $z$-dependent curvature is introduced in the infrared region of AdS which leads to conformal symmetry breaking in QCD,  but its form is left largely unspecified.  One can impose from the onset a viable phenomenological confining structure to determine the effective IR modification of AdS space. For example, one can adjust the dilaton background to reproduce the correct linear and equidistant Regge behavior of the hadronic mass spectrum $M^2$~\cite{Karch:2006pv},  a form supported by semiclassical arguments~\cite{Shifman:2007xn}.  One can also consider models where the dilaton field is dynamically coupled to gravity~\cite{Csaki:2006ji, Shock:2006gt, Gursoy:2007cb, Gursoy:2007er, Batell:2008zm, dePaula:2008fp}. In one approach to AdS/QCD~\cite{Erlich:2005qh, DaRold:2005zs, DaRold:2005vr}, bulk fields are introduced to match the $SU(2)_L \times SU(2)_R$ chiral symmetries of QCD and its spontaneous breaking, but without explicit connection with the internal constituent structure of hadrons~\cite{Brodsky:2003px}.  Instead, axial and vector currents become the primary entities as in effective chiral theory.  Following this bottom-up approach only a limited number of operators is introduced, and consequently only a limited number of fields is required to construct  phenomenologically viable five-dimensional gravity duals.

\section{Light-front holographic QCD} 

Light-front quantization is the ideal relativistic, frame independent framework to describe the internal constituent structure of hadrons. The simple structure of the light-front (LF) vacuum allows an unambiguous definition of the partonic content of a hadron in QCD and of hadronic light-front wave functions (LFWFs), the underlying link between large distance hadronic states and the constituent degrees of freedom at short distances.   The QCD light-front Hamiltonian $H_{LF}$  is constructed from the QCD Lagrangian using the  standard methods of quantum field theory~\cite{Brodsky:1997de}.  The spectrum and light-front wave functions of relativistic bound states are obtained from the eigenvalue equation  $H_{LF} \vert  \psi \rangle  = M^2 \vert  \psi \rangle$. It becomes an infinite set of coupled integral equations for the LF components  $\psi_n = \langle n \vert \psi\rangle$  in a Fock-state expansion, {\it i.~e.} in a complete basis of non-interacting $n$-particle states $\vert n \rangle$, with an infinite number of components. This provides a quantum-mechanical probabilistic interpretation of the structure of hadronic states in terms of their constituents at the same light-front time  $x^+ = x^0 + x^3$, the time marked by the front of a light wave~\cite{Dirac:1949cp}. The constituent spin and orbital angular momentum properties of the hadrons are also encoded in the LFWFs.  Unlike instant time quantization, the Hamiltonian eigenvalue equation  in the light front is frame independent.   In practice, the matrix diagonalization~\cite{Brodsky:1997de} of the LF Hamiltonian eigenvalue equation in four-dimensional space-time has proven to be a daunting task because of the large size of the matrix representations.  Consequently, alternative methods and approximations  are necessary to better understand the nature of relativistic bound states in the strong-coupling regime of QCD.

To a first semiclassical approximation, where quantum loops and quark masses are not included, the relativistic bound-state equation for  light hadrons can be reduced to an effective LF Schr\"odinger equation by identifying as a key dynamical variable the invariant mass of the constituents, which is the measure of the off-shellness in  the LF kinetic energy,  and it is thus the natural variable to characterize the hadronic wave function.  In conjugate position space, the relevant dynamical variable  is an invariant impact kinematical variable $\zeta$, which measures the separation of the partons within the hadron at equal light-front  time~\cite{deTeramond:2008ht}.  Thus, by properly identifying the key dynamical variable, one can reduce, to a first semi-classical approximation, the multiparton problem  in QCD to an effective one dimensional quantum field theory. As a result, all the complexities of the strong interaction dynamics are hidden in an effective potential $U(\zeta)$,  but the central question -- how to derive the confining  potential from QCD, remains open.

It is remarkable that in the semiclassical approximation described above,  the light-front Hamiltonian  has a structure which matches exactly the eigenvalue equations in AdS space.  This offers the unique possibility to make an explicit connection of the AdS wave function $\Phi(z)$ with the internal constituent structure of hadrons. In fact, one can obtain the AdS wave equations by starting from the semiclassical approximation to light-front QCD in physical space-time -- an emergent property of this framework.  This connection yields a  relation between the coordinate  $z$ of AdS space with the impact LF variable $\zeta$~\cite{deTeramond:2008ht},  thus giving  the holographic variable $z$ a precise definition and intuitive meaning in light-front QCD.

Light-front  holographic methods  were originally introduced~\cite{Brodsky:2006uqa, Brodsky:2007hb} by matching the  electromagnetic current matrix elements in AdS space~\cite{Polchinski:2002jw} with the corresponding expression  derived from light-front  quantization in physical space-time~\cite{Drell:1969km, West:1970av}. It was also shown that one obtains  identical holographic mapping using the matrix elements of the energy-momentum tensor~\cite{Brodsky:2008pf} by perturbing the AdS metric (\ref{AdSm}) around its static solution~\cite{Abidin:2008ku}, thus establishing a precise relation between wave functions in AdS space and the light-front wave functions describing the internal structure of hadrons.

The description of higher spin in AdS space is a notoriously difficult problem~\cite{Fronsdal:1978vb, Fradkin:1986qy, Buchbinder:2001bs,  Metsaev:1999ui,
 Metsaev:2003cu, Metsaev:2011uy, Metsaev:2013kaa, Metsaev:2014iwa}, and thus  there is much interest in finding a simplified approach which can describe higher-spin hadrons using the  gauge/gravity duality. In the  framework of  Ref.~\cite{Karch:2006pv}  the recurrences of the $\rho$ and its daughter trajectories are obtained from a gauge invariant AdS Lagrangian. In the light-front holographic approach, where the internal structure, and notably the orbital angular momentum of the constituents,  is reflected in the AdS  wave functions by the LF mapping, wave equations with arbitrary integer and half-integer spin can be derived from an invariant effective action in  AdS space~\cite{deTeramond:2013it}.  Remarkably, the pure AdS  equations correspond to the light-front kinetic energy of  the partons inside a hadron, whereas the light-front interactions  which build confinement correspond to the modification of AdS space in an effective dual gravity approximation~\cite{deTeramond:2008ht}. From this point of view, the non-trivial geometry of pure AdS space encodes  the kinematical aspects and additional deformations of AdS space  encode dynamics, including confinement,  and determine the form of the effective potential $U$ from the precise holographic mapping to light-front physics. It can also be shown that  the introduction of a dilaton profile is equivalent to a modification of the AdS metric, even for arbitrary spin~\cite{deTeramond:2013it}.

It is important to notice that the construction of higher-spin modes given in Ref. \cite{Karch:2006pv} starts from a gauge invariant action in AdS and uses the gauge invariance of the model to construct a higher-spin effective action. However, this approach which is based in gauge invariance in the higher dimensional theory,  is not applicable to  light-front  mapping to physical space-time which incorporates LF partonic physics in the holographic approach.  In contrast, for light-front mapping the identification of orbital angular momentum of the constituents with the fifth dimensional AdS mass, in principle an arbitrary parameter, is a key element in the description of the internal structure of hadrons using light-front holographic principles, since hadron masses depend crucially on it.

\section{Confinement and conformal algebraic structures}

In principle, LF Hamiltonian theory provides a rigorous, relativistic and frame-independent framework for solving nonperturbative QCD and understanding the central problem of hadron physics -- color confinement.   For QCD(1+1) the mass of the mesons and baryon eigenstates at zero quark mass is determined   in units of its dimensionful coupling using the Discretized Light Cone Quantization (DLCQ) method~\cite{Pauli:1985pv, Pauli:1985ps}.  However, in the case of 3+1 space-time, the QCD coupling is dimensionless, so the physical mechanism that sets the hadron mass scale for zero quark mass is not  apparent.  Since our light-front semiclassical approximation~\cite{Brodsky:1997de} is effectively a one-dimensional quantum field theory, it is natural to apply the framework developed  by de Alfaro, Fubini and Furlan (dAFF)~\cite{deAlfaro:1976je} which can generate a mass scale and a confinement potential  without affecting the conformal invariance of the action. In their remarkable paper,  published  some 40 years ago, a hint to the possible appearance  of  scale in nominally conformal theories was given~\cite{deAlfaro:1976je}.  This remarkable result is based on the isomorphism of the algebra of the one-dimensional conformal group ${\it Conf}\!\left(R^1\right)$ to the algebra of  generators of the group $SO(2,1)$ and the isometries  of AdS$_2$ space. In fact, one of the generators of this group, the rotation in the 2-dimensional space,  is compact and has therefore a discrete spectrum with normalizable eigenfunctions. As a result, the form of  the evolution operator  is fixed and includes  a confining harmonic oscillator potential, and the time variable has a finite range. Since the generators of the conformal group have different dimensions their relations with generators of $SO(2,1)$ imply a scale, which here plays a fundamental role, as already conjectured in~\cite{deAlfaro:1976je}.    These considerations have led to the realization that  the form of the effective LF confining potential can be obtained by extending the results found by dAFF to light-front dynamics and  to the embedding space~\cite{Brodsky:2013ar}~\footnote{Harmonic confinement also follows from the covariant Hamiltonian description of mesons given in Ref.~\cite{Leutwyler:1977pv}.}.  These results become particularly relevant, since it was also shown recently that an effective harmonic potential in the light-front form of dynamics corresponds, for light quark masses, to a linear potential in the usual instant-form~\cite{Trawinski:2014msa, Eichten:1978tg}.  Thus, these results also lead to the prediction of linear Regge trajectories in the hadron mass square for small quark masses  in agreement with the observed spectrum for light hadrons.

\begin{figure}[ht]
\begin{center}
\includegraphics[width=7.0cm]{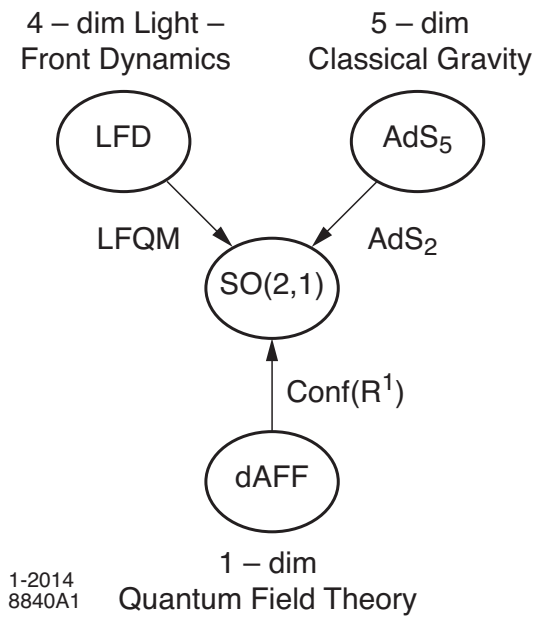}\hspace{0pt}
\caption{\label{connections} \small An effective light-front theory for QCD endowed with an  $SO(2,1)$  algebraic structure  follows from the one-dimensional semiclassical approximation to light-front dynamics in physical space-time, higher dimensional gravity in AdS$_5$ space and the extension of conformal quantum mechanics to light-front dynamics. The result is a relativistic light-front quantum mechanical wave equation which  incorporates essential spectroscopic and dynamical features of hadron physics.  The emergence of a mass scale and the effective confining potential has its origins in the isomorphism of the one-dimensional conformal group  ${\it Conf}\!\left(R^1\right)$  with the group $SO(2,1)$, which is also the isometry group of AdS$_2$.}
\end{center}
\end{figure}

The remarkable  connection between the  semiclassical approximation to light-front dynamics  in physical four-dimensional space-time  with gravity in a higher dimensional AdS space,  and the constraints imposed by the invariance properties under  the full conformal group in one dimensional quantum field theory, is depicted in Fig. \ref{connections} and is central to this report.  We shall describe  how to construct a light-front effective theory which encodes the fundamental conformal symmetry of the four-dimensional classical QCD Lagrangian.  This construction is endowed with and $SO(2,1)$ underlying symmetry, consistent with the emergence of a mass scale. We will also describe how to obtain effective wave equations for any spin in the higher dimensional embedding space, and how to map these results to light-front physics in physical space-time. The end result is a  semiclassical relativistic light-front  bound-state  equation,  similar to the Schr\"odinger equation in atomic physics,  which describes essential spectroscopic and dynamical features of hadron physics.

\section{Other approaches and applications}

A completely different approach to an effective treatment of nonperturbative QCD, which,  however turns out to be closely related to holographic QCD is a ``meromorphization procedure'' of perturbative QCD. In fact, it has been shown~\cite{Erlich:2006hq} that the hard wall model corresponds to a procedure proposed by Migdal \cite{Migdal:1977nu, Migdal:1977ut}, whereas the soft wall model has been related to the QCD sum rule method~\cite{Shifman:1978bx}  in Refs. \cite{Cata:2006ak, Jugeau:2013zza} (See also Appendix \ref{generating}).  Other approaches to emergent holography are discussed in~\cite{Koch:2010cy, Jevicki:2014mfa, Koch:2014mxa, Koch:2014aqa, Mintun:2014gua, Glazek:2013jba, Qi:2013caa, Dietrich:2013kza}.

We also briefly review other holographic approaches to  QCD, in particular top-down and bottom-up models based on chiral symmetry breaking.  Top-down models, such as the Sakai-Sugimoto model, are derived from brane configurations in string theory, whereas bottom-up models, such as the hard or soft-wall models, are more phenomenological and are not derived from string theory.  Each of the models discussed in this review include degrees of freedom which are identified with Standard Model hadrons via their quantum numbers, and predictions of  holographic models for QCD observables may be compared to experiment and to other models, often with remarkable quantitative success~\cite{Erlich:2008gp, Erlich:2009me}. The domain of small coupling in QCD would require, however, quantum corrections beyond the semiclassical approximation~\cite{Csaki:2008dt}.

A particularly interesting application of the holographic ideas is to high-energy small-angle scattering  in QCD, usually described by pomeron exchange~\cite{Forshaw:1997dc} which carries the vacuum quantum numbers.  The gauge/string duality provides a unified framework for the description of the soft Regge regime and hard BFKL Pomeron~\cite{Brower:2006ea}. The gauge/string framework can also be used to compute strong coupling high-energy oderon exchange~\cite{Brower:2008cy}, which distinguish particle anti-particle cross sections and thus carries   $C=-1$ vacuum quantum numbers. The gauge/gravity duality has also been applied to deep inelastic scattering (DIS), first discussed in this context in Ref.~\cite{Polchinski:2002jw}. We will not discuss in this report these interesting applications, but refer the reader to the original articles cited here~\footnote{Other interesting applications of the gauge/gravity correspondence include, but are not limited to,  high-energy $p p$ and $p \bar p$ scattering~\cite{Domokos:2009hm, Domokos:2010ma},  high-energy photon-hadron scattering~\cite{Nishio:2011xz}, compton scattering~\cite{Gao:2009se, Marquet:2010sf, Costa:2012fw} and polarized DIS~\cite{Gao:2009ze}.}.  Neither shall we discuss in this report  applications of the gauge/gravity duality to strongly coupled quark-gluon plasma observed in heavy ion collisions at RHIC and CERN, also an important subject which has attracted much attention~\footnote{For a review  see Ref. \cite{CasalderreySolana:2011us} and references therein.}.

Another interesting topic which we only touch upon  in this report is holographic renormalization~\cite{deBoer:1999xf}: the relation between the flow in the holographic coordinate in AdS space and the renormalization group flow of the dual quantum field theory~\cite{Wilson:1973jj}~\footnote{For a review of holographic renormalization, see for example~\cite{Skenderis:2002wp, Bianchi:2001kw}}. Thus, the description of the large-scale  behavior should be independent of the of the ``microscopic'' degrees of freedom (quarks and gluons) of the ultraviolet boundary theory and expressed in terms of  ``macroscopic''  infrared degrees of freedom (hadrons). As a result, the interaction potential of the effective infrared theory should retain universal characteristics from the renormalization group flow. For example in hadronic physics the universality of the Regge trajectories, but this universal behavior should also be relevant to other areas.

A number of  excellent reviews on the AdS/CFT correspondence are already available.  We refer the reader to the Physics Report by Aharony, {\em et al.}~\cite{Aharony:1999ti}, the TASI lectures by Klebanov \cite{Klebanov:2000me} and by D'Hoker and Freedman \cite{D'Hoker:2002aw} for some early reviews.  For more recent discussions of holographic QCD see the reviews in Refs.~\cite{Erdmenger:2007cm, Peeters:2007ab, Kim:2012ey, Ramallo:2013bua}.

\section{Contents of this review}

The report   is organized as follows:  in Chapter  \ref{ch2}  we  describe important aspects of light-front quantization and its multi-parton semiclassical approximation.  This leads to a relativistic invariant light-front  wave equation to compute hadronic bound states in terms of an effective potential   which is {\it a priori} unknown.  We also discuss how the semiclassical results are modified by the introduction of light quark masses. We show in Chapter \ref{ch3}  how a specific introduction of a scale  determines uniquely the form of the  light-front effective confinement potential, while leaving the action  conformally  invariant.   We also describe in this chapter the relation of the one-dimensional conformal group with the group $SO(2,1)$, and the extension of conformal quantum mechanics  to light-front physics described in Chapter \ref{ch2}.  In Chapter \ref{ch4} we derive  hadronic AdS wave equations for arbitrary integer and half-integer spin.  We give particular care  to the separation of kinematic and dynamical effects  in view of the mapping to LF  bound-state equations.  We perform  the  actual light-front mapping  in Chapter \ref{ch5}, and  and we compare the theoretical results  with the observed light meson and baryon spectra. In Chapter \ref{ch6}  we carry out the actual  LF mapping of amplitudes in AdS to their corresponding expressions in light-front QCD.  We  describe  form factors and transition amplitudes of hadrons in holographic QCD.  We also give a comparison with data  and we discuss present limitations of the model. In Chapter \ref{ch7} we present  other approaches to holographic QCD, including bottom-up and top-down gauge/gravity models.   We present our  conclusions and final remarks  in Chapter \ref{ch8}.   We include a discussion of open problems and future applications. In particular, we point out a possible connection of our effective light-front approach with holographic renormalization flows to AdS$_2$ geometry in the infrared and its  one-dimensional conformal dual theory.  In Appendix \ref{RG} we give a brief introduction to Riemanian geometry and  maximally symmetric Riemannian spaces. In particular  we exhibit  the connection between the conformal group in one dimension,  $SO(2,1)$, and AdS$_2$.  In Appendix \ref{metric} we give  a short collection of notations and conventions. We  present in the Appendices  \ref{cfqm} and \ref{HSWEAdS} several more technical derivations, relevant for Chapter \ref{ch3} and \ref{ch4} respectively.   We describe  in Appendix \ref{EMT} the light-front holographic mapping of the gravitational form factor of composite hadrons. In Appendix \ref{generating}   we discuss the relation of the generating functional of the boundary conformal field theory and the classical action in the 5-dimensional gravity theory \cite{Gubser:1998bc, Witten:1998qj} for fields with arbitrary integer spin, both in the soft- and the hard-wall models.   In Appendix \ref{formulae} some useful {formul\ae}  are listed.  In Appendix \ref{LFint} we describe an algebraic procedure  to construct the holographic light-front Hamiltonians corresponding to the hard and soft-wall models discussed in this report for bosons and fermions~\cite{Brodsky:2008pg}.  Finally in Appendix \ref{pform} we describe the equations of motion of $p$-form fields in AdS.

\chapter{A Semiclassical Approximation to Light-Front Quantized QCD  \label{ch2}}

Light-front quantization is the natural framework for the description  of the QCD nonperturbative relativistic bound-state structure in quantum field theory in terms of a frame-independent $n$-particle Fock expansion.  The central idea is due to Dirac who demonstrated the remarkable advantages  of using light-front time $x^+ = x^0+x^3$ (the ``front-form") to quantize a theory versus the standard time $x^0$ (the ``instant-form'').  As Dirac showed~\cite{Dirac:1949cp}, the front-form has the maximum number of kinematic generators of the Lorentz group, including the boost operator.  Thus the description of a hadron at fixed $x^+$ is independent of the observer's frame, making it ideal for addressing dynamical processes in quantum chromodynamics. An extensive review of light-front quantization is given in Ref.~\cite{Brodsky:1997de}.  As we shall discuss  in this  and in the next two chapters, a semiclassical approximation to light-front quantized field theory in physical four-dimensional space-time has a holographic dual with dynamics of  theories in five-dimensional anti-de Sitter space. Furthermore, its confining dynamics follows from the mapping to a one-dimensional conformal quantum field theory~\cite{Brodsky:2013ar}.

Quantization in the light-front provides  a rigorous field-theoretical realization of the intuitive ideas of the parton model~\cite{Feynman:1969ej, Feynman:1973xc} formulated at fixed time $x^0$ in the infinite-momentum frame~\cite{Fubini:1965xx, Weinberg:1966jm}.  Historically, the prediction of Bjorken scaling in deep inelastic scattering~\cite{Bjorken:1968dy} followed from a combination of the high energy limit $q_0 \to i \infty$ with the infinite momentum frame $P \to \infty$, introduced in~\cite{Fubini:1965xx}, using the usual definition of time; {\it i.e.}, the instant-form.   The same results are obtained in the front-form but with complete rigor; {\it e.g.}, the structure functions and other probabilistic parton distributions measured in deep inelastic scattering are obtained from the squares of the light-front wave functions, the eigensolution of the light-front Hamiltonian. Unlike the instant-form, the front-form results are independent of the hadron's Lorentz frame. A measurement in the front form is analogous to taking a flash photograph. The image in the resulting picture  records the state of the object as the front of  a light wave from the flash illuminates it,   consistent with observations within the space-like causal horizon $\Delta x_\mu^2 < 0$.  Similarly, measurements such as deep inelastic electron-proton scattering, determine  the structure of the  target proton  at fixed light-front time.

In the constituent quark model~\cite{GellMann:1964nj, Zweig:1981pd} the minimum quark content required by the hadronic quantum numbers is included in the  wave functions, which describe how hadrons are built  of their constituents.  In the conventional interpretation of the quark model, the main contribution to the hadron masses is supposed to arise from the explicit breaking of chiral symmetry by constituent quark masses. Typical computations of the hadron spectrum generally include a spin-independent confining interaction and a spin-dependent interaction, usually modeled from one-gluon-exchange in QCD~\cite{Godfrey:1985xj}. The parton model and the constituent quark model provide, respectively,  a good intuitive understanding of many high- and low-energy phenomena. In practice, however, it has been proven  difficult to reconcile the constituent quark model with QCD, and the best hope to make a connection between both approaches is provided by light-front dynamics. In fact, the original formulation of QCD was given in light-front coordinates~\cite{Fritzsch:1972jv, Fritzsch:1973pi} and the idea to derive a light-front constituent quark model~\cite{Casher:1973vh} also dates to the same time. The physical connections of the front-form with the constituent model is  a reason to hope that  light-front quantization will eventually  provide an understanding of the most challenging dynamical problems in QCD, such as color confinement~\cite{Wilson:1994fk}.

Just as in non-relativistic quantum mechanics, one can obtain bound-state light-front wave functions in terms of the hadronic constituents from solving the  light-front Hamiltonian eigenvalue problem.   The eigenstates of the light-front Hamiltonian are defined at fixed light-front time $x^+$ over all space within the causal horizon, so that causality is maintained without normal-ordering.  In fact, light-front physics is a fully relativistic field theory but its structure is similar to non-relativistic theory~\cite{Dirac:1949cp}, and the bound-state equations are relativistic Schr\"odinger-like equations at equal light-front time. Because of Wick's theorem, light-front time-ordered perturbation theory is equivalent to the covariant Feynman perturbation theory.  Furthermore, since boosts are kinematical, the light-front wave functions are frame independent.

In principle, one can  solve QCD by diagonalizing the light-front QCD Hamiltonian $H_{LF}$ using, for example, the discretized light-cone quantization method~\cite{Brodsky:1997de} or the Hamiltonian transverse lattice formulation introduced in~\cite{Bardeen:1976tm}. The spectrum and light-front  wave functions are then obtained from the eigenvalues and eigenfunctions of the Heisenberg problem $H_{LF} \vert \psi \rangle = M^2 \vert \psi \rangle$, which becomes an infinite set of coupled integral equations for the light-front components $\psi_n = \langle n \vert \psi \rangle$ in a Fock expansion~\cite{Brodsky:1997de}. This nonperturbative method has the advantage that it is frame-independent, is defined in physical Minkowski space-time, and has no fermion-doubling problem.  It has been applied successfully in lower space-time dimensions~\cite{Brodsky:1997de}, such as QCD(1+1)~\cite{Pauli:1985pv, Pauli:1985ps}.  In practice, solving the actual eigenvalue problem is a formidable computational task   for a non-abelian quantum field theory  in four-dimensional space-time. An analytic approach to nonperturbative  relativistic bound-states is also vastly difficult because of the unbound particle number  with arbitrary momenta and helicities. Consequently, alternative methods and approximations  are necessary to better understand the nature of relativistic bound-states in the strong-coupling regime.

Hadronic matrix elements and form factors are computed from simple overlaps of the boost invariant light-front wave functions as in the Drell-Yan West formula~\cite{Drell:1969km, West:1970av}. In contrast,  at ordinary fixed time $x^0$,  the hadronic states must be boosted  from the hadron's rest frame to a moving frame -- an intractable dynamical problem which involves changes in particle number.   Moreover, the form factors at fixed time $x^0$  also require computing off-diagonal matrix elements and the contributions of currents which arise from the instant vacuum fluctuations in the initial state and which connect to the hadron in the final state. Thus, the knowledge of wave functions alone is not sufficient to compute covariant current matrix elements in the usual instant form.  When a hadron is examined in the light front in the Drell-Yan frame~\cite{Drell:1969km, Brodsky:1980zm}, for example,  a virtual photon couples only to forward moving quarks and only processes with the same number of initial and final partons are allowed. A quantum-mechanical probabilistic  constituent interpretation in terms of wave functions is thus an important property of light-front dynamics required for both the constituent quark model and the parton model.

In axiomatic quantum  field theory the vacuum state is defined as the unique state invariant under Poincar\'e  transformations~\cite{SW64}.  Conventionally it is defined as the lowest energy eigenstate of the instant-form Hamiltonian. Such an eigenstate is defined at a single time $x^0$ over all space $x$. It is thus acausal and frame-dependent.  In contrast, in the front form, the vacuum state is defined as the eigenstate of lowest invariant mass $M^2$ at fixed light-front time $x^+ \! = x^0 + x^3$. It is frame-independent and only requires information within the causal horizon.  Thus, an important advantage of the LF framework is the apparent simplicity and Lorentz invariance of the LF vacuum.  In contrast, the equal-time vacuum  contains quantum loop graphs and thus an infinite sea of quarks and gluons.

\section{The Dirac forms of relativistic dynamics}

According to Dirac's classification of the forms of relativistic dynamics~\cite{Dirac:1949cp}, the fundamental generators of the Poincar\'e group can be separated into kinematical and dynamical generators. The kinematical generators act along the initial hypersurface where the initial conditions (the quantization conditions) are imposed.   The kinematical generators leave invariant the initial surface and are thus independent of the dynamics; therefore they contain no interactions. The dynamical generators are responsible for the evolution of the system  (mapping one initial surface into another surface) and depend consequently on the interactions.

\begin{figure}[ht] 
\begin{centering}
\includegraphics[width=14.6cm]{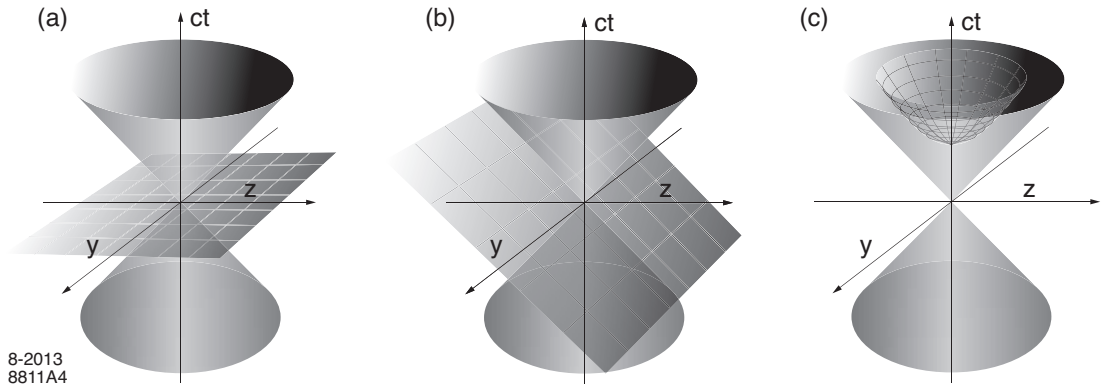}
\caption{\label{forms} \small Dirac forms of relativistic dynamics: (a) the instant form,  (b) the front form and ({c}) the point form. The initial surfaces are defined respectively by $x^0 = 0, \, x^0 + x^3 = 0$ and
$x^2 = \kappa^2 >0, \, x^0 >0.$}
\end{centering}
\end{figure}

In his original paper Dirac~\cite{Dirac:1949cp}  found three forms of relativistic dynamics\footnote{Subsequently Leutwyler and Stern~\cite{Leutwyler:1977vy} found two additional forms for a total of five inequivalent forms which correspond to the number of subgroups of the Poincar\'e group.} which correspond to different parameterizations of space-time and which cannot be transformed into each other by a Lorentz transformation. The  three  forms of Dirac are illustrated  in Fig. \ref{forms}.  The  {\it instant form} is the usual form where the initial surface is the surface defined at $x^0=0$. In the {\it front form}  (discussed above)  the initial surface is the tangent plane to the light-cone $x^0 + x^3 = 0$ -- the null plane, thus without reference to a specific Lorentz frame. According to Dirac~\cite{Dirac:1949cp}, it is the ``three dimensional surface in space-time formed by a plane wave front advancing with the velocity  of light.'' In the third form, the {\it point form}, the initial surface is  the hyperboloid defined by $x^2 = \kappa^2 > 0, \, x^0 >0$, which is left invariant by the Lorentz generators. Each front has its own Hamiltonian and evolves with a different time, but the results  computed in any front should be identical, since physical observables cannot depend on how space-time is parameterized.

The Poincar\'e group is the full symmetry group of any form of relativistic dynamics. Its Lie algebra is given by the well known commutation relations
\begin{eqnarray}
\left[P^\mu, P^\nu\right]  &\!=\!&0 ,\\ \nn
\left[M^{\mu \nu}, P^\rho\right] &\!=\!& i \left(g^{\mu \rho} P^\nu - g^{\nu \rho} P^\mu \right), \\ \nn
\left[M^{\mu \nu}, M^{\rho \sigma} \right] &\!=\!& i \left(g^{\mu \rho} M^{\nu \sigma} - g^{\mu \sigma} M^{\nu \rho}
                                                             +      g^{\nu \sigma} M^{\mu \rho} - g^{\nu \rho} M^{\mu \sigma} \right) ,
\end{eqnarray}
where the $P^\mu$ are the generators of space-time translations and the antisymmetric tensor $M^{\mu \nu}$ of the generators of the Lorentz transformations.

In the instant form the Hamiltonian $P^0$ and the three components of the boost vector  $K^i = M^{0 i}$ are dynamical generators, whereas the momentum $\mbf{P}$ and the three components of angular momentum $J^i = \half \epsilon^{i j k} M^{j k}$ are kinematical.   In the front form~\cite{Kogut:1969xa, Srivastava:1999js}, the dynamical generators are the ``minus" component generators, the Hamiltonian $P^- $ and  the generator  $M^{-1}  = K^1 - J^2$ and $M^{- 2} = K^2 + J^1$, which correspond to LF rotations along the $x$ and $y$-axes~\footnote{The $\pm$ components of a tensor are defined by $a^{\pm} =  a^0 \pm a^3$, and the metric follows from the scalar product $a \cdot b = \half \left(a^+ b^- + a^- b^+\right) - a^1 b^1 - a^2 b^2$ (See Appendix  \ref{metric}).}. The kinematical generators are the longitudinal ``plus" momentum  $P^+ $ and transverse momentum $P^i$. The boost operators are also kinematical in the light-front: $M^{+1} = K^1 + J^2$ and  $M^{+ 2}  = K^2 - J^1$, which boost the system in the $x$- and $y$-direction respectively, as well as the generator $\half M^{+ -} = K^3$ which boost the system in the longitudinal direction. Finally, the $z$-component of angular momentum $M^{12} = J^3$, which rotates the system in the $x-y$ plane is also a kinematical operator, and labels the angular moment states in the light front. In the point-form the four generators $P^\mu$ are dynamical and the six Lorentz generators $M^{\mu \nu}$ kinematical. The light-front frame has the maximal number of kinematical generators.

\section{Light-front dynamics}
\label{LFD}

For a hadron with four momentum  $P^\mu =  (P^+, P^-, \mbf{P}_{\!\perp})$, $P^\pm = P^0 \pm P^3$, the mass-shell relation  $P^2 = M^2$,  where $P_\mu P^\mu = P^+ P^-  - \mbf{P}_\perp^2$, leads to the  dispersion relation for the LF Hamiltonian $P^-$
\beq \label{LFDR}
P^- = \frac{\mbf{P}_\perp^2 + M^2}{P^+},   \hspace{20pt} P^+ > 0.
\enq
This LF relativistic dispersion relation has several remarkable properties.  The square root operator does not appear in \req{LFDR}, and thus the dependence on the transverse momentum $\mbf{P}_{\!\perp}$ is similar to the non-relativistic dispersion relation.  For massive physical hadronic states $P^2 > 0$ and $P^0$ are positive, thus $P^+$ and $P^-$ are also positive.  Furthermore, since  the longitudinal momentum $P^+$ is  kinematical,  it is given by the sum of the single-particle longitudinal momentum of the constituents  of the bound state.  In fact, for an $n$-particle bound state with particle  four momentum $p_i^\mu= \left(p^+_i, p^-_i, \mbf{p}_{\perp i}\right)$,  where $p_i^2 =   p^+_i p^-_i - \mbf{p}_{\perp i}^2 = m_i^2$, we have
\beq 
p^-_i  = \frac{\mbf{p}^2_{\perp i}+ m^2_i}{p^+_i} ,    \quad p^+_i > 0,
\enq
for each constituent $i$.  Thus
\beq
P^+ = \sum_i^n p^+_i,  \quad p^+_i > 0.
\enq
On the other hand, since the bound-state is arbitrarily off the LF energy shell we have the inequality
\beq
P^- - \sum_i^n p^-_i  <  0,
\enq
for the LF Hamiltonian which contains the interactions.

The LF Hamiltonian $P^-$ is the momentum conjugate to the  LF-time coordinate, $x^+ = x^0 + x^3$. Thus, the
evolution of the system is given by the relativistic light-front  Schr\"odinger-like equation
\beq \label{LFSE}
i \frac{\partial}{\partial x^+} \vert \psi{(P}) \rangle =  P^- \vert \psi({P}) \rangle ,
\enq
where $P^-$ is given by \req{LFDR}. Since the generators  $P^+$ and $\mbf{P}_\perp$ are kinematical, we can construct the LF Lorentz-invariant Hamiltonian  $H_{LF} = P^2  =  P^+ P^- -  \mbf{P}_\perp^2$ with eigenvalues corresponding to  the invariant mass $P_\mu P^\mu = M^2$
\beq \label{HLF}
H_{LF} \vert \psi({P}) \rangle = M^2 \vert \psi({P}) \rangle .
\enq
As one could expect,  the eigenstates of the LF Hamiltonian $H_{LF}$ are invariant since the LF boost generators are kinematical. Consequently, if the eigenstates are projected onto an $n$-particle Fock component $\vert n \rangle$  of the free LF Hamiltonian,  the resulting light-front wave function $\psi_n = \langle n \vert \psi \rangle$ only depends on the relative coordinates of the constituents. Thus, an additional important property of the light-front frame  for a bound state is the separation of relative and overall variables.

Since $p^+_i >0$ for every particle, the vacuum is the unique state with $P^+ = 0$ and contains no particles. All other states have $P^+ > 0$.   Since plus momentum is kinematic, and thus conserved at every vertex,  loop graphs with constituents with positive $p^+_i$ cannot occur in the light-front vacuum. Because this also holds in presence of interactions, the vacuum of the interacting theory is also the trivial vacuum of the non-interacting theory. However,  one cannot discard the presence of zero modes, possible background fields with  with $p^+ = 0$, which also lead to $P^+ = 0$ and  thus can mix with the  trivial vacuum. The light-front  vacuum  is defined at fixed LF time $x^+ = x^0 + x^3$ over all $x^- = x^0 - x^3$  and  $\mbf{x}_\perp$, the expanse of space  that can be observed within the speed of light. Thus the frame independent definition of the vacuum 
 \beq
P^2 \vert  0 \rangle = 0.
\enq
Causality is maintained since the LF vacuum only requires information within the causal horizon. Since the LF vacuum is causal and frame independent, it can provide a representation of the empty universe for quantum field theory~\cite{Brodsky:2009zd, Brodsky:2012ku}. In fact, the front form is a natural basis for cosmology because the  universe is observed along the front of a light wave.

\section{Light-front quantization of QCD \label{LFquant}}

We can now proceed to relate the LF generators to the underlying QCD Lagrangian in terms of the dynamical fields of the theory. In the light-front, the Dirac equation is written as a pair of coupled equations for plus and minus components, $\psi_\pm = \Lambda_\pm \psi$, with the projection operator $\Lambda_\pm = \gamma^0 \gamma^\pm$. One of the equations does not have a derivative with respect to the LF evolution time $x^+$, and it is therefore a constraint equation which determines the minus component $\psi_-$ in terms of the dynamical field $\psi_+$~\cite{Brodsky:1997de, Burkardt:1995ct}.  Likewise, the dynamical transverse field $\mbf{A}_\perp$ in the light-cone gauge $A^+ = 0$ has no ghosts  nor unphysical
negative metric gluons.

Our starting point is the $SU(3)_C$ invariant Lagrangian of QCD
\begin{equation} \label{LQCD}
\mathcal{L}_{\rm QCD} = \bar \psi \left( i \gamma^\mu D_\mu - m\right) \psi
- \tfrac{1}{4} G^a_{\mu \nu} G^{a \, \mu \nu} ,
\end{equation}
where $D_\mu = \partial_\mu - i g_s A^a_\mu T^a$ and
$G^a_{\mu \nu} = \partial_\mu A^a_\nu - \partial_\nu A^a_\mu +
g_s c^{abc} A_\mu^b A_\nu^c$, with $\left[T^a, T^b\right] = i c^{abc} T^c$ and
$a, b ,c$ are $SU(3)_C$ color indices.

One can express the  hadron four-momentum  generator $P^\mu =  (P^+, P^-,\mbf{P}_{\!\perp})$  in terms of the
dynamical fields $\psi_+$  and
$\mbf{A}_\perp$ quantized on the light-front at fixed light-front time $x^+ $, $x^\pm = x^0 \pm x^3$~\cite{Brodsky:1997de}
\begin{eqnarray} \nonumber \label{Pm}
P^- &\!\!= &  \half \int \! dx^- d^2 \mbf{x}_\perp \left( \bar \psi_+ \, \gamma^+
\frac{m^2 + \left( i \mbf{\nabla}_{\! \perp} \right)^2}{ i \partial^+} \psi_+
 -  A^{a \mu}   \left( i \mbf{\nabla}_{\! \perp} \right)^2 \! A^a_\mu\right) \\ \nonumber
 &\! +\! &  g_s \int \! dx^- d^2 \mbf{x}_\perp \, \bar \psi_+ \gamma^\mu T^a \psi_+ A^a_\mu +
 \\ \nonumber
&\!+\!& \frac{g_s^2}{4} \int \! dx^- d^2 \mbf{x}_\perp \,   c^{a b c}  c^{a d e} A_\mu^b A_\nu^c  A^{d \mu} \! A^{e \nu} \\ \nonumber
&\!+\!& \frac{g_s^2}{2} \int \! dx^- d^2 \mbf{x}_\perp \, \bar \psi_+ \gamma^+ T^a \psi_+
\, \frac{1}{\left(i \partial^+\right)^2} \,
\bar \psi_+ \gamma^+ T^a \psi_+  \\
&\!+\!& \frac{g_s^2}{2} \int \! dx^- d^2 \mbf{x}_\perp \, \bar \psi_+ \gamma^\mu T^a A^a_\mu
\, \frac{\gamma^+}{i \partial^+}
\left(T^b A^b_\nu \gamma^\nu \psi_+ \right) ,
\end{eqnarray}
The first term in (\ref{Pm}) is the kinetic energy of quarks and gluons; it is the only non-vanishing term in the limit $g_s \to 0$. The second term is the three-point vertex interaction. The third term is the four-point gluon interaction. The fourth term represents the instantaneous gluon interaction which originates from the imposition of light-cone gauge, and the last term is the instantaneous fermion interaction~\cite{Brodsky:1997de}. The integrals in \req{Pm} are over the null plane $x^+ = 0$, the initial surface,  where the commutation relations for the fields are fixed.  The LF Hamiltonian $P^-$ generates LF time translations
\begin{equation}
\left[\psi_+(x), P^-\right] = i \frac{\partial}{\partial x^+} \psi_+(x),  \hspace{30pt}  \left[\mbf{A_\perp}, P^-\right] = i \frac{\partial}{\partial x^+} \mbf{A_\perp}(x),
\end{equation}
which evolve the initial conditions for the fields to all space-time.

The light-front longitudinal momentum  $\mbf{P}^+$
\begin{equation}
P^+ = \int \! dx^- d^2 \mbf{x}_\perp \left(
 \bar \psi_+ \gamma^+   i \partial^+ \psi_+
 - A^{a \mu} \, (i \partial^+)^2  A^a_\mu \right),
 \end{equation}
and the light-front transverse momentum  $\mbf{P}_\perp$
\begin{equation}
\mbf{P}_{\! \perp}  =  \half \int \! dx^- d^2 \mbf{x}_\perp \left(
\bar \psi_+ \gamma^+   i \mbf{\nabla}_{\! \perp} \psi_+
- A^{a \mu} \, i\partial^+ \, i\mbf{\nabla}_{\! \perp}  A^a_\mu \right),
\end{equation}
are kinematical generators and  do not involve interactions.

The Dirac field $\psi_+$ and the transverse gluon field $\mbf{A}_\perp$ are expanded in terms of particle creation and annihilation operators as~\cite{Brodsky:1997de}
\begin{equation} \label{eq:psiop}
\psi_+(x^- \!,\mbf{x}_\perp)_\alpha = \sum_\lambda \int_{q^+ > 0} \frac{d q^+}{\sqrt{ 2
 q^+}}
\frac{d^2 \mbf{q}_\perp}{ (2 \pi)^3}
\left[b_\lambda (q)
u_\alpha(q,\lambda) e^{-i q \cdot x} + d_\lambda (q)^\dagger
v_\alpha(q,\lambda) e^{i q \cdot x}\right],
\end{equation}
and
\begin{equation} \label{eq:Aop}
\mbf{A}_\perp(x^- \!,\mbf{x}_\perp) = \sum_\lambda \int_{q^+ > 0}
\frac{d q^+}{\sqrt{2
 q^+}} \frac{d^2 \mbf{q}_\perp}{(2 \pi)^3}
\left[a(q)
\vec\epsilon_\perp(q,\lambda) e^{-i q \cdot x} + a(q)^\dagger
\vec\epsilon_\perp^{\,*}(q,\lambda) e^{i q \cdot x } \right],
\end{equation}
with $u$ and $v$ LF spinors~\cite{Lepage:1980fj}
and commutation relations
\begin{equation} \label{eq:crpsi}
\left\{b(q), b^\dagger(q')\right\} = \left\{d(q), d^\dagger(q')\right\} =
(2 \pi)^3 \,\delta (q^+ - {q'}^+) ~
\delta^{(2)}\negthinspace\left(\mbf{q}_\perp - \mbf{q}'_\perp\right).
\end{equation}
\begin{equation} \label{eq:crA}
\left[a(q), a^\dagger(q')\right] =
(2 \pi)^3 \,\delta (q^+ - {q'}^+) ~
\delta^{(2)}\negthinspace\left(\mbf{q}_\perp - \mbf{q}'_\perp\right).
\end{equation}

Using the  LF commutation relations given above  and the properties of the light-front spinors given in  Appendix \ref{metric}, we obtain the expression of the light-front Hamiltonian $P^-$ in the particle number representation
\begin{equation} \label{Pmp}
P^- \!  = \!  \sum_\lambda \int \!  \frac{dq^+ d^2 \mbf{q}_\perp}{(2 \pi)^3 }   \,
\Big( \frac{\mbf{q}_\perp^2 + m^2 }{q^+} \! \Big) \,
 b_\lambda^\dagger(q) b_\lambda(q) + { \rm (interactions)} ,\\
\end{equation}
where, for simplicity, we have omitted from \req{Pmp} the terms corresponding to antiquarks and gluons. We recover the LF dispersion relation $q^- = ({\mbf{q}_\perp^2 \! + m^2})/{q^+}$ for a quark or antiquark in absence of interactions and the dispersion relation for the gluon quanta
$q^- = \mbf{q}_\perp^2/q^+$, which follows
from the on shell relation $q^2 = m^2$ and $q^2 =0$ respectively.  The LF time evolution operator $P^-$ is thus conveniently written as a term which represents the sum of the kinetic energy of all the partons plus a sum of all the interaction terms.  The longitudinal and transverse kinematical  generators are
\begin{eqnarray} \label{Pplus}
P^+ &\! =\!&  \sum_\lambda \int \!  \frac{dq^+ d^2 \mbf{q}_\perp}{(2 \pi)^3 }   \, q^+\,\,
 b_\lambda^\dagger(q) b_\lambda(q), \\ \label{Pperp}
 \mbf{P}_\perp  &\! =\!&  \sum_\lambda \int \!  \frac{dq^+ d^2 \mbf{q}_\perp}{(2 \pi)^3 }   \,\,
\mbf{q}_\perp  \,
 b_\lambda^\dagger(q) b_\lambda(q),
\end{eqnarray}
and contain no interactions. For simplicity we have also omitted from \req{Pplus} and \req{Pperp} the contribution of the kinetic terms from antiquarks and gluons.

\subsection{Representation of hadrons in the light-front Fock basis}

An important advantage of light-front quantization is that a particle 
Fock expansion can be used as the basis for representing the physical
states of QCD.  The light-front Fock representation is thus an interpolating basis projecting the hadronic eigenstate onto the Fock basics of free on-shell partonic constituents.
The complete basis of Fock-states $\vert n\rangle$ is constructed by applying
free-field creation operators to
the vacuum state $\vert 0 \rangle$ which has no particle content,
$P^+ \vert 0 \rangle =0$, $\mbf{P}_{\! \perp} \vert 0 \rangle = 0$.
A one-particle state is defined by
$\vert q \rangle = \sqrt{2 q^+} \,b^\dagger(q) \vert 0 \rangle$,
so that its normalization has the Lorentz invariant form
\beq \label{normq}
\langle q | q' \rangle =  2 {q}^+  (2 \pi)^3  \delta ({q}^+ -
{q'}^+)\, \delta^{(2)} (\mbf{q}_{\perp} - \mbf{q}_{\perp}'),
\enq
and this fixes our normalization. Each $n$-particle Fock state
$|p_i^+ \!, \mbf{p}_{\perp i}, \la_i \rangle$ is an eigenstate of  $P^+$,
$\mbf{P}_{\! \perp}$ and $J^3$ and it is normalized according to
\begin{equation}  
\left\langle  p_i^+, \mbf{p}_{\perp i},\lambda_i
 \big|{p'}_i^+, \mbf{p'}_{\negthinspace\perp i},\lambda'_i \right\rangle
= 2 p_i^+ (2 \pi)^3 \, \delta \bigl(p_i^+ - {p'}_i^+\bigr) \,
\delta^{(2)} \negthinspace \bigl(\mbf{p}_{\perp i} - \mbf{p'}_{\negthinspace\perp
    i}\bigr) \, \delta_{\lambda_i,\lambda'_i}.
\label{eq:normFC}
\end{equation}

We now proceed to the separation of relative and overall kinematics by introducing the  partonic variables $k_i^\mu= \left(k^+_i, k^-_i, \mbf{k}_{\perp i}\right)$ according to ~$k_i^+ =  x_i P^+$, ~  $\mbf{p}_{\perp i} = x_i \mbf{P}_{\perp i} + \mbf{k}_{\perp i}$,   where the longitudinal momentum fraction for each constituent is $x_i = k_i^+/P^+$   (See Fig. \ref{npartons}).  Momentum conservation requires that $P^+\!  = \sum_{i=1}^n k^+_i$, $k^+_i > 0$, or equivalently
$\sum_{i=1}^n x_i =1$, and $\sum_{i=1}^n \mbf{k}_{\perp i}=0$. The light-front momentum coordinates $x_i$ and  $\mbf{k}_{\perp i}$  are actually relative coordinates; {\it i.e.}, they are independent of the total momentum $P^+$  and $\mbf{P}_\perp$ of the bound state.

\begin{figure}[ht] 
\begin{centering}
\includegraphics[width=6.8cm]{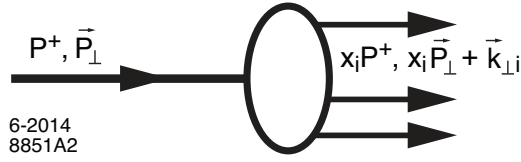}
\caption{\label{npartons} \small  Overall and relative partonic variables in a hadronic bound state.}
\end{centering}
\end{figure}

The hadron state is an eigenstate of the total momentum $P^+$
and $\mbf{P}_{\! \perp}$ and the  total spin projection $S^z$.
Each hadronic eigenstate $\vert \psi \rangle$  is expanded in a complete
Fock-state  basis of noninteracting $n$-particle states
$\vert n \rangle$ with an infinite number of components. For example, a proton with four-momentum $P^\mu =  (P^+, P^-,\mbf{P}_{\!\perp})$
is described by the expansion
\begin{equation}
\left\vert \psi(P^+,\mbf{P}_{\! \perp}, S_z) \right\rangle
= \sum_{n,\lambda_i}
 \int \!  \int \big[d x_i\big]  \left[d^2 \mbf{k}_{\perp i}\right]
\frac{1}{\sqrt{x_i}} \, \psi_n(x_i,\mbf{k}_{\perp i},\lambda_i)
\bigl\vert n: x_i P^+\!, x_i \mbf{P}_{\! \perp} \! +  \mbf{k}_{\perp i},\lambda_i \bigr\rangle,
\label{eq:LFWFexp}
\end{equation}
where the sum is over all Fock states and helicities, 
beginning with the valence state; {\it e.g.}, $n \ge 3$ for baryons.
The measure of the constituents phase-space momentum integration  is
\begin{equation} \label{fsx}
\int \big[d x_i\big] \equiv
\prod_{i=1}^n \int dx_i \,\delta \Bigl(1 - \sum_{j=1}^n x_j\Bigr) ,
\end{equation}
\vspace{-10pt}
\begin{equation} \label{fsk}
\int \left[d^2 \mbf{k}_{\perp i}\right] \equiv \prod_{i=1}^n \int
\frac{d^2 \mbf{k}_{\perp i}}{2 (2\pi)^3} \, 16 \pi^3 \,
\delta^{(2)} \negthinspace\Bigl(\sum_{j=1}^n\mbf{k}_{\perp j}\Bigr).
\end{equation}
The coefficients of the  Fock expansion
\begin{equation} \label{eq:LFWF}
\psi_n(x_i, \mbf{k}_{\perp i},\lambda_i)
= \bigl\langle n:x_i,\mbf{k}_{\perp i},\lambda_i \big\vert \psi \bigr\rangle ,
\end{equation}
are frame independent; {\it i.e.}, the form of the LFWFs is independent of the total longitudinal and transverse momentum $P^+$ and $\mbf{P}_{\! \perp}$ of
the hadron and depend only on the partonic coordinates: the longitudinal momentum fraction $x_i$,  the transverse momentum $\mbf{k}_{\perp i}$, and $\lambda_i$, the projection of the constituent's spin along the $z$ direction. The
wave function $\psi_n(x_i, \mbf{k}_{\perp i},\lambda_i) $ represents the probability amplitudes to find on-mass-shell constituents $i$ in a specific light-front Fock state $\vert n \rangle$ with longitudinal momentum $ x_i P^+$, transverse momentum $x_i \mbf{P}_{\! \perp} + \mbf{k}_{\perp i}$ and
helicity $\lambda_i$ in a given hadron.  
Since  the boundary conditions are specified in the null-plane,
the ability to specify wave functions simultaneously in any frame is a
special feature of light-front quantization.

Each constituent of the light-front wave function  $\psi_{n}(x_i, \mbf{k}_{\perp i}, \lambda_i)$  of a hadron is on its respective mass shell  $k^2_i= m^2_i$, where $k_{\mu i} \, k_i^\mu = k^+_i k^-_i - \mbf{k}^2_{\perp i}$,   thus each single particle state has four-momentum
\beq
k_i^\mu =  \left( k_i^+, k^-_i, \mbf{k}_i\right)= \left( k_i^+, \frac{\mbf{k}^2_{\perp i} + m_i^2}{k^+_i}, \mbf{k}_i \right), ~~~ {\rm for} ~~i = 1, 2 \cdots n.
\enq
However,  the light-front wave function represents a state which is off the light-front energy shell, $P^-  - \sum_i^n k^-_i < 0$, for a stable hadron.  In fact, the invariant mass of the constituents in each $n$-particle Fock state is given by
\begin{equation} \label{M2n}
M_n^2= \Big(\sum_{i=1}^n k^\mu_i\Big)^2=\Big(\sum_{i=1}^n k_i^+\Big)\,\Big(\sum_{i=1}^n k_i^-\Big)
-\Big(\sum_{i=1}^n \mbf{k}_{\perp i}\Big)^2= \sum_{i=1}^n\frac{\mbf{k}_{\perp i}^2 + m_i^2}{x_i},
\end{equation}
and is a measure of the off-energy shell of the bound state,  with   $M_n^2$ in general  different from the hadron bound-state mass $P_\mu P^\mu = M^2$.

The hadron state is normalized according to
\begin{equation}
\bigl\langle \psi(P^+,\mbf{P}_{\! \perp}, S^z) \big\vert
\psi(P'^+,\mbf{P}'_\perp, S^{z'}) \bigr\rangle
= 2 P^+ (2 \pi)^3 \,\delta_{S^z\!, S^{z'}} \,\delta \bigl(P^+ - P'^+ \bigr)
\,\delta^{(2)} \negthinspace \bigl(\mbf{P}_{\! \perp} - \mbf{P}'_\perp\bigr).
\label{eq:Pnorm}
\end{equation}
Thus, the  normalization of the  LFWFs is determined by
\begin{equation}
\sum_n  \int \big[d x_i\big] \left[d^2 \mbf{k}_{\perp i}\right]
\,\left\vert \psi_n(x_i, \mbf{k}_{\perp i}) \right\vert^2 = 1,
\label{eq:LFWFknorm}
\end{equation}
where the internal-spin indices have been suppressed.

The constituent spin and orbital angular momentum properties of the
hadrons are also encoded in the LFWFs  $\psi_n(x_i, \mbf{k}_{\perp i},\lambda_i)$  which obey the
total orbital angular momentum sum rule~\cite{Brodsky:2000ii}
\beq \label{Jz}
J^z = \sum_{i=1}^n  S^z_i + \sum_{i=1}^{n-1} L^z_i,
\enq
since there are only $n-1$ relative angular momenta in an $n$-particle light-front Fock state in the sum \req{Jz}. The internal spins $S^z_i$ are denoted as $\lambda_i$. The  orbital angular momenta have the operator form
\begin{equation}
L^z_i =-i \left(\frac{\partial}{\partial k^x_i}k^y_i -
\frac{\partial}{\partial k^y_i}k^x_i \right).
\end{equation}
Since the total angular momentum projection $J^z$ in the light front is a kinematical operator, it
is conserved Fock state by Fock state and by every interaction in the LF Hamiltonian.
In the light-cone gauge $A^+ = 0$,  the gluons only have physical angular momentum projections
$S^z= \pm 1$ and the orbital angular momentum of quark and gluons is defined unambiguously~\cite{Brodsky:1997de}.

\section{Semiclassical approximation to QCD in the light front \label{LFQM}}

 Our goal is to find a semiclassical approximation to strongly coupled QCD dynamics and derive a simple relativistic wave equation to compute hadronic bound states and other hadronic properties. To this end it is necessary to reduce the  multiple particle eigenvalue problem of the LF Hamiltonian (\ref{HLF}) to an effective light-front  Schr\"odinger equation, instead of diagonalizing the full Hamiltonian. The central problem then becomes the derivation of the effective interaction, which acts only on the valence sector of the theory and has, by definition, the same eigenvalue spectrum as the initial Hamiltonian problem.   For carrying out this program  in the front from, one must systematically express the higher-Fock components as functionals of the lower ones. This method has the advantage that the Fock space is not truncated and the symmetries of the Lagrangian are preserved~\cite{Pauli:1998tf}. The method is similar to the methods used in many-body problems in nuclear physics to reduce the great complexity of a dynamical problem with a large number of degrees of freedom to an effective model with fewer degrees of freedom~\cite{Brown:2010}.  The same method is used in QED;  for example, the reduction of the higher Fock states of  muonium $\mu^+ e^-$ to an effective $\mu^+ e^-$ equation introduces interactions which yield the hyperfine splitting, Lamb shift, and other corrections to the Coulomb-dominated potential.

In principle one should determine the effective potential  from the two-particle irreducible  $ q \bar q \to q \bar q $ Greens' function for a pion.  In particular, the reduction from higher Fock states in the intermediate states would lead to an effective interaction $U$  for the valence $\vert q \bar q \rangle$ Fock state of the pion~\cite{Pauli:1998tf}.  However, in order to capture the nonperturbative dynamics  one most  integrate out all higher Fock states, corresponding to an infinite number of degrees of freedom --  a formidable problem. This is apparent, for example, if one identifies the sum of infrared sensitive ``H'' diagrams as the source of the effective potential, since the horizontal rungs correspond to an infinite number of higher gluonic Fock states~\cite{Appelquist:1977tw, Appelquist:1977es}.  A related approach for determining the valence light-front wave function and studying the effects of higher Fock states without truncation has been given in Ref.~\cite{Chabysheva:2011ed}.

We will describe below a simple procedure  which allows us to reduce the strongly correlated multi-parton bound-state problem in light-front QCD into an effective one-dimensional problem~\cite{deTeramond:2008ht}.  To follow this procedure, it is crucial to identify as the key dynamical variable,  the invariant mass  $M_n^2$ \req{M2n},  $M_n^2 = \left(k_1 + k_2 + \cdots k_n\right)^2$,   which controls the bound state. In fact, the LFWF is of-shell in $P^-$ and consequently in the invariant mass. Alternatively, it is useful to consider its canonical conjugate invariant variable in impact space. This choice of variable will also allow us  to separate  the dynamics of quark and gluon binding from the kinematics of constituent spin and internal orbital angular momentum~\cite{deTeramond:2008ht}.

For an  $n$-Fock component $\psi \left(k_1, k_2, \cdots, k_n\right)$ we make the substitution
\beq
\psi_n \left(k_1, k_2, \cdots, k_n\right)  ~\to ~  \phi_n\left( \left(k_1 + k_2 + \cdots k_n\right)^2\right), ~~~m_q \to 0.
\enq
Using this semiclassical approximation, and in the limit of zero quark masses, the $n$-particle bound-state problem is reduced effectively to a single-variable LF quantum mechanical wave equation~\cite{deTeramond:2008ht},  which describes the bound-state dynamics of light hadrons in terms of an effective confining interaction $U$  defined at equal LF time.  In this semiclassical approximation there is no particle creation or absorption.

Let us outline how this reduction is actually carried out in practice. We  compute $M^2$ from the hadronic matrix element
$
\langle \psi(P') \vert P_\mu P^\mu \vert\psi(P) \rangle  =
M^2  \langle \psi(P' ) \vert\psi(P) \rangle,
$
expanding  the initial and final hadronic states in terms of their Fock components using (\ref{eq:LFWFexp}).
The computation is  simplified in the frame $P = \big(P^+, M^2/P^+, \vec{0}_\perp \big)$ where $H_{LF} =  P^2 \! = P^+ P^-$.
Using the normalization condition (\ref{eq:normFC}) for each individual constituent and after integration over the internal coordinates of the $n$ constituents for each Fock state in
the $\mbf{P}_\perp = 0$ frame, one finds~\cite{deTeramond:2008ht}
 \begin{equation} \label{eq:MKk}
 M^2  =  \sum_n  \! \int \! \big[d x_i\big]  \! \left[d^2 \mbf{k}_{\perp i}\right]
 \sum_{a=1}^n \left(\frac{ \mbf{k}_{\perp a}^2 + m_a^2 }{x_a} \right)
 \left\vert \psi_n (x_i, \mbf{k}_{\perp i}) \right \vert^2  + ({\rm interactions}) ,
 \end{equation}
plus similar terms for antiquarks and gluons ($m_g = 0)$. The integrals in (\ref{eq:MKk}) are over the internal coordinates of the $n$ constituents for each Fock state  with the phase space normalization  (\ref{eq:LFWFknorm}). Since the LF kinetic energy has a finite  value in each Fock state, it follows that the  LFWFs of bound states $\psi_n (x_i, \mbf{k}_{\perp i})$ have the small momentum fraction-$x$ boundary conditions  $\psi_{n} (x_i, \mbf{k}_{\perp i}) \to (x_i)^\al$ with $\al \ge \half$ in the limit $x_i \to 0$~\cite{Antonuccio:1997tw}.

It is useful to express (\ref{eq:MKk}) in terms of  $n\! - \! 1$ independent transverse impact variables $\mbf{b}_{\perp j}$, $j = 1,2,\dots,n-1$,
conjugate to the relative coordinates $\mbf{k}_{\perp i}$ using the Fourier expansion~\cite{Soper:1976jc}
\begin{equation} \label{eq:LFWFb}
\psi_n(x_j, \mathbf{k}_{\perp j}) =  (4 \pi)^{(n-1)/2}
\prod_{j=1}^{n-1}\int d^2 \mbf{b}_{\perp j}
\exp{\left(i \sum_{k=1}^{n-1} \mathbf{b}_{\perp k} \cdot \mbf{k}_{\perp k}\right)} \,
{\psi}_n(x_j, \mathbf{b}_{\perp j}),
\end{equation}
where $\sum_{i=1}^n \mbf{b}_{\perp i} = 0$.  We find
\begin{equation}
 M^2  =  \sum_n  \prod_{j=1}^{n-1} \int d x_j \, d^2 \mbf{b}_{\perp j} \,
\psi_n^*(x_j, \mbf{b}_{\perp j})
 \sum_{a=1}^n \left(\frac{ \mbf{- \nabla}_{ \mbf{b}_{\perp a}}^2 + m_a^2 }{x_a} \right)
 \psi_n(x_j, \mbf{b}_{\perp j})
  + ({\rm interactions}) ,
  \label{eq:MKb}
 \end{equation}
where the normalization in impact space is defined by
\begin{equation}  \label{eq:LFWFbnorm}
\sum_n  \prod_{j=1}^{n-1} \int d x_j d^2 \mbf{b}_{\perp j}
\left\vert \psi_n(x_j, \mbf{b}_{\perp j})\right\vert^2 = 1.
\end{equation}

The simplest example is a two-parton hadronic bound state. If we want to  reduce further the dynamics to a single-variable problem, we must take the limit of quark masses to zero.
 In the limit $m_q \to 0$ we find
\begin{eqnarray} \nonumber
M^2  &\!\!=\!\!&  \int_0^1 \! d x \! \int \!  \frac{d^2 \mbf{k}_\perp}{16 \pi^3}   \,
  \frac{\mbf{k}_\perp^2}{x(1-x)}
 \left\vert \psi (x, \mbf{k}_\perp) \right \vert^2  + ({\rm interactions}) \\    \label{eq:Mb}
  &\!\!=\!\!& \int_0^1 \! \frac{d x}{x(1-x)} \int  \! d^2 \mbf{b}_\perp  \,
  \psi^*(x, \mbf{b}_\perp)
  \left( - \mbf{\nabla}_{\mbf{b}_ \perp}^2 \right)
  \psi(x, \mbf{b}_\perp)   +  ({\rm interactions}),
 \end{eqnarray}
 with normalization
\beq\label{N}
\int_0^1 \! d x \! \int \!  \frac{d^2 \mbf{k}_\perp}{16 \pi^3}   \,  \left\vert \psi (x, \mbf{k}_\perp) \right \vert^2  
=  \int_0^1 \! d x \int  \! d^2 \mbf{b}_\perp  \, \left\vert\psi(x, \mbf{b}_\perp)\right\vert^2 = 1.
\enq

\begin{figure}[ht] 
\begin{centering}
\includegraphics[width=6.0cm]{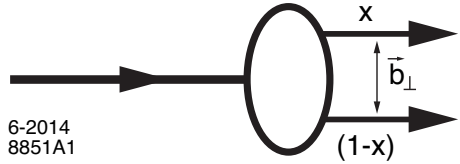}
\caption{\label{qbarq} \small Relative $q-\bar q$ variables in impact space for a pion bound state.}
\end{centering}
\end{figure}

For $n=2$, the invariant mass \req{M2n} is $M_{q \bar q}^2 = \frac{\mbf{k}_\perp^2}{x(1-x)}$.
Similarly, in impact space the relevant variable for a two-parton state is  $\zeta^2= x(1-x)\mbf{b}_\perp^2$,  the invariant separation between the quark and  antiquark (See Fig. \ref{qbarq}).  Thus, to first approximation,  LF dynamics  depend only on the boost invariant variable $M_n$ or $\zeta$,  and the dynamical properties are encoded in the hadronic LF wave function
$\phi(\zeta)$
\begin{equation} \label{eq:psiphi}   
\psi(x,\zeta, \varphi) = e^{i L \varphi} X(x) \frac{\phi(\zeta)}{\sqrt{2 \pi \zeta}} , 
\end{equation}
where we have factored out the longitudinal $X(x)$ and orbital dependence from the LFWF $\psi$.  This is a natural factorization in the light front since the
corresponding canonical generators, the longitudinal and transverse generators $P^+$ and $\mbf{P}_\perp$ and the $z$-component of the orbital angular momentum
$J^z$ are kinematical generators which commute with the LF Hamiltonian generator $P^-$.
From \req{N} the normalization of the transverse and longitudinal modes is given by
\beqa\label{NT}
 \langle\phi\vert\phi\rangle &\!=\!& \int \! d \zeta  \, \phi^2(\zeta) = 1, \\ \label{NL}
 \langle X \vert X \rangle &\!=\!&  \int_0^1 dx \, x^{-1} (1-x)^{-1} X^2(x) =1 .
 \enqa

To proceed, we  write the Laplacian operator in (\ref{eq:Mb}) in  polar coordinates
$(\zeta, \varphi)$
\begin{equation} \label{eq:Lzeta}
\nabla_\zeta^2 = \frac{1}{\zeta} \frac{d}{d\zeta} \left( \zeta \frac{d}{d\zeta} \right)
+ \frac{1}{\zeta^2} \frac{\partial^2}{\partial \varphi^2},
\end{equation}
and factor out the angular dependence of the modes in terms of the $SO(2)$ Casimir representation $L^2$ of orbital angular momentum in the transverse plane.
Using  (\ref{eq:psiphi}) we find~\cite{deTeramond:2008ht}
\beq \label{eq:KV}
M^2  =  \int \! d\zeta \, \phi^*(\zeta) \sqrt{\zeta}
\left( -\frac{d^2}{d\zeta^2} -\frac{1}{\zeta} \frac{d}{d\zeta}
+ \frac{L^2}{\zeta^2}\right)
\frac{\phi(\zeta)}{\sqrt{\zeta}}
+ \int \! d\zeta \, \phi^*(\zeta) U(\zeta) \phi(\zeta) ,
\enq
where $L = \vert L^z \vert$.  In writing the above equation we have summed up all the complexity of the interaction terms in the QCD Hamiltonian \req{Pm}  in the introduction of the effective potential $U(\zeta)$ which acts in the valence state, and which should enforce confinement at some IR scale, which determines the QCD mass gap. The light-front eigenvalue equation $P_\mu P^\mu \vert \phi \rangle = M^2 \vert \phi \rangle$ is thus a light-front wave equation for $\phi$
\begin{equation} \label{LFWE}
\left(-\frac{d^2}{d\zeta^2}
- \frac{1 - 4L^2}{4\zeta^2} + U(\zeta) \right)
\phi(\zeta) = M^2 \phi(\zeta),
\end{equation}
a relativistic single-variable LF  quantum-mechanical wave equation.   Its eigenmodes $\phi(\zeta)$ determine the hadronic mass spectrum and represent the probability amplitude to find the partons at transverse impact separation $\zeta$, the invariant separation between pointlike constituents within the hadron~\cite{Brodsky:2006uqa} at equal LF time.   This equation is an effective two-particle wave equation where  an infinite number of  higher Fock states~\cite{Pauli:1998tf}  and retarded interactions are  incorporated in the light-front effective potential, which acts on the valence states.   In practice,  computing the effective potential from QCD is a formidable task and other methods have to be devised to incorporate the essential dynamics from confinement. The effective interaction potential in \req{LFWE} is instantaneous in LF time $x^+$, not instantaneous in ordinary time $x^0$.  The LF potential thus satisfies causality, unlike the instantaneous Coulomb interaction appearing in atomic physics.

If $L^2 < 0$, the LF Hamiltonian  defined in Eq. \req{HLF}  is unbounded from below $\langle \phi \vert P_\mu P^\mu \vert \phi \rangle <0$,  and the spectrum contains an infinite number of unphysical negative values of $M^2 $ which can be arbitrarily large. As $M^2$ increases in absolute value, the particle becomes localized within a very small region near $\zeta = 0$, if the effective potential  vanishes at small  $\zeta$. For $M^2 \to - \infty$ the particle is localized at $\zeta = 0$, the particle ``falls towards the center''~\cite{LL:1958}. The critical value  $L=0$  corresponds to the lowest possible stable solution, the ground state of the light-front Hamiltonian. It is important to notice that in the light front, the $SO(2)$ Casimir for orbital angular momentum $L^2$ is a kinematical quantity, in contrast to the usual $SO(3)$ Casimir $L(L+1)$ from non-relativistic physics which is rotational, but not boost invariant. The  $SO(2)$ Casimir form $L^2$  corresponds to the group of rotations in the transverse LF plane. Indeed, the Casimir operator for $SO(N)$  is $L(L+N-2)$.

If we compare the invariant mass  in the instant form in the hadron center-of-mass system, $\mbf{P} = 0$, $M^2_{q \bar q} = 4 \, m_q^2 +  4 \mbf{p}^2$, with the invariant mass in the front form in the constituent rest frame, $\mbf{p}_q +  \mbf{p}_{\bar q} = 0$ for equal quark-antiquark masses~\footnote{Notice that the hadron center-of-mass frame and the constituent rest frame are not identical in the front form of dynamics since the third component of momentum is not conserved in the light front (See Ref.~\cite{Trawinski:2014msa} and references therein.}, we obtain the relation~\cite{Trawinski:2014msa} 
\beq \label{pots}
U =  V^2  + 2 \sqrt{\mbf{p}^2 + m_q^2} \,  V +  2 \,  V \sqrt{\mbf{p}^2 + m_{q}^2},
\enq
where we identify $\mbf{p}_\perp^2 = \frac{\mbf{k}_\perp^2}{4 x (1-  x)}$, $p_3 = \frac{m_q (x - 1/2)}{\sqrt{ x(1-x)}}$, and  $V$ is the effective potential in the instant form. Thus, for small quark masses a  linear instant-form potential $V$ implies a harmonic  front-form potential $U$ at large distances.    One can also show   how the two-dimensional front-form harmonic oscillator potential for massless quarks takes on a three-dimensional form when the quarks have mass since the third space component is conjugate to $p_3$, which has an infinite range for $m \ne 0$~\cite{Trawinski:2014msa}.

Extension of the results to arbitrary $n$ follows from the $x$-weighted definition of the
transverse impact variable of the $n-1$ spectator system~\cite{Brodsky:2006uqa} given by the LF cluster decomposition
\begin{equation} \label{zetan}
\zeta = \sqrt{\frac{x}{1-x}} \Big\vert \sum_{j=1}^{n-1} x_j \mbf{b}_{\perp j} \Big\vert,
\end{equation}
 where $x = x_n$ is the longitudinal
momentum fraction of the active quark. One can also
generalize the equations to allow for the kinetic energy of massive
quarks using Eqs. (\ref{eq:MKk}) or (\ref{eq:MKb}) as discussed in Sec. \ref{FQM} below. In this case, however,
the longitudinal mode $X(x)$ does not decouple from the effective LF bound-state equations.

\subsection{Inclusion of light quark masses \label{FQM}}

The noticeable simplicity of the transverse single-variable light-front wave equation derived from the bound-state Hamiltonian equation of motion in Sec. \ref{LFQM} is lost when we consider massive quarks, as longitudinal LF variables have to be taken into account as well. In the limit of massless quarks the scheme is very concise and unique: in the semiclassical approximation the underlying conformal symmetry of QCD determines the dynamics (Chapter \ref{ch3}) and there is an exact agreement of the AdS equations of motion and the light-front Hamiltonian (Chapter \ref{ch4}). However, as we will discuss here, the inclusion of small quark masses can still  be treated in a simple way following the semiclassical approximation described above.

As it is clear from \req{pots}, the effective light-front confining interaction $U$ has a strong dependence on heavy quark masses  and consequently in the longitudinal variables~\footnote{This connection was used in Ref.~\cite{Gutsche:2014oua} to construct a light-front  potential for heavy quarkonia.}. However,  for small quark masses -- as compared to the hadronic scale,   one expects that the effective confinement interaction  in the quark masses is unchanged to first order. In this approximation  the confinement potential only depends on the transverse invariant variable $\zeta$, and the transverse dynamics are  unchanged.  The partonic shift in the hadronic mass is computed straightforwardly from  \req{eq:MKk} or  \req{eq:MKb}
\beqa \label{eq:DelM2}
\Delta M^2  &=& 
\sum_n  \int \big[d x_i\big] \left[d^2 \mbf{k}_{\perp i}\right] \sum_{a=1}^n \frac{m_a^2}{x_a}
\,\left\vert \psi_n(x_i, \mbf{k}_{\perp i}) \right\vert^2   \nn  \\
&=&  \sum_n  \prod_{j=1}^{n-1} \int d x_j \, d^2 \mbf{b}_{\perp j} \,
 \sum_{a=1}^n \frac{m_a^2 }{x_a}  \,
\left\vert\psi_n(x_j, \mbf{b}_{\perp j})\right\vert^2 \nn \\
&\equiv & \left\langle \psi  \left\vert \sum_a \frac{m_a^2}{x_a} \right \vert \psi  \right \rangle, 
\enqa
where $\Delta M^2 = M^2 - M_0^2$. Here $M_0^2$ is the value of the hadronic mass computed in the limit of zero quark masses. This expression is identical to the Weisberger result for a partonic mass shift~\cite{Weisberger:1972hk}. Notice that this result is exact to first order in the light-quark mass if the sum in \req{eq:DelM2} is over all Fock states $n$.  

For simplicity, we consider again the case of a meson bound-state of a quark and an antiquark with longitudinal momentum $x$ and $1-x$ respectively.  To lowest order in the quark masses  we find
\beq  \label{MKbmq}  
\Delta M^2  =   \int_0^1 \frac{d x}{x(1-x)} 
\left( \frac{m_q^2}{x} + \frac{m_{\bar q}^2}{1-x} \right)  X^2(x),
\enq
using the normalization  \req{NL}. The quark masses $m_q$ and $m_{\bar q}$  in \req{MKbmq} are effective  quark masses  from the renormalization due to the reduction of higher Fock states as functionals of the valence state~\cite{Pauli:1998tf},  not ``current" quark masses, {\it i.e.}, the quark masses appearing in the QCD Lagrangian.

As it will be shown in Chapter \ref{ch6}, the factor $X(x)$ in the LFWF in \req{eq:psiphi} can be determined in the limit of massless quarks from the precise mapping of light-front amplitudes for arbitrary momentum transfer $Q$. Its form is $X(x) = x^\half (1-x)^\half$ \cite{Brodsky:2006uqa}. This expression of the LFWF gives a divergent expression for the partonic mass shift \req{MKbmq}, and, evidently,  realistic effective two-particle wave functions have to be additionally suppressed at the end-points $x = 0$ and $x = 1$. As pointed out in \cite{Brodsky:2008pg}, a key for this modification is suggested by the construction of the light-front wave  functions discussed above. It relies on the assumption that the essential dynamical variable which controls the bound state  wave function in momentum space is the invariant mass \req{M2n}, which determines the off-energy shell behavior of the bound state. For the effective two-body bound state the inclusion of  light quark masses amounts to the replacement
\beq \label{Mqbarq}
M^2_{q \bar q} = \frac{\mbf{k}_\perp^2}{x(1-x)} \to \frac{\mbf{k}_\perp^2}{x(1-x)} +\frac{m_q^2}{x} + \frac{m_{\bar q}^2}{1-x},
\enq
in the LFWF in momentum space, $\psi_{q \bar q}\left(x, \mbf{k}_\perp\right)$, which is then Fourier transformed to impact space.  We will come back to this point in Chapter \ref{ch5} for the specific models dictated by conformal invariance. The longitudinal dynamics in presence of quark masses has also been discussed in Ref.~\cite{Chabysheva:2013pia}.

In the next chapter  we will show how the form of the effective confinement potential is uniquely determined from the algebraic structure of an effective one-dimensional quantum field theory, which encodes the underlying conformality of the classical QCD Lagrangian. In subsequent chapters we will  discuss the connection of effective gravity theories in AdS space with the light-front results presented in this section.

\chapter{Conformal Quantum Mechanics and Light-Front Dynamics \label{ch3}}
As  we have emphasized in the introduction, conformal symmetry plays a special,  but somehow hidden,  role in QCD. The classical Lagrangian is, in the limit of massless quarks, invariant under conformal transformations~\cite{Parisi:1972zy, Braun:2003rp}.    The symmetry, however,  is broken by quantum corrections. Indeed, the need for renormalization of the theory introduces a scale $\La_{\rm QCD}$ which leads to the ``running coupling" $\al_s\left(Q^2\right)$ and asymptotic freedom~\cite{Gross:1973id, Politzer:1973fx} for large values of $Q^2$, $Q^2 \gg \La^2_{\rm QCD}$, a mechanism conventionally named ``dimensional transmutation".  But there are theoretical and phenomenological indications that at large distances,  or small values of $Q^2$,  $Q^2 \le \La^2_{\rm QCD}$, where the string tension has formed,  the QCD $\beta$-function vanishes and scale invariance is in some sense restored  (See for example Ref.~\cite{Brodsky:2007hb} and references therein). Since we are interested in a semiclassical approximation to nonperturbative QCD, analogous to the quantum mechanical wave equations  in atomic physics, it is natural to have a closer look at conformal quantum mechanics, a conformal field theory in one dimension. De Alfaro, Fubini and Furlan~\cite{deAlfaro:1976je} have obtained remarkable results which,  extended to light-front holographic QCD~\cite{deTeramond:2008ht, Brodsky:2013ar}, give important insights into the  QCD confining mechanism.  It turns out that it is possible to introduce a scale in the light-front Hamiltonian, by modifying the variable of dynamical evolution and  nonetheless the underlying action  remains conformally invariant. Remarkably this procedure determines uniquely the form of the light-front effective potential and correspondingly the  modification of AdS space.

\section{One-dimensional conformal field theory \label{alg}}

Our aim is to incorporate in a one-dimensional quantum field theory  -- as an effective theory, the fundamental conformal symmetry of the four-dimensional classical QCD Lagrangian in the limit of massless quarks. We will require that the corresponding one-dimensional effective action, which encodes the chiral symmetry of QCD, remains conformally invariant.  De Alfaro {\it et al.}~\cite{deAlfaro:1976je} investigated in detail the simplest scale-invariant  model, one-dimensional field theory, namely
 \beq \label{AQ} 
A[Q] = \half \int dt  \left(\dot Q^2 - \frac{g}{Q^2} \right),
\enq
 where  $\dot Q \equiv d Q / dt$. Since the action is dimensionless, the dimension of the field $Q$ must be half the dimension of the ``time'' variable  $t$, dim[$Q$] = $\half$dim$[t]$,  and  the constant $g$  is dimensionless. The translation operator in $t$, the Hamiltonian, is
 \beq \label{HtQ}
H = 
 \half \left(\dot Q^2 + \frac{g}{Q ^2}\right),
 \enq
 where the field momentum operator is $P = \dot Q$,  and therefore the equal time commutation relation is
\beq \label{CCR}
[Q(t), \dot Q(t)] = i.
\enq
The equation of motion for the field operator $Q(t)$ is  given by the usual quantum mechanical evolution
\beq  \label{QME}
 i \left[H, Q(t) \right]  = \frac{d Q(t)}{d t}.
 \enq

In the Schr\"odinger picture with $t$-independent operators and $t$-dependent state vectors  the evolution is given by
\beq  \label{QEf}
H  \vert \psi(t) \rangle = i \frac {d}{d t} \vert \psi(t)\rangle.
\enq
Using  the representation of  the field operators $Q$ and  $P = \dot Q$ given by the substitution $Q(0) \to x,  ~ \dot Q(0) \to - i {d}/{d x}$ we obtain  the usual quantum mechanical evolution
\beq  \label{Sp}
i \frac{\partial}{\partial t} \psi(x,\tau) = H \Big(x,   - i \frac{d}{d x} \Big) \psi(x, t),
\enq
with the Hamiltonian 
\beq  \label{Htxf}
H  = \half \left(- \frac{d^2}{dx^2} + \frac{g}{x^2} \right) .
\enq
It has the same structure as the LF Hamiltonian \req{LFWE} with a vanishing light-front potential,  as expected for a conformal theory. The dimensionless constant $g$ in the action \req{AQ} is now related to the Casimir operator of rotations in the light-front wave equation \req{LFWE}.

However, as emphasized by dAFF, the absence of dimensional constants in \req{AQ} implies that the action action $A[Q]$ is invariant under  a larger group of transformations, the full conformal group in one dimension, that is, under translations, dilatations, and special conformal transformations. For one dimension these can be expressed by the transformations of the variable $t$
 \beq
\label{GC} 
t'=\frac{\al t + \be}{\ga t + \de}; \quad \quad  \al \de - \be \ga =1\;,  
\enq
 and the corresponding field transformation
\beq
 Q'(t') = \frac{Q(t)}{\ga t + \de}.
 \enq
As we show in Appendix \ref{cfqm}, the action $A[Q]$ \req{AQ} is indeed, up to a surface term, invariant under conformal transformations.

The constants of motion of the action are obtained by applying Noether's theorem. The three conserved generators corresponding to the invariance of the action \req{AQ} under  the full conformal group in one dimension are (Appendix \ref{cfqm}):

\begin{enumerate}

\item
Translations in the variable  $t$:
\beq
H  = \half \left( \dot Q^2 + \frac{g}{Q^2} \right),
\enq

\item
Dilatations: 
\beq
D = \half \left( \dot Q^2 + \frac{g}{Q^2} \right) t - {\textstyle{\frac{1}{4} }}  \left( \dot Q Q + Q \dot Q \right),
\enq

\item
Special conformal transformations:
\beq
 K= \half \left( \dot Q^2 + \frac{g}{Q^2}\right) t^2 -\half \left(\dot Q Q+ Q \dot Q\right) \,t + \half Q^2 ,
 \enq
\end{enumerate}
where we have taken the symmetrized product of the classical expression $\dot Q Q$  because  the operators have to be Hermitean. Using the commutation relations \req{CCR} one can check that the  operators  $H,  D ~ {\rm and} ~ K$ do indeed fulfill the algebra of the generators of the  one-dimensional conformal
group ${\it Conf}\!\left(R^1\right)$  as it shown in Appendix \ref{cfqm}:
 \beq 
 [H,D]= i\,H,  \quad [H ,K]=2\, i \, D, \quad [K,D]=- i\,  K .
\enq

The conformal group in one dimension is locally isomorphic to the group $SO(2, 1)$,  the Lorentz  group in 2+1 dimensions.  In fact, by introducing the combinations
 \beq  \label{a}
 L^{0,-1} =  \half\left(  a \, H +\frac{1}{a}  \,K \right) ,\quad L^{1,0}= \half\left( - a \, H +\frac{1}{a}  \,K \right) ,\quad  L^{1,-1} = D,
 \enq
one sees that that the generators $L^{0,-1}$, $L^{1,0}$ and $L^{1,-1}$ satisfy the commutation relations of the algebra of the generators  of the group $SO(2,1)$
\beq \label{SO2}
[L^{0,-1},L^{1,0}]= i\, L^{1,-1}, \quad [L^{0,-1},L^{1,-1}]=- i\, L^{1,0},
\quad [L^{1,0},L^{1,-1}]= - i\, L^{0,-1},
 \enq
where $L^{1,i},\;  i=-1,\,0$ are the boosts in  the space direction 1 and $L^{0,-1}$ the rotation in the $(-1,0)$ plane (See  Sec.~\ref{group-relation}). The rotation operator $L^{0,-1}$ is compact and has thus a discrete spectrum with normalizable eigenfunctions. Since the dimensions of $H$ and $K$ are different, the constant $a$ has the dimension of $t$. In fact, the relation between the generators of the conformal group and the generators of $SO(2, 1)$ suggests that the scale $a$ may play a fundamental role~\cite{deAlfaro:1976je}. This superposition of different invariants of motion, which implies the introduction of  a scale, opens the possibility to construct  a confining semiclassical theory  based on an underlying conformal symmetry.

Generally one can construct a  new ``Hamiltonian'' by any superposition of the three constants of motion
\beq \label{G}
G = u \, H + v\,D + w\,K.
\enq
The new Hamiltonian acts on the state vector, but its evolution involves a new ``time'' variable.
To determine the action of the  generator $G$  \req{G} on the state vector, we consider
 the infinitesimal  transformation properties of the generators  $H$, $D$ and $K$ given by \req{HDKSev} in  Appendix \ref{cfqm}
\beq
e^{ - i \, \ep \, G}  |\psi(t)\rangle = |\psi(t)\rangle +   \ep (u+v  t + w t^2) \frac{d}{dt}|\psi(t)\rangle + O(\ep^2).
\label{HDKev} 
\enq
Thus, we recover the usual quantum mechanical evolution for the state vector
\beq \label{EG}
G \vert \psi(\tau)\rangle  = i \frac{d}{d \tau}  \vert \psi(\tau) \rangle ,
\enq
provided that we introduce a new time variable $\tau$ defined through~\cite{deAlfaro:1976je}
\beq   \label{dtau} 
d\tau= \frac{d t} {u+v\,t + w\,t^2}. 
\enq

Likewise, we can find the evolution of the field $Q(t)$ from the combined action of the generators $H$, $D$ and $K$. Using the equations \req{HDKHev} in Appendix \ref{cfqm} we find 
\beq
i \left[G, Q(t) \right]  =  \frac{d Q(t)}{d t} \left(u + v t + w t^2\right)   -  Q(t) \half \frac{d}{d t} \left(u + v t + w t^2\right)   .
\enq
and thus the Heisenberg equation of motion
\beq  \label{HE}
i \left[G, q(\tau) \right]  = \frac{d q(\tau)}{d \tau},
\enq
where the rescaled field $q(\tau)$ is given by
 \beq \label{qtau}
 q(\tau) = \frac{Q(t)}{(u + v\, t + w \,t^2)^{1/2}}.
 \enq
 From \req{CCR}  it follows   that  the new field also satisfies  the usual quantization condition
\beq
[q(\tau),\dot q(\tau)]= i,
\enq
where $\dot q = dq/d\tau$.

If one expresses the action $A$  \req{AQ} in terms of the transformed fields $q(\tau)$, one finds 
\beqa \label{Aq}
A[Q] &=& \half \int  d \tau \Big( \dot q^2  - \frac{g}{q^2}  - \frac{4 u \omega - v^2}{4} q^2 \Big) + A_{\rm surface} \nn \\
         &=& A[q] + A_{\rm surface},
\enqa
where  $A_{\rm surface}$ is given through \req{AQq}.   Thus, up to a surface term, which does not modify the equations of motion, the action \req{AQ} remains unchanged under the transformations \req{dtau} and \req{qtau}.   However, the Hamiltonian derived from $A[q]$,
\beq \label{Htauq}
G =  \half \Big( \dot q^2 + \frac{g}{q^2} +\frac{4\,uw - v^2}{4} q^2\Big),
\enq
contains the factor $4 u w-v^2$ which breaks the scale invariance. It is a compact operator  for $4 u w - v^2 > 0$.  It is important to notice that the appearance of the generator of special conformal transformations $K$ in \req{G} is essential for confinement. This stresses the importance of the total derivative modifying the Lagrangian under the special conformal transformation (See \req{r1a}).

We use the Schr\"odinger picture from the representation of $q$ and $p = \dot q$:  $q \to y,  ~ \dot q \to - i {d}/{d y}$,
\beq  \label{Sp}
i \frac{\partial}{\partial \tau} \psi(y,\tau) = G \Big(y,   - i \frac{d}{d y} \Big) \psi(y,\tau),
\enq
with the corresponding Hamiltonian
\beq  \label{Hx}
G  = \half \Big(- \frac{d^2}{dy^2} + \frac{g}{y^2} + \frac{4 u \omega - v^2}{4} y^2 \Big),
\enq
in the Schr\"odinger representation~\cite{deAlfaro:1976je}. For $g \geq -1/4$ and $4\, u w -v^2 > 0$ the operator \req{Hx} has a discrete spectrum.  It is remarkable  that it is indeed possible to construct a compact operator  starting from the conformal action  \req{AQ}, without destroying the scale invariance  of the action itself.

We go now back to the original field operator $Q(t)$ in \req{AQ}.  From  \req{qtau} one obtains  the relations 
\beq \label{Qq}
q(0) = \frac{Q(0)}{\sqrt{u}},  \quad \quad \dot q(0)= \sqrt{u}\;  \dot Q(0) - \frac{ v}{2 \sqrt{u}}\,Q(0),
\enq
and thus from \req{Htauq} we obtain
\beqa \label{HtauQ}
G (Q, \dot Q ) \! &= \! & \frac{1}{2} u \left(\dot Q^2 + \frac{g}{Q^2} \right)  - \frac{1}{4} v \left( Q \dot Q + \dot Q Q\right) + \frac{1}{2} w Q^2\\  \nn
            \! &= \! & u H + v D + w K,
            \enqa
at $t=0$.  We thus recover the evolution operator \req{G} which describes, like \req{Htauq}, the evolution in the variable $\tau$ \req{EG},  but expressed in terms of the original field $Q$. The change in the time variable \req{dtau}, required by the conformal invariance of the action, implies the change of the original Hamiltonian \req{HtQ} to \req{HtauQ}.

With the realization of the operator $Q(0)$ in the state space  with wave functions $\psi(x, \tau)$ and the substitution $Q(0) \to x$ and $\dot Q(0) \to - i \frac{d}{d x}$ we obtain
\beq  \label{Sp}
i \frac{\partial}{\partial \tau} \psi(x,\tau) = G \Big(x,   - i \frac{d}{d x} \Big) \psi(x,\tau),
\enq
and  from \req{HtauQ}  the Hamiltonian
\beq \label{Htaux}
G = \frac{1}{2} u \left(- \frac{d^2}{d x^2} + \frac{g}{x^2}\right)  + \frac{i}{4} v  \left(x \, \frac{d}{d x}+ \frac{d}{d x} \, x  \right) +\frac{1}{2}  w x^2.
\enq

The field $q(\tau)$ was only introduced as an intermediate step in order  to recover the evolution equation \req{HE} and the Hamiltonian \req{Htauq} from the action \req{AQ}  with variational methods.  This shows that the essential point  for confinement  and the emergence of a mass gap  is indeed the change from $t$ to $\tau$ as evolution parameter.

\section{Connection to light-front dynamics \label{connectionLFD}}

We can now apply the group theoretical results from the conformal algebra to the front-form ultra-relativistic bound-state wave equation obtained in Chapter \ref{ch2}. Comparing the Hamiltonian \req{Htaux}  with the light-front wave equation \req{LFWE} and identifying the variable $x$ with the light-front invariant variable $\zeta$,  we have to choose $u=2, \; v=0$ and relate the dimensionless constant $g$ to the LF orbital angular momentum, $g=L^2-1/4$,  in order to reproduce the light-front kinematics. Furthermore  $w = 2 \lambda^2$ fixes the confining light-front  potential to a quadratic $\la^2 \, \zeta^2$ dependence.

For the Hamiltonian $G$ \req{Htaux} mapped to the light-front Hamiltonian in \req{LFWE},  {\it i.e.},  $u=2$, $v=0$ and $w =  2 \la^2$ one has
 \beqa \label{HtauLF}
 G
  & \! = \!&  - \frac{d^2}{d x^2} + \frac{g}{x^2} + \la^2 x^2 \nn \\
 & \! = \! & 2 \left( H + \lambda^2 K\right),
 \enqa
 and the relation with the algebra of the group $SO(2,1)$ becomes particularly compelling.  From the relations \req{a}  follows the connection of the free Hamiltonian $H$, \req{HtQ} with the group generators of $SO(2,1)$
\beq \label{free}
L^{0, -1} - L^{1, 0}= a H.
\enq
The Hamiltonian  \req{HtauLF}  can be expressed as a generalization of \req{free} by replacing $L^{0, -1} - L^{1, 0}$ by $ L^{0, -1} - \chi L^{1, 0}$. This generalization yields indeed
\beq \label{conf}
 L^{0, -1} - \chi L^{1, 0}
= \frac{1}{4} {a}(1+\chi) G,
\enq
with
\beq \label{aw}
\la =\frac{1}{a^2} \, \frac{1-\chi}{1+\chi},
\enq
in \req{HtauLF}.   Thus the confining LF Hamiltonian
\beq  \label{HchiLF}
H_{LF} =  - \frac{d^2}{d \ze^2} + \frac{g}{\ze^2}   +  \la \ze^2.
\enq
For $\chi=1$ we recover the free case \req{free}, whereas for $ -1 <\chi < 1$ we obtain a confining LF potential. For $\chi$ outside this region, the Hamiltonian is not bounded from below.

This consideration based on the isomorphism of  the conformal group in one dimension with the group $SO(2,1)$ makes  the appearance of a dimensionful constant in the Hamiltonian \req{HchiLF}, derived from a conformally invariant action, less astonishing. In fact,  as mentioned below Eqs. \req{a} and \req{SO2}, one has to introduce the dimensionful constant $a \ne 0$ in order to relate the generators of the conformal group with those of the group $SO(2,1)$.  This constant $a$  sets the scale for the confinement strength $\la^2$ but does not determine its magnitude, as can be seen from \req{aw}. This value depends on  $a$ as well as on the relative weight of the two generators $L^{0, -1}$ and $L^{1, 0}$ in the construction of the Hamiltonian \req{conf}.

Since the  invariant light-front Hamiltonian is the momentum square operator, $H_{LF} = P^2 = P^+ P^-  - \mbf{P}_\perp^2$ with the LF evolution operator $P^- = \frac{\pa}{\pa x^+}$,   it follows from the identification of $G$ \req{Htaux} with the LF  wave equation \req{LFWE}} and the  Hamiltonian evolution equation \req{LFSE},   that  in the LF frame  $\bf{P}_\perp=0$  the evolution parameter $\tau$ is proportional to the LF time $x^+ = x^0 + x^3$, namely $\tau = x^+ /P^+$, where $P^+= P^0+P^3$ is the hadron longitudinal momentum. Therefore,   the dimension of $\tau$ is that of an inverse mass squared, characteristic of the fully relativistic treatment.

In the original paper of de Alfaro {\it et al.}~\cite{deAlfaro:1976je} and subsequent investigations~\cite{Fubini:1984hf, Chamon:2011xk}, the aim was not so much to obtain a confining  model, but rather to investigate conformal field theories. The use of the compact operator $L^{0,-1}$, constructed inside the algebra of generators of the conformal group, served mainly to obtain a normalizable vacuum state in order to make contact with quantum fields operating in a Fock space. In this sense, the parameters $u, \, v$ and $w$ played only  an auxiliary role. In the context of the present review, one is  however primarily interested in the dynamical evolution in terms  of the new variable $\tau$, which turns out to be proportional to the light front time $x^+ = x^0 + x^3$, therefore the dimensioned parameter $w$ plays a physical role~\footnote{This possibility was also shortly discussed in general terms by dAFF~\cite{deAlfaro:1976je}.}; namely, that of setting the hadronic  scale in the LF potential.

In their discussion of the evolution operator $G$ de Alfaro {\it et. al.} mention a critical point, namely that ``the time evolution is quite different from a stationary one''.  By this statement they refer to the fact that the variable $\tau$ is related to the variable $t$  by 
\beq
 \tau= \frac{1}{\sqrt{  2 w}} \, \arctan \left(t \sqrt{\frac{w}{2}}\right),
\enq
a quantity which is of finite range.  Thus $\tau$ has the natural interpretation in a meson as the difference of light-front times between events involving the quark and antiquark, and in principle could be measured in double parton-scattering processes~\cite{Brodsky:2013ar}.     Thus the finite range of $\tau$ corresponds to the finite size of hadrons due to confinement.

To sum up, the dAFF mechanism for introducing a scale makes use of the algebraic structure of one dimensional conformal field theory.  A new Hamiltonian with a mass scale $\sqrt{\la}$ is constructed from the generators of the conformal group and its  form is  therefore fixed uniquely: it is, like the original Hamiltonian with unbroken dilatation symmetry, a constant of motion~\cite{deAlfaro:1976je}.  The essential point of this procedure is the introduction of a new evolution parameter $\tau$.  The theory defined in terms of the new evolution Hamiltonian $G$ has a well-defined vacuum, but the dAFF procedure breaks Poincar\'e and scale invariance~\cite{Fubini:1984hf}.  The symmetry breaking in this procedure is reminiscent of spontaneous symmetry breaking, however, this is not the case  since, in contrast with current algebra,  there are no degenerate vacua~\cite{Fubini:1984hf} (the vacuum state is chosen ${\it ab \, initio}$) and thus a massless scalar $0^{++}$ state is not required. The dAFF mechanism is also different from  usual explicit breaking by just adding a  mass  term to the Lagrangian~\cite{Kharzeev:2008br}.

\section{Conformal quantum mechanics, $SO(2,1)$ and AdS$_2$ \label{CCMAdS2}}

The local isomorphism between the conformal  group in one-dimension and the  group $SO(2,1)$ is fundamental for introducing the  scale for confinement in the light-front Hamiltonian.   In fact, the conformal group in one dimension $Conf  \! \left(R^1\right)$ is locally isomorphic  not only  to the group $SO(2,1)$, but also to the isometries of AdS$_2$~\footnote{The isomorphism of  the algebra of  generators of the group $SO(2,1)$ and the isometries  of AdS$_2$ space is the basis of the AdS$_2$/CFT$_1$ correspondence~\cite{Chamon:2011xk}.}. 
Using Table \ref{tabiso}, one  can see explicitly the equivalence of the generators of AdS$_2$ isometries at  the AdS$_2$ boundary, $z=0$, with  the representation of the conformal generators $H$, $D$ and $K$ in conformal quantum mechanics given by \req{HDKSev} in Appendix \ref{cfqm}. In the limit $z \to 0$ we have:
\beqa
H &=& \frac{i R}{a}  \frac{\pa}{\pa t}, \\
D &=& i \left(t  \frac{\pa}{\pa t} + z  \frac{\pa}{\pa z} \right) \to  i t  \frac{\pa}{\pa t}, \\
K &=& \frac{i a}{R} \left( \left(t^2+z^2 \right)  \frac{\pa}{\pa t} + 2 z t  \frac{\pa}{\pa z} \right) \to  \frac{i a}{R} \, t^2   \frac{\pa}{\pa t} ,
\enqa
In the equation above   $a$ is the dimensionfull constant required to establish the isomorphism between the generators of the conformal group $Conf\!(R^1)$ and the generators of $SO(2,1)$ given by \req{a}, and $R$ is the AdS$_2$ radius; a priori two independent scales. Comparing with Eq.  \req{HDKSev}) we find that the generators are indeed identical, if the scales are  also identical. Thus $H = i \pa_t$, $D = i t \pa_t$ and  $H = i t^2 \pa_t$ provided that  $a = R$.

In the next chapter, we shall derive bound-state equations for hadronic states from  classical gravity in AdS space, where the confinement arises from distortion of this  higher dimensional space. We will see there that the method discussed in this chapter fixes uniquely the modification of AdS space and not only  the form of the light-front confinement potential.

\chapter{Higher-Spin Wave Equations and AdS Kinematics and Dynamics \label{ch4}}
In this chapter we derive hadronic bound-state wave equations  with arbitrary  spin  in a higher-dimensional space asymptotic to  anti-de Sitter  space.  We give first an introductory derivation of the wave equations for  scalar and vector fields. We then extend our treatment to higher spin using AdS tensors or generalized Rarita-Schwinger spinor fields in AdS for all integer and half-integer spins respectively.  Our procedure takes advantage of the Lorentz frame (the local inertial frame). Further simplification is brought by the fact that physical hadrons  form tensor representations in  3 + 1 dimensions.  In the present approach, the subsidiary conditions required to eliminate the lower-spin states from the symmetric tensors follow from the higher-dimensional Euler-Lagrange equations of motion and are not imposed. It turns out that  the AdS geometry fixes the kinematical features of the theory, whereas the breaking of maximal symmetry  from additional deformations of AdS space determines the dynamical features,  including confinement. It will be shown that a strict separation of the two is essential in light-front  holographic QCD~\footnote{A more detailed discussion of the procedures discussed here is given in Ref. \cite{deTeramond:2013it}.}.

We  briefly review in  Appendix \ref{RG} the relevant elements of Riemannian geometry useful in the discussion of anti-de Sitter space and applications of the 
 gauge/gravity correspondence, and in Appendix \ref{HSWEAdS} we discuss technical details useful for the derivation of integer and half-integer wave equations in holographic QCD.

\section{Scalar  and vector fields \label{scal}}

The derivation of the equation of motion  for a scalar field in AdS is a particularly simple example. As mentioned in the introduction, {in order to describe hadronic states using holographic methods,} one has to break the maximal symmetry of the AdS metric, which is done  by introducing   a scale via a dilaton profile depending explicitly on the holographic variable $z$. For the sake of generality we mostly work in a $(d+1)$-dimensional curved space, and for all direct physical applications  we take $d=4$. In Sec. \ref{warpmetric} we shall also consider the breaking of maximal symmetry by warping the AdS metric.

The coordinates of AdS$_{d+1}$ space are the $d$-dimensional Minkowski  coordinates $x^\mu$ and the holographic variable $z$. The combined coordinates are   labeled $x^M = \left(x^\mu, z = x^d \right)$ with $M, N = 0, \dots , d$ the indices of the higher dimensional $d+1$ curved space, and $\mu, \nu= 0, 1, \dots, d-1$ the Minkowski flat space-time indices. In Poincar\'e coordinates (Sec. \ref{A22}),  the conformal AdS metric is
\beqa 
\label{AdSmetric} 
ds^2 &=&  g_{M N} dx^M dx^N \\ \nn
          &=& \frac{R^2}{z^2}  \left( \eta_{\mu \nu} dx^\mu dx^\nu - dz^2\right).
\enqa
Here $g^{MN}$ is the  full space metric tensor \req{ge} and $\eta_{\mu \nu}= diag[1,-1 \cdots -1]$ the metric tensor of Minkowski space. For a scalar field in AdS space $\Phi(x^0, \cdots x^{d-1}, z)$  the invariant action (up to bilinear terms) is 
\beq \label{SA}
S = \half \int \! d^d x \, dz  \,\sqrt{g} \,e^{\varphi(z)}
  \left( g^{M N} \partial_M \Phi \partial_N \Phi -  \mu^2   \Phi^2 \right)  ,
\enq
where  $g= \left(\frac{R}{z}\right)^{2d+2}$ is the modulus of the determinant of the metric tensor  $g_{MN}$.  At this point, the AdS$_{d+1}$  mass $\mu$ in (\ref{SA}) is not a physical observable and  it is {\it a priori} an arbitrary parameter. The  integration measure $d^d x \, dz \,\sqrt{g}$ is AdS invariant, and the action \req{SA} is written in terms of simple derivatives, since the derivative of a scalar field transforms as a covariant vector field. For $\varphi \equiv 0$ the action is also AdS invariant, but the  dilaton background $\varphi(z)$  effectively breaks the maximal symmetry of AdS (See \ref{A2}). It is a function of the holographic variable $z$ which vanishes in the conformal limit $z \to 0$.  As we will show below, it is crucial that the dilaton profile is only a function of the variable $z$. This  allows the separation  of the overall movement of the hadron from  its internal dynamics. In AdS$_5$, this unique $z$-dependence of the dilaton allows the description of the bound-state dynamics in terms of a one-dimensional wave equation.  It also enable us to to establish a map to the semiclassical one-dimensional approximation to light-front QCD given by the frame-independent light-front Schr\"odinger equation obtained in Chapter \ref{ch2}.

The equations of motion  for the field $\Phi(x,z)$ are obtained from the variational principle
\beq
\frac{\de S}{\de \Phi}= { \frac{1}{ \sqrt{g} \,e^\varphi} }  \partial_M \left(\sqrt{g} \, e^\varphi g^{M
N}\frac{\pa \cL}{\pa(\pa_N \Phi)}\right) -\frac{ \pa \cL}{\pa \Phi}= 0,
  \label{SAdSA}
\enq
from which one obtains the wave equation for the scalar field
\begin{equation} \label{WES}
\left[
 \partial_\mu \partial^\mu
- \frac{ z^{d-1}}{e^{\varphi(z)}}   \partial_z
\left(\frac{e^{\varphi(z)}}{z^{d-1}} \partial_z\right)
+ \frac{{( \mu} R)^2}{z^2}\right] \Phi = 0,
\end{equation}
where $\pa_\mu \pa^\mu \equiv \eta ^{\mu \nu} \pa_\mu \pa_\nu$.

A free hadronic state in holographic QCD  is described by a plane wave in physical space-time and a $z$-dependent profile function:
\beq \label{J0}
\Phi(x, z) = e ^{ i P \cdot x} \,  \Phi_{J=0}(z) ,
\enq 
with invariant hadron mass $P_\mu P^\mu \equiv \eta^{\mu \nu} P_\mu P_\nu = M^2$.  Inserting  \req{J0} into the wave equation (\ref{WES}) we obtain the bound-state
eigenvalue equation  
\beq  \label{WESM}
 \left[ 
   -  \frac{ z^{d-1}}{e^{\varphi(z)}}   \partial_z \left(\frac{e^{\varphi(z)}}{z^{d-1}} \partial_z   \right) 
  +  \frac{(\mu  R )^2}{z^2}  \right]  \Phi_{J=0} = M^2 \Phi_{J=0}.
  \enq
for spin $J=0$ hadronic states.

As a further example, we also derive  the equations of motion for a vector field $\Phi_M(x,z)$. We start  with the generalized Proca action in AdS$_{d+1}$ space
 \beq
S =   \int \! d^d x \, dz  \,\sqrt{g} \,e^{\varphi(z)}
  \left( \frac{1}{4} g^{M R} g^{N S} F_{M N} F_{R S} - \half \, \mu^2 g^{M N} \Phi_M \Phi_N \right),
\label{SV}
 \enq
where $F_{MN} = \partial_M \Phi_N -\partial_N \Phi_M$. The antisymmetric tensor $F_{MN}$ is covariant, since the parallel transporters in the covariant derivatives \req{Dco} cancel. Variation of the action  leads  to  the equation of motion
\beq
\frac{1}{\sqrt{g} \, e^\varphi} \partial_M \left(\sqrt{g} \, e^\varphi g^{M R} g^{N S} F_{R S} \right) +  \mu^2 g^{N R} \Phi_R = 0,
\label{VADS2}
\enq
together with the supplementary condition 
\begin{equation} \label{Lorentz}
\pa_M \left( \sqrt{g} e^\varphi g^{M N} \Phi_N \right) = 0.
\end{equation}

One obtains from \req{VADS2} and the  the condition (\ref{Lorentz}), the system of coupled differential  equations~\footnote{$\Phi_z$ denotes the $d$-th coordinate, $\Phi_z \equiv \Phi_d$.}
\beqa \label{WEV1}
\left[ \partial_\mu \partial^\mu  - \frac{z^{d-1}}{e^{\varphi(z)}}   \partial_z \left(\frac{e^{\varphi(z)}}{z^{d-1}} \partial_z\right)  
  -  \partial_z^2 \varphi + \frac{(\mu R)^2}{z^2} - d + 1\right] \Phi_z &\! = \!& 0 , \\ \nn
\left[ \partial_\mu \partial^\mu - \frac{z^{d-3}}{e^{\varphi(z)}}   \partial_z \left(\frac{e^{\varphi(z)}}{z^{d-3}} \partial_z\right) 
+ \frac{(\mu R)^2}{z^2}\right] \Phi_\nu & \! =  \! & - \frac{2}{z} \partial_\nu \Phi_z .
\enqa
In the conformal limit $\varphi \to 0$ we recover the results given in Ref. \cite{Muck:1998iz}.  In the gauge defined by $\Phi_z = 0$, the equations \req{WEV1} decouple and we find the wave equation~\footnote{Technically we impose the condition $\Phi_z=0$ since physical hadrons have no polarization in the $z$ direction. If $\mu=0$ in the Proca action \req{SV} this can be viewed as a gauge condition.}
\beq
  \label{WEV} \left[
\partial_\mu \partial^\mu-\frac{ z^{d-3}}{e^{\varphi(z)}}
\partial_z \left(\frac{e^{\varphi(z)}}{z^{d-3}} \partial_z\right)
+ \left(\frac{\mu R}{z}\right)^2\right] \Phi_\nu  =  0.
 \enq

A physical spin-1 hadron has physical polarization components   $ \epsilon_\nu({P})$ along the  physical coordinates. We thus write
\beq \label{J1}
\Phi_\nu(x, z) = e ^{ i P \cdot x} \,  \Phi_{J=1}(z) \epsilon_\nu({P}),
\enq 
with invariant mass $P_\mu P^\mu  = M^2$. Substituting \req{J1} in \req{WEV} we find the eigenvalue equation
\beq
  \label{WEVJ1} 
  \left[-\frac{ z^{d-3}}{e^{\varphi(z)}}
\partial_z \left(\frac{e^{\varphi(z)}}{z^{d-3}} \partial_z\right)
+ \left(\frac{\mu R}{z}\right)^2\right] \Phi_{J=1}  =  M^2 \Phi_{J=1},
 \enq
describing a spin-1  hadronic  bound-state.

\section{Arbitrary integer spin \label{arbitrary}}

The description of higher-spin fields in pure AdS space is a complex, but relatively well known problem~\cite{Fronsdal:1978vb, Fradkin:1986qy, Buchbinder:2001bs,  Metsaev:1999ui, Metsaev:2003cu, Metsaev:2011uy, Metsaev:2013kaa, Metsaev:2014iwa} \footnote{The light-front  approach can be used advantageously to describe arbitrary spin fields in AdS. See Refs.~\cite{Metsaev:1999ui, Metsaev:2013kaa}.}. The treatment of higher-spin states in the ``bottom-up'' approach to holographic QCD is an important touchstone for this procedure~\cite{deTeramond:2013it}, but it requires a simplified and well defined framework to extend the  computations to warped spaces \footnote{With warped spaces we denote curved spaces which deviate from AdS spaces either by a dilaton term or by a modified metric.} asymptotic to AdS. For example,  the approach of Ref.~\cite{ deTeramond:2008ht} relies on rescaling the solution of a scalar field $\Phi(z)$, $\Phi(z) \to \Phi_J(z) =  \left(\frac{z}{R}\right)^J \Phi(z)$, thus introducing a spin-dependent factor~\cite{deTeramond:2008ht, deTeramond:2012rt}. The approach of Karch, Katz, Son and Stephanov (KKSS)~\cite{Karch:2006pv}   starts from a gauge-invariant action in a  warped AdS space, and uses the gauge invariance of the model to construct explicitly an effective action in terms  of higher-spin fields with only the physical degrees of freedom. However, this approach is not applicable to pseudoscalar particles and their trajectories, and their angular excitations do not lead to a relation with light-front quantized QCD, which is the main subject of this report. The treatment described in this report relies on the approach of Ref. \cite{deTeramond:2013it}, which starts from a manifestly covariant action for higher spin states in a warped space asymptotic to AdS.

Fields with integer spin  $J$ are represented  by a totally symmetric rank-$J$ tensor field {  $\Phi_{N_1  \dots N_J}$.}  Such a symmetric tensor also contains  lower spins, which have to  be eliminated by imposing  subsidiary conditions, as will be discussed below. The action for a spin-$J$ field in AdS$_{d+1}$ space in the presence of a dilaton background field $\varphi(z)$  is given by
\begin{multline}
\label{action1}
 S = \int d^{d} x \,dz \,\sqrt{g}  \; e^{\vp(z)} \,g^{N_1 N_1'} \cdots  g^{N_J N_J'}   \Big(  g^{M M'} D_M \Phi^*_{N_1 \dots N_J}\, D_{M'} \Phi_{N_1 ' \dots N_J'}  \\
 - \mu^2  \, \Phi^*_{N_1 \dots N_J} \, \Phi_{N_1 ' \dots N_J'} + \cdots \Big),
 \end{multline}
where  $\sqrt{g} = (R/z)^{d+1}$ and $D_M$ is the covariant derivative which includes the affine connection (Sec. \ref{A11}) and   $\mu$ is the AdS mass. The omitted terms  in the action indicated by $\cdots$,  refer to  additional  terms with different contractions of the indices.

Inserting the covariant derivatives in the  action leads to a rather complicated expression. Furthermore the additional terms from different contractions in (\ref{action1}) bring an enormous complexity.  A considerable simplification in (\ref{action1}) is due to the fact that one has  only to consider the  subspace of tensors   $\Phi_{\nu_1 \nu_2 \cdots \nu_J}$ which has no  indices along the $z$-direction. In fact,  a physical hadron has  polarization indices along the $3+1$ physical coordinates, $\Phi_{\nu_1 \nu_2 \cdots \nu_J}$, all other components  must vanish identically 
\beq\label{no}
\Phi_{z N_2 \cdots N_J}  = 0.
\enq   
As will be  seen later, the constraints imposed by the mapping of the AdS equations of motion to the LF Hamiltonian in physical space-time for the hadronic bound-state system at fixed LF time, will give further insights  in the description of higher spin states, since it allows an explicit distinction between kinematical and dynamical aspects.

As a practical procedure, one starts from an effective action, which includes   a $z$-dependent effective AdS mass $\mu_{\it
eff}(z)$~\cite{deTeramond:2013it}
\begin{multline}
\label{action2}
S_{\it eff} = \int d^{d} x \,dz \,\sqrt{g}  \; e^{\vp(z)} \,g^{N_1 N_1'} \cdots  g^{N_J N_J'}   \Big(  g^{M M'} D_M \Phi^*_{N_1 \dots N_J}\, D_{M'} \Phi_{N_1 ' \dots N_J'}  \\
 - \mu_{\it eff}^2(z)  \, \Phi^*_{N_1 \dots N_J} \, \Phi_{N_1 ' \dots N_J'} \Big).
 \end{multline}
Again, for $\vp\equiv 0$ and a constant mass term $\mu$,  the action is AdS invariant. The  function $\mu_{\it eff}(z)$, which can absorb  the contribution from different contractions in (\ref{action1}), is {\it a priori} unknown.  But, as  shall   be shown  below, the additional symmetry breaking due to the $z$-dependence of the effective AdS mass allows a clear separation of kinematical and dynamical effects. In fact, its $z$-dependence can be determined either by the precise mapping of AdS to light-front physics, or by eliminating interference terms between kinematical and dynamical effects~\cite{deTeramond:2013it}. The agreement between the two methods shows how the light-front mapping and the explicit separation of kinematical and dynamical effects are intertwined.

The equations of motion are  obtained from the Euler-Lagrange equations in the subspace defined by (\ref{no})
 \beq \label{ELJ}
 \frac{ \de S_{\it eff}}{ \de \Phi^*_{\nu_1 \nu_2 \cdots \nu_J}} = 0,
\enq
 and
 \beq \label{ELz}
  \frac{\de S_{\it eff}}{\de \Ph^*_{ z N_2 \cdots N_J}} = 0.
 \enq
The wave equations for hadronic modes follow from the Euler-Lagrange equation \req{ELJ}, whereas \req{ELz} will yield the kinematical constraints required to eliminate the lower-spin states.

The appearance of covariant derivatives in the action for higher spin fields, \req{action1} and \req{action2}, leads to multiple sums and rather complicated expressions. As shown  in \cite{deTeramond:2013it}  (See also Appendix  \ref{HSWEAdS}), these expressions simplify considerably if one does not use generally covariant tensors but goes intermediately  to a local inertial frame with Lorentz  (tangent) indices. The final expression for the equation of motion for AdS fields with all polarizations in the physical directions is derived from \req{ELJ} and one obtains   (Sec. \ref{inertial})
\beq
\label{PhiJ}
 \left[  \pa_\mu \pa^\mu
   -  \frac{ z^{d-1 - 2J}}{e^{\vp(z)}}   \partial_z \left(\frac{e^{\varphi(z)}}{z^{d-1 - 2 J}} \partial_z   \right)
  +  \frac{(mR)^2 }{z^2}  \right]   \Phi_{\nu_1 \dots \nu_J} = 0,
  \enq
with
  \beq \label{muphi}
  (m\, R)^2 =(\mu_{\it eff}(z) R)^2  - J z \, \vp'(z) + J(d - J +1) ,
  \enq
  which agrees with the result found in Refs.~\cite{deTeramond:2008ht, deTeramond:2012rt} by rescaling the wave equation for a scalar field. As will be shown in Chapter \ref{ch5} mapping to the light front implies that the quantity $m$ is independent of the variable $z$ (See Eq. \req{muRJL}).

Terms in the action which are linear in tensor fields, with one or more indices along the holographic direction, $\Phi_{z N_2 \cdots N_J}$ (See Appendix  \ref{HSWEAdS}), yield from (\ref{ELz}) the results~\cite{deTeramond:2013it}
 \beq 
 \label{scPhi} \eta^{\mu \nu} \pa_\mu  \Phi_{\nu \nu_2 \cdots
\nu_J}=0, \quad \eta^{\mu \nu}   \Phi_{ \mu \nu \nu_3  \cdots
\nu_J}=0. 
\enq 
These are just the kinematical constraints required to eliminate the states with spin lower than $J$ from the symmetric tensors $ \Phi_{\nu_1 \nu_2 \cdots \nu_J}$.

The conditions \req{scPhi} are independent of the conformal symmetry breaking terms $\vp(z)$ and $\mu(z)$ in the effective action \req{action2}; they are a consequence of the purely kinematical aspects encoded in the AdS metric. It is remarkable that, although one  has started in AdS space with unconstrained symmetric spinors, the non-trivial affine connection of AdS geometry gives precisely the subsidiary conditions needed to eliminate the lower spin states $J-1, \,J-2, \cdots$ from the fully symmetric  AdS  tensor field $\Phi_{\nu_1 \dots \nu_J}$.

In order to make contact with the LF Hamiltonian, one considers hadronic states with momentum $P$ and a  $z$-independent spinor $\ep_{\nu_1 \cdots \nu_J}(P)$. In holographic QCD such a state is described by a $z$-dependent wave function and a plane wave propagating in physical space-time  representing a free hadron
\beq \label{scalarcov} \Phi_{\nu_1 \cdots \nu_J}(x, z) = e ^{ i P
\cdot x} \,  \Phi_J(z) \,  \ep_{\nu_1 \cdots \nu_J}({P}), \enq
with invariant hadron mass $P_\mu P^\mu \equiv \eta^{\mu \nu} P_\mu P_\nu = M^2$.  Inserting  (\ref{scalarcov}) into the wave equation (\ref{PhiJ}) one  obtains the bound-state eigenvalue equation \beq  \label{PhiJM}
 \left[
   -  \frac{ z^{d-1- 2J}}{e^{\varphi(z)}}   \partial_z \left(\frac{e^{\varphi(z)}}{z^{d-1-2J}} \partial_z   \right)
  +  \frac{(m\,R )^2}{z^2}  \right]  \Phi_J(z) = M^2 \Phi_J(z),
  \enq
  where the  normalizable solution of (\ref{PhiJM}) is normalized according to
\beq  \label{Phinorm} R^{d - 1 - 2 J} \int_0^{\infty} \!
\frac{dz}{z^{d -1 - 2 J}} \, e^{\varphi(z)} \Phi_J^2 (z)  =1. \enq
 One  also recovers from (\ref{scPhi}) and (\ref{scalarcov}) the { usual} kinematical constraints
\beq \label{sub-spin}
 \eta^{\mu \nu } P_\mu \,\ep_{\nu \nu_2 \cdots \nu_J}=0, \quad
\eta^{\mu \nu } \,\ep_{\mu \nu \nu_3  \cdots \nu_J}=0.
 \enq

One sees that the  wave equation \req{WES} and \req{WEV} for the scalar and vector field, respectively, are special cases of the  equation for general spin \req{PhiJ} with the mass \req{muphi}. In the case of a scalar field, the covariant derivative is  the usual partial derivative, and there are no additional contractions in the action;  thus $\mu_{\it eff} = \mu = m$ is a constant. For a spin-1 wave equation, there is one additional term from the antisymmetric contraction, and the contribution from the parallel transport cancels out.  It is also simple in this case to determine the effective mass $\mu_{\it eff}$ in (\ref{action2}) by the comparison with the full expression for the action of a vector field which includes the antisymmetric contraction   (See Eq. \req{WEV}).  Thus for spin-1, one  has $\mu = m$ and  $(\mu_{\it eff}(z) R)^2 = ( \mu R)^2   +  z \, \vp'(z) - d$.

In general, the AdS mass $m$ in the wave equation (\ref{PhiJ})  or (\ref{PhiJM}) is determined from the mapping  to the light-front Hamiltonian (Chaper \ref{ch5}).  Since  $m$ maps  to the Casimir operator of the orbital angular momentum in the light front (a kinematical quantity) it follows that $m$  should be a constant. Consequently, the $z$-dependence of the effective mass (\ref{muphi}) in the AdS action (\ref{action2}) is determined {\it a posteriori} by kinematical constraints in the light front, namely  by the requirement that the mass $m$ in \req{PhiJ}  or (\ref{PhiJM}) must be a constant. 

 The relation \req{muphi} can also be derived independently {\it a priori}, if one demands that  the kinematical effects from AdS  and the dynamical  effects due to the breaking of maximal symmetry are clearly separated in the equations of motion~\cite{deTeramond:2013it}. In general, the presence of a dilaton in the effective action \req{action2} and the quadratic appearance of covariant derivatives lead  to a mixture of kinematical and dynamical effects. However,  by  choosing the appropriate $z$ dependence of the effective mass $\mu_{\rm eff}(z)$ the interference terms cancel. This requirement determines  $\mu_{\rm eff}(z)$ completely and one recovers (\ref{muphi}).

In the case where the maximal symmetry of AdS is not broken by a dilaton,  $\vp(z)=0$, no $z$-dependent AdS mass is necessary, and one can start with a constant mass term  in \req{action2}. This is the case in the hard-wall model, where   dynamical effects are introduced by  the boundary conditions and indeed no mixing between kinematical and dynamical aspects does occur.

\subsection{Confining interaction and warped metrics \label{warpmetric}}

As an alternative to introducing  a dilaton term into the action \req{action2} in order  to break the maximal symmetry  of AdS space, the approach presented in this section allows one to equivalently modify the AdS metric by a  $J$-independent warp factor
 \beq 
 \label{gw} \tilde g_{MN}= e^{2 \tilde \vp(z)} g_{MN} ,
 \enq 
 where $g_{MN}$ is the metric tensor of AdS in Poincar\'e coordinates (Sec. \ref{A22}). The effective action is then given by
\begin{multline}
\label{action3} \tilde S_{\it eff} = \int d^{d} x \,dz
\,\sqrt{\tilde g}
\, \tilde g^{N_1 N_1'} \cdots  \tilde g^{N_J N_J'}   \Big(  \tilde g^{M M'} D_M \Phi^*_{N_1 \dots N_J}\, D_{M'} \Phi_{N_1 ' \dots N_J'}  \\
 - \tilde \mu_{\it eff}^2(z)  \, \Phi^*_{N_1 \dots N_J} \, \Phi_{N_1 ' \dots N_J'} \Big),
 \end{multline}
where  $\sqrt{\tilde g} = (R\,  e^{\tilde \vp(z)}/z)^{d+1}$ and the effective mass $\tilde \mu_{\it eff}(z)$ is again an  {\it a priori} unknown function.

The use of warped metrics is useful for visualizing the overall confinement behavior by  following  an object in warped AdS space as it falls to the infrared region by the effects of gravity. The gravitational potential energy for an object of mass $M$ in general relativity is given in terms of  the time-time component of the metric tensor $g_{00}$
\begin{equation} \label{V}
V = M c^2 \sqrt{\tilde g_{00}} = M c^2 R \, \frac{e^{ \tilde \vp(z)}}{z} ;
\end{equation}
thus, one may expect a potential that has a minimum at the hadronic scale $z_0 \sim 1/ \Lambda_{\rm QCD}$ and grows fast for larger values of $z$ to confine effectively a particle in a hadron within distances $z \sim z_0$. In fact, according to Sonnenschein~\cite{Sonnenschein:2000qm}, a background dual to a confining theory should satisfy the conditions for the metric component $g_{00}$
\begin{equation}\label{Scond}
\partial_z (g_{00}) \vert_{z=z_0} = 0 , ~~~~ g_{00} \vert_{z = z_0} \ne 0,
\end{equation}
to display the Wilson loop-area law for confinement of strings.

\begin{figure}[h]
\begin{center}
\includegraphics[width=7.2cm]{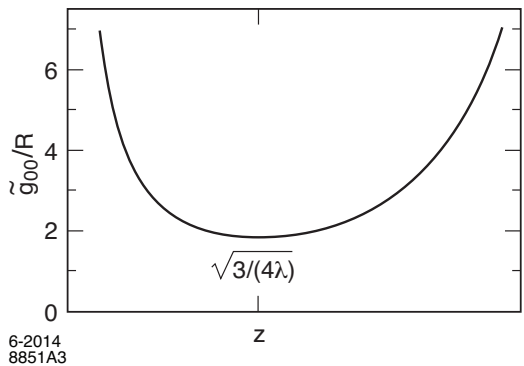}
\caption{\label{adspot} \small $\tilde g_{00}(z)/ R$ for a positive dilaton profile   $\tilde \vp(z) = \la z^2 / 3$.}
\end{center}
\end{figure}

The action with a warped metric (\ref{action3}) and the effective action with a dilaton field (\ref{action2}) lead to identical results for the equations of motion for arbitrary spin
(\ref{PhiJM}), provided that one identifies the metric warp factor $\tilde \vp(z)$ in (\ref{gw}) and the effective mass  $\tilde \mu_{\it eff}(z)$ with the dilaton profile $\vp(z)$ and the mass
$m$ in \req{PhiJM} according to
 \beqa \label{mueffw}
\tilde \vp(z) &=& \frac{\vp(z)}{d-1},\\
 \left(\tilde
\mu_{\it eff}(z) R\right)^2 &=&  \left( m^2 + J z \frac{\tilde \vp
'(z)}{d-1} - J z^2 \tilde \Omega^2(z) - J(d-J)  \right) e^{-2
\tilde \vp(z)}, \enqa
where $\tilde \Om(z)$ is the warp factor of the affine connection for the metric (\ref{gw}), $\tilde \Om(z) = 1/z - \pa_z \tilde \vp$ (For more details see Ref. \cite{deTeramond:2013it} and Sec. \ref{warp}).  As an example, we show in Fig. \ref{adspot}  the metric component $\tilde g_{00}/R$ for the positive profile $  \tilde \vp(z) =  \la \, z^2/3 $.  The corresponding potential \req{V} satisfies Eq. \req{Scond} which leads to the Wilson loop-area condition for confinement. This type of solution is also expected from simple arguments based on stability considerations, since the potential energy should display a deep minimum as a function of the holographic variable $z$~\cite{Klebanov:2009zz}.

\subsection{Higher spin in a gauge invariant AdS model \label{KKSS}}

 In their seminal paper~\cite{Karch:2006pv}  Karch {\it et. al.}, introduced the soft wall model for the treatment of higher spin states in AdS/QCD. They also started from the  covariant  action \req{action1}, in $d=4$ dimensions, but in addition demanded that it is invariant under gauge transformations in AdS$_5$
 \beqa \label{gaugeKKSS}  
\Phi_{N_1 \cdots N_J} & \to & \Phi_{N_1 \cdots N_J} +\de\Phi_{N_1 \cdots N_J} , \\
\de\Phi_{N_1 \cdots N_J} &=& D_{ N_1}\, \xi_{N_2 \cdots N_J }+  D_{ N_2}\, \xi_{N_1 \cdots N_J } + \cdots \nn , 
\enqa
where $\xi_{N_2 \cdots N_J }$ is a symmetric tensor of rank $J-1$. In order to achieve gauge invariance, the genuine AdS mass $\mu$ in  \req{action1} must be $z$-dependent in order to cancel terms   in the gauge-transformed action, arising from the affine connection in the covariant derivatives $D_N$.

The assumed gauge invariance of the  equations of motion  allows  one  to simplify the action \req{action1} considerably: one first chooses a gauge in which
 $\Phi_{z \cdots } =0$.
This choice does not fix the gauge uniquely, but there still exist nontrivial  tensors $\tilde \xi_{N_2 \cdots N_J }$ which leave this condition invariant, {\it i.e.},  for which
\beq
 \de \Phi_{z,\mu_2 \cdots \mu_J} =D_z \tilde \xi_{N_2 \cdots N_J } + \dots =0.
\enq
One can show that invariance of the action under these gauge transformations demands for the rescaled fields
\beq
\tilde \Phi_{\mu_1 \cdots \mu_J} =  \left(\frac{z}{R}\right)^{2J - 2}\Phi_{\mu_1 \cdots \mu_J},
 \label{KKSS1}
 \enq
and the following form of the action~\cite{Karch:2006pv}  
\beq
S_{\rm KKSS} = \half \int d^4x \, dz \, e^{\la z^2} \left(\frac{R}{z}\right)^{2 J-1} \eta^{N N'} \eta^{\mu_1 \mu'_1}
\cdots \eta^{\mu_J \mu'_J} \pa_N \tilde \Phi_{\mu_1 \cdots \mu_J} \,  \pa_{N'} \tilde \Phi_{\mu'_1 \cdots \mu'_J}.
\enq

Applying the variational principle \req{ELJ} and introducing $z$-independent spinors, such as the spinors in \req{scalarcov}, one obtains the { KKSS} equation of motion 
\beq \label{PhiKKSS}
- \frac{z^{2J-1}}{e^{\la z^2}}\pa_z\, \left(\frac{e^{\la z^2}}{ z^{2 J -1}} \tilde \Phi_J (z)\right)= M^2 \tilde \Phi_J(z).
\enq
The comparison of \req{PhiKKSS} with \req{PhiJM} for $\vp(z)=\la z^2$ and $d=4$ shows that the structure is the same, but  $m=0$ and the $J$ dependence of the exponents of $z$ are completely different.  The phenomenological consequences of the differences, especially concerning the sign of $\la $ will be discussed in the next Chapter. In contrast with the method presented  in Sec. \ref{warpmetric}, the dilaton cannot be absorbed into an  additional  warp factor in the AdS metric in the treatment  based on gauge invariance~\cite{Karch:2006pv}  discussed in this section.

\section{Arbitrary half-integer spin \label{half}}

Fields with half-integer spin, $J = T + \half$, are conveniently described by Rarita-Schwinger spinors~\cite{Rarita:1941mf}, $\left[ \Psi_{N_1 \cdots N_T}\right]_\al$,  objects which
transform as symmetric tensors of rank $T$ with indices $N_1 \dots N_T$, and as  Dirac spinors with index $\alpha$. The Lagrangian of fields with arbitrary half-integer spin in a higher-dimensional space is  in general more complicated than the integer-spin case.  General covariance allows for a superposition of terms  of the form
$$\bar \Psi_{N_1 \dots N_T} \Ga^{[N_1 \dots N_T M N_1' \dots N_T']} D_M  \Psi_{N'_1\ \dots N_T'}, $$
and mass terms
$$\mu \bar \Psi_{N_1 \dots N_T} \Ga^{[N_1 \dots N_T N_1' \dots N_T']}   \Psi_{N_1 \dots N_T'}, $$
where the tensors $\Ga^{[\cdots]}$ are antisymmetric products of Dirac  matrices and a sum over spinor indices is  understood. The maximum number of independent Dirac matrices depends on the dimensionality of space.  As a specific example, we present in Sec.~\ref{threehalf} the case of spin $\frac{3}{2}$.

In flat space-time, the equations describing a free particle with spin $T+\frac{1}{2}$ are~\cite{Rarita:1941mf} \beq \label{RS}
 \left(i \ga^\mu \pa_\mu - M\right) \Psi_{\nu_1 \cdots \nu_T} =0,  \qquad {\ga^\nu} \Psi_{\nu \nu_2 \cdots \nu_T}=0.
  \enq
 Because of the symmetry of the tensor indices of  the spinor $\Psi_{\nu_1 \cdots \nu_T}$ and the anti-commutation relation $\ga^\mu \ga^\nu +\ga^\nu \ga^\mu = 2 \eta^{\mu \nu}$, the relations in \req{RS} imply the   subsidiary conditions of the integral spin theory for the $T$ tensor indices  \req{scPhi}
 \beq \label{scPsi}
 \eta^{\mu \nu} \pa_\mu  \Psi_{\nu \nu_2 \cdots \nu_T}=0, \quad
\eta^{\mu \nu}   \Psi_{ \mu \nu \nu_3  \cdots \nu_T}=0. 
\enq

The actual form of the Dirac equation for Rarita-Schwinger spinors  (\ref{RS}) in flat space-time motivates  us to start with a simple  effective action for arbitrary
half-integer spin in AdS space which, in the absence of dynamical terms, preserves maximal symmetry of AdS  in order to describe the correct kinematics  and constraints in physical space-time. 

We start with the effective action in AdS$_{d+1}$
 \begin{multline} \label{af}
~~~  S_{F \it  eff} = \half  \int  \!
d^{d} x \,dz\,  \sqrt{g} \, e^{\vp(z)}  g^{N_1\,N_1'} \cdots g^{N_T\,N_T'}  \\
\left[ \bar  \Psi_{N_1 \cdots N_T}  \Big( i \, \Ga^A\, e^M_A\, D_M
-  \mu - \rho(z)\Big)
 \Psi_{N_1' \cdots N_T'} + h.c. \right] , ~~~~~
 \end{multline} 
including a dilaton term $\vp(z)$ and an effective interaction $\rho(z)$ (See also Refs. \cite{Abidin:2009hr} and \cite{Gutsche:2011vb}). In \req{af} $\sqrt{g} = \left(\frac{R}{z}\right)^{d+1}$ and $e^M_A$ is the inverse vielbein, $e^M_A = \left(\frac{z}{R}\right)
\delta^M_A$. The covariant derivative $D_M$ of a Rarita-Schwinger spinor  includes the affine connection and the spin connection (Sec. \ref{A13}) and the tangent-space Dirac matrices obey the usual anticommutation relation $\left\{\Gamma^A, \Gamma^B\right\} = 2 \eta^{A B}$. For $\vp(z) = \rho(z) = 0$ the effective action  (\ref{af}) preserves the maximal symmetry of AdS space. The reason why one has to introduce  an additional  symmetry breaking term $\rho(z)$  in \req{af} will become clear soon. Similarly to the integer-spin case, where the subsidiary conditions  follow from the simple AdS effective action \req{action2}, it will be shown below that the action  \req{af} indeed implies  that the Rarita-Schwinger condition given in \req{RS}, and thus the subsidiary conditions \req{scPsi}, follow from the non-trivial kinematics of the higher  dimensional gravity theory~\cite{deTeramond:2013it}.

Since only  physical polarizations orthogonal to the holographic dimension are relevant for fields corresponding to hadron  bound states we put  
 \beq 
 \Psi_{z N_2 \dots N_T}=0 \label{orthf}.
 \enq

The equations of motion are  derived in a similar way from the effective action \req{af} as in the case of integer spin \cite{deTeramond:2013it}. Since the covariant derivatives occur only linearly, the expressions are considerably simpler~\cite{deTeramond:2013it}. The equations of motion follow from the variation of the effective action 
 \beq \label{ELJf}
 \frac{ \de S_{F  \it eff}}{ \de \bar \Psi_{\nu_1 \nu_2 \cdots \nu_J}} = 0,
\enq
 and
 \beq \label{ELzf}
  \frac{\de S_{F \it eff}}{\de \bar \Psi_{ z N_2 \cdots N_J}} = 0.
 \enq
One then obtains the AdS Dirac-like wave equation (Appendix \ref{HSWEAdS})
 \beq \label{DEz} \left[ i \left( z \eta^{M N}
\Gamma_M \partial_N + \frac{d - z \vp'-2 T}{2} \Gamma_z \right) -
\mu R - R \, \rho(z)\right] \Psi_{\nu_1 \dots \nu_T}=0,
\enq 
and  the Rarita-Schwinger condition in physical space-time \req{RS}
 \beq \label{dirac-SE1}
 \gamma^\nu \Psi_{\nu  \nu_2 \, \dots \,\nu_T} =0. 
 \enq

Although the dilaton term $\vp'(z)$ shows up in the equation of motion  \req{DEz}, it actually does not lead to dynamical effects, since it  can be absorbed by rescaling  the Rarita-Schwinger spinor according to $ { \Psi}_{\nu_1 \dots \nu_T} \to e^{\vp(z)/2}
\Psi_{\nu_1 \dots \nu_T} $. Thus, from \req{DEz} one  obtains
\beq
\label{DE2} \left[ i \left( z \eta^{M N} \Gamma_M
\partial_N + \frac{d -2 T}{2} \Gamma_z \right) - \mu R - R \,
\rho(z)\right]    \Psi_{\nu_1 \dots \nu_T}=0.
 \enq
Therefore, for fermion fields in AdS a dilaton term has no dynamical effects on the spectrum since it can be rotated away~\cite{Kirsch:2006he}. This is a consequence of the linear covariant derivatives in the fermion action, which also prevents a mixing between dynamical and kinematical effects, and thus, in contrast to the effective action for integer spin fields \req{action2}, the AdS mass $\mu$  in Eq. \req{af} is constant. As a result, one must introduce an effective confining interaction $\rho(z)$ in the fermion action to break conformal symmetry and generate a baryon spectrum~\cite{Brodsky:2008pg, Abidin:2009hr}. This interaction can be constrained by the condition that the `square' of the Dirac equation leads to a potential which matches the dilaton-induced potential for integer spin.

The Rarita-Schwinger condition  \req{dirac-SE1} in flat four-dimensional space also entails, with the extended Dirac equation \req{DE2},  the subsidiary conditions for the tensor indices required to eliminate the lower spins~\cite{deTeramond:2013it}. The results from the effective action \req{af} for spin-$\frac{3}{2}$  are in agreement with the results from Refs. \cite{Volovich:1998tj} and \cite{Matlock:1999fy} (Sec.  \ref{threehalf}).

\subsubsection{Warped metric}

 Identical results for the equations of motion for arbitrary half-integer spin are obtained if one starts with the  modified metric \req{gw}.  One finds that the effective fermion action with a dilaton field \req{af} is equivalent to the  corresponding fermion action with modified AdS metrics, provided that one identifies the dilaton profile according to $\tilde \vp(z) = \vp(z)/d$ and the effective mass  $\tilde \mu(z)$ with  the mass $\mu$ in \req{af} according to $\tilde \mu(z) = e^{- \tilde \vp(z)} \mu$. Thus, one cannot introduce confinement in the effective AdS action  for fermions either by a dilaton profile or by additional warping of the AdS metrics in the infrared. In  both cases one requires  an additional effective interaction, as introduced in the effective action \req{af}, with $\rho(z) \ne 0$.

\chapter{Light-Front Holographic Mapping and Hadronic Spectrum \label{ch5}}
The study of the  internal structure and excitation spectrum of mesons and baryons is one of the most challenging aspects of hadronic physics. In fact, an important goal of computations in lattice QCD is the reliable extraction of the excited hadron mass spectrum. Lattice calculations of the ground state of hadron masses agree well with experimental values~\cite{Dudek:2011zz}. However, the excitation spectrum of the light hadrons, and particularly nucleons,  represent a formidable challenge to lattice QCD due to the enormous computational complexity required for the extraction of meaningful data beyond the leading ground state configuration~\cite{Edwards:2011jj}. In addition to the presence of continuous thresholds, a large basis of interpolating operators is required  since excited hadronic states are classified according to  irreducible representations of the lattice, not the total angular momentum. Furthermore, it is not simple to distinguish the different radial excitations by following the propagation of modes in the euclidean lattice because of the identical short distance behavior of radial states.  In contrast, the semiclassical light-front holographic wave equation (\ref{LFWE})  describes relativistic bound states  at equal light-front time with an analytic simplicity comparable to the Schr\"odinger equation of atomic physics at equal instant time, where the excitation  spectrum  follows from the solution of an eigenvalue equation. Also, it is simple to identify the radial excitations in the spectrum as they corresponds to the number of nodes in the eigenfunctions.

The structure of the QCD light-front Hamiltonian   (\ref{HLF}) for the hadron bound state  $\vert \psi(P) \rangle$ formulated at equal light-front time is similar to the structure of the wave equation \req{PhiJM} for the $J$-mode $\Phi_{\mu_1 \cdots \mu_J}$ in AdS space; they are both frame independent and have identical eigenvalues $M^2$, the mass spectrum of the color-singlet states of QCD. This provides the basis for a profound connection between physical QCD formulated in the light front  and the physics of hadronic modes in AdS space. However, important differences are also apparent:  Eq. (\ref{HLF}) is a linear quantum-mechanical equation of states in Hilbert space, whereas Eq. (\ref{PhiJM}) is a classical gravity equation; its solutions describe spin-$J$ modes propagating in a higher-dimensional warped space.  In order to establish a connection between both approaches,  it is important to realize that physical hadrons are inexorably endowed with orbital angular momentum since they are composite. Thus, the identification of orbital angular momentum is of primary interest in making such a connection. As we discuss below, this identification follows from the precise mapping between the one-dimensional semiclassical approximation to light-front dynamics found in Chapter \ref{ch2} and the equations of motion of hadronic spin modes described in Chapter \ref{ch4}.  Furthermore, if one imposes the requirement that the action of the corresponding one-dimensional effective theory  remains conformal invariant  (See Chapter \ref{ch3}), this fixes uniquely the form of the effective potential; and, as we will show below,  the dilaton profile  to have the specific form $\varphi(z) = \lambda z^2$.  This remarkable result follows from the dAFF construction of conformally invariant quantum mechanics~\cite{deAlfaro:1976je, Brodsky:2013ar} and the mapping of AdS to light-front physics.  The resultant effective theory possess an $SO(2,1)$ algebraic structure, which, as we shall discuss in the present chapter, encodes fundamental dynamical aspects of confinement and reproduces quite well the systematics of the light-hadron excitation spectrum.

In the usual AdS/CFT correspondence the baryon is an $SU(N_C)$ singlet bound state of $N_C$ quarks in the large-$N_C$ limit. Since  there are no quarks in this theory, quarks are introduced as external sources at the AdS asymptotic boundary~\cite{Gross:1998gk, Witten:1998xy}.  The baryon is constructed  as an $N_C$ baryon vertex located in the interior of AdS. In this top-down string approach baryons are usually described as solitons or Skyrmion-like objects~\cite{Nawa:2006gv, Hong:2007kx, Hata:2007mb}.   In contrast, the light-front holographic approach  is based on the precise mapping of AdS expressions to the light front in physical space-time.  Consequently, we will describe in this review physical baryons corresponding to $N_C=3$ not $N_C \to  \infty$.   In fact,  the light-front approach  to  relativistic bound-state dynamics corresponds to  strongly correlated multiple-particle states in the Fock expansion, and we may expect that the large number of degrees of freedom, required to have a valid description in terms of a semiclassical gravity theory, would correspond to the very large number of components in the large $n$-Fock expansion~\cite{Brodsky:2013ar}.   The enormous complexity arising as a result of the strong interaction dynamics of an infinite number of components and Fock states is encoded in the effective potential $U$.   To a first semiclassical approximation, this confining potential  is determined by the underlying conformal symmetry of the classical QCD Lagrangian incorporated in the effective one-dimensional effective theory.  As it turns out, the analytical exploration of the baryon spectrum using gauge/gravity duality ideas is not as simple, or as well understood yet, as the meson case, and further work beyond the scope of the present review is required.  However, as we shall discuss below,   even a relatively simple approach provides a framework for a useful analytical exploration of the strongly-coupled dynamics of baryons  which gives important insights into the systematics of the light-baryon spectrum using simple analytical methods.

\section{Integer spin}

An essential step is the mapping of the equation of motion describing a hadronic mode in a warped AdS space to the light front.  To this end, we compare the relativistic one-dimensional light-front wave equation  \req{LFWE} with the spin-$J$ wave equation in AdS \req{PhiJM}, and factor out the scale  $(1/z)^{ J - (d-1)/2}$ and  dilaton factors from the AdS field  as follows
\begin{equation} \label{Phiphi}
\Phi_J(z)   = \left(\frac{R}{z} \right)^{ J - (d-1)/2 } e^{- \varphi(z)/2} \, \phi_J(z) .
\end{equation}
Upon the substitution of the holographic variable $z$ by the light-front invariant variable $\zeta$ and replacing \req{Phiphi}
into the AdS eigenvalue equation  (\ref{PhiJM}), we find for $d=4$ the QCD light-front frame-independent wave equation (\ref{LFWE})
\begin{equation} \label{LFWEbis}
\left(-\frac{d^2}{d\zeta^2}
- \frac{1 - 4L^2}{4\zeta^2} + U(\zeta) \right)
\phi(\zeta) = M^2 \phi(\zeta),
\end{equation}
with the effective potential  in the front form of dynamics~\cite{deTeramond:2010ge}
\begin{equation} \label{U}
U(\zeta, J) = \half \varphi''(\zeta) +\frac{1}{4} \varphi'(\zeta)^2  + \frac{2J - 3}{2 \zeta} \varphi'(\zeta) .
\end{equation}
The  AdS mass $m$ in (\ref{PhiJM}) is related to the light-front internal orbital angular momentum $L$  and the total angular momentum $J$ of the hadron according to
\begin{equation} \label{muRJL}
(m R)^2 = - (2-J)^2 + L^2,
\end{equation}
where the critical value  $L=0$  corresponds to the lowest possible stable solution. Light-front holographic mapping thus implies that the  AdS mass  $m$ in (\ref{PhiJM})  is not a free parameter  but scales according to (\ref{muRJL}), thus giving a precise expression for the AdS effective mass $\mu_{\it eff}(z)$ in the AdS effective action \req{action2}. For $J = 0$ the five dimensional AdS mass $m$ is related to the orbital  momentum of the hadronic bound state by  $(m R)^2 = - 4 + L^2$ and thus  $(m R)^2 \ge - 4$. The quantum mechanical stability condition $L^2 \ge 0$ is thus equivalent to the Breitenlohner-Freedman stability bound in AdS~\cite{Breitenlohner:1982jf}.

\subsection{A light-front holographic model for mesons \label{Mesons}}

The simplest holographic example is the truncated model  of Polchinski and Strassler  whereas the confinement dynamics is included by the boundary conditions at $1/\Lambda_{\rm QCD}$~\cite{Polchinski:2001tt}.  This ``hard-wall''  model was introduced  to study high-energy fixed-angle hard scattering of glueballs using the gauge/gravity duality in confining gauge theories~\cite{Polchinski:2001tt}. It was then realized by Boschi-Filho and Braga~\cite{BoschiFilho:2002ta, BoschiFilho:2002vd} that this simple model could be used advantageously to compute the glueball mass spectrum and obtain results which compare favorably with more elaborated computations based, for example, on lattice QCD~\cite{Morningstar:1997ff, Teper:1997am} or supergravity in an AdS blackhole geometry background~\cite{Csaki:1998qr, deMelloKoch:1998qs, Brower:2000rp} using the AdS/CFT correspondence.  The hard-wall model was extended by two of the authors of this report to  light hadrons in Ref.~\cite{deTeramond:2005su}, where it was shown that the pattern of orbital excitations  of light mesons and baryons is well described in terms of a single parameter, the QCD scale $\Lambda_{\rm QCD}$.  In Refs.  \cite{Erlich:2005qh, DaRold:2005zs, DaRold:2005vr} it was shown by different authors, including another of the authors of this report, how to construct a five-dimensional holographic model which incorporates chiral symmetry and other properties of light mesons, including  quark masses, decay rates and couplings (See Sect. \ref{abum}).

In the light-front version of the hard-wall model~\cite{Brodsky:2006uqa},  the holographic variable $z$ corresponds exactly with the impact variable $\zeta$, which represents the  invariant measure of transverse separation of the constituents within the hadrons, and quarks propagate freely in the hadronic interior up to the confinement scale. This model provides an  analog of the MIT bag model~\cite{Chodos:1974je} where quarks are permanently confined inside a finite region of space. In contrast to bag models, boundary conditions are imposed on the boost-invariant variable $\zeta$, not on the bag radius at fixed time.  The resulting model is a manifestly Lorentz invariant model with confinement at large distances, while incorporating conformal behavior at small physical separation.  The eigenvalues of the  LF wave equation (\ref{LFWE}) for the hard-wall model ($U = 0$) are determined by the boundary conditions $\phi(z = 1/\Lambda_{\rm QCD}) = 0$, and are given in terms of the roots  $\beta_{L,k}$ of the Bessel functions: $M_{L,k} = \beta_{L,k} \Lambda_{\rm QCD}$. By construction, the hard wall model  has  a simple separation of kinematical and dynamical aspects, but it has shortcomings  when trying to describe the observed meson spectrum~\cite{deTeramond:2005su, deTeramond:2012rt}.  The model fails to account for the pion as a chiral $M=0$ state and it is degenerate with respect to the orbital quantum number $L$, thus leading to identical trajectories for pseudoscalar and vector mesons. It also fails to account for the important   splitting for the $L=1$ $a$-meson states for different values of $J$. Furthermore,  for higher quantum excitations the spectrum behaves as $M \sim 2n + L$, in contrast to the Regge dependence $M^2 \sim n + L$ found experimentally~\cite{Afonin:2007jd, Klempt:2007cp}. As a consequence, the radial modes are not well described in the truncated-space model. For example the first radial AdS eigenvalue has a mass 1.77 GeV, which is too high compared to the mass of the observed first  radial excitation of the meson, the $\pi(1300)$.

The shortcomings of the hard-wall model are evaded with the ``soft-wall'' model~\cite{Karch:2006pv} where the sharp cutoff is modified by a a dilaton profile $\vp(z) \sim z^2$. The soft-wall model leads to linear Regge trajectories~\cite{Karch:2006pv} and avoids the ambiguities in the choice of boundary conditions at the infrared wall. From the relation \req{U} it follows that the harmonic potential is holographically related to a unique dilaton profile, $\varphi(z) = \la z^2$ provided that $\varphi(z) \to 0$ as $z \to 0$. This choice  follows  from the requirements  of conformal invariance  as described in Chapter \ref{ch3}, and  leads through \req{U} to the  effective LF potential~\footnote{The notation $\la = \ka^2$ is used in Ref. \cite{deTeramond:2012rt}.}
\beq \label{UJ} 
U(\ze, J) =   \la^2 \ze^2 + 2 \la (J - 1),
\enq
which  corresponds  to a transverse oscillator in the light-front.  The term  $\la^2 \ze^2$ is determined uniquely by the underlying conformal invariance incorporated in the one-dimensional effective theory, and the constant term $2 \la (J-1)$ by the spin representations in the embedding space.

For the effective potential (\ref{UJ}) equation  (\ref{LFWE}) has eigenfunctions (Appendix \ref{solutions})
\beq \label{phi}
\phi_{n, L}(\zeta) = \vert \la\vert^{(1+L)/2} \sqrt{\frac{2 n!}{(n\!+\!L\!)!}} \, \zeta^{1/2+L}
e^{- \vert \la \vert \zeta^2/2} L^L_n(\vert \la \vert \zeta^2) ,
\enq
and eigenvalues
\beq \label{M2SFM}
M^2 = \left( 4n + 2 L + 2\right) \vert \la \vert + 2 \la(J-1).
\enq
The LF wave functions $\phi(\zeta) = \langle \zeta \vert \phi \rangle$ are normalized as $\langle \phi \vert \phi \rangle = \int d \zeta \, \phi^2(\zeta) = 1$ in accordance with (\ref{Phinorm}).

\begin{figure}[!]
\centering
\includegraphics[angle=0,width=6.8cm]{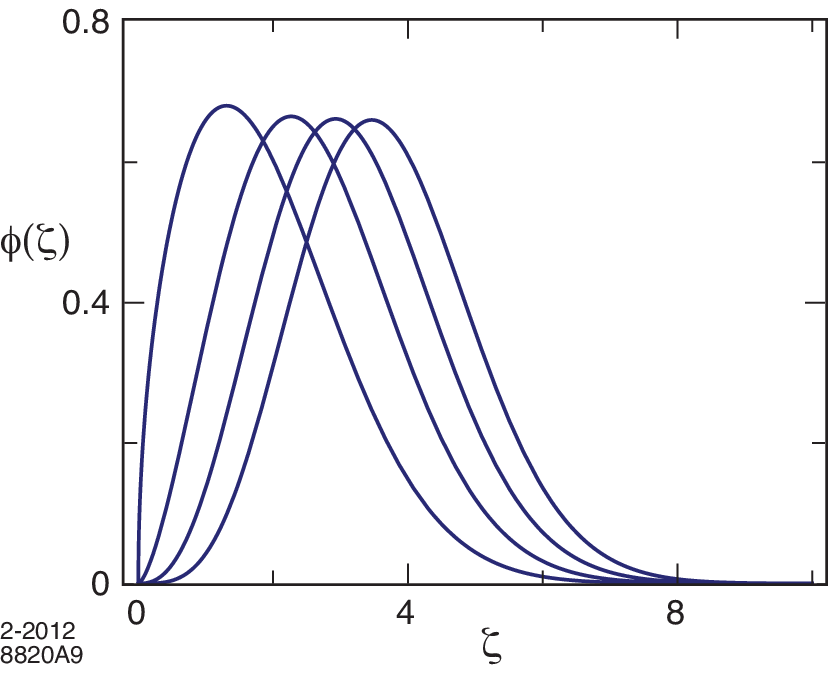} \hspace{10pt}
\includegraphics[angle=0,width=6.8cm]{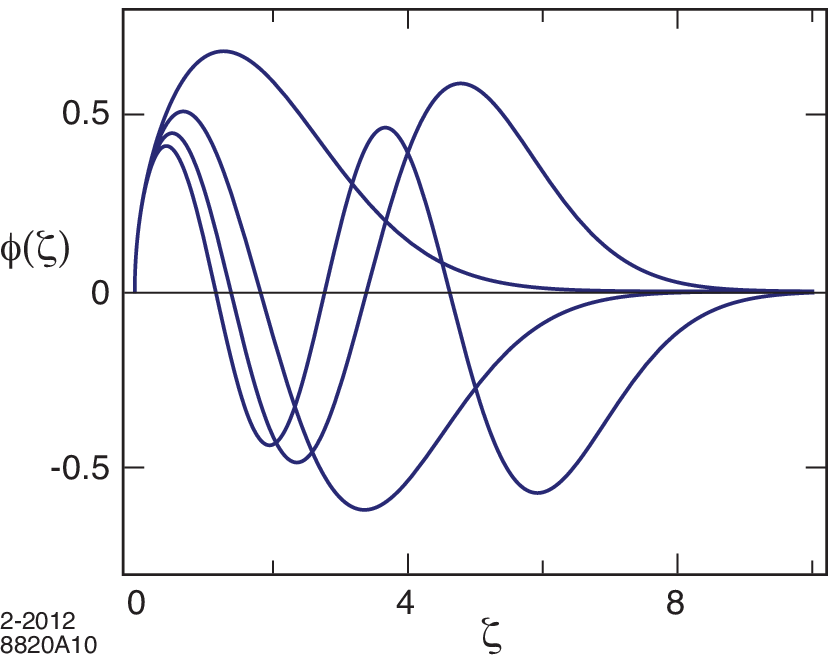}
\caption{\label{LFWFs}
\small Light-front wave functions $\phi_{n,L}(\zeta)$  in physical space-time corresponding to a dilaton profile $\exp(\la z^2)$: (left) orbital modes ($n=0$) and (right) radial modes ($L=0$).}
\end{figure}

Except for $J=1$ the spectrum predictions are significantly different for $\la > 0$ or  $\la < 0$.   The predicted spectrum for  $\la>0$ is
\beq\label{M2SFM} 
M_{n, J, L}^2 = 4 \la \left(n + \frac{J+L}{2} \right).
\enq
For the lowest possible solution $n = L = J =0$ there is an exact cancellation of the LF kinetic and potential energy and the ground state eigenvalue turns out to be $M^2 = 0$~\footnote{One can easily write the LF eigenvalue equation for general dimension $d$. The dimension has  no influence on the confining term $\la^2 z^2$  but it determines the constant term in the  potential. For $J=0$ this term is $(2-d)\la$. Only in  $d=4$ dimensions the vacuum energy is exactly compensated by a constant term in the potential and the $J=0$, $L=0$  state is massless.}.  This is a  bound state of two massless quarks and scaling dimension 2, which we identify with the lowest state, the pion. This result not only implies linear Regge trajectories and a massless pion but also the relation between the $\rho$ and $a_1$ mass usually obtained from the Weinberg sum rules~\cite{Weinberg:1967kj}\footnote{A discussion of chiral symmetry breaking in the light-front is given in Ref. \cite{Beane:2013oia}.}
\beq
m_\pi^2 = M_{0,0,0}^2 = 0, \,
m_\rho^2 = M_{0,1,0}^2 = 2 \la, \,
m_{a_1}^2 = M_{0,1,1}^2 = 4 \la.
\enq

The meson spectrum (\ref{M2SFM}) has a string-model Regge form~\cite{Nambu:1978bd}.   In fact, the linear dependence of the squared masses in both the angular momentum $L$ and radial quantum number $n$,  $M^2 \sim n + L$, and thus the degeneracy in the quantum numbers $n + L$,  was first predicted using  semiclassical quantization of effective strings in Ref.~\cite{Baker:2002km}.  The LFWFs (\ref{phi}) for different orbital and radial excitations are depicted in Fig. \ref{LFWFs}. Constituent quark and antiquark separate from each other as the orbital and radial quantum numbers increase. The number of nodes in the light-front wave function depicted in Fig. \ref{LFWFs} (right) correspond to the radial excitation quantum number $n$.  The  result \req{M2SFM} was also found in Ref. \cite{Gutsche:2011vb}.

 \begin{table}[htdp] 
  \begin{center} 
 {\begin{tabular}{@{}cccccc@{}}
 \hline\hline \\[-3.0ex]
 $L$ &    $S$ & $ n$ &  $J^{PC}$ &   Meson State & 
 \\[0.2ex]
 \hline
 \multicolumn{6}{c}{}\\[-3.0ex]
 0 & 0 & 0  &  $0^{-+}$   & $\pi(140)~$    \\[-0.5ex]
 0 & 0 & 1  &  $0^{-+}$   & $\pi(1300)$    \\[-0.5ex]
 0 & 0 & 2  &  $0^{-+}$   & $\pi(1800)$    \\[-0.5ex]
 0 & 1 & 0  &  $1^{- -}$  &  $\rho(770)~$   \\[-0.5ex] 
 0 & 1 & 0  &  $1^{- -}$  &  $\om(782)~$   \\[-0.5ex] 
 0 & 1 & 1  &  $1^{- -}$  &  $\om(1420)~$   \\[-0.5ex] 
 0 & 1 & 1  &  $1^{- -}$  &  $\rho(1450)~$   \\[-0.5ex] 
 0 & 1 & 2  &  $1^{- -}$  &  $\om(1650)~$   \\[-0.5ex]
 0 & 1 & 2  &  $1^{- -}$  &  $\rho(1700)~$   \\[0.5ex]   \hline \\  [-3.0ex] 
 1 & 0 & 0  &  $ 1^{+-}$  & $b_1(1235)$  \\[-0.5ex]
 1  & 1 & 0 &  $0^{++}$  & $a_0(980)~$ \\[-0.5ex]
 1  & 1 & 1 &  $0^{++}$  & $a_0(1450)~$ \\[-0.5ex]
 1  & 1 & 0 &  $1^{++}$  & $a_1(1260)$ \\[-0.5ex]
 1  & 1 & 0 &  $2^{++}$  & $f_2(1270)$  \\[-0.5ex]
 1  & 1 & 0 &  $2^{++}$  & $a_2(1320)$   \\[-0.5ex]  
 1  & 1 & 2 &  $2^{++}$  & $f_2(1950)$   \\[-0.5ex] 
 1  & 1 & 3 &  $2^{++}$  & $f_2(2300)$   \\[0.5ex] \hline \\  [-3.0ex] 
 2  & 0 & 0 &  $2^{-+}$   & $\pi_2(1670)$ \\[-0.5ex]
 2  & 0 & 1 &  $2^{-+}$   & $\pi_2(1880)$ \\[-0.5ex]
 2  & 1 & 0 &  $3^{--}$    &$\om_3(1670)$ \\[-0.5ex]
 2  & 1 & 0 &  $3^{--}$    &$\rho_3(1690)$ \\[0.5ex] \hline \\  [-3.0ex] 
 3  & 1 & 0 &  $4^{++}$  & $ a_4(2040)$\\[-0.5ex]
 3  & 1 & 0 &  $4^{++}$  & $ f_4(2050)$ \\[0.4ex] 
 \hline\hline
 \end{tabular}}
  \caption{\label{mesons} \small Confirmed  mesons listed by PDG~\cite{PDG:2014}. The labels $L$,  $S$ and  $n$ refer to assigned internal orbital angular momentum, internal spin and radial quantum number respectively.  For a $q \bar q$ state $P = (-1)^{L+1}$, $C = (-1)^{L+S}$.  For the pseudoscalar sector only the $I=1$ states are listed.}
 \end{center}
  \end{table}

To compare the LF  holographic model predictions with experiment,  we list in Table \ref{mesons} confirmed (3-star and 4-star) meson states  corresponding to the  light-unflavored meson sector from the  Particle Data Group (PDG)~\cite{PDG:2014}. The Table includes isospin   $I = 0$ and $I =1$ vector mesons and the $I = 1$  pseudoscalar mesons. We have included the assigned internal spin, orbital angular momentum and radial quantum numbers from the  quark content $\vert u \bar d \rangle$, $\frac{1}{\sqrt{2}} \vert u \bar u - d \bar d \rangle$ and  $\vert d \bar u \rangle$. The $I=1$ mesons are the $\pi$, $b$, $\rho$ and $a$ mesons.  We have not listed in Table  \ref{mesons} the $I=0$ mesons for the pseudoscalar sector  which are a mix of $u \bar u$, $d \bar d$ and $ s \bar s$, thus more complex entities. The light $I = 0$ mesons  are $\eta$, $\eta'$, $h$, $h'$ $\omega$, $\phi$, $f$ and $f'$. This list comprises the puzzling $I=0$ scalar $f$-mesons~\cite{Parganlija:2013xsa}, which may be interpreted as a superposition of tetra-quark  states with a $q \bar q$  configuration with $L=1$, $S=1$,  which couple to a $J=0$ state~\cite{Klempt:2007cp}~\footnote{The interpretation of the $\pi_1(1400)$ is not very clear~\cite{Klempt:2007cp} and is not included in Table  \ref{mesons}. Similarly, we do not include the $\pi_1(1600)$ in the present analysis.}.  We also include in Table \ref{mesons}  the $I=0$ vector-mesons the  $f$ and $\om$, which are well described as a $q \bar q$ system with no mixings with the strange sector.  Likewise, in Sec.~\ref{LQM} we compute the masses of the $\phi$ mesons which are well described as an $s \bar s$ bound state. We will also discuss in Sec.~\ref{LQM} light hadronic  bound states composed of  $u$ or $d$  with an $s$-quark: the $K$ meson and $K^*$  vector-meson families.

\begin{figure}[h]
\centering
\includegraphics[width=14.6cm]{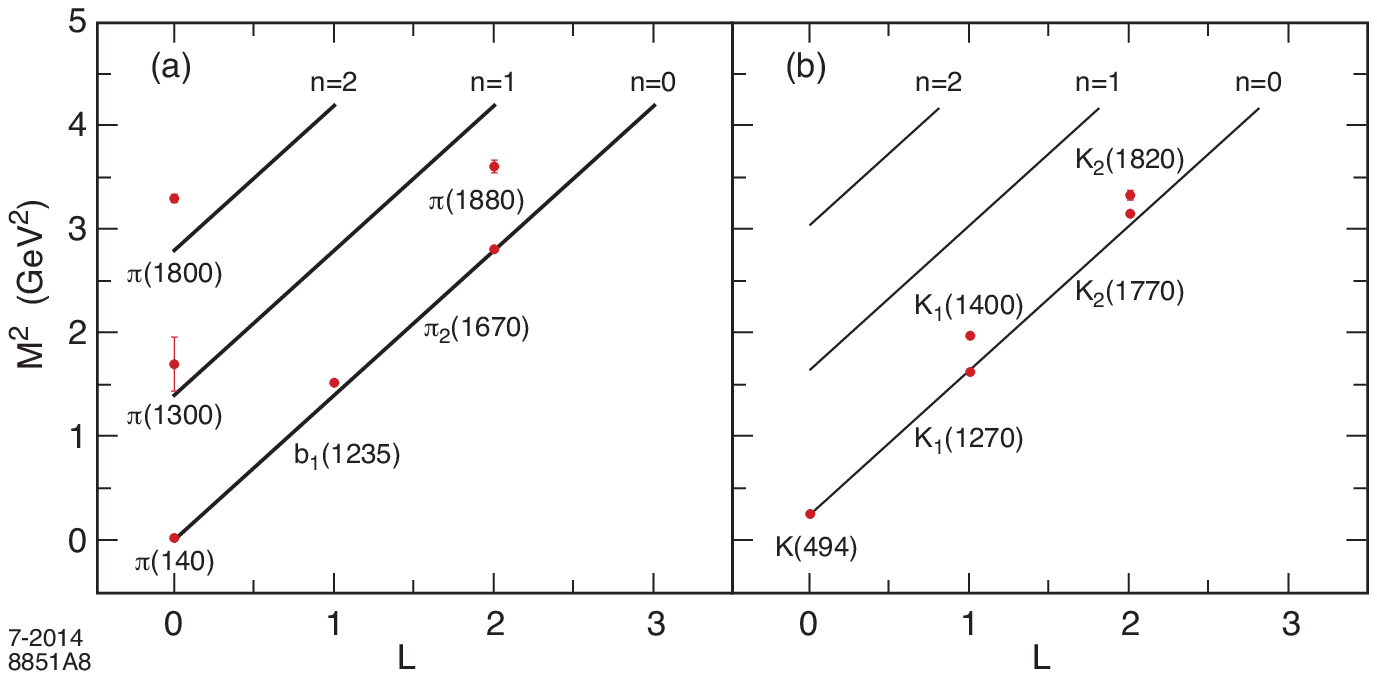} 
\caption{\small   Orbital and radial excitation spectrum for the light pseudoscalar mesons:  (a) $I = 0$ unflavored mesons and (b)  strange mesons, for $\sqrt \la = 0.59$ GeV.}
\label{pionspec}
\end{figure} 

\begin{figure}[h]
\centering
\includegraphics[width=14.6cm]{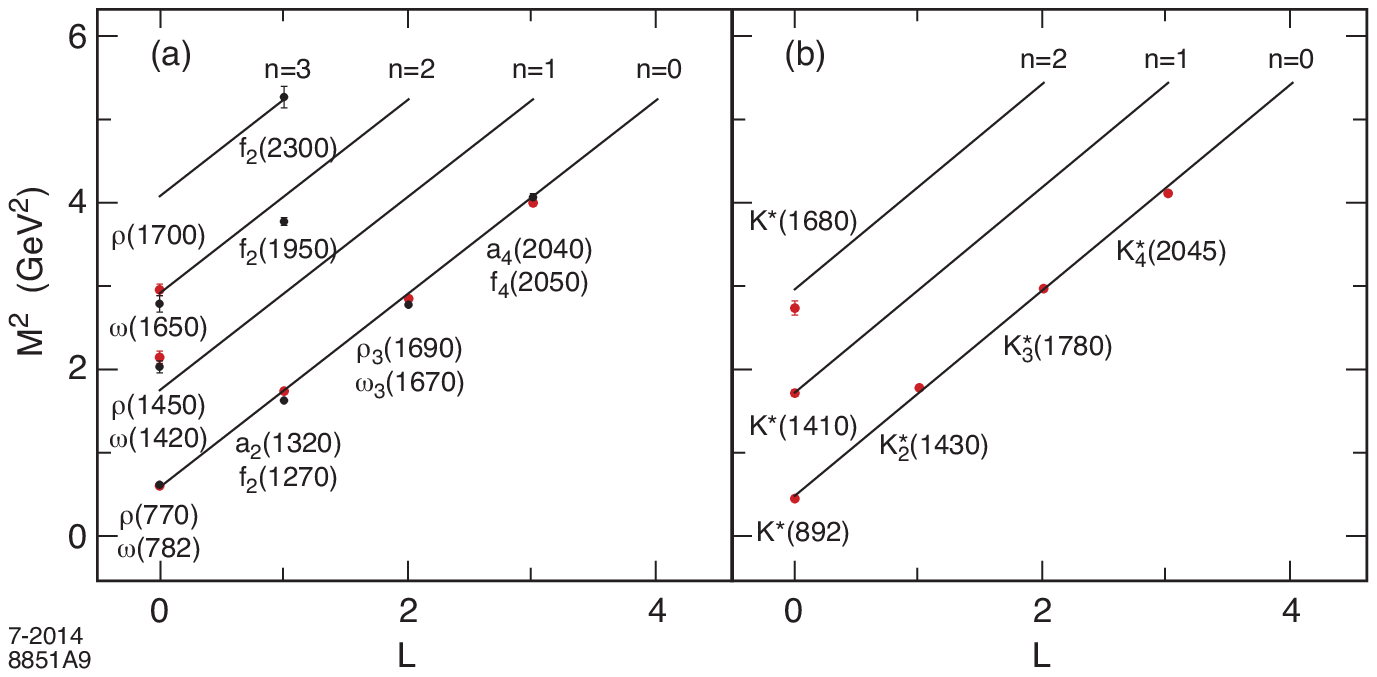}  
 \caption{\small  Orbital and radial excitation spectrum for the light vector mesons:  (a) $I = 0$ and  $I=1$ unflavored mesons and (b)  strange mesons, for $\sqrt \la = 0.54$ GeV.}
 \label{VMspec}
\end{figure}

 For the $J = L + S$ meson families Eq. (\ref{M2SFM}) becomes
 \begin{equation} \label{M2SFMLS}
 M_{n,L,S}^2 = 4 \la  \left(n + L + \frac{S}{2}\right).
 \end{equation}
 The lowest possible stable solution for $n = L = S = 0$, the pion, has eigenvalue $M^2 = 0$. Thus one can compute the full  $J = L + S$,  mass spectrum $M^2$  in Table \ref{mesons}, by simply adding  $4 \la$ for a unit change in the radial quantum number, $4 \la$ for a change in one unit in the orbital quantum number and $2 \la$ for a change of one unit of spin to the ground state value.  The spectral predictions  for the $J = L + S$ light  unflavored meson  states,  listed in  Table. \ref{mesons}, are  compared with experimental data 
in Fig. \ref{pionspec} (a) and Fig. \ref{VMspec} (a) for the positive sign dilaton model discussed here.  The data is from PDG~\cite{PDG:2014}.

In contrast to the hard-wall model, the soft-wall model with positive dilaton accounts for the mass pattern observed in radial excitations, as well as for the triplet splitting for the $L=1$, $J = 0,1,2$ observed for the vector meson $a$-states.  As we will discuss in the next section, a spin-orbit effect is only predicted  for mesons not baryons, as observed in experiment~\cite{Klempt:2007cp, Klempt:2009pi};  it thus becomes a crucial test for any model which aims to describe the systematics of the light hadron spectrum. Using the spectral formula (\ref{M2SFM}) we find~\cite{deTeramond:2012rt}
 \begin{equation}
 M_{a_2(1320)} >  M_{a_1(1260)} >   M_{a_0(980)}.
 \end{equation}
The predicted values are 0.76, 1.08 and 1.32 GeV for the masses of the  $a_0(980)$, $a_1(1260)$ and   $a_2(1320)$ vector mesons, compared with the experimental values 0.98, 1.23 and 1.32  GeV respectively. The prediction for the mass of the $L=1$, $n=1$  state $a_0(1450)$ is  1.53 GeV, compared with the observed value 1.47 GeV. Finally, we would like to mention  the recent precision measurement at COMPASS~\cite{COMPASS:2013xra}  which found a new resonance named the $a_1(1420)$ with a mass 1.42 GeV, the origin of which remains unclear. In the present framework the $a_1(1420)$ is interpreted as a  $J = 1, S = 1, L = 1, n = 1$ vector-meson state with a predicted mass of 1.53 GeV.  For other calculations of the hadronic spectrum in the framework of AdS/QCD, see Refs.~\cite{BoschiFilho:2005yh, Katz:2005ir, Evans:2006ea, Hong:2006ta,  Colangelo:2007pt, Forkel:2007cm, Forkel:2007ru, Vega:2008af, Nawa:2008xr,   Colangelo:2008us, Forkel:2008un, Gherghetta:2009ac, Ahn:2009px, dePaula:2009za, Sui:2009xe, Kapusta:2010mf, Zhang:2010bn, Zhang:2010sb,  Kirchbach:2010dm, Iatrakis:2010zf,  Branz:2010ub, Kelley:2010mu, Sui:2010ay, Afonin:2011ff, Zhang:2011cm, Liu:2012vda, Capossoli:2013kb, Li:2013oda, Afonin:2013npa}~\footnote{For recent reviews see, for example, Refs.~\cite{Erlich:2009me, Kim:2011ey}. One can also use the AdS/QCD framework to study hadrons at finite temperature (See, for example Refs.~\cite{Colangelo:2009ra, Li:2013lfa, Mamani:2013ssa}  and references therein) or in a hadronic medium~\cite{Park:2011qq}.}.

The LF holographic model with $\la>0$ accounts for the mass pattern observed in the radial and orbital excitations of the light mesons, as well as for the triplet splitting for the $L=1$, $J = 0, 1, 2$, vector meson $a$-states~\cite{deTeramond:2012rt}. The slope of the Regge trajectories gives a value $\sqrt \la \simeq 0.5 ~\rm GeV$,  but  the value of $\la$ required for describing the pseudoscalar sector is slightly higher that the value of $\la$ extracted from the vector sector. In general the description of the vector sector is better than the pseudoscalar sector. However, the prediction for  the observed spin-orbit splitting for the $L=1$  $a$-vector mesons  is overestimated by the model.

The solution for $\la<0$  leads to a pion mass heavier than the $\rho$ meson and a meson spectrum given by $M^2 = 4 \lambda \, (n + 1 + (L - J)/2$, in clear disagreement with the observed spectrum.  Thus  the solution $\la<0$ is incompatible with the light-front  constituent interpretation of hadronic states.   We also note that the solution with $\la>0$ satisfies the stability requirements  from the  Wilson loop area condition for confinement~\cite{Sonnenschein:2000qm} discussed in Sec. \ref{warpmetric}.

\subsection{Meson spectroscopy in a gauge invariant AdS model}
 
Like  the AdS wave equation for arbitrary spin \req{PhiJM}, the AdS wave equation which follows from a gauge invariant construction described in Sec. \ref{KKSS} (See  Ref. \cite{Karch:2006pv})  can be brought into a Schr\"odinger-like form by rescaling the AdS field in  \req{PhiKKSS}  according to $\tilde \Phi_J(z) =
z^{J-1/2} e^{-\la z^2/2} \tilde  \phi_J(z)$. The result is
\beq \label{SchKKSS}
\left( -\frac{d^2}{d z^2} - \frac{1 - 4 J^2}{4} +\la^2 z^2 - 2 J \la \right) \tilde \phi_J(z) = M^2\, \tilde \phi_J(z),
\enq
and yields  the spectrum
\beq
M^2=(4n + 2 J + 2)|\la| -2 J\la.
\enq

Besides the difference in sign in the dilaton profile,  there are conceptual differences in the treatment of higher spin given by KKSS~\cite{Karch:2006pv} in Sec. \ref{KKSS}, as compared with the treatment given in Sec. \ref{arbitrary}.  The mapping of the AdS equation of motion  \req{PhiKKSS} onto the Schr\"odinger equation \req{SchKKSS} reveals that $J=L$ and therefore an essential kinematical degree of freedom is missing in the light-front interpretation of the KKSS AdS wave equation. In particular the $\rho$ meson would be an $L=1$ state. Furthermore the method of treating higher spin, based on gauge invariance, can only be applied to the vector meson trajectory, not pseudoscalar mesons.
Generally speaking, one can say that insisting on gauge invariance in AdS$_{d+1}$ favours a negative dilaton profile
$(\la<0)$, whereas the mapping onto the LF equation demands an AdS mass $\mu \neq 0$ and a positive profile $(\la >0)$.

\subsection{Light quark masses and meson spectrum \label{LQM}}

In general, the effective  interaction depends on quark masses and the longitudinal momentum fraction $x$ in addition to the transverse invariant variable $\zeta$. However, if the confinement potential is unchanged for small quark masses it then only depends on the transverse invariant variable $\zeta$, and the transverse dynamics are unchanged (See Sect. \ref{FQM}).  This is consistent with the fact that the potential is determined from the  conformal symmetry of the effective one-dimensional quantum field theory, which is not badly broken for small quark  masses.

In the limit of zero quark masses  the effective LFWF for a  two-parton ground state in impact space is
\beq   \label{LFWFb}
\psi_{\bar q q/\pi}(x, \zeta)  \sim  \sqrt{x(1-x)}~e^{-\half \la \zeta^2},
\enq
where the invariant transverse variable $\zeta^2 = x(1-x) \mbf{b}_\perp^2$ and  $\la > 0$. The factor $\sqrt{x(1-x)}$  is determined from the precise holographic mapping of transition amplitudes in the limit of massless quarks (Chapter \ref{ch6}). The  Fourier transform of \req{LFWFb} in momentum-space is
\beq  \label{LFWFk}
\psi_{\bar q q/\pi}(x, \mbf{k}_\perp) \sim  \frac{1}{\sqrt{x(1-x)}}
~e^{- \frac{\mbf{k}_\perp^2}{2 \la x(1-x)}},
\enq
where the explicit dependence of the wave function in the LF off shell-energy is evident.

For the effective two-body bound state the inclusion of  light quark masses~\footnote{The light quark masses introduced here are not the constituent masses of the nonrelativistic quark model, but effective quark masses from  the reduction of higher Fock states as functionals of the valence state (See Sec. \ref{FQM}). In the chiral limit, however, these masses should be zero.} amounts to the replacement in \req{LFWFk} of the $q \! - \!  \bar q$  invariant  mass \req{Mqbarq}, the key dynamical variable which describes the off energy-shell behavior of the bound state~\cite{Brodsky:2008pg},
\beq  \label{LFWFkm}
\psi_{\bar q q/\pi}(x, \mbf{k}_\perp) \sim  \frac{1}{\sqrt{x(1-x)}}
~e^{- \frac{1}{2 \la} \left( \frac{\mbf{k}_\perp^2}{x(1-x)} + \frac{m_q^2}{x} + \frac{m_{\bar q}^2}{1-x}\right) },
\enq
which has the same exponential form of the succesful phenomenological LFWF introduced in Ref.  \cite{Brodsky:1980vj}.
The Fourier transform of  \req{LFWFkm} gives the LFWF in impact space including light-quark masses
\beq   \label{LFWFbm}
\psi_{\bar q q/\pi}(x, \zeta)  \sim  \sqrt{x(1-x)}~ e^{-  \frac{1}{2 \la} \big(\frac{m_q^2}{x} + \frac{m_{\bar q}^2}{1-x} \big) }  e^{ -\half \la \, \zeta^2},
\enq
which  factorizes neatly  into transverse and longitudinal components. The holographic LFWF \req{LFWFbm} has been successfully used in the description of diffractive vector meson electroproduction at HERA~\cite{Forshaw:2012im} by extending the LF holographic approach to the longitudinal component of the $\rho$ LFWF, in $B \to \rho \gamma$~\cite{Ahmady:2012dy} and $B \to K^* \gamma$~\cite{Ahmady:2013cva} decays as well as in the prediction of $B \to \rho$~\cite{Ahmady:2013cga},  $B \to K^*$ form factors~\cite{Ahmady:2014sva} and $B \to K^* \mu^+ \mu^-$ decays~\cite{Ahmady:2014cpa}.  This LFWF has also been used to study the spectrum~\cite{Branz:2010ub} and  the distribution amplitudes~\cite{Hwang:2012xf}  of  light and heavy mesons.

For excited meson states we can follow the same procedure by replacing the key invariant mass variable in the polynomials in the LFWF using \req{Mqbarq}.  An explicit calculation shows, however,  that the essential modification in the hadronic  mass, from small quark masses, comes from the shift in the exponent of the LFWF. The corrections from the shift in the polynomials accounts for less than 3 \%. This can be understood from the fact that to first order  the transverse dynamics is unchanged, and consequently the transverse LFWF is also unchanged to first order. Thus our expression for the  LFWF 
\beq   \label{LFWFmnl}
\psi_{n, L}(x, \zeta)  \sim  \sqrt{x(1-x)}~e^{-  \frac{1}{2 \la} \big(\frac{m_q^2}{x} + \frac{m_{\bar q}^2}{1-x} \big)}   \zeta^2 e^{-\half \la \, \zeta^2}  L^L_n(\la \zeta^2),
\enq
and from \req{MKbmq} the hadronic mass shift $\Delta M^2$ for small quark masses~\cite{Teramond:2009ghp}
\beq
\De M_{m_q,m_{\bar q}}^2 =\frac{\int_0^1 dx\,   e^{- \frac{1}{\la}\big(\frac{m_q^2}{ x} +\frac{m_{\bar q}^2 }{1- x}\big)}  \left(\frac{m_q^2}{x} + \frac{m_{\bar q}^2}{1-x}\right)}
{\int_0^1 dx \, e^{- \frac{1}{\la}\big(\frac{m_q^2}{ x} +\frac{m_{\bar q}^2 }{1- x}\big)}} ,
\enq
which is independent of $L$, $S$ and $n$.  For light quark masses, the hadronic mass is  the longitudinal  $1/x$ average of the square of the effective quark masses,
{\it i. e.}, the effective  quark masses from  the reduction of higher Fock states as functionals of the valence state~\cite{Pauli:1998tf}. The final result for the hadronic spectrum of mesons modified by light quark masses is thus
\beq \label{M2mq} 
M_{n, J, L, m_q, m_{\bar q}}^2 = \De M_{m_q,m_{\bar q}}^2 + 4 \la \left(n + \frac{J+L}{2} \right),
\enq
with identical slope $4 \la$ from the limit of zero quark masses.  In particular, we obtain from \req{M2mq}, respectively, the spectral prediction for the unflavored meson and strange meson mass spectrum. We have
\beq \label{M2pi}
M^2_{n,L,S} = M^2_{\pi^\pm} +  4 \la \left(n +   \frac{J + L}{2}\right) ,
\enq
for the $\pi$ and $b$ pseudoscalar and  $\rho, \omega, a, f$ vector mesons,
and
\beq \label{M2S}
M^2_{n,L,S} = M^2_{K^\pm} +  4 \la \left(n +  \frac{J + L}{2}\right) ,
\enq
for the $K$ and $K^*$ meson spectrum. The PDG values are~\cite{PDG:2014} $M_{\pi^\pm} \cong 140$ MeV and $M_{K^\pm} \cong 494$ MeV.

 \begin{table}[htdp]
  \begin{center} 
 {\begin{tabular}{@{}cccccc@{}}
 \hline\hline \\[-3.0ex]
 $L$ &    $S$ & $ n$ &  $J^{P}$ &   Meson State & 
 \\[0.2ex]
 \hline
 \multicolumn{6}{c}{}\\[-3.0ex]
  0 & 0 & 0& $ 0^{-}$ & $ K(494)~$\\  [-0.5ex]
  0 & 1 & 0& $ 1^{-}$ & $ K^*(892)$\\  [-0.5ex]
  0 & 1 & 1& $ 1^{-}$ & $ K^*(1410)$\\  [-0.5ex]
  0 & 1 & 2& $ 1^{-}$ & $ K^*(1680)$  \\  [0.3ex]  \hline \\  [-3.0ex] 
  1 & 0 & 0 & $ 1^+ $ & $ K_1(1270) $\\  [-0.5ex]
  1 & 0 & 1 & $ 1^+ $ & $ K_1(1400) $\\  [-0.5ex]
  1 & 1 & 0 & $ 2^+ $ & $ K^*_2(1430) $\\  [-0.5ex]
  1 & 1 & 1 & $ 0^+ $ & $ K^*_0(1430) $\\ [0.3ex]  \hline \\  [-3.0ex] 
  2 & 0 & 0 & $ 2^- $ & $ K_2(1770)$\\  [-0.5ex]
  2 & 0 & 1 & $ 2^- $ & $ K_2(1820)$\\   [-0.5ex]
  2 & 1 & 0 & $ 3^- $ & $ K^*_3(1780)$\\[0.3ex]  \hline \\  [-3.0ex] 
  3 & 1 & 0 & $ 3^- $ & $ K^*_4(2045)$ \\[0.4ex] 
 \hline\hline
 \end{tabular}}
  \caption{ \label{Kmesons} \small Confirmed  strange mesons listed by PDG~\cite{PDG:2014}. The labels $L$,  $S$ and  $n$ refer to assigned internal orbital angular momentum, internal spin and radial quantum number respectively.  For a $q \bar q$ state $P = (-1)^{L+1}$.}
 \end{center}
 \end{table}

We list in Table \ref{Kmesons} the confirmed (3-star and 4-star) strange mesons from the  Particle Data Group~\cite{PDG:2014}. The  predictions  for the $J = L + S$ strange pseudoscalar and vector mesons  are  compared with experimental data in Figs. \ref{pionspec} (b) and \ref{VMspec} (b) respectively. The data is from PDG~\cite{PDG:2014}.  The spectrum is well reproduced with  the same values of the mass scale as for the massless case,  $\sqrt \la = 0.59$ GeV for the light  pseudoscalar meson sector and  $\sqrt \la = 0.54$  GeV for the light vector sector.  It is clear from Table \ref{Kmesons} or Fig. \ref{pionspec} (b) that  the interpretation of the states $K_1(1400)$ and $K_2(1820)$ as $n=1$ radial excitations is not straightforward as their masses are very close to the $n=0$ states. As in the case of the light  unflavored $q \bar q$ mesons, the model predictions are much better for the vector sector.  In fact, the model predictions for the $K^*$ sector shown in  Fig. \ref{VMspec} (b) are very good~\cite{Dosch:2014wxa}.   We note, however, from Table \ref{Kmesons} that the states $K^*_0(1430)$ and  $K^*_2(1430)$ -- which belong to the  $J = 0$, $J=1$ and $J =2$ triplet for $L=1$, are degenerate. This result  is in contradiction with the spin-orbit coupling predicted by the LF holographic model for mesons; a possible indication of mixing of the $K^*_0$  with states which carry the vacuum quantum numbers.

Fitting the quark masses to the observed masses of the $\pi$ and $K$ we obtain for $\sqrt{\la} = 0.54 ~{\rm MeV}$ the average effective light quark  mass $m_q= 46$ MeV, $q = u,d$,  and $m_s = 357$ MeV, values between the current $\overline {MS}$ Lagrangian masses normalized at 2 GeV and typical constituent masses.  With these values one obtains $\De M^2_{m_q, m_{\bar q}} = 0.067\,  \la , ~ \De M^2_{m_q, m_{\bar s}} = 0.837 \, \la,  ~ \De M^2_{m_s,m_{\bar s}} = 2.062 \, \la$,  for $\sqrt{\la} = 0.54~ {\rm MeV}$.

 \begin{table}[htdp]   
  \begin{center} 
 {\begin{tabular}{@{}cccccc@{}} 
 \hline\hline \\[-3.0ex]
 $L$ &    $S$ & $ n$ &  $J^{P}$ &   Meson State & 
 \\[0.2ex]
 \hline
 \multicolumn{6}{c}{}\\[-3.0ex]
  0 & 1 & 0& $ 1^{-}$ & $ \phi(1020)$\\  [-0.5ex]
  0 & 1 & 1& $ 1^{-}$ & $ \phi(1680)$\\  [-0.5ex]  
  0 & 1 & 3& $ 1^{-}$ & $ \phi(2170)$\\  [-0.0ex]   \hline \\  [-3.0ex] 
  2 & 1 & 0 & $ 3^- $ & $ \phi_3(1850) $\\  [-0.0ex]
  \hline\hline
 \end{tabular}}
  \caption{ \label{phimesons} \small Confirmed  $\phi$ mesons listed by PDG~\cite{PDG:2014}. The labels $L$,  $S$ and  $n$ refer to assigned internal orbital angular momentum, internal spin and radial quantum number respectively.  The parity assignment is given by $P = (-1)^{L+1}$.}
 \end{center}
 \end{table}

\begin{figure}[h]
\centering
\includegraphics[width=7.6cm]{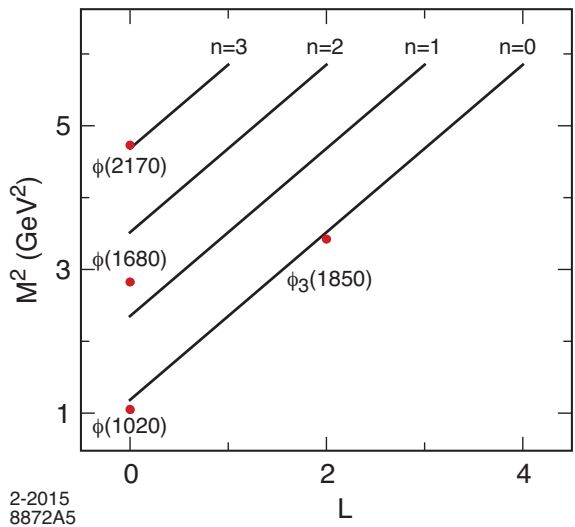}  
 \caption{\small  Orbital and radial excitation spectrum for the $\phi$ vector mesons for $\sqrt \la = 0.54$ GeV.}
 \label{phispec}
\end{figure}

Since the $\phi$ vector mesons are essentially $s \bar s$ bound-states, we can use our previous results to predict the $\phi$  spectrum  without introducing  any additional parameter. To this end we list in Table \ref{phimesons} the confirmed $\phi$ mesons from PDG~\cite{PDG:2014}.   The model predictions shown in  Fig. \ref{phispec} follow from \req{M2mq} with $\De M^2_{m_s,m_{\bar s}} = 2.062 \, \la$.

For heavy mesons conformal symmetry is strongly broken and there is no reason to assume that the LF potential in that case is similar to the massless one. Indeed, a simple computation shows that the model predictions for heavy quarks (without introducing additional elements in the model) is not satisfactory.  In fact,  the data for heavy mesons can only be reproduced at the expense of introducing vastly different values for the scale $\la$~\cite{Branz:2010ub, Gutsche:2012ez, Fujita:2009ca}. Another important point are the leptonic decay widths. For light quarks the quark masses have little influence on the result, only about 2 \% for the $\pi$ meson and 5 \% for the $K$ meson, but using the formalism also for the $B$ and $D$ mesons leads to widely different values when compared with experiment. For large quark masses the form of the LF confinement potential $U$ cannot be obtained  from the  conformal symmetry of the effective one-dimensional quantum field theory~\cite{Brodsky:2013ar}. In this case an important dependence on the heavy quark mass is expected, as  suggested by the relation given by Eq. \req{pots} between the effective potentials in the front form and instant form of dynamics~\cite{Trawinski:2014msa}.

\section{Half-integer spin \label{half-integer-map}}

One can also take as starting point the construction of light-front wave equations in physical space-time for baryons by studying  the LF transformation properties of spin-$\half$ states~\cite{Brodsky:2008pg, Teramond:2007conf}. The light-front wave equation describing baryons is a matrix eigenvalue equation $D_{LF} \vert \psi \rangle = M \vert \psi \rangle$ with $H_{LF} = D_{LF}^2$. In a $2 \times 2$ chiral spinor component representation,  the light-front equations are given by the coupled linear differential equations  (See Appendix \ref{LFint})
\begin{eqnarray} \label{LFDE}  \nonumber
- \frac{d}{d\zeta} \psi_-  - \frac{\nu+\half}{\zeta}\psi_-  -  V(\zeta) \psi_-&=& M \psi_+ , \\
 \frac{d}{d\zeta} \psi_+ - \frac{\nu+\half}{\zeta}\psi_+  - V(\zeta) \psi_+ &=& M \psi_- , 
\end{eqnarray}
where the invariant variable $\zeta$ for an $n$-parton bound state  is the transverse impact variable of the $n-1$ spectator system given by  \req{zetan}.

As we will consider below, we can identify $\nu$ with the light-front orbital angular momentum $L$,  the relative  angular momentum  between the active and the spectator cluster, but this identification is less straightforward than the relation for mesons, since it involves the internal spin and parity of the 3-quark baryon configuration. Note that $L$ is the maximal value of $|L^z|$ in a given LF Fock state.  An important feature of  bound-state relativistic theories  is that hadron eigenstates have in general Fock components with different $L$ components.  By convention one labels the eigenstate with its minimum value of $L$.  For example, the symbol $L$ in the light-front holographic spectral  prediction for mesons (\ref{M2SFM}) refers to the {\it minimum } $L$  (which also corresponds to the leading twist) and $J$ is the total angular momentum of the hadron.

A physical baryon has plane-wave solutions  with four-momentum $P_\mu$,  invariant mass $P_\mu P^\mu = M^2$, and polarization indices along the physical coordinates. It thus  satisfies the Rarita-Schwinger equation for spinors in physical space-time \req{RS}
 \beq
  \left(i \ga^\mu \pa_\mu - M \right) u_{\nu_1 \cdots \nu_T}({P}) =0,  \qquad {\ga^\nu} u_{\nu \nu_2 \cdots \nu_T}({P})=0.
  \enq
 Factoring out  from the AdS spinor field $\Psi$ the four-dimensional plane-wave and spinor dependence, as well as the scale factor $(1/z)^{T-d/2}$, we write
\beq \label{Psipsi} 
\Psi^\pm_{\nu_1 \cdots \nu_T}(x, z)   = e^{ i P \cdot x}  u^\pm_{\nu_1 \cdots \nu_T} ({P})  \left(\frac{R}{z} \right)^{T-d/2}   \psi^\pm_T(z) ,
\enq
where $T = J - \half$ and the chiral spinor  $u^\pm_{\nu_1 \dots \nu_T} =
\half (1 \pm \gamma_5)u_{\nu_1 \dots \nu_T}$  satisfies the four-dimensional chirality equations
 \beq
 \gamma \cdot P \, u^\pm_{\nu_1 \dots \nu_T}({P}) = M  u^\mp_{\nu_1 \dots \nu_T}({P}), \qquad
\gamma_5 u^\pm_{\nu_1 \dots \nu_T}({P}) = \pm \, u^\pm_{\nu_1 \dots \nu_T}({P}).
\enq

Upon replacing  the holographic variable $z$ by the light-front invariant variable $\zeta$ and substituting \req{Psipsi} into  the AdS wave equation \req{DE2} for arbitrary spin $J$, we recover the LF expression (\ref{LFDE}), provided that $ \vert \mu R \vert = \nu + \half $ and $\psi^\pm_T = \psi_\pm$, independent of the value of $J = T + \half$.  We also find that the effective LF potential in the light-front Dirac equation \req{LFDE} is determined by the effective interaction $\rho(z)$ in the effective action \req{af},
\begin{equation}
V(\zeta) = \frac{R}{\zeta} \rho(\zeta),
\end{equation}
 which is a $J$-independent potential. This is a remarkable result, since it implies that independently of the specific form of the potential, the value of the baryon masses along a given Regge trajectory depends only on the LF orbital angular momentum $L$, and thus, in contrast with the vector mesons, there is no spin-orbit coupling, in agreement with the observed near-degeneracy in the baryon spectrum~\cite{Klempt:2007cp, Klempt:2009pi}.   Equation (\ref{LFDE})  is equivalent to the system of second order equations
 \begin{equation} \label{LFWEA}
\left(-\frac{d^2}{d\zeta^2}
- \frac{1 - 4 \nu^2}{4\zeta^2} + U^+(\zeta) \right) \psi_+ = \mathcal{M}^2 \psi_+,
\end{equation}
and
\begin{equation} \label{LFWEB}
\left(-\frac{d^2}{d\zeta^2}
- \frac{1 - 4(\nu + 1)^2}{4\zeta^2} + U^-(\zeta) \right) \psi_- = \mathcal{M}^2 \psi_-,
\end{equation}
where
\beq \label{Upm}
U^\pm(\zeta) = V^2(\zeta) \pm V'(z) + \frac{1 + 2 \nu}{\zeta} V(\zeta).
\enq

 \subsection{A light-front holographic model for baryons \label{Baryons}}

 As for the case of mesons,  the simplest holographic model of baryons is the hard-wall model, where confinement dynamics is introduced by the boundary conditions  at $z \simeq 1/\La_{\rm QCD}$.  To determine the boundary conditions we integrate (\ref{af}) by parts  for $\varphi(z) = \rho(z) = 0$  and use the equations of motion.  We then find
\begin{equation}
S_F= - \lim_{\epsilon \to 0} \, R^d \int \frac{d^dx}{2 z^d} \Big ( \bar \Psi_+ \Psi_- - \bar \Psi_- \Psi_+ \Big) \Big \vert_\epsilon^{z_0},
\end{equation}
where  $\Psi_\pm = \half \left(1 \pm \gamma_5 \right) \Psi$.  Thus in a truncated-space holographic model, the light-front modes $\Psi_+$ or $\Psi_-$ should vanish at the boundary $z = 0$ and $z_0 = 1/\La_{\rm QCD}$. This condition fixes the boundary conditions and determines the baryon spectrum in the truncated hard-wall model~\cite{deTeramond:2005su}:
$M^+ = \beta_{\nu,k} \, \Lambda_{\rm QCD}$, and $M^- = \beta_{\nu+1,k} \, \Lambda_{\rm QCD}$,
 with a scale-independent  mass ratio determined by the zeros of Bessel functions $\beta_{\nu, k}$. Equivalent results follow from the hermiticity of the LF Dirac operator $D_{LF}$ in the eigenvalue equation  $D_{LF} \vert \psi \rangle = \mathcal{M} \vert \psi \rangle$. The orbital excitations of baryons in this model are approximately aligned along  two trajectories corresponding to even and odd parity states~\cite{deTeramond:2012rt, deTeramond:2005su}. The spectrum shows a clustering of states with the same orbital $L$, consistent with a strongly suppressed spin-orbit force.  As for the case for mesons, the hard-wall model predicts $M \sim 2n + L$, in contrast to the usual Regge behavior $M^2 \sim n + L$ found experimentally~\cite{Klempt:2007cp, Klempt:2009pi}.  The radial modes are also not well described in the truncated-space model.

Let us now examine a model similar to the soft-wall dilaton model  for mesons by introducing an effective potential, which also leads to linear Regge trajectories in both the orbital and radial quantum numbers for baryon excited states.  As we have discussed in Sec. \ref{half}, a dilaton factor in the  fermion action can be scaled away by a field redefinition.   We thus choose instead an effective  linear confining potential $V = \lambda_F \zeta$ which reproduces the linear Regge behavior for baryons~\cite{Brodsky:2008pg, Abidin:2009hr}. Choosing $V = \lambda_F \zeta$ we find from  \req{Upm}
\beqa  \label{Uplus}
U^+(\zeta) &=& \lambda_F^2 \zeta^2 + 2 (\nu + 1) \lambda_F ,\\  \label{Uminus}
U^-(\zeta) &=& \lambda_F^2 \zeta^2 + 2 \nu  \lambda_F ,
\enqa
and from  \req{LFWEA} and \req{LFWEB} the two-component solution
\beqa \label{psip}
\psi_+(\zeta) &\sim& 
  \zeta^{\frac{1}{2} + \nu} e^{-\vert \lambda_F\vert \zeta^2/2}
  L_{n}^{\nu}\left(\vert \la_F \vert \zeta^2\right),   \\ \label{psim}
\psi_-(\zeta) &\sim&  \zeta^{\frac{3}{2} + \nu} e^{-\vert \la_F \vert  \zeta^2/2}
 L_{n}^{\nu + 1}\left(\vert \la_F\vert \zeta^2\right).
\enqa

We can compute separately the eigenvalues for the wave equations \req{LFWEA} and \req{LFWEB} for arbitrary $\la_F$ and compare the results  for consistency, since the eigenvalues determined from both equations should be identical.  For the potential \req{Uplus} the eigenvalues of \req{LFWEA} are 
\beq \label{Mplus}
M_+^2 = \left(4n + 2 \nu + 2\right ) \vert \la_F  \vert + 2 \left( \nu  +1 \right) \la_F,
\enq
whereas for the potential \req{Uminus} the eigenvalues of \req{LFWEB} are
\beq
M_-^2 = \left(4n + 2(\nu +1) +2 \right) \vert \la_F \vert + 2 \nu \la_F.
\enq
For $\la_F>0$ we find $M_+^2 = M_-^2 = M^2$ where
\beq \label{M2F}
M^2 = 4 \, \lambda_F  \left(  n +  \nu + 1 \right),
\enq
identical for plus and minus eigenfunctions. For $\la_F<0$ it follows that $M_+^2 \ne M_-^2$ and no solution is possible. Thus the solution $\la_F <0$ is discarded. Note that, as expected, the oscillator form  $\lambda_F^2 \zeta^2$ in the second order equations  \req{LFWEA} and \req{LFWEB}, matches the soft-wall potential for mesons prescribed by the underlying conformality of the classical QCD Lagrangian  as discussed in Chapter \ref{ch3}. We thus set  $\la_F = \la$  reproducing the universality of the Regge slope for mesons and baryons. Notice that in contrast with the meson spectrum \req{M2SFM}, the baryon spectrum  \req{M2F} in the soft wall does not depend on $J$,  an important result also found in Ref. \cite{Gutsche:2011vb}.

\begin{table}[!]    
\begin{center} 
 {\begin{tabular}{@{}ccccc@{}}
 \hline\hline  \\[-3.0ex]
 $SU(6)$ &  $S$ & $ L$ &  $n$ &   Baryon State
 \vspace{1pt}
 \\[0.5ex]
 \hline
 \multicolumn{5}{c}{}\\[-2.5ex]
 ${\bf 56}$  & $\half$ & 0  &  0  & $N{\half^+}(940)$\\[0.0ex]
 {}    &   $\threehalf$& 0 &   0  &$\Delta{\threehalf^+}(1232)$ \\[1.0ex]
 \hline  \\[-2.5ex]
  ${\bf 56} $     &   $\half$ & 0  &  1  & $N{\half^+}(1440)$\\[0.0ex]
  {}  &   $\threehalf$& 0 &   1  &$\Delta{\threehalf^+}(1600)$\\[1.0ex]   
  \hline\\[-2.5ex]
 ${\bf 70}$ & $\half$    & 1 & 0  & $N{\half^-}(1535)~~ N{\threehalf^-}(1520)$ \\[0.0ex]
 {}  & $\threehalf$ & 1 &  0  & $N{\half^-}(1650)~~ N{\threehalf^-}(1700)~~N{\fivehalf^-}(1675)$\\[0.0ex]
 {}  & $\half$ & 1  &  0  &$\Delta{\half^-}(1620)~~ \Delta{\threehalf^-}(1700)$ \\[1.0ex]
 \hline  \\[-2.5ex]
  ${\bf 56}$      &  $\half$ & 0  &  2  & $N{\half^+}(1710)$\\[0.0ex]
 ${}$  & $\half$ & 2  &  0  &$N{\threehalf^+}(1720)~~ N{\fivehalf^+}(1680)$ \\[0.0ex]
 {}    & $\threehalf$ & 2  & 0  & $\Delta{\half^+}(1910)~~ \Delta{\threehalf^+}(1920)
                     ~~ \Delta{\fivehalf^+}(1905)~~\Delta{\sevenhalf^+}(1950)$\\[1.0ex]
 \hline \\[-2.5ex]
{\bf 70}   & $\threehalf$ & 1 &  1  & $N{\half^-}~~~~~~~~ ~~ N{\threehalf^-}(1875)~~N{\fivehalf^-}~~~~~~~~$\\[1.0ex]
 \hline \\[-2.5ex]
  ${}$      & $\threehalf$ & 1  &  1  &$\Delta{\fivehalf^-(1930)} $  \\[1.0ex]
 \hline   \\[-2.5ex]
  ${\bf 56}$      & $\half$ & 2  &  1  &$N{\threehalf^+(1900)}  ~~ N{\fivehalf^+}$~~~~~~~~ \\[1.0ex]
   \hline \\[-2.5ex]
 ${\bf 70}$ &    $\half$ & 3 &  0  & $N{\fivehalf^-}~~ ~~~~~~ N{\sevenhalf^-}$ \\[0.0ex]
 {}   &$\threehalf$ & 3  & 0  &   $N{\threehalf^-}~~~~~~ ~~ N{\fivehalf^-}~~~~~~ ~~
                     N{\sevenhalf^-}(2190)~~ N{\ninehalf^-}(2250)$\\[0.0ex]
 {}  &   $\half$ & 3  & 0   & $\Delta{\fivehalf^-}~~~~~ ~~ \Delta{\sevenhalf^-}$ \\[1.0ex]
 \hline   \\[-2.5ex]
 ${\bf 56}$ &    $\half$ & 4 &  0  & $N{\sevenhalf^+}~~~~~~ ~~ N{\ninehalf^+}(2220)$ \\[0.0ex]
 {}    &$\threehalf $ & 4 &  0  & $\Delta{\fivehalf^+}~~~~~~ ~~ \Delta{\sevenhalf^+} ~~~~~~ ~~
                       \Delta{\ninehalf^+}~~~~~~ ~~\Delta{\elevenhalf^+}(2420)$
 \\[1.0ex]                      
  \hline   \\[-2.5ex]
 ${\bf 70}$   &  $\half$ & 5 &  0  &$N{\ninehalf^-}~~~~~~ ~~ N{\elevenhalf^-}~~~~~~~$ \\[0.0ex]
 {}     &   $\threehalf$ & 5 &  0  & $N{\sevenhalf^-}~~~~~~~ ~~ N{\ninehalf^-}~~~~~~~
      N{\elevenhalf^-}(2600) ~~N{\thirteenhalf^-} $\\[1.5ex]
 \hline\hline
 \end{tabular}}
 \caption{\label{baryons}\small Classification of confirmed baryons listed by the PDG~\cite{PDG:2014}. The labels $L$,  $S$ and  $n$ refer to the internal  orbital angular momentum, internal spin and radial quantum number respectively. The even-parity baryons correspond to the $\bf 56$ multiplet of $SU(6)$ and the odd-parity to the ${\bf 70}$.}
 \end{center}
 \end{table}

 It is important to notice that the solutions \req{psip} and \req{psim} of the second-order differential equations \req{LFWEA} and \req{LFWEB} are not independent
since the solutions must also obey the linear Dirac equation \req{LFDE} \cite{Mueck:1998iz}. This fixes the relative normalization. Using the relation 
$L_{n-1}^{\nu+1}(x) + L_n^\nu(x) = L_n^{\nu+1}(x)$ between the associated Laguerre functions we find for $\la = \la_F >0$
\beqa
\psi_+(\zeta) &=& \la ^{(1+\nu)/2}\sqrt{\frac{2 n!}{(n+\nu - 1)!}} \,
  \zeta^{\frac{1}{2} + \nu} e^{-\la  \zeta^2/2}
  L_{n}^{\nu}\left(\la \zeta^2\right),   \\
\psi_-(\zeta) &=&  \la^{(2+\nu)/2} \frac{1}{\sqrt{n + \nu + 1}} \sqrt{\frac{2 n!}{(n+\nu - 1)!}} \,
 \zeta^{\frac{3}{2} + \nu} e^{- \la   \zeta^2/2}
 L_{n}^{\nu + 1}\left(\la \zeta^2\right),
\enqa
with equal probability
\beq
\int d\zeta \, \psi_+^2(\zeta) =  \int d\zeta \, \psi_-^2(\zeta) =1.
\enq
Equation  (5.41) implies that the spin $S^z_q$ of the quark in the proton has equal probability to be aligned or anti-aligned with the proton's spin $J^z$.   Thus there is equal probability for states  with $L^z_q=0$ and $L^z_q =\pm 1$.  This remarkable equality means that in the chiral limit the proton's spin $J^z$ is carried by the quark orbital angular momentum:  $J^z= \langle L^z_q \rangle = \pm 1/2$. This equality also holds for the hard-wall model.

We list in Table \ref{baryons} the confirmed (3-star and 4-star) baryon states from  updated Particle Data Group~\cite{PDG:2014} \footnote{A recent exploration of the properties of baryon resonances derived from a multichannel partial wave analysis~\cite{Anisovich:2011fc} report additional resonances not included in the Review of Particle Properties~\cite{PDG:2014}.}. To determine the internal spin, internal orbital angular momentum and radial quantum number assignment of the $N$ and $\Delta$ excitation spectrum from the total angular momentum-parity PDG assignment, it is  convenient to use the conventional $SU(6) \supset SU(3)_{flavor} \times SU(2)_{spin}$ multiplet structure~\cite{Lichtenberg:1978pc}, but other model choices are also possible~\cite{Klempt:2009pi} \footnote{In particular the $\Delta\fivehalf^-(1930)$ state (also shown in Table \ref{baryons}) has been given the non-$SU(6)$ assignment $S = 3/2$, $L =1$, $n=1$  in Ref.~\cite{Klempt:2009pi}. This assignment will be further discussed  below.}.

\begin{figure}[htdp]
\centering
\includegraphics[width=15.6cm]{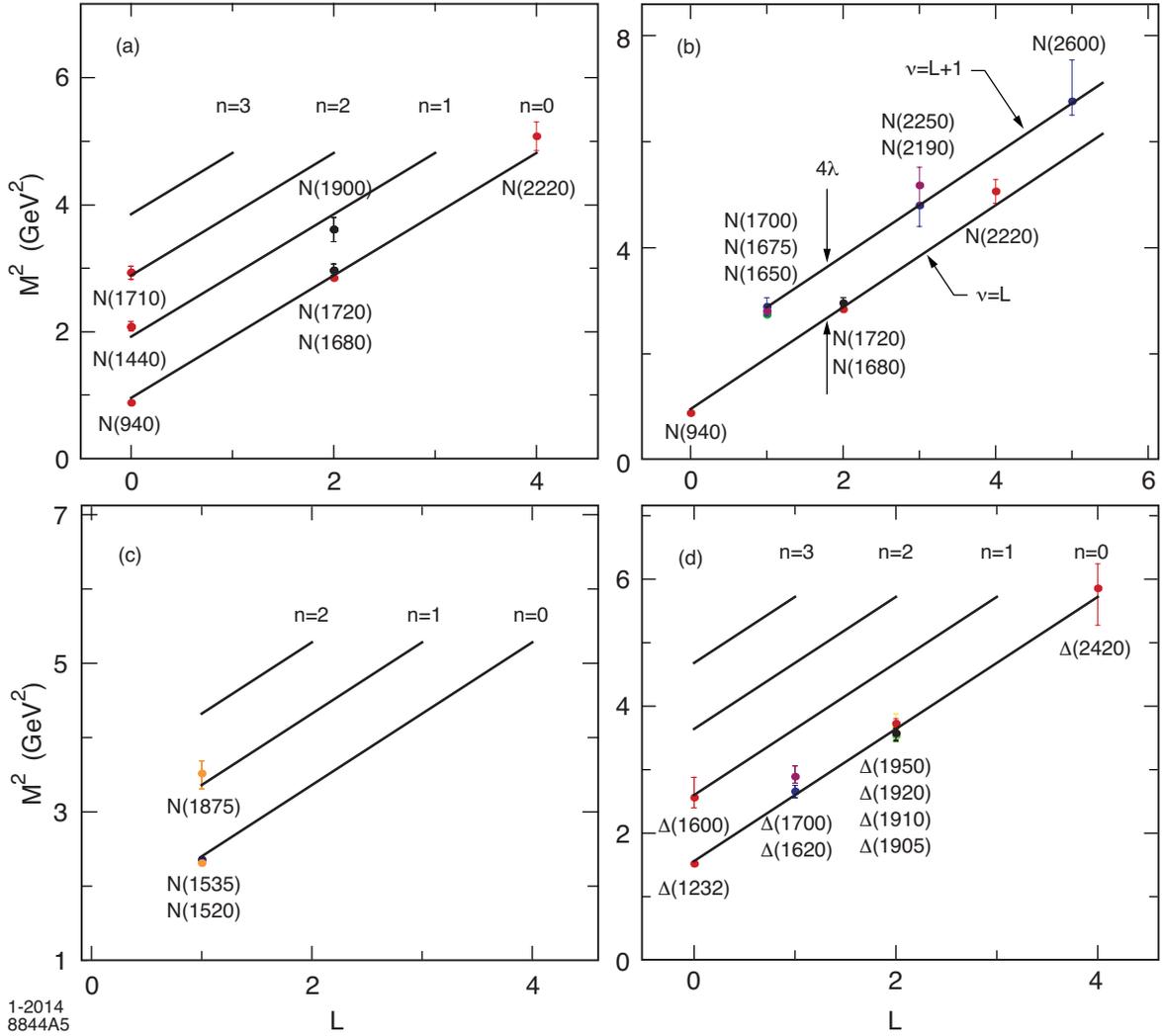}  
\caption{\small Orbital and radial baryon excitation spectrum.  Positive-parity spin-$\half$ nucleons (a) and spectrum gap between the negative-parity spin-$\threehalf$ and the positive-parity spin-$\half$ nucleons families (b). Negative parity $N$ ({c}) and positive and negative parity $\Delta$  families (d).  The values of $\sqrt \la$ are $\sqrt{\la} = 0.49$ GeV (nucleons) and  0.51 GeV (deltas).}
 \label{baryonspec}
\end{figure} 

As for the case of mesons, our first task is to identify the lowest possible stable state, the proton, which corresponds to $n=0$ and $\nu = 0$. This fixes the scale $\sqrt{\lambda} \simeq 0.5$ GeV. The resulting predictions for the spectroscopy of the positive-parity spin-$\half$ light nucleons are shown in  Fig. \ref{baryonspec} (a) for  the parent Regge trajectory for $n =0$ and  $\nu = 0, 2, 4, \cdots,  L $, where $L$ is the relative LF angular momentum between the active quark and the spectator cluster. The predictions for the daughter trajectories for $n=1$, $n = 2, \cdots $ are also shown in this figure. Only confirmed PDG \cite{PDG:2014} states are shown. The Roper state $N(1440)$ and the $N(1710)$ are well accounted for in this model as the first and second radial excited states of the proton. The newly identified state, the $N(1900)$~\cite{PDG:2014} is depicted here as the first radial excitation of the $N(1720)$. The model is successful in explaining the parity degeneracy observed in the light baryon spectrum, such as the $L=2$, $N(1680) - N(1720)$ degenerate pair in Fig. \ref{baryonspec} (a).

In Fig.  \ref{baryonspec} (b) we compare the positive parity  spin-$\half$ parent nucleon trajectory with the negative parity spin-$\threehalf$ nucleon trajectory.  It is remarkable that the gap scale $4 \la$ determines not only the slope of  the trajectories, but also the spectrum gap between the positive-parity spin-$\half$  and the negative-parity spin-$\threehalf$ nucleon families, as indicated by arrows in this figure. This means the respective assignment $\nu = L$ and $\nu = L+1$ for the lower and upper trajectories in Fig. \ref{baryonspec} (b). We  also note that the degeneracy of states with the same orbital quantum number $L$ is also well described, as for example the degeneracy of the $L=1$ states  $N(1650)$, $N(1675)$ and $N(1700)$ in Fig. \ref{baryonspec} (b).

We have also to take into account baryons  with negative parity and internal spin $S = \half$, such as the $N(1535)$, as well as baryon states with positive parity and internal spin $S = \threehalf$, such as the $\Delta(1232)$.  Those states are well described by the assignment $\nu = L + \half$. This means, for example, that  $M^{2 \,(+)}_{n, L, S = \frac{3}{2}} = M^{2 \,(-)}_{n, L, S =  \frac{1}{2}}$ and consequently the positive and negative-parity $\Delta$ states lie in the same trajectory consistent with the experimental results, as depicted in Fig. \ref{baryonspec} (d).  The newly found state, the $N(1875)$ \cite{PDG:2014}, depicted in Fig. \ref{baryonspec} ({c}) is well described as the first radial excitation of the $N(1520)$, and the near degeneracy of the $N(1520)$ and $N(1535)$ is also well accounted. Likewise, the  $\Delta(1660)$ corresponds to the first radial excitation of the $\Delta(1232)$ as shown in Fig. \ref{baryonspec} (d).  The model explains the important degeneracy of  the $L=2$, $\Delta(1905), $ $ \Delta(1910),$ $\Delta(1920), $ $ \Delta(1950)$ states which are degenerate within error bars. The parity degeneracy of the light baryons    is also a property of the  hard-wall model, but in that case the radial states are not well described~\cite{deTeramond:2005su}. Our results for the $\Delta$ states agree with those of Ref.~\cite{Forkel:2007cm}. ``Chiral partners"~\cite{Glozman:1999tk}  such as the $N(1535)$ and the $N(940)$ with the same total angular momentum $J = \half$, but with different orbital angular momentum are non-degenerate from the onset~\footnote{Since our approach is based on a semiclassical  framework,  the Regge trajectories remain linear and there is no chiral symmetry restoration for highly excited states~\cite{Shifman:2007xn}.}. To recapitulate, the parameter $\nu$, which is related to the fifth dimensional AdS mass by the relation $ \vert \mu R \vert  = \nu + 1$,  has the internal spin $S$ and parity $P$ assignment  given in  Table \ref{nuasig} shown below~\cite{deTeramond:2014yga}.

\begin{table}[ht] \label{nuT}
\centering 
\vspace{5pt}
\resizebox{6.5cm}{!} {
\begin{tabular}{ l | c r } 
 & $S = \half$ & \,$S = \threehalf$ \ ~~~~~\\ [1.0ex]
 \hline
P = + & $\nu = L$ &  $ \nu = L + \half$ \\ [1.0ex]
 P = \ --  &  $\nu = L + \half$ &  $\nu = L+1$ \\ [0.6ex]
 \hline
\end{tabular}
}
\caption{\label{nuasig}
\small Orbital assignment for baryon trajectories  according to parity and internal spin.}
\end{table}

The assignment     $\nu = L$ for the lowest trajectory, the proton trajectory,  is straightforward and follows from the mapping of AdS to light-front physics. The assignment for other spin and parity baryons states  given in Table \ref{nuasig} is phenomenological. It is expected that further analysis of the different quark configurations and symmetries  of the baryon wave function, as suggested by the model discussed in Ref.~\cite{Forkel:2008un}, will indeed explain the actual assignment given in this table. This particular assignment successfully describes the full light baryon orbital and radial excitation spectrum, and in particular the gap between trajectories with different parity and internal spin~\cite{deTeramond:2014yga}.   If we follow the non-$SU(6)$ quantum number assignment for the $\Delta(1930)$ given in Ref.~\cite{Klempt:2009pi}, namely  $S = 3/2$, $L =1$, $n=1$  we find  the value $\mathcal{M}_{\Delta(1930)} = 4 \sqrt \la \simeq 2$ GeV, consistent with the experimental result 1.96 GeV~\cite{PDG:2014}.

An important feature of light-front holography is that it predicts a similar multiplicity of states for mesons and baryons,  consistent with what is 
observed experimentally~\cite{Klempt:2007cp}. This remarkable property could have a simple explanation in the cluster decomposition of the
holographic variable \req{zetan}, which labels a system of partons as an active quark plus a system of $n-1$ spectators. From this perspective, a baryon with $n=3$ looks in light-front holography as a quark--diquark system. It is also interesting to notice that in the hard-wall model the proton mass is entirely due to the kinetic energy of the light quarks, whereas in the soft-wall model described here, half of the invariant mass squared $M^2$ of the proton is due to the kinetic energy of the partons, and half is due  to the confinement potential.

\chapter{Light-Front Holographic Mapping and Transition Amplitudes  \label{ch6}}
A form factor in QCD is defined by the transition matrix element of a local quark current between hadronic states.  The great advantage of the front form  -- as emphasized in Chapter \ref{ch2} --  is that boost operators are kinematical.  Unlike in the instant form,  the boost operators in the front form have no interaction terms.  The calculation of a current matrix element $\langle P + q \vert J^\mu \vert P \rangle$ requires boosting the hadronic  eigenstate from $\vert P \rangle $ to $\vert P + q \rangle $, a task which becomes hopelessly complicated in the instant form which includes  changes even in particle number for the boosted state~\cite{Dietrich:2012iy, Hoyer:2014gna}. In fact, the boost of a composite system at fixed time $x^0$ is only known at weak binding~\cite{Brodsky:1968xc,  Brodsky:1968ea}.  In addition, the virtual photon couples to connected currents which arise from the instant-form vacuum.

In AdS space form factors are computed from the overlap integral of normalizable modes with boundary currents which propagate in AdS space. The AdS/CFT duality incorporates the connection between the twist-scaling dimension of the  QCD boundary interpolating operators with the falloff of the normalizable modes in AdS near its conformal boundary~\cite{Polchinski:2001tt}. If both quantities represent the same physical observable, a precise correspondence can be established at any momentum transfer  between   the string modes $\Phi$ in AdS space and the light front wave functions of hadrons $\psi$ in physical four-dimensional space-time~\cite{Brodsky:2006uqa}. In fact, light-front holographic methods were originally  derived by observing the correspondence between matrix elements obtained in AdS/CFT with the corresponding formula using the light-front representation~\cite{Brodsky:2006uqa}. As shown in Chapter \ref{ch4} the same results follow from comparing the relativistic light-front Hamiltonian equation describing bound states in QCD with the wave equations describing the propagation of modes in a warped AdS space for arbitrary spin~\cite{deTeramond:2008ht, deTeramond:2013it}.

Form factors are among the most basic observables of hadrons, and thus central for our understanding of hadronic structure and dynamics.  The physics includes the important interplay of perturbative and nonperturbative elements, which if properly taken into account, should allow us to study the transition from perturbative dynamics at large momentum transfer $q^2$ to non-perturbative dynamics at moderate and small $q^2$. Thus,  the transition from quark and gluon degrees of freedom to hadronic degrees of freedom, which is not a simple task.

As will become clear from our discussion in this Chapter, holographic QCD  incorporates important elements for the study of hadronic form factors which encompasses perturbative and nonperturbative elements, such as the connection between the twist of the hadron to the fall-off of its current matrix elements for large $q^2$, and essential aspects of vector meson dominance which are relevant at lower energies.  This framework is also useful for analytically continuing the space-like results to the time-like region using simple analytic formulas expressed in terms of vector meson masses.

\section{Meson electromagnetic form factor \label{MFF}}

\subsection{Meson form factor in AdS space}

In the higher dimensional gravity theory, the hadronic transition matrix element corresponds to the  coupling of an external electromagnetic (EM) field $A^M(x,z)$,  for a photon propagating in AdS space, with the extended field $\Phi_P(x,z)$ describing a hadron in AdS~\cite{Polchinski:2002jw}.  To simplify the discussion we treat here first the electromagnetic form factor for a spinless particle in a  model with a wall at  a finite distance  $z = 1/\Lambda_{\rm QCD}$ -- the hard wall model, which limits the propagation of the string modes in AdS space beyond  the IR boundary, and also sets the hadronic mass scale~\cite{Polchinski:2001tt}.  The coupling of the EM field  $A^M(x,z)$ follows from minimal coupling by replacing in \req{action1} or \req{action2} the covariant derivative $D_M$ by $D_M - i {\rm e}_5 A_M$, where ${\rm e}_5$ is the charge in the bulk theory.  To first order in the EM field the interaction term is
\beq
S_{int} = {\rm e}_5  \int \! d^4x \, dz  \sqrt{ g} \,  g^{M M'} \Phi^*(x,z) i \overleftrightarrow\partial_{\! \! M} \Phi(x,z) \,  A_{M'} (x,z),
\enq
where $g \equiv \vert \rm {det } \, g_{MN}  \vert$.
 We recall from Chapter \ref{ch4} that the coordinates of AdS$_5$ are the Minkowski coordinates $x^\mu$ and $z$ labeled $x^M = (x^\mu, z)$,
 with $M, N = 0, \cdots 4$,  and $g$ is the determinant of the metric tensor.  The hadronic transition matrix element has thus the form~\cite{Polchinski:2002jw}
\beq
\label{MFF}
 {\rm e}_5 \! \int \! d^4x \, dz  \sqrt{g} \,   \Phi^*_{P'}(x,z) \overleftrightarrow\partial_{ \! \!  M}\Phi_P(x,z)  A^M(x,z)
  \sim
  (2 \pi)^4 \delta^4 \left( P'  \! - P - q\right) \epsilon_\mu  (P + P')^\mu {\rm e}  F(q^2) ,
 \enq
where, the pion has initial and final four momentum $P$ and $P'$ respectively and $q$ is the four-momentum transferred to the pion by the photon with polarization $\epsilon_\mu$. The expression on the right-hand side of (\ref{MFF}) represents the space-like QCD electromagnetic transition amplitude in physical space-time
\begin{equation}
\langle P' \vert J^\mu(0) \vert P \rangle = \left(P + P' \right)^\mu F(q^2).
\end{equation}
It is the EM matrix element of the quark current  $J^\mu = e_q \bar q \gamma^\mu q$, and represents a local coupling to pointlike constituents. Although the expressions for the transition amplitudes look very different, one can show  that a precise mapping of the matrix elements  can be carried out at fixed light-front time for an arbitrary number of partons in the bound-state~\cite{Brodsky:2006uqa, Brodsky:2007hb}.

The propagation of the pion in AdS space is described by a normalizable mode $\Phi_P(x^\mu, z) = e^{i P  \cdot x} \Phi(z)$ with invariant  mass $P_\mu P^\mu = M^2$ and plane waves along the physical coordinates $x^\mu$.   The physical incoming electromagnetic probe (no physical polarizations along the AdS coordinate $z$) propagates in AdS according to 
\beq \label{Az}
A_\mu(x^\mu ,z) = e^{ i q \cdot x} V(q^2, z)  \epsilon_\mu(q),   \quad  A_z = 0,
\enq
where $\ep$ is  the EM polarization vector in 4 dimensions, with $q \cdot \ep=0$. The function $V(q^2,z)$ -- the bulk-to-boundary propagator, has the value 1 at zero momentum transfer, since we are normalizing the solutions to the total charge operator. It also has the value 1 at  $z = 0$, since the boundary limit is the external current:  $A_\mu(x^\mu ,z \to 0) = e^{ i q \cdot x}  \epsilon_\mu(q)$. Thus
\beq \label{BCV}
V(q^2 = 0, z ) = V(q^2, z = 0) = 1.
\enq
Extracting the overall factor  $(2 \pi)^4 \delta^4 \left( P'  \! - P - q\right)$ from momentum conservation at the vertex, which arises from integration over Minkowski variables in (\ref{MFF}), we find~\cite{Polchinski:2002jw} 
\beq  \label{FFHW}
F(q^2) = R^3 \int_0^{1/\Lambda_{\rm QCD}} \frac{dz}{z^3} \, V(q^2, z)  \, \Phi^2(z),
\enq
where $F( 0) = 1$.  The pion form factor in AdS is the overlap of the normalizable modes corresponding to the incoming and outgoing hadrons $\Phi_P$ and $\Phi_{P'}$ with the non-normalizable mode $V(q^2,z)$, corresponding to the external EM current~\cite{Polchinski:2002jw}~\footnote{The equivalent expression for a spin-$J$ meson is $F(q^2) = R^{3- 2J} \int \frac{dz}{z^{3 - 2 J}} \, V(q^2, z)  \, \Phi_J^2(z)$, where the hadronic mode $\Phi_J$ is normalized according to \req{Phinorm} for $d = 4$.}.

\subsection{Meson form factor in light-front QCD}

The light-front formalism provides an exact Lorentz-invariant representation of current  matrix elements in terms of the overlap of light-front wave functions. The electromagnetic current has elementary couplings to the charged constituents since the full Heisenberg current can be replaced in the interaction picture by the free quark current $J^\mu(0)$, evaluated at fixed light-cone time $x^+=0$ in the $q^+=0$ frame~\cite{Drell:1969km}. In contrast to the covariant Bethe-Salpeter equation, in the light front Fock expansion one does not need to include the contributions to the current from an infinite number of irreducible kernels, or the interactions of the electromagnetic current with vacuum fluctuations~\cite{Brodsky:1997de}.

In the front-form, the EM form factor is most conveniently computed  from the matrix elements of the plus component of the current component  $J^+$ at LF time $x^+ = 0$
\beq \label{Jplus}
\langle P' \vert J^+(0) \vert P \rangle = \left(P + P' \right)^+ F(q^2).
\enq
This component of the current does not couple to Fock states with different numbers of constituents in the $q^+=0$ frame~\cite{Drell:1969km}. We express the plus component of the current operator 
\beq
J^+(x) = \sum_q e_q\bar\psi(x) \gamma^+ \psi(x),
\enq
in the particle number representation  from the momentum expansion of $\psi(x)$ in terms of creation and annihilation operators (\ref{eq:psiop})~\footnote{Notice that $\gamma^+$ conserves the spin component of the struck quark (Appendix \ref{metric}), and thus the current $J^+$ only couples Fock states with the same number of constituents.}. The matrix element \req{Jplus} is then computed by expanding the initial and final meson states $\vert \psi_M(P^+ \! , \mbf{P}_\perp)\rangle$  in terms of its Fock components \req{eq:LFWFexp}. Using the normalization condition \req{eq:normFC} for each individual constituent,  and after integration over the intermediate variables in the $q^+ = 0$ frame  we obtain the  Drell-Yan-West expression~\cite{Drell:1969km, West:1970av} 
 \begin{equation} \label{eq:DYW}
F_M(q^2) = \sum_n  \int \big[d x_i\big] \left[d^2 \mbf{k}_{\perp i}\right]
\sum_j e_j \psi^*_{n/M} (x_i, \mbf{k}'_{\perp i},\lambda_i)
\psi_{n/M} (x_i, \mbf{k}_{\perp i},\lambda_i),
\end{equation}
where the phase space momentum integration $[d x_i\big] \left[d^2 \mbf{k}_{\perp i}\right]$ is given by \req{fsx} and \req{fsk}, and the variables of the light-front  Fock components in the final state are given by $\mbf{k}'_{\perp i} = \mbf{k}_{\perp i} + (1 - x_i)\, \mbf{q}_\perp $ for a struck  constituent quark and
$\mbf{k}'_{\perp i} = \mbf{k}_{\perp i} - x_i \, \mbf{q}_\perp$ for each spectator. The formula is exact if the sum is over all Fock states $n$.

The form factor can also be conveniently written in impact space as a sum of overlap of LFWFs of the $j = 1,2, \cdots, n-1$ spectator constituents~\cite{Soper:1976jc}.  Suppose that the charged parton $n$ is the active constituent struck by the current, and the quarks $i = 1,2, \cdots ,n-1$ are spectators. We substitute (\ref{eq:LFWFb}) in the DYW formula (\ref{eq:DYW}). 
Integration over $k_\perp$ phase space gives us $n - 1$ delta functions to integrate over the $n - 1$ intermediate transverse variables with the result
\beq \label{eq:FFb} 
F_M(q^2) =  \sum_n  \prod_{j=1}^{n-1}\int d x_j d^2 \mbf{b}_{\perp j} 
\exp \! {\Bigl(i \mbf{q}_\perp \! \cdot \sum_{j=1}^{n-1} x_j \mbf{b}_{\perp j}\Bigr)} 
\left\vert  \psi_{n/M}(x_j, \mbf{b}_{\perp j})\right\vert^2,
\enq
corresponding to a change of transverse momentum $x_j \mbf{q}_\perp$ for each of the $n-1$ spectators. This is a convenient form for comparison with  AdS results, since the form factor
is expressed in terms of the product of light-front wave functions  with identical variables.

\subsection{Light-front holographic mapping \label{EMLFHM}}

We now have all the elements to establish a connection of the AdS and light-front formulas. For definiteness we shall consider  the $\pi^+$  valence Fock state $\vert u \bar d\rangle$ with charges $e_u = \frac{2}{3}$ and $e_{\bar d} = \frac{1}{3}$. For $n=2$, there are two terms which contribute to Eq. (\ref{eq:FFb}). Integrating over angles and exchanging  $x \leftrightarrow 1 \! - \! x$ in the second integral  we find 
\begin{equation}  \label{eq:PiFFb}
 F_{\pi^+}(q^2)  =  2 \pi \int_0^1 \! \frac{dx}{x(1-x)}  \int \zeta d \zeta \,
J_0 \! \left(\! \zeta q \sqrt{\frac{1-x}{x}}\right) 
\left\vert \psi_{u \bar d/ \pi}\!(x,\zeta)\right\vert^2,
\end{equation}
where $\zeta^2 =  x(1  -  x) \mathbf{b}_\perp^2$ and $F_{\pi^+}(0)=1$.

We now compare this result with the electromagnetic  form factor in  AdS.   Conserved currents are not renormalized and correspond to five-dimensional massless  fields propagating in AdS$_5$ space according to the relation  $(\mu R)^2 = (\Delta - p) (\Delta - 4 + p)$ for a $p$-form field in AdS space \req{Deltamup}. This  corresponds   for $\mu = 0$ and $p = 1$ to either $\Delta = 3$ or 1,  the canonical dimensions of an EM current and the massless gauge field respectively. The equation of motion describing the propagation of the electromagnetic field   in AdS space is obtained from the action
\beq \label{sem}
S_{em}=\int d^dx \, dz \,\sqrt{g}   \; g^{MM'}\,g^{NN'} F_{MN}\,F_{M'N'} ,
\enq with the covariant field  tensor $F_{MN} = \pa_M A_N- \pa_N A_M$. It gives for  $V(Q^2,z)$ (Eq. \req{Az}) the wave equation
\beq
\left[ \frac{d^2}{d z^2} - \frac{1}{z} \frac{d}{d z} - Q^2 \right] V \! \left(Q^2, z\right) = 0,
\enq
where $Q^2 = - q^2 > 0$. Its solution,  subject to the boundary conditions \req{BCV}, is
\beq 
\label{eq:V}
V(Q^2, z) = z Q K_1(z Q),
\enq
which decays exponentially for large values of $Q^2$: here $K_1(Q z) \sim \sqrt{\frac{\pi}{2 Q z}\,}e^{-Q z}$~\footnote{This solution corresponds to a ``free" EM current in physical space.  Confined EM currents in AdS correspond to ``dressed" currents in QCD. This will be discussed in the next section.}.    Using the integral representation of $V(Q^2,z)$ from \req{JKint}
\begin{equation} \label{eq:intJ}
V(Q^2, z) = \int_0^1 \! dx \, J_0 \! \left(\! z  Q \sqrt{\frac{1-x}{x}}\right) ,
\end{equation} we can write the AdS electromagnetic form-factor as
\begin{equation} 
F(Q^2)  =    R^3 \! \int_0^1 \! dx  \! \int \frac{dz}{z^3} \, 
J_0\!\left(\!z Q\sqrt{\frac{1-x}{x}}\right)  \Phi^2(z) .
\label{eq:AdSFx}
\end{equation}

To compare with  the light-front QCD  form factor expression (\ref{eq:PiFFb})  we  use the expression of the  light-front wave function  (\ref{eq:psiphi})
\begin{equation} \label{e:psiphi}   
\psi(x,\zeta, \varphi) = e^{i L \varphi} X(x) \frac{\phi(\zeta)}{\sqrt{2 \pi \zeta}} , 
\end{equation}
which we use to  factor out the longitudinal  and transverse modes $\phi(\zeta)$ and $X(x)$  in \req{eq:PiFFb}. Both expressions for the form factor have to  be identical for arbitrary values of $Q$.  We obtain the result~\cite{Brodsky:2006uqa} 
\beq
\phi(\zeta) =  \left(\frac{R}{\zeta}\right)^{-3/2} \Phi(\zeta)  \quad\quad  {\rm and}   \quad\quad  X(x) = \sqrt{x(1-x)},
\enq
where we identify the transverse impact LF variable $\zeta$ with the holographic variable $z$,
$z \to \zeta = \sqrt{x(1-x)} \vert \mbf b_\perp \vert$.~\footnote{Extension of the results to arbitrary $n$ follows from the $x$-weighted definition of the transverse impact variable of the $n-1$ spectator system given by Eq. (\ref{zetan}). In general the mapping relates the AdS density  $\Phi^2(z)$ to an effective LF single particle transverse density~\cite{Brodsky:2006uqa}. }
Thus, in addition of recovering the expression found in Chapter \ref{ch5}, which relates the transverse mode $\phi(\zeta)$ in physical space-time to the field $\Phi(z)$ in AdS space from the  mapping to the LF Hamiltonian equations, we find a definite expression for the longitudinal LF mode $X(x)$~\footnote{It is interesting to notice that  computations based on lattice QCD and rainbow-ladder truncation of  Dyson-Schwinger equations of twist-two parton distribution amplitudes give similar results for the longitudinal component $X(x)$~\cite{Segovia:2013eca, Gao:2014bca}.}. The identical result follows from mapping the matrix elements of the energy-momentum tensor~\cite{Brodsky:2008pf} (See Appendix \ref{EMT}).

Although the expression for the form-factor \req{FFHW} is derived in the simple hard-wall model, the power falloff for large $Q^2$ is model independent.  This follows from the fact that the leading large-$Q^2$ behavior of form factors in AdS/QCD arises from the small $z \sim 1/Q$  kinematic domain in AdS space. According to the AdS/CFT duality (See Chapter \ref{ch1}), this corresponds to small distances $x_\mu x^\mu \sim 1/Q^2$ in physical space-time, the domain where the current matrix elements are controlled by the conformal twist-dimension $\tau$ of the hadron's interpolating operator. In the case of the front form, where $x^+= 0$, this corresponds to small transverse separation $x_\mu x^\mu  =  -  {\bf x}_\perp^2$.  In general, the short-distance behavior of a hadronic state is characterized by its twist  (dimension minus spin)  $\tau = \Delta - \sigma$, where $\sigma$ is the sum over the constituent's spin $\sigma = \sum_{i = 1}^n \sigma_i$. Twist is also equal to the number of partons $\tau = n$~\footnote{For a hadronic state with relative orbital angular momentum $L$ the twist is $\tau = n + L$.}.

In a high-energy electron-proton elastic collision experiment, for example,  the photon propagation is near to the light-cone, and thus its short space-like interval maps to the boundary of AdS near $z=0$ (Chapter \ref{ch1}). This means that the photon propagation function $V(Q^2,z)$ is strongly suppressed in the AdS interior. At large enough $Q^2$ the important contribution to the integral in \req{FFHW} is from the asymptotic boundary region near $z \sim 1/Q$ where the function $V(Q^2,z)$ has its important support. At small $z$ the string modes scale as $\Phi \sim z^\De$, and the ultraviolet point-like power-scaling behavior (instead of a soft collision amplitude)  is recovered~\cite{Polchinski:2001tt}
\beq \label{asymFF}
F(Q^2)  \to \left[ \frac{\La^2_{\rm QCD}}{Q^2} \right]^{\De - 1},
\enq
upon the substitution $\De \to n$.

It is remarkable that  the QCD dimensional counting rules~\cite{Brodsky:1973kr, Brodsky:1974vy, Matveev:ra} are also a key feature of nonperturbative models~\cite{Polchinski:2001tt} based on the gauge/gravity duality.  If fact, the phenomenological success of dimensional scaling laws implies that QCD is a strongly coupled conformal theory at moderate, but not asymptotic, energies~\footnote{For small $z$  it follows that $\phi_J \sim z^{1/2 + L}$ and thus from \req{Phiphi}  $\Phi_J \sim z^{3/2 - J} \phi_J \sim z^{2 + L -J}$, in agreement with the pion twist $\tau = n + L$ for $n=2$.}.  In hard exclusive scattering there is little sign of the logarithmic running of the QCD coupling from QCD perturbative predictions~\cite{Lepage:1980fj}. For example, the measured proton Dirac form factor $F_1$  scales as $Q^4 F_1(Q^2) \simeq$ constant, up to $Q^2 \le 35\, {\rm GeV}^2$~\cite{Diehl:2005wq}. This puzzling behavior could have an explanation in the fact that in exclusive reactions the virtualty of the gluons exchanged in the hard QCD processes is typically much less than the momentum transfer scale $Q$, since several gluons share the total momentum transfer, and thus the $Q^2$-independence of the strong coupling is tested in the conformal IR window. Since the simple nonperturbative counting rules \req{asymFF} encode the conformal aspects of the theory, the holographic predictions seem to explain quite well the exclusive data in this large, but not asymptotically large energy range.

\begin{figure}[h]
\centering
\includegraphics[angle=0,width=8.6cm]{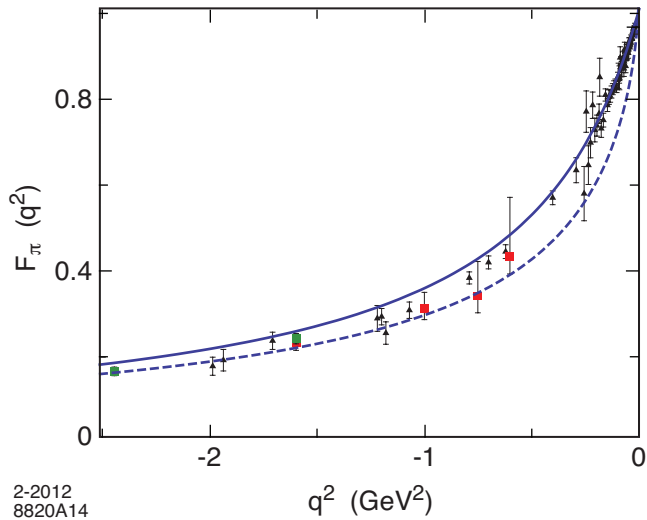} 
\caption{\label{PionFF} \small Space-like electromagnetic pion form factor $F_\pi(q^2)$. Continuous line: confined current, dashed  line: free current. Triangles are the data compilation  from Baldini~\cite{Baldini:1998qn}, boxes  are JLAB data~\cite{Tadevosyan:2007yd, Horn:2006tm}. }
\end{figure} 

The results for the elastic form factor described above correspond to a `free' current propagating on AdS space. It is dual to the electromagnetic
point-like current in the Drell-Yan-West light-front formula~\cite{Drell:1969km, West:1970av} for the pion form factor.  
The DYW formula is an exact expression for the form factor. It is written as an infinite sum of an overlap of LF Fock components with an arbitrary number of constituents. This allows one to map state-by-state to the effective gravity theory in AdS space. However, this mapping has the shortcoming that the nonperturbative pole structure of the time-like form factor does not appear in the time-like region unless an infinite number of Fock states is included. Furthermore, the moments of the form factor at  $Q^2 = 0$ diverge term-by-term; for example one obtains an infinite charge radius~\cite{deTeramond:2011yi} as shown in Fig. \ref{PionFF}~\footnote{This deficiency is solved by taking into account finite quark masses. In this case the charge radius becomes finite for free currents in the DYW formula.}  In fact, infinite slopes also occur in chiral theories when coupling to a massless pion.

\subsubsection{Pion form factor with confined AdS current}

The description of form factors in AdS has the feature that the time-like pole structure is incorporated in the EM current when the current is `confined', {\it i. e.}, the EM current is modified as it propagates in an IR modified AdS space to incorporate confinement. In this case, the confined current in AdS is dual to a hadronic EM current which includes any number of virtual $q \bar q$ components.  The confined EM  current  also leads to finite moments at $Q^2=0$, since a hadronic scale is incorporated in the EM current. This is  illustrated in  Fig. \ref{PionFF} for the EM pion form factor. 

As a specific  example,  consider the hard-wall model with a wall at a finite distance $z  = 1/ \La_{\rm QCD}$~\footnote{A logarithmically divergent result for the pion radius does not appear in the hard-wall model if one uses Neumann boundary conditions for the EM current. In this case the EM current is confined and $\langle r^2_\pi \rangle \sim 1 / \Lambda^2_{\rm QCD}$.}. The gauge-invariant boundary conditions for the confined EM field lead to the expression~\cite{Hong:2004sa} 
\beq  \label{VLa}
V(Q^2 \!, z) = z\, Q \left[ K_1(z \, Q) + I_1(z \, Q)  \, \frac{K_0(Q/\La_{\rm QCD})}{I_0(Q/\La_{\rm QCD})} \right],
\enq
for the bulk-to-boundary propagator, where an infinite series of  time-like poles in the confined AdS current corresponds to the zeros of the Bessel function $I_0(Q/\La_{\rm QCD})$. This is conceptually very satisfying,   since by using the relation  $J_\al(i x) = e^{ i \al \pi /2} I_\al(x)$ for the modified Bessel function $I_\al(x)$, it follows that the poles in  \req{VLa} are determined by the  dimension 2 solution of the hadronic wave equation for $L = 0$, even if the EM current itself scales with dimension 3.  Thus, the poles in the current correspond to the  mass spectrum of radial excitations computed in the hard-wall model. Notice that the scaling dimension 2 corresponds   to leading twist $\tau = 2$ quark-antiquark bound state, whereas the scaling dimension 3 corresponds to the naive conformal dimension of the EM conserved  current.  The downside of the hard-wall model, however, is that the spectrum of radial excitations in this model behaves as $M \sim 2 n$, and thus this model is not able to describe correctly  the radial vector-meson excitations, and consequently neither the time-like form factor data. The observed $\rho$ vector-meson radial trajectory has instead the Regge behavior $M^2 \sim n$. In the limit for large $Q^2$ we recover the `free' current propagating in AdS, given by \req{eq:V}.  A discussion of  pion and vector form factors  in the hard-wall model is given in Refs. \cite{Hong:2004sa, Grigoryan:2007vg, Grigoryan:2007wn, Grigoryan:2008cc, Kwee:2007dd, Kwee:2007nq}.

\subsection{Soft-wall form factor  model \label{SWFF}}

We can extend the computation of form factors with a confined current for the soft wall model.  For a general dilaton profile one need to introduce a $z$-dependent AdS effective  coupling $e_5(z)$.  This procedure does not affect  gauge symmetry  in  asymptotic physical Minkowski space, and generally for any fixed value of the holographic variable  $z$. As it turns out,  the functional dependence on $z$ is determined by the requirements of charge conservation in Minkowski space at $Q^2 = 0$.  This is analogous to the  introduction of a $z$-dependent mass in \req{action2},  which was fixed by the requirement of a separation of dynamical and kinematical features.  Following the same steps as for the hard-wall model  discussed above, we find for the pion form factor
\beq  \label{FFSW}
{\rm e} F(Q^2) = R^3 \int  \frac{dz}{z^3} \, e^{\vp(z)} {\rm e}_5(z) V(Q^2, z)  \, \Phi^2(z),
\enq
with boundary conditions
\beq \label{bcdil}
 \frac{1}{\rm e} \, \lim_{Q^2 \to 0}  \, e_5(z) V(Q^2, z) = \frac{1}{\rm e} \,   \lim_{z \to 0}  e_5(z) V(Q^2, z) = 1.
\enq

To find the behavior of the bulk-to-boundary propagator we consider the dilaton-modified action for the EM field in AdS
\beq \label{sem}
S_{em}=\int d^dx \, dz \,\sqrt{g} e^{\vp(z)}  \; g^{MM'}\,g^{NN'} F_{MN}\,F_{M'N'}.
\enq 
Its variation gives  the wave equation
\beq
\left[ \frac{d^2}{d z^2} - \left(\frac{1}{z} - \vp'(z)\right) \frac{d}{d z} - Q^2 \right] V \! \left(Q^2, z\right) = 0.
\enq

For the harmonic dilaton profile $\vp(z) = \la z^2$ its non-normalizable solution is the EM bulk-to-boundary propagator~\cite{Brodsky:2007hb, Grigoryan:2007my}
(See Appendix \ref{solutions})
\beq \label{soldressed}
V(Q^2,z) = e^{(-|\la| - \la)\,z^2/2}\; \Ga\!\left[1 +\frac{Q^2}{4 |\la|}\right] \;  U\!\left[ \frac{Q^2}{4 |\la|},0,|\la| z^2\right]
\enq
where $U(a,b,c)$ is the Tricomi confluent hypergeometric function
\begin{equation}
\Gamma(a)  U(a,b,z) =  \int_0^\infty \! e^{- z t} t^{a-1} (1+t)^{b-a-1} dt.
\end{equation}
The current \req{soldressed} has the limiting  values  $V(0,z) = e^{(-|\la| - \la)\,z^2/2}$  and  $V(Q^2,0)=1$.   

We can determine the $z$-dependence of the AdS coupling ${\rm e}_5$ from charge conservation, $F(0) = 1$, in the limit $Q \to 0$ using \req{bcdil}.   This requirements fixes  for $\la <0$ the $z$-dependence of  $e_5(z) =  \rm e$ independent of $z$, and for $\la >0$  to ${\rm e}_5 = {\rm e} \,e^{\la z^2}$. Thus the effective current is 
\beq \label{Vmod}
 \tilde V (Q^2, z) =  \Ga\!\left[1 +\frac{Q^2}{4 |\la|}\right] \;  U\!\left[ \frac{Q^2}{4 |\la|},0,|\la| z^2\right],
\enq
where $\tilde V(z) = \frac{1}{\rm e} {\rm e}_5 V(z)$.   The modified current $\tilde V(Q^2,z)$, Eq. (\ref{Vmod}),  has the same boundary conditions (\ref{BCV}) as the free current (\ref{eq:V}),
and reduces to (\ref{eq:V}) in the  limit $Q^2 \to \infty$~\cite{Brodsky:2007hb}.

The soft-wall model of confinement~\cite{Karch:2006pv} also has important analytical properties which are particularly useful for the study of transition amplitudes. As shown in Ref.~\cite{Grigoryan:2007my} the bulk-to-boundary propagator  (\ref{Vmod}) has  the integral representation
\begin{equation}  \label{Vx}
\tilde V(Q^2,z) = \vert \la \vert  z^2 \int_0^1 \! \frac{dx}{(1-x)^2} \, x^{Q^2 / 4 \vert \la \vert}  e^{-\vert \la \vert z^2 x/(1-x)}.
\end{equation} 
Since the integrand in \req{Vx} contains the generating function of the associated Laguerre polynomials
\begin{equation}
\frac{e^{-\vert \la \vert  z^2 x/(1-x)}}{(1-x)^{k+1}} =   \sum_{n=0}^{\infty} L_n^k(\vert \la \vert  z^2) x^n,
\end{equation}
$\tilde V(Q^2, z)$ can thus be expressed as a sum of poles~\cite{Grigoryan:2007my}
\begin{equation} \label{Vsum}
\tilde V(Q^2, z) = 4 \la^2 z^2 \sum_{n=0}^{\infty}  \frac{L_n^1(\vert \la \vert  z^2)}{ M_n^2 + Q^2},
\end{equation}
with $M_n^2 = 4 \vert \la \vert (n+1)$. 

For negative values of $Q^2$ (time-like), the poles of the dressed current \req{Vsum} occur at $-Q^2 =  4  \vert \la \vert (n + 1)$. On the other hand, the poles  of the observed vector mesons with  quantum numbers $J=1, \; L=0$ according to the  bound-state equation \req{M2SFM},  should occur at $-Q^2 =  4 |\la| (n+\half)$~\footnote{For a negative dilaton profile~\cite{Karch:2006pv} the vector meson radial trajectory corresponds to the quantum numbers $J = L = 1$ and thus the poles are located at $-Q^2 =  4  \vert \la \vert (n + 1)$. This identification, however,   is not compatible with  light-front QCD.}.   From here on, we will shift the vector mesons mass poles to their physical twist-2 location to obtain a meaningful comparison with measurements. When this is done, the agreement with data is very good~\cite{deTeramond:2012rt}.

Let us now compute the elastic EM form factor  corresponding to the lowest radial $n = 0$ and orbital $L=0$  state for an arbitrary twist $\tau$ described by the hadronic state
\begin{equation}  \label{Phitau}
\Phi_\tau(z) =   \sqrt{\frac{2 }{\Gamma(\tau \! - \! 1)} } \, \kappa^{\tau -1} z ^{\tau} e^{- \kappa^2 z^2/2},
\end{equation} with normalization.
\begin{equation} \label{PhitauNorm}
\langle\Phi_\tau\vert\Phi_\tau\rangle = \int \frac{dz}{z^3} \,  \Phi_\tau(z)^2  = 1.
\end{equation}
This agrees with the fact that the field $\Phi_\tau$ couples to a local hadronic interpolating operator of twist $\tau$ defined at the asymptotic boundary of AdS space, and thus the scaling dimension of $\Phi_\tau$ is $\tau$.  For convenience we have redefined the wave function to absorb the dilaton profile.  To compute the form factor
\beq  \label{FFtau}
F_\tau(Q^2) =  R^3 \int  \frac{dz}{z^3}  \tilde V(Q^2, z)  \, \Phi_\tau^2(z),
\enq
we substitute in \req{FFtau} the field \req{Phitau}  and the bulk-to-boundary propagator (\ref{Vx}).  Upon integration over the variable  $z$ we find the result~\cite{Brodsky:2007hb} 
\begin{equation} \label{eq:FFxSW}
F_\tau(Q^2) =  \int_0^1 dx \, \rho_\tau(x,Q),
\end{equation}
where
\begin{equation}  \label{rhotau}
\rho_\tau(x,Q) = (\tau \!-\!1) \, (1 - x)^{\tau-2} \, x^{\frac{Q^2}{4 \kappa^2}} .
\end{equation}
The integral (\ref{eq:FFxSW}) can be expressed in terms of Gamma functions
\begin{equation} \label{eq:FFSWtau}
F_\tau(Q^2) = \Gamma(\tau)  
\frac{\Gamma\left(1\! + \! \frac{Q^2}{4 \kappa^2}\right)}{\Gamma\left(\tau \! + \! \frac{Q^2}{4 \kappa^2}\right)}.
\end{equation}
For integer  twist-$\tau$ (the number of constituents $N$) for a given Fock component
 we find~\cite{Brodsky:2007hb} 
 \begin{equation} \label{Ftau}   
 F_\tau(Q^2) =  \frac{1}{{\Big(1 + \frac{Q^2}{\mathcal{M}^2_\rho} \Big) }
 \Big(1 + \frac{Q^2}{\mathcal{M}^2_{\rho'}}  \Big)  \cdots 
       \Big(1  + \frac{Q^2}{\mathcal{M}^2_{\rho^{\tau-2}}} \Big)} ,
\end{equation}
which is  expressed as a $\tau - 1$ product of poles along the vector meson Regge radial trajectory.  For a pion, for example, its lowest Fock state -- the valence state -- is a twist-2 state, and thus the form factor is the well known monopole form~\cite{Brodsky:2007hb}. For the proton, the minimal Fock state is a twist-3 state, and the corresponding form factor is the product of two monopoles, corresponding to the two lowest vector meson states.  It is important to notice that even if  the confined EM dressed current (\ref{Vsum}) contains an infinite number of poles, the actual number of poles appearing in the expression for the elastic form factor (\ref{Ftau})  is determined by the twist of the Fock component: the resulting form factor is given by a product of $N-1$ poles for an $N$-component Fock state.
The remarkable analytical form of (\ref{Ftau}), expressed in terms of the lowest vector meson mass and its radial excitations, incorporates not only the correct leading-twist scaling behavior expected from the constituent's hard scattering with the photon but also  vector meson dominance (VMD)  al low energy~\cite{Sakurai:1960ju}  and a finite mean-square charge radius  $\langle r^2\rangle \sim \frac{1}{\la}$~\footnote{In contrast, the computation with a free current gives the logarithmically divergent result $\langle r^2\rangle \sim  \frac{1}{\la}  \ln\left(\frac {4 \kappa^2}{Q^2}\right)\Big\vert_{Q^2 \to 0}$.}. The light-front holographic approach extends the traditional Sakurai form of vector meson dominance~\cite{Sakurai:1960ju} to a product of vector poles.    Since the LF holographic amplitude encodes the power-law behavior for hard-scattering~\cite{Brodsky:1973kr, Brodsky:1974vy, Matveev:ra}, the result \req{Ftau} can be extended naturally to high-energies, thus overcoming the limitations of the original VMD model~\cite{Friedman:1991nq} \footnote{Other extensions are discussed for example in Ref.~\cite{Masjuan:2012sk}.}.

\subsubsection{Effective wave function from holographic mapping of a current}

It is also possible to find a precise  mapping of a confined EM current propagating in a warped AdS space to the light-front QCD Drell-Yan-West expression for the form factor. In this case we  find an effective LFWF, which corresponds to a superposition of an infinite number of Fock states generated by the ``dressed'' confined current. For the soft-wall model this mapping can be done analytically.

The form factor in light-front  QCD can be expressed in terms of an effective single-particle density~\cite{Soper:1976jc}
\begin{equation} 
F(Q^2) =  \int_0^1 dx \, \rho(x,Q),
\end{equation}
where
$\rho(x, Q) = 2 \pi \int_0^\infty \!  b \,  db \, J_0(b Q (1-x)) \vert \psi(x,b)\vert^2$, for a two-parton state ($b = \vert \mbf{b}_\perp \vert$). By direct comparison with \req{rhotau} for arbitrary values of $Q^2$    we find the effective two-parton  LFWF~\cite{Brodsky:2011xx}
\begin{equation} \label{ELFWF}
\psi_{\it eff}(x, \mbf{b}_\perp) = \kappa \frac{ (1-x)}{\sqrt{\pi \ln(\frac{1}{x})}} \,
e^{- \half \kappa^2 \mbf{b}_\perp^2  (1-x)^2 / \ln(\frac{1}{x})},
\end{equation}
in impact space.  The momentum space expression follows from the Fourier transform of  \req{ELFWF} and it is given by~\cite{Brodsky:2011xx}  
\beq 
\psi_{\it eff}(x, \mbf{k}_\perp) = 4 \pi \, \frac{ \sqrt{\ln\left(\frac{1}{x}\right)}}{\kappa (1-x)} \,
e^{ - \mbf{k}_\perp^2/2 \kappa^2 (1-x)^2 \ln \left(\frac{1}{x}\right)}.
\enq
The effective LFWF  encodes  nonperturbative dynamical aspects that cannot be determined from a term-by-term holographic mapping, unless one includes an infinite number of terms.  However, it has the correct analytical properties to reproduce the bound state vector meson pole in the pion time-like EM form factor. Unlike the ``true'' valence LFWF, the effective LFWF, which represents a sum of an infinite number of Fock components  in the EM current, is not symmetric in the longitudinal variables $x$ and $1-x$ for the active and spectator quarks,  respectively.

\subsubsection{Higher Fock components and form factors}

One can extend the formalism in order to examine the contribution of higher-Fock states in the nonperturbative analytic structure of time-like hadronic form factors. In fact, as we have shown above for the soft-wall model,   there is a precise non-trivial relation between the twist (number of components)  of each Fock state in a hadron and the number of poles from the hadronized $q \bar q$ components in the electromagnetic current inside the hadron.  In general, the pion state  is a superposition of an infinite number of Fock components  $\vert N \rangle$, $\vert \pi\rangle = \sum_N \psi_N \vert N \rangle$, and thus 
the full pion form factor is  given by
\begin{equation} \label{Fpitau}
F_\pi(q^2) = \sum_\tau P_\tau F_\tau(q^2),
\end{equation}
since the charge  is a diagonal operator. Normalization at $Q^2 =0$,  $F_\pi(0) = 1$, implies that 
 $\sum_\tau P_\tau = 1$ if all possible states are included.

 Conventionally the analysis of form factors is based on the generalized vector meson dominance model
 \begin{equation} \label{Fpilambda}
F_\pi(q^2) = \sum_\lambda C_\lambda \frac{M^2_\lambda}{M^2_\lambda - q^2},
\end{equation}
 with a dominant contribution from the  $\rho$ vector meson  plus contributions from the higher resonances  $\rho'$, $\rho''$, $\rho'''$, \dots, etc~\cite{Kuhn:1990ad}. Comparison with
\req{Fpitau} and \req{Ftau} allow us to determine the coefficients  $C_\lambda$ in terms of the of the probabilities $P_\tau$ for each Fock  state and the vector meson masses $M_n^2$.  However, no fine tuning of the coefficients $C_\la$ is necessary in the holographic LF framework, since the correct scaling is incorporated in the model.

\subsection{Time-like form factors in holographic QCD \label{TLFF}}

The computation of form factors in the time-like region is in general a complex task. For example, lattice results of Minkowski observables, such as  time-like hadronic form factors cannot be obtained directly from numerical Euclidean-space lattice simulations,  and Dyson-Schwinger computations are often specific to the space-like region.  Typically models  based on hadronic degrees of freedom involve sums over a large number of intermediate states and thus require a  large number of hadronic parameters~\cite{Kuhn:1990ad, Bruch:2004py, Hanhart:2012wi, Dumm:2013zh, Achasov:2013usa}.  In contrast, as we shall show below for an specific example, the analytical extension of the holographic model to $s> 4 m_\pi^2$  encodes  most relevant dynamical aspects of the time-like form factor, including the relative phases between different vector-meson resonance contributions.

In the strongly coupled semiclassical gauge/gravity limit, hadrons have zero widths and are stable,  as  in the  $N_C \to \infty$ limit of QCD~\footnote{In Refs.~\cite{Brodsky:2007hb, Brodsky:2011xx} the computation of  the pion leptonic decay constant in LF holography is examined. A computation of decay constants in the framework of bottom-up AdS/QCD models  will be given in Chapter \ref{ch5}.  See also \cite{Ballon-Bayona:2014oma} for a recent computation of the decay constant of the pion and its excited states.}. In a realistic theory, the resonances have widths due to their mixing with the continuum, e.g., two pions. As a practical approach, we modify  (\ref{Ftau})  by introducing  finite widths  in the expression for $F_\pi(s)$ according to
\begin{equation} \label{FpiGamma}
F_\tau(s) = \frac{M^2_{\rho} M^2_{\rho'} \cdots M^2_{\rho^{\tau-2}} }{{\left(M^2_{\rho} - s-  i \sqrt{s} \, \Gamma_\rho(s) \right) }
 \left(M^2_{\rho'} - s - i \sqrt{s} \, \Gamma_{\rho'}(s)\right)  \cdots 
  \left(M^2_{\rho^{\tau-2}}  - s - i \sqrt{s} \, \Gamma_{\rho^{\tau-2}}(s)\right)},
  \end{equation}
 with normalization   $F_\tau(0) = 1$.

The effect of multiparticle states coupled to  $\rho$-resonances is to introduce an $s$-dependent width $\Gamma(s)$. The modified width include the kinematical factors from the mixing of the vector mesons with the continuum -- which occurs mainly to pion pairs, although higher multiparticle states also occur at large $s$~\footnote{An alternative form given by Gounaris and Sakurai is  often used~\cite{Gounaris:1968mw}.}.

\subsubsection{A simple holographic model \label{hologmodel}}

In holographic QCD, the Fock states of hadrons can have any number of extra $q \bar q$ pairs created by the confining potential;  however, there are no constituent dynamical gluons~\cite{Brodsky:2011pw}.  This result is consistent with the flux-tube interpretation of QCD~\cite{Isgur:1984bm}  where soft gluons interact strongly to build a color confining potential for quarks.  Gluonic degrees of freedom only arise at high virtuality and gluons with smaller virtuality are sublimated in terms of the effective confining potential. This unusual property of QCD may explain the dominance of quark interchange~\cite{Gunion:1973ex}  over quark annihilation or gluon exchange contributions in large-angle elastic scattering~\cite{Baller:1988tj}.  In fact, empirical evidence confirming gluonic degrees of freedom at small virtualities or constituent gluons is lacking~\cite{Klempt:2007cp, Brodsky:2012rw}.

 In a complete treatment the unstable hadron eigenvalues and widths should emerge due to mixing with the continuum.  This is a formidable dynamical problem;  thus to  illustrate the relevance of higher Fock states in the analytic  structure of the pion form factor, we will consider a simple phenomenological model where the widths are constant and basically taken from the Particle Data Group  and the probabilities are taken as adjustable parameters. We  will consider a simplified model~\cite{deTeramond:2010ez} where we only include the first two components in a Fock expansion of the LF pion wave function
\beq
\vert \pi \rangle  = \psi_{q \bar q /\pi} \vert q \bar q   \rangle_{\tau=2} 
+  \psi_{q \bar q q \bar q} \vert q \bar q  q \bar q  \rangle_{\tau=4}  +   \cdots,
\enq
 and no constituent dynamical gluons~\cite{Brodsky:2011pw}.  The $J^{PC} = 0^{- +}$ twist-2 and  twist-4 states  are created by the interpolating operators $\mathcal{O}_2 = \bar q \gamma^+ \gamma_5  q$ and  $\mathcal{O}_4 = \bar q \gamma^+ \gamma_5  q  \bar q q$.  Up to twist-4  the corresponding expression for the pion form factor is
\begin{equation} \label{Fpi4}
F_\pi(q^2) = (1 - \gamma ) F_{\tau = 2}(q^2) + \gamma F_{\tau = 4}(q^2) ,
\end{equation}
where we have labeled the twist-4  probability  $P_{q \bar q q \bar q} = \gamma$, the admixture of the $\vert q \bar q q \bar q \rangle$  state.

\begin{figure}[h] 
\centering
\includegraphics[width=12.6cm]{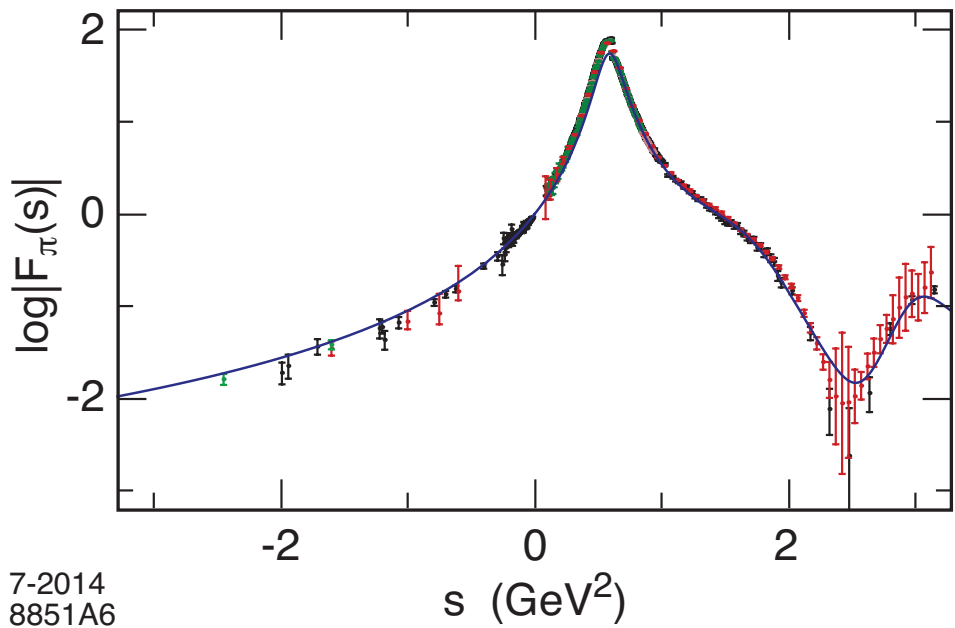}  
\caption{\label{PionFFSLTL} \small The structure of the space-like ($s = - Q^2 <0$)  and time-like ($s = q^2 > 4 m_\pi^2$) pion  form factor in light-front holographic QCD for a truncation of the pion wave function up to twist four.
The space-like data are taken  from the compilation  from Baldini  {\it et al.}~\cite{Baldini:1998qn} (black)  and JLAB data~~\cite{Tadevosyan:2007yd, Horn:2006tm} (red and green). The time-like data are from the precise measurements from KLOE~\cite{Ambrosino:2008aa, Ambrosino:2010bv, Babusci:2012rp} (dark green and dark red),  BABAR~\cite{Aubert:2009ad, Aubert:20012} (black) and BELLE~\cite{Fujikawa:2008ma} (red).}
\label{pionFF}
\end{figure} 

The predictions of the light-front holographic model up to twist-4 \req{Fpi4} for the space-like and time-like pion elastic form factor  are shown in Fig. \ref{pionFF}.  We choose the values $\Gamma_\rho =  149$ MeV,   $\Gamma_{\rho'} =  360$ MeV and $\Gamma_{\rho''} =  160$ MeV.    The chosen values for the width of the $\rho'$ and $\rho''$ are on the lower side of the PDG  values  listed in Ref. \cite{PDG:2014}. The results correspond to the probability $P_{q \bar q q \bar q}$ = 12.5 \%.  The values of $P_{q \bar q q \bar q}$  (and the corresponding widths) are inputs in the model. We use the value of $\sqrt \la= 0.5482$ GeV determined from the $\rho$ mass: $\sqrt \la = M_\rho/ \sqrt 2$, and the masses of the radial excitations follow from setting the poles at their physical locations, $M^2 \to 4 \la(n + 1/2)$. The main features of the pion form factor in the space-like and time-like regions are well described by the same physical picture with a minimal number of parameters.  The value for the pion radius is  $\langle r_\pi \rangle =  0.644$ fm, compared with the experimental value  $\langle r_\pi \rangle =
0.672 \pm 0.008$ fm from Ref. \cite{PDG:2014}.  Since we are interested in the overall behavior of the model, we have not included $\om-\rho$ mixing and kinematical threshold effects. This simple model, however,  reproduces quite well the space- and  time-like structure in the momentum transfer regime where the model is valid: this is, up to the second radial excitation of the $\rho$, $s \simeq  M^2_{\rho''} \simeq 3$ GeV$^2$ in the time like region (which covers the Belle data). Above this energy interference with higher twist contributions and a detailed study of the effects of the mixing of the vector mesons with the continuum  should be incorporated, as well as the effects of $s$-dependent widths from multiparticle states coupled to the $\rho$-resonances.

\begin{figure}[h]  
\centering
\includegraphics[width=7.6cm]{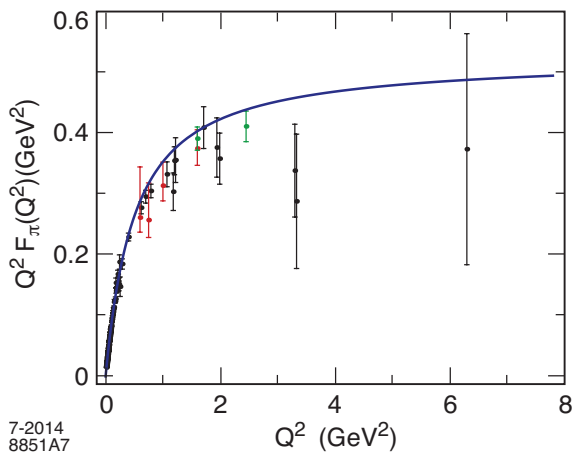}  
\caption{\label{Q2Fpi} \small Scaling predictions for $Q^2 F_\pi(Q^2)$. The space-like data is  from the compilation of Baldini  {\it et al.} (black)~\cite{Baldini:1998qn}  and JLAB data~~\cite{Tadevosyan:2007yd, Horn:2006tm} (red and green). }
\end{figure} 

The analytical structure of the holographic model encodes essential  dynamical aspects of the pion form factor, including its  pole and relative phase dependence as derived from  multiple vector meson resonances, leading-twist scaling at high virtuality,  as well as the transition from the hard-scattering domain to the long-range confining hadronic region. The scaling behavior in the space-like region is illustrated in Fig. \ref{Q2Fpi} where we plot $Q^2 F_\pi(Q^2)$.

 At very  large time-like momentum transfer  we  expect that higher twist contributions and the effects of the mixing with the continuum are not important.  This follows from the structure of the hadronic form-factor $F_\tau$ \req{Ftau}, which is the product of  ($N$-1)-poles for the  twist $\tau=N$ component in the light-front Fock expansion of the wave function. As a result, the resonant contributions for twist $N +1$ is decoupled by a factor $1/Q^2$ for large $Q^2$, compared to the twist $N$  contribution. Thus at large time-like momentum transfer, the resonant structure should be less and less visible and melts to a smooth curve.  This is indeed the case for the Babar data which is well reproduced above $s  \simeq 6$ GeV$^2$ by the leading twist-2 amplitude. However, a recent  measurement from CLEO~\cite{Seth:2012nn}  at $s = 14.2$ and 17.4 GeV$^2$ gives results significantly higher than those expected from QCD scaling considerations.  One can also extend the LF holographic approach to describe other processes, as for example the  photon-to-meson transition form factors, such as $\ga^* \ga \to \pi^0$,  a reaction which  has been of intense experimental and theoretical interest~\footnote{See for example Ref.~\cite{Brodsky:2011xx} and references therein.}.

\section{Nucleon electromagnetic form factors}

Proton and neutron electromagnetic form factors are among the most basic observables of the nucleon, and thus central for our
understanding of the nucleon's structure and dynamics~\footnote{For a recent review see Ref. \cite{Pacetti:2015iqa}.}. In general two form factors are required to describe the elastic scattering of 
electrons by spin-$\half$ nucleons, the Dirac and  Pauli form factors, $F_1$ and  $F_2$
\begin{equation} \label{NFF}
\langle P' \vert J^\mu(0) \vert P \rangle = u(P') \left[ \gamma^\mu F_1(q^2) +  \frac{i \sigma^{\mu \nu} q^\nu}{2 \mathcal{M}} F_2(q^2)\right] u({P}),
\end{equation}
where $q = P' - P$.
In the light-front formalism one can identify
the Dirac and Pauli form factors from the LF spin-conserving and spin-flip current matrix elements of the $J^+$ current~\cite{Brodsky:1980zm}:

\beq
\label{LFFF1}
\left \langle P', \uparrow \left \vert \frac{J^+(0)}{ 2 P^+}  \right \vert P, \uparrow \right \rangle = F_1\left( q^2 \right),
\enq
and 
\beq
\label{LFFF2}
\left \langle P', \uparrow \left \vert \frac{J^+(0)}{ 2 P^+}  \right \vert P, \downarrow \right \rangle = - \frac{\left(q^1 - i q^2\right)}{2 M} F_2\left( q^2 \right).
\enq

On the higher dimensional gravity theory on the bulk, the spin-non-flip amplitude for the  EM transition  corresponds to
the  coupling of an external EM field $A^M(x,z)$  propagating in AdS with a  fermionic mode
$\Psi_P(x,z)$, given by the left-hand side of the equation below 
 \begin{multline} \label{FF}
 \int d^4x \, dz \,  \sqrt{g}   \,  \bar\Psi_{P'}(x,z)
 \,  e_M^A  \, \Gamma_A \, A^M(x,z) \Psi_{P}(x,z) \\  \sim 
 (2 \pi)^4 \delta^4 \left( P'  \! - P - q\right) \epsilon_\mu u(P') \gamma^\mu F_1(q^2) u({P}),
 \end{multline} 
 where $e^A_M = \left(\frac{R}{z}\right) \delta_M^A$ is the vielbein with curved space indices  $M, N = 0, \cdots 4$ and tangent indices
 $A, B = 0, \cdots, 4$. The expression on the right-hand side  represents the Dirac EM form factor in physical space-time. It is the EM  spin-conserving matrix element  \req{LFFF1} of the local quark current  $J^\mu = e_q \bar q \gamma^\mu q$  with local coupling to the constituents. In this case one can also show  that a precise mapping of the $J^+$ elements  can be carried out at fixed LF time, providing an exact correspondence between the holographic variable $z$ and the LF impact variable $\zeta$ in ordinary space-time with the 
result~\cite{Brodsky:2008pg}
 \begin{equation} \label{pmFFAdS}
G_\pm(Q^2)  =  g_\pm R^4 \int \frac{dz}{z^4} \, V(Q^2, z)  \, \Psi^2_\pm(z), 
\end{equation}
for the components $\Psi_+$ and $\Psi_-$ with angular momentum $L^z = 0$ and $L^z = +1$ respectively.
The effective charges $g_+$ and $g_-$ are determined from the spin-flavor structure of the theory.

A precise mapping for the Pauli form factor using light-front holographic methods has not been carried out. To study the spin-flip nucleon form factor $F_2$ \req{LFFF2} using holographic methods, Abidin and Carlson~\cite{Abidin:2009hr} 
 propose to introduce a non-minimal electromagnetic coupling with the  `anomalous' gauge invariant term 
\begin{equation} \label{F2AdS}
\int d^4x~dz~  \sqrt{g}  ~ \bar\Psi
 \,  e_M^A\,  e_N^B \left[\Gamma_A, \Gamma_B\right] F^{M N}\Psi,
 \end{equation}
 in the five-dimensional action, since the structure of (\ref{FF}) can only account for $F_1$.  Although this is a practical   avenue, the overall strength of the new term has to be fixed by the static quantities and thus some predictivity is lost.

Light-front holographic QCD methods have also been used to obtain hadronic momentum densities and  generalized parton distributions (GPDs) of  mesons and nucleons in the  zero skewness limit~\cite{Dahiya:2007is, Abidin:2008sb, Vega:2010ns, Vega:2012iz}.  GPDs  are nonperturbative, and thus holographic methods are well suited to explore their analytical structure~\footnote{A computation of GPDs and nucleon structure functions at small $x$ using gravity duals has been carried out in Refs.~\cite{Nishio:2011xa}  and  \cite{Watanabe:2013spa}  respectively.}. LF holographic methods have been used to model transverse momentum dependent (TMD) parton distribution functions and fragmentation functions~\cite{Aghasyan:2014zma}.  LF holography can also be used to study the flavor separation of the elastic nucleon form factors~\cite{deTeramond:2012X} which have been determined recently up to $Q^2 = 3.4 ~{\rm GeV}^2$~\cite{Cates:2011pz}. One can also use the holographic framework to construct light-front wave functions and parton distribution functions (PDFs) by matching quark counting rules~\cite{Gutsche:2013zia}.  Recently, models of nucleon and flavor form factors and GPDs has been discussed using LF holographic ideas and AdS/QCD~\cite{Chakrabarti:2013gra, Chakrabarti:2013dda, Kumar:2014coa, Sharma:2014voa}. The Dirac and Pauli weak neutral nucleon form factors have also been examined using the framework of light-front holographic QCD in Ref. \cite{Lohmann:2014}. LF holography has also been used to describe nucleon transition form factors, such as $ \ga^* N \to N^*$~\cite{deTeramond:2011qp}~\footnote{A computation of nucleon transition form factors has been carried out in the framework of the Sakai-Sugimoto model in Refs.~\cite{Grigoryan:2009pp} and \cite{BallonBayona:2012jy}.  Baryon form factors have also been computed using the SS framework in Refs.~\cite{Kim:2008pw, Bayona:2011xj}.}.

\subsection{Computing nucleon form factors in light-front holographic QCD \label{FFLFH}}

In computing nucleon form factors we should impose the asymptotic boundary conditions by the leading fall-off of the form factors to match the twist of the hadron's interpolating operator, {\it i. e.} $\tau = 3$, to represent the fact that at high energies the nucleon is essentially a system of 3 weakly interacting partons. However, as discussed at the end of Chapter \ref{ch5}, at low energies the strongly correlated bound state of $n$ quarks behaves as a system of an active quark {\it vs.} the $n-1$ spectators. This means, for example,  that for a proton the nonperturbative bound state behaves as a quark-diquark system, {\it i. e.}, a twist-2 system. In this simple  picture,  at large momentum transfer, or at small distances, where the cluster is resolved into its individual constituents,  the baryon is governed by twist-3, whereas in the long-distance nonperturbative region by twist-2. Thus,  at the transition region the system should evolve from twist-3 to twist-2.  In practice, since the behavior of the form factors at very low energy is much constrained by its normalization, we will  use a simple approximation where the nucleon form factor is twist-3 at all momentum transfer scales (In fact, twist-3 for the Dirac form factor, and twist-4 for the Pauli form factor to account for the $L=1$ orbital angular momentum). As in the case of the pion form factor described in the previous section, the vector-meson poles should be shifted to their physical locations for a meaningful comparison with data. With this limitations in mind, we describe below a simple  model approximation  to describe the space-like nucleon form factors.

In order to compute the individual features of the proton and neutron form factors  one needs to incorporate the spin-flavor structure of the nucleons,  properties  which are absent in models of the gauge/gravity correspondence. The spin-isospin symmetry can be readily included in light-front holography by weighting the different Fock-state components  by the charges and spin-projections of the quark constituents;  {\it e.g.}, as given by the $SU(6)$  spin-flavor symmetry. We label by $N_{q \uparrow}$  and $N_{q \downarrow}$ the probability to find the constituent $q$ in a nucleon with spin up or down respectively. For the  $SU(6)$ wave function~\cite{Lichtenberg:1978pc} we have 
\begin{equation}
N_{u \uparrow} = \frac{5}{3}, \hspace{20pt}  N_{u \downarrow} = \frac{1}{3}, \hspace{20pt}  
N_{d \uparrow} = \frac{1}{3},  \hspace{20pt}  N_{d \downarrow} = \frac{2}{3},
\end{equation}
for the proton and
\begin{equation}
N_{u \uparrow} = \frac{1}{3}, \hspace{20pt}  N_{u \downarrow} = \frac{2}{3}, \hspace{20pt}  
N_{d \uparrow} = \frac{5}{3},  \hspace{20pt}  N_{d \downarrow} = \frac{1}{3},
\end{equation}
for the neutron.  The effective charges $g_+$ and $g_-$ in (\ref{pmFFAdS})  are computed by the sum of the charges of the struck quark convoluted by the corresponding probability for the $L^z = 0$ and $L^z = + 1$ components $\Psi_+$ and $\Psi_-$ respectively. We find $g^+_p = 1$, $g^-_p = 0$, $g_+^n = - \frac{1}{3}$ and $g_-^n = \frac{1}{3}$. The nucleon Dirac form factors in the $SU(6)$ limit are thus given by
\begin{eqnarray}
F_1^p(Q^2) &\!=\!& R^4 \int \frac{dz}{z^4} \, V(Q^2, z)  \, \Psi_+^2(z), \\
F_1^n(Q^2) &\!=\!& - \frac{1}{3}R^4 \int \frac{dz}{z^4} \,   V(Q^2, z) \left[  \Psi_+^2(z) - \Psi_-^2(z)  \right],
\label{pnF1AdS}
\end{eqnarray}
where $F_1^p(0) = 1$ and  $F_1^n(0) = 0$.

\begin{figure}[h]
\begin{center}
 \includegraphics[width=7.6cm]{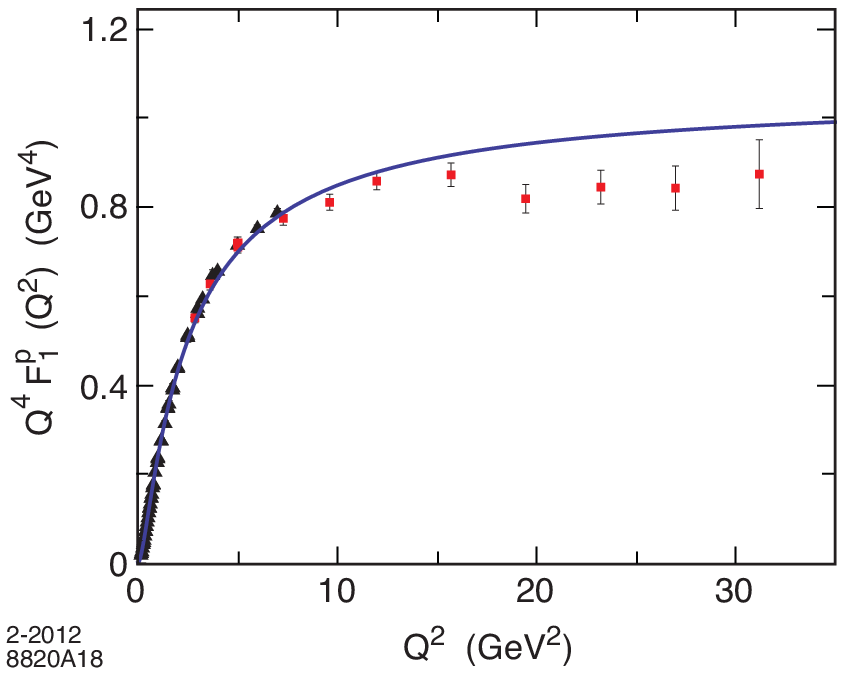}   \hspace{0pt}
\includegraphics[width=7.6cm]{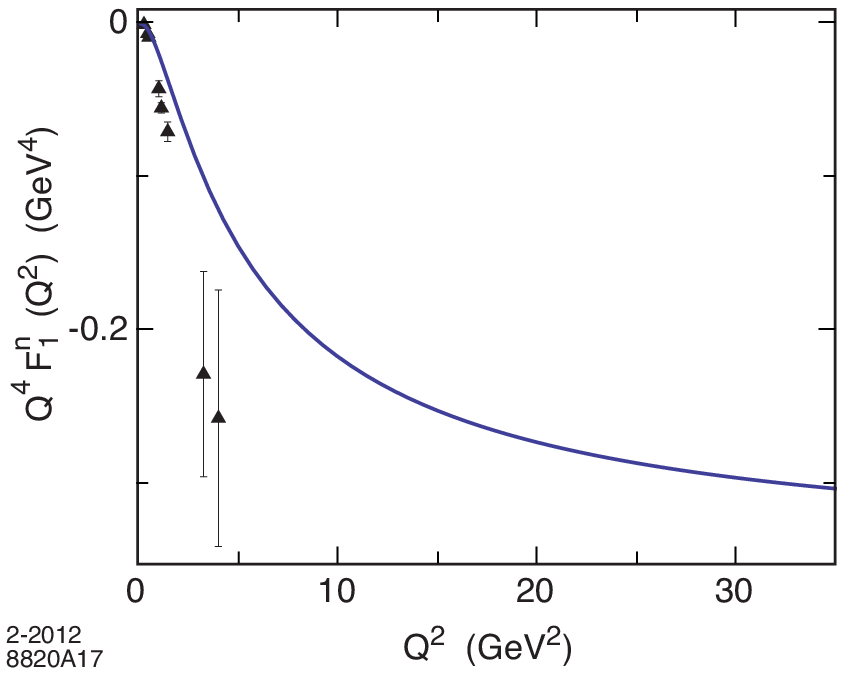}
 \caption{\small {Light-front holographic predictions  for  $Q^4 F_1^p(Q^2)$ (left) and   $Q^4 F_1^n(Q^2)$ (right) in the
$SU(6)$ limit}. Data compilation  from Diehl~\cite{Diehl:2005wq}.}
\label{fig:nucleonFF1}
\end{center}
\end{figure}

In the soft-wall model the plus and minus  components of the  leading twist-3 nucleon wave function are
\begin{equation} \label{AdSNWF}
\Psi_+(z) = \frac{ \sqrt{2} \kappa^2}{R^2} z^{7/2} e^{- \kappa^2 z^2/2}, \hspace{20pt}
\Psi_-(z) =                \frac{\kappa^3}{R^2} z^{9/2} e^{- \kappa^2 z^2/2} ,
\end{equation}
where we have absorbed the dilaton exponential dependence by a redefinition of the AdS wave function,  and the bulk-to-boundary propagator $V(Q^2, z)$ is given by (\ref{Vx}). The results for $F_1^{p,n}$ follow from the analytic form (\ref{Ftau}) for any twist $\tau$. We find
\begin{equation} \label{protonF1p}
F_1^p(Q^2) =  F_+(Q^2),
 \end{equation}
 and
 \begin{equation} \label{neutronF1n}
 F_1^n(Q^2) = -\frac{1}{3} \left(F_+(Q^2) - F_-(Q^2)\right),
 \end{equation}
where we have, for convenience,  defined  the twist-3 and twist-4 form factors
\begin{equation} \label{Fp}
F_+(Q^2) =  \frac{1}{{\Big(1 + \frac{Q^2}{\mathcal{M}^2_\rho} \Big) }
 \Big(1 + \frac{Q^2}{\mathcal{M}^2_{\rho'}}  \Big) },
 \end{equation}
and
  \begin{equation} \label{Fm}
 F_-(Q^2) =  
  \frac{1}{\Big(1 + \frac{Q^2}{\mathcal{M}^2_\rho} \Big) 
 \Big(1 + \frac{Q^2}{\mathcal{M}^2_{\rho'}}  \Big)
       \Big(1  + \frac{Q^2}{\mathcal{M}^2_{\rho^{''}}} \Big)} ,
\end{equation}
with the multiple pole structure  derived from the soft-wall dressed EM current propagating in AdS space. The results for $Q^4 F_1^p(Q^2)$ and $Q^4 F_1^n(Q^2)$  are shown in Fig. \ref{fig:nucleonFF1}.   The value $\sqrt \la = 0.548$ GeV  is determined from the $\rho$ mass.

\begin{figure}[h]
\begin{center}
 \includegraphics[width=7.6cm]{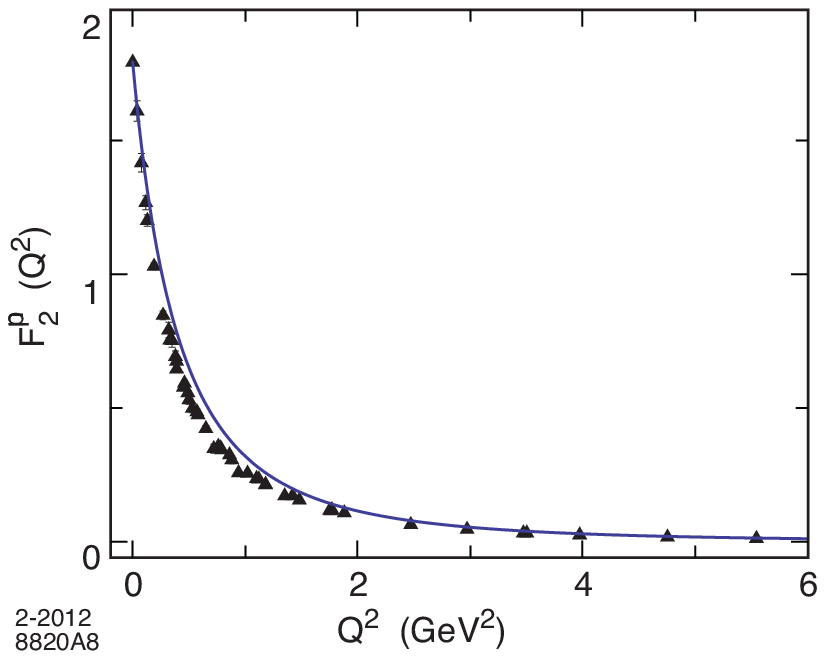}   \hspace{0pt}
\includegraphics[width=7.6cm]{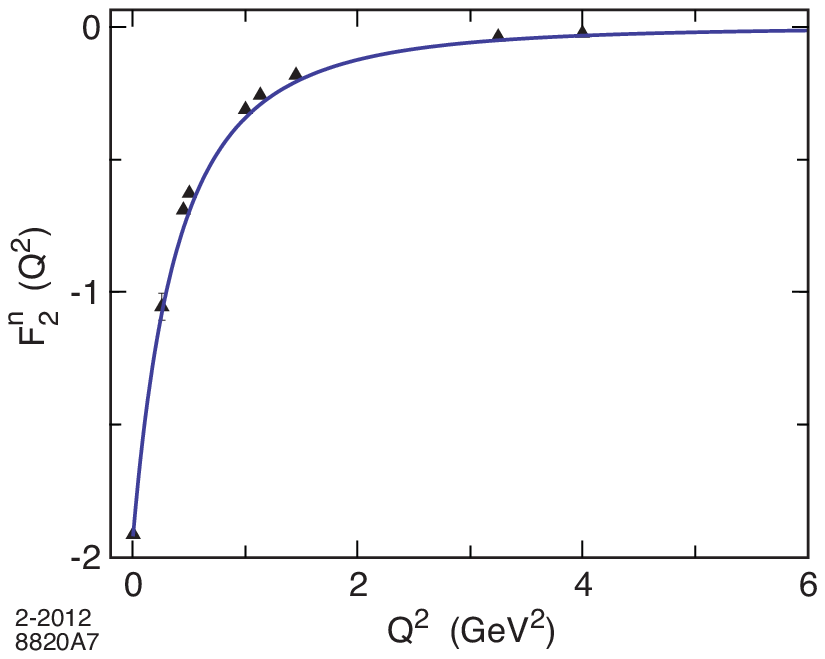}
 \caption{\small {Light-front holographic predictions for  $F_2^p(Q^2)$ (left)  and  $F_2^n(Q^2)$ (right). The value of the nucleon anomalous magnetic moment $\chi$ is taken from experiment}. Data compilation  from Diehl~\cite{Diehl:2005wq}.}
\label{fig:nucleonFF2}
\end{center}
\end{figure}

The  expression for the elastic nucleon form factor $F_2^{p,n}$  follows  from (\ref{NFF}) and (\ref{F2AdS}). \begin{equation} \label{F2}
F_2^{p,n}(Q^2) \sim \int \frac{d z}{z^3} \Psi_+(z) V(Q^2,z) \Psi_-(z).
\end{equation}
Using the twist-3 and twist-4  nucleon soft-wall wave functions $\Psi_+$ and $\Psi_-$ (\ref{AdSNWF})
we find
\begin{equation}
 F_2^{p,n}(Q^2) =  
  {\chi_{p,n}} F_-(Q^2),
\end{equation}
where the amplitude (\ref{F2}) has been normalized to the static quantities $\chi_p$ and $\chi_n$ and $F_-(Q^2)$ is given by (\ref{Fm}). The experimental values $\chi_p = 1.793$ and $\chi_n = -1.913$ are consistent with the $SU(6)$ prediction~\cite{Beg:1964nm} $\mu_P/\mu_N = -3/2$. In fact $(\mu_P/\mu_N)_{\rm exp} = - 1.46$, where $\mu_P = 1 + \chi_p$ and $\mu_N = \chi_n$. The results for $F_2^p(Q^2)$ and $F_2^n(Q^2)$ are shown in Fig. \ref{fig:nucleonFF2}.  The vector meson masses are given by   $M^2 = 4 \la \left(n + \half\right)$ with the value   $\sqrt \la = 0.548$ GeV obtained from the $\rho$ mass.

We compute the charge and magnetic root-mean-square (rms) radius from the usual electric and magnetic nucleon form factors
\begin{equation}
G_E(q^2) = F_1(q^2) + \frac{q^2}{4 \mathcal{M}^2} F_2(q^2)
\end{equation}
and
\begin{equation}
G_M(q^2) = F_1(q^2) + F_2(q^2).
\end{equation}
Using the definition
\begin{equation}
\langle r^2 \rangle = -  \frac{6}{F(0)} \frac{d F(Q^2)}{d Q^2} \Big|_{Q^2 =0},
\end{equation}
we find the values $\sqrt{\langle r_E \rangle_p} = 0. 802 ~{\rm fm}$, $\sqrt{\langle r^2_M \rangle_p} = 0. 758 ~{\rm fm}$,
 $\langle r^2_E \rangle_n  = -0.10 ~{\rm fm^2}$ and $\sqrt{\langle r^2_M \rangle_n} = 0.768 ~{\rm fm}$, compared with the
 experimental values   $\sqrt{\langle r_E \rangle_p} = (0.877 \pm 0.007) ~{\rm fm}$, 
 $\sqrt{\langle r^2_M \rangle_p} = (0.777 \pm 0.016) ~{\rm fm}$,
 $\langle r^2_E \rangle_n  = (- 0.1161 \pm 0.0022) ~{\rm fm^2}$ and 
 $\sqrt{\langle r^2_M \rangle_n} = (0.862 \pm 0.009) ~{\rm fm}$ from electron-proton scattering experiments~\cite{PDG:2014}.~\footnote{The neutron charge radius is defined by $\langle r_E^2 \rangle_n = -  6 \frac{d G_E(Q^2)}{d Q^2} \Big|_{Q^2 =0}$. }
 The muonic hydrogen measurement gives  $\sqrt{\langle r_E \rangle_p} = 0.84184(67)~{\rm fm}$ from  Lamb-shift measurements~\cite{Pohl:2010zz}.~\footnote{Other soft and hard-wall model predictions of the
 nucleon  rms radius are given, for example,  in Refs. \cite{Abidin:2009hr, Vega:2010ns, Vega:2012iz}.}

Chiral effective theory predicts that the slopes of the form factors are singular for zero pion mass. For example, the slope of the Pauli form factor of the proton at $q^2=0$ computed by Beg and Zepeda diverges as $1/ m_\pi$~\cite{Beg:1973sc}. This result comes from the  triangle diagram $\gamma^* \to \pi^+ \pi^- \to p \bar p.$ One can also argue from dispersion theory that the singular behavior of the form factors as a function of the pion mass comes from the two-pion cut.  Lattice theory computations of nucleon form factors require a strong dependence at small pion mass to extrapolate the predictions to the physical pion mass~\cite{Collins:2011mk}. � The two-pion calculation~\cite{Beg:1973sc} is a Born computation which probably does not exhibit vector dominance. To make a reliable computation in the hadronic basis of intermediate states one evidently has to include an infinite number of states. On the other hand, divergences do not appear in light-front holographic QCD even for massless pions when we use the dressed current.  In fact, the holographic analysis with a dressed EM current in AdS generates  a nonperturbative multi-vector meson pole structure.~\footnote{In the case of a free propagating current in AdS, we obtain logarithmic divergent results in the chiral limit.}

\chapter{Other Bottom-Up and Top-Down Holographic Models \label{ch7}}
 Here we review some of the other holographic approaches to hadronic physics, and after discussing generic features of these models we focus on particular top-down and bottom-up models based on chiral symmetry breaking.  Top-down models are derived from brane configurations in string theory, while bottom-up models, like Light Front Holographic QCD, are more phenomenological and are not restricted by the constraints of string theory-based models.  All holographic QCD models include
degrees of freedom which are identified with Standard Model hadrons via their quantum numbers, 
and predictions of   QCD observables may be compared to experiment and to other models, often with remarkable quantitative success accurate to within the 10-15\% level.

By now there are enough examples of gauge theories with established gravity duals that a 
dictionary exists between features of the gauge theory and
corresponding properties of the higher-dimensional gravity dual.  
The AdS/CFT correspondence relates operators in the lower-dimensional theory
to fields in the higher-dimensional dual theory \cite{Maldacena:1997re}.  The quantum numbers and
conformal dimensions of the gauge theory operators dictate the nature of
the corresponding fields \cite{Gubser:1998bc, Witten:1998qj}. 
Global symmetries in the 3+1 dimensional theory become gauge invariances
of the 4+1 dimensional theory.  Hence, symmetry currents are related to gauge
fields in the gravity dual.  In the case of ${\cal N}$=4 Yang-Mills theory,
the $SO(6)$ $R$-symmetry, which is a global symmetry of the theory, 
is associated with the $SO(6)$ isometry of the five-sphere in
the AdS$_5\times S^5$ supergravity background.  
A Kaluza-Klein decomposition of the gravitational 
fluctuations on the 5-sphere include spin-1
$SO(6)$ gauge fields in the effective 4+1 dimensional theory on AdS$_5$.  In addition
to $R$-symmetries, there may be flavor 
symmetries  due to the addition of probe branes \cite{Karch:2002sh}, on which
gauge fields propagate as fluctuations of open strings ending on the
probe branes.  The 3+1 dimensional interpretation of the theory follows either from the behavior of the solutions to the equations of motion near the boundary of AdS$_5$, or from 
a further Kaluza-Klein decomposition of the 4+1 dimensional theory on AdS$_5$.
The boundary conditions  do not allow massless
Kaluza-Klein modes of the gauge fields to propagate, so the gauge invariance is absent in the 3+1
dimensional effective theory while a global symmetry typically remains.  
In  AdS/QCD models, 
isospin is the remnant of a 4+1 dimensional gauge invariance with the same
gauge group as  the isospin symmetry group.  The
notion that boundary conditions in an extra dimension
may be responsible for breaking of gauge
invariances is also the basis of Higgsless models of electroweak
symmetry breaking \cite{Csaki:2003dt} and holographic technicolor models
\cite{Agashe:2004rs,Hong:2006si,Hirn:2006nt}.

In order to describe a confining theory the conformal invariance encoded in the isometries of 
AdS$_5$
must be broken.  In bottom-up models the breaking of conformal symmetry is most easily modeled by way of a hard wall
\cite{Polchinski:2000uf},
so that the space-time geometry becomes a slice of AdS$_5$. The location of the hard wall determines
the typical scale of hadronic masses in hard-wall AdS/QCD models.
While phenomenological AdS/QCD models often take the slice of AdS$_5$ as
the background space-time, in models based on string theory 
the space-time geometry
is dictated by the corresponding
brane configuration and may be quite different from AdS$_5$
\cite{Kruczenski:2003uq,Sakai:2004cn,Sakai:2005yt}.  
A comparison of the predictions
of AdS/QCD models in various space-time backgrounds
appears in Ref.~\cite{Becciolini:2009fu}.  The result of such
a comparison is that, for a wide class of space-time geometries, 
naive predictions of the
classical 5D models (ignoring quantum corrections) agree with
experiment at the 10-30\% level, with the slice of AdS$_5$ often
producing among the most accurate predictions.  It is not known whether there is an underlying
reason for the phenomenological success of models based on the AdS geometry, though there
are arguments based on decoupling of high dimension operators at low energies \cite{Fitzpatrick:2013twa} and suggestions of scale invariance in QCD at low energies  \cite{Brodsky:2008be} that may help to understand this.

\section{Bottom-up models}

Bottom-up holographic QCD models, like Light-Front Holographic QCD described in previous chapters, are loosely related to string-theory examples of holographic dualities, but are phenomenological and not constrained by the restrictions of string theory.  The flexibility in bottom-up model building allows for matching of various aspects of QCD, for example the pattern of
explicit and spontaneous chiral symmetry breaking and certain aspects of 
QCD at high energies ({\em e.g.} operator product expansions \cite{Erlich:2005qh, Hirn:2005vk}). Bottom-up models are motivated by the AdS/CFT dictionary relating properties of a non-gravitational theory to a gravitational theory in one additional dimension.  Interpolating operators for QCD states map into fields in the higher-dimensional model.  For example, as described in earlier sections,  the rho mesons are created by the $SU(2)$ vector isospin current, so in order to model the spectrum of rho mesons the higher-dimensional model would include a field with the appropriate quantum numbers to couple to the vector current, namely an $SU(2)$ gauge field.  For simplicity, and motivated by conformal symmetry, the higher-dimensional space-time in bottom-up holographic models is typically chosen to be Anti-de Sitter space, with metric that we repeat here: \begin{equation}
ds^2=\frac{R^2}{z^2}\left(\eta_{\mu\nu}dx^\mu dx^\nu-dz^2\right), \label{eq:AdSmetric}
\end{equation}
where the $z$ coordinate describes the extra spatial dimension.  In order to break the conformal symmetry the AdS space-time may be chosen to end at some value of $z$, as in hard-wall models, or else some field may have a non-vanishing background with a dimensionful parameter, as in the soft-wall model.  Both hard-wall and soft-wall models maintain the 3+1 dimensional Poincar\'e invariance of the space-time.

Consider an $SU(2)$ gauge theory on the slice of AdS$_5$ with metric
Eq.~(\ref{eq:AdSmetric}) between  the boundary at $z=0$ and an infrared cutoff at $z=z_m$. The 5-dimensional $SU(2)$ gauge fields are related via the AdS/CFT correspondence to a 4-dimensional $SU(2)$ current which may be identified with the isospin current of QCD.
The action for this theory is \begin{eqnarray}
S&=&
-\frac{1}{4g_5^2}\int d^4x\,dz \sqrt{g}\,F_{MN}^a\,F_{PQ}^a\, g^{MP} g^{NQ}
\nonumber \\
&=&-\frac{1}{4g_5^2}\int d^4x\,dz \frac{R}{z}
\left(F_{\mu\nu}^a\,F_{\rho\sigma}^a\, \eta^{\mu\rho} \eta^{\nu\sigma}
-2 F_{\mu z}^a \,F_{\nu z}^a\, \eta^{\mu\nu}\right) \nonumber \\
&=&-\frac{1}{4g_5^2}\int d^4x\,dz \frac{R}{z}
\left(F_{\mu\nu}^a\,F^{a\,\mu\nu}+2 F_{z\mu}^a\,F^{a\,z\mu} \right) 
, \label{eq:SSU2}\end{eqnarray}
where $F_{MN}^a=\partial_MV_N^a-\partial_NV_M^a+\epsilon^{abc}V_M^bV_N^c$
is the field strength tensor for the gauge fields $V_M^a$,
$g_5$ is the 5-dimensional gauge coupling, and the $SU(2)$ gauge index $a$ runs from 1 to 3.  As usual,
Greek indices are contracted with the 4-dimensional Minkowski tensor
$\eta_{\mu\nu}$ as in the last line of Eq.~(\ref{eq:SSU2}), 
while capital Latin indices run from 0 to 4 and 
contractions of capital Latin indices with $g_{MN}$ are made explicit so as to avoid confusion.

The linearized equations of motion are \begin{eqnarray}
\frac{1}{z}\,\partial_\mu F^{a\,\mu\nu}+\partial_z\,\left(\frac{1}{z}\,
F^{a\,z\nu}\right)&=&0, \label{eq:linEOM} \\
\frac{1}{z}\partial_\mu F^{a\,\mu z}=0, \end{eqnarray}
where for the present discussion the field strengths are to be linearized.
Note that $R$ and $g_5$ factor out of the linearized equations of motion,
and otherwise they  appear in the dimensionless combination $R/g_5^2$.
At the infrared boundary $z=z_m$ there is freedom in the choice of boundary
conditions.  We will impose the simplest gauge-invariant boundary conditions,
$F_{\mu z}^a(x,z_m)=0$.  The Kaluza-Klein modes must also be normalizable as
$z\rightarrow 0$, {\em i.e.} the integral over $z$ in the 
action Eq.~(\ref{eq:SSU2}) must remain finite
upon replacing the gauge fields by a Kaluza-Klein mode.  As we will review, a Weinberg sum rule is an automatic consequence of the choice of Dirichlet boundary conditions on the gauge fields in a UV-regulated boundary at some $z=\epsilon$, where $\epsilon\rightarrow0$.

Suppose we fix a gauge $V_z^a=0$.  In that case the linearized equations of
motion become, \begin{equation}
\frac{1}{z}\,\partial_\mu F^{a\,\mu\nu}-\partial_z\,\left(\frac{1}{z}\,
\partial_z V^{a\,\nu}\right)=0, \label{eq:VEOMlin}\end{equation}
together with a transverseness condition $\partial_z
\partial_\mu V^{a\,\mu}$=0.
From the transverseness condition follows that $\partial_\mu V^{a\,\mu}$ is independent of $z$,
and then by a $z$-independent gauge transformation condition consistent with our choice $V^{a\,z}=0$, we may also set $\partial_\mu V^{a\,\mu}=0$.
The 4-dimensional-transverse
Kaluza-Klein modes are solutions to the linearized equations of motion of 
the form \begin{equation}
V_n^{a\,\mu}(x,z)=\varepsilon_n^{a\,\mu}\,e^{iq\cdot x}\,\psi_n(z), \end{equation}
where $q^2=m_n^2$, $q\cdot\varepsilon^a=0$.  The wavefunction 
$\psi_n(z)$ then satisfies the equation \begin{equation}
\frac{d}{dz}\,\left(\frac{1}{z}\,\frac{d\psi_n(z)}{dz} \right)+\frac{m_n^2}{z}
\,\psi_n(z)=0, \label{eq:psin}\end{equation}
with boundary conditions $\psi'(z_m)=0$ and normalizability, with normalization
\begin{equation}
 \int_\epsilon^{z_m} dz\, \psi_n(z)^2/z=1. \label{eq:psinorm}\end{equation}  
The subscript $n$ denotes the discrete set of eigenvalues 
$m_n^2$ of the Sturm-Liouville system.  In practice we
impose  Dirichlet ($\psi_n(\epsilon)=0$)
boundary conditions on the modes at an unphysical ultraviolet cutoff $z=\epsilon
\ll z_m$.  As $\epsilon\rightarrow0$ the spectrum becomes independent
of the specific form of the Sturm-Liouville boundary condition except for
the addition of a zero mode in the Neumann case.  The extra zero mode 
decouples from the rest
of the spectrum in the $\epsilon\rightarrow0$ limit, as can be seen from the
fact that the integral over $z$ in the action for the zero mode diverges in 
this limit.

The equations of motion Eq.~(\ref{eq:VEOMlin}) imply that the Kaluza-Klein
modes satisfy the Proca equation, \begin{equation}
\partial_\mu F^{a\,\mu\nu}=-m_n^2\, V^{a\,\nu}, \end{equation}
so the eigenvalue $m_n$ is identified with the mass of the corresponding
3+1 dimensional field.

By construction the Kaluza-Klein modes have quantum numbers conjugate to the 
corresponding 3+1 dimensional current $J_\mu^a=\overline{q}\gamma^\mu T^a q$,
where $T^a$ are the $SU(2)$ isospin generators normalized such
that ${\rm Tr}\,T^aT^b=\frac{1}{2}\delta^{ab}$.  Both the gauge
fields $V_\mu^a$ and the current $J_\mu^a$ transform in the adjoint 
representation of  $SU(2)$.  Under charge conjugation $V_\mu^a$ is odd, and 
parity acts as $V_i^a(t,\mathbf{x},z)\rightarrow -V_i^a(t,-\mathbf{x},z)$,
while $V_0^a$ and $V_z^a$ are even under parity.
These charge assignments identify the Kaluza-Klein modes with states obtained
by acting with the current $J_\mu^a$ on the vacuum.

A useful object to consider is the bulk-to-boundary propagator, which
is related to
the solution of the equations of motion subject to
the boundary condition
$V^{a\,\mu}(x,z)\rightarrow V^{a\,\mu}_0(x)$ as $z\rightarrow 0$
for arbitrary fixed $V^{a\,\mu}_0(x)$  \cite{Gubser:1998bc,Witten:1998qj}.  We have already encountered this object in Chapter \ref{ch6}, where it was used in the calculation of meson and nucleon electromagnetic form factors.  The boundary profile of the field, $V_0^{a\,\mu}(x)$, plays the role of the source of the current $J^a_\mu(x)$ in the AdS/CFT correspondence.  Fourier transforming in 3+1 dimensions,
with a slight abuse of notation we equivalently have
$V^{a\,\mu}(q,z)\rightarrow V^{a\,\mu}_0(q)$ as $z\rightarrow 0$.
With \begin{equation}
V^{a\,\mu}(x,z)= \int \frac{d^4q}{(2\pi)^4}\,e^{-iq\cdot x} V^{a\,\mu}_0(q) \,V(q,z), \end{equation}
the bulk-to-boundary propagator $V(q,z)$ satisfies, \begin{equation}
\partial_z\,\left(\frac{1}{z}\,\partial_z V(q,z)\right)
+\frac{q^2}{z}\,V(q,z)=0, \end{equation}
with $V(q,\epsilon)=1$ and $\left.\partial_z V(q,z)\right|_{z=\epsilon}=0$.
The bulk-to-boundary propagator determines solutions to the equations
of motion that have a prescribed form on the UV boundary of the space-time,
$z=\epsilon$.

The bulk-to-boundary propagator 
can be decomposed in terms of the normalizable eigensolutions
as follows \cite{Hong:2004sa}:  Define the Green function 
$G(q,z,z')$ with Dirichlet boundary conditions at $z=\epsilon$, 
\begin{equation}
\left(\partial_z\frac{1}{z}\partial_z+\frac{q^2}{z}
\right)G(q,z,z')= \delta(z-z'), \label{eq:Geom}\end{equation}
\begin{equation}
G(q,\epsilon,z')=0,\ \ \left.\partial_zG(q,z,z')\right|_{z=z_m}=0. \label{eq:Gbc}
\end{equation}
Now consider the integral
\begin{eqnarray}
\int_\epsilon^{z_m} dz\,V(q,z)\left(\partial_z\frac{1}{z}\partial_z+\frac{q^2}{z}
\right)G(q,z,z')=V(q,z'). \end{eqnarray}
Integrating by parts twice, we have, \begin{eqnarray}
V(q,z')&=&\int_\epsilon^{z_m} dz \,G(q,z,z')\left(\partial_z\frac{1}{z}\partial_z
+\frac{q^2}{z}\right)V(q,z) \nonumber \\
&&+\left.\left(V(q,z)\frac{1}{z}\partial_z G(q,z,z')-
G(q,z,z')\frac{1}{z}\partial_z V(q,z)\right)\right|_{z=\epsilon}^{z_m}
\nonumber \\
&=&-\left.\frac{1}{z}\partial_zG(q,z,z')\right|^{z=\epsilon}. \end{eqnarray}
By  Eqs.(\ref{eq:Geom}) and (\ref{eq:Gbc}),  the Green function  can be decomposed
in the complete set of normalizable solutions $\psi_n(z)$ defined earlier:
\begin{equation}
G(q,z,z')=\sum_n \frac{\psi_n(z)\psi_n(z')}{q^2-m_n^2}, \end{equation}
where the wavefunctions are normalized as in Eq.~(\ref{eq:psinorm}).
Hence, we obtain \begin{equation}
V(q,z')=-\sum_n\,\frac{\psi_n'(\epsilon)}{\epsilon}\frac{\psi_n(z')}{
q^2-m_n^2},
\label{eq:B2B}\end{equation}
as long as $z'\neq\epsilon$.  A consequence of the hard-wall boundary condition on the Kaluza-Klein modes is the Gibbs phenomenon
at $z'=\epsilon$, which is the discontinuity in the Fourier series of a function with a jump discontinuity.  In our case the discontinuity can be seen by comparing Eq.~(\ref{eq:B2B}) with 
$\psi_n(\epsilon)=0$ and the condition
$V(q,\epsilon)=1$.    However, limits as $z'\rightarrow
\epsilon$ are well defined.
This situation is analogous to the problem in Fourier
transforming a square waveform in terms of modes which vanish at the boundaries
of the square.  In that case, the Fourier transform strictly vanishes at
the discontinuity, and for any sum over a finite number of modes there is
a deviation in the function from the desired waveform in the neighborhood of
the discontinuity.

In the context of the hard-wall model, 
the precise statement of the AdS/CFT correspondence in the gravity limit
is that the generating functional of connected
correlation functions of products of currents, 
$W[V^{a\,\mu}_0(x)]$ in 3 + 1 physical space-time, is identified with the 
4+1 dimensional action $S$ evaluated on solutions to the equations of motion
such that 
$V^{a\,\mu}(x,z)\rightarrow V^{a\,\mu}_0(x)$ as $z\rightarrow \epsilon$~\cite{Gubser:1998bc, Witten:1998qj, Muck:1998iz}.
The gravity limit is the limit of infinite $N$ and large 't Hooft
coupling \cite{Maldacena:1997re}, but in the context of the hard-wall model the same analysis is equivalent to studying the effective 3+1
dimensional theory derived from the 4+1 dimensional model
in the classical limit, {\em i.e.} having absorbed quantum corrections into the definition of the model.

Integrating the action by parts, the action vanishes by the equations of
motion except for a boundary term at $z=\epsilon$.  Hence, the 
generating functional has the form, \begin{eqnarray}
&&W[V^{a\,\mu}_0(x)]=S[V^{a\,\mu}_0(x)]
=-\frac{R}{2g_5^2}\int d^4x \left.\left[ V^{a\,\mu}(x,z)
\left(g_{\mu\nu}-\frac{\partial_\mu \partial_\nu}{\partial^2}\right)
\frac{1}{z}\partial_z V^{a\,\nu}(x,z)\right]
\right|_{z=\epsilon} +\dots
\nonumber \\
&&= -\frac{R}{2g_5^2}
\int d^4x \int \frac{d^4\bar{q}}{(2\pi)^4} \,\frac{d^4\bar{q}'}{(2\pi)^4}
 \left.\left[V_{0}^{a\,\mu}(\bar{q}') 
e^{-i(\bar{q}'+\bar{q})\cdot x}
V(\bar{q}',z) \left(g_{\mu\nu}-\frac{\bar{q}_\mu \bar{q}_\nu}{\bar{q}^2}\right)
\,V^{a\,\nu}_0(\bar{q})\frac{1}{z}\partial_z
V(\bar{q},z)\right]\right|_{z=\epsilon} +\dots \nonumber \\
&&=-\frac{R}{2g_5^2}
\int \frac{d^4\bar{q}}{(2\pi)^4}  \left.\left[ V^{a\,\mu}_0(-\bar{q})\left(
g_{\mu\nu}-\frac{\bar{q}_\mu \bar{q}_\nu}{\bar{q}^2}\right)V^{a\,\nu}(\bar{q})
\,\frac{1}{\epsilon}\partial_z V(\bar{q},z)\right]\right|_{z=\epsilon}+ \dots
\nonumber \\
&&=-\frac{R}{2g_5^2}
\int \frac{d^4\bar{q}}{(2\pi)^4}
\int d^4\bar{x}\,d^4\bar{x}'  \left.\left[ e^{i\bar{q}\cdot (\bar{x}-\bar{x}')}\,V_0^{a\,\mu}(\bar{x}')
\left(g_{\mu\nu}-\frac{\bar{q}_\mu \bar{q}_\nu}{\bar{q}^2}\right)
V_0^{a\,\nu}(\bar{x})
\frac{1}{z}\partial_z V(\bar{q},z)\right]\right|_{z=\epsilon}+\dots ,\label{eq:W}
\end{eqnarray}
where the ellipsis 
represents terms more than quadratic in $V^{a\,\mu}_0$. The transverse projection operator  enforces the transverseness of field-theory correlators derived from this generating functional. 

The AdS/CFT prediction for the current-current correlator 
follows: \begin{eqnarray}
\langle J_\mu^a(x)\,J_\nu^b(0)\rangle&\!\!=\!\!&
\frac{\delta^2W}{\delta V^{a\,\mu}(x)\,
\delta V^{b\,\nu}(0)} \nonumber \\
&\!\!=\!\!& 
-\delta^{ab}\,\frac{R}{g_5^2}\,
\int \frac{d^4\bar{q}}{(2\pi)^4}
\int d^4\bar{x}\,d^4\bar{x}' \,e^{i\bar{q}\cdot (\bar{x}-\bar{x}')}
\delta^4(\bar{x}'-x)\,\left(g_{\mu\nu}-\frac{\bar{q}_\mu \bar{q}_\nu}{\bar{q}^2}\right) \delta^4(\bar{x})\left.\frac{1}{z}\partial_z V(\bar{q},z)
\right|_{z=\epsilon} \nonumber \\
&\!\!=\!\!& -\delta^{ab}\,\frac{R}{g_5^2}\,
\int \frac{d^4\bar{q}}{(2\pi)^4}
\, e^{-i\bar{q}\cdot x}\,\left(g_{\mu\nu}-\frac{\bar{q}_\mu \bar{q}_\nu}{\bar{q}^2}\right)
\left.\frac{1}{z}\partial_z V(\bar{q},z)\right|_{z=
\epsilon} . \end{eqnarray}
Fourier transforming, \begin{equation}
\int d^4x \,e^{iq\cdot x}\,\langle J_\mu^a(x)\,J_\nu^b(0)\rangle
= -\delta^{ab}\,\frac{R}{g_5^2}\,\left(g_{\mu\nu}-\frac{q_\mu q_\nu}{q^2}\right)
\left.\frac{1}{z}\partial_z V(q,z)\right|_{z=\epsilon} .
\end{equation}
Expressing the bulk-to-boundary propagator $V(\bar{q},z)$ in terms of the
Kaluza-Klein modes as in Eq.~(\ref{eq:B2B}), we obtain the decomposition
\begin{equation}
\int d^4x\, e^{iq\cdot x}\,\langle J_\mu^a(x)\,J_\nu^b(0)\rangle
= \delta^{ab}
\,\frac{R}{g_5^2}\,\left(g_{\mu\nu}-\frac{q_\mu q_\nu}{q^2}\right)
\lim_{z\rightarrow\epsilon}
\sum_n\,\frac{\left(\psi_n'(\epsilon)/\epsilon\right)
\left(\psi_n'(z)/z\right)}{q^2-m_n^2}. \end{equation}
The polarization $\Pi_V(q^2)$ is defined in terms of the two-point function of
currents via \begin{equation}
\int d^4x \,e^{iq\cdot x}\,\langle J_\mu^a(x)\,J_\nu^b(0)\rangle
= - \delta^{ab}\left(g_{\mu\nu}q^2-q_\mu q_\nu\right)\Pi_V(q^2),
\end{equation}
so we identify \begin{eqnarray}
\Pi_V(q^2)&=&\frac{R}{g_5^2}\frac{1}{q^2}
\left.\frac{\partial_z V(q,z)}{z}\right|_{z=
\epsilon} \label{eq:PiV0} \\
&=&-\frac{R}{g_5^2}\,\lim_{z\rightarrow\epsilon}
\sum_n\,\frac{\left(\psi_n'(\epsilon)/\epsilon\right)
\left(\psi_n'(z)/z\right)}{q^2(q^2-m_n^2)} \label{eq:PiV1}\\
&=&-\frac{R}{g_5^2}\,\sum_n\,\frac{\left[\psi_n'(\epsilon)/\epsilon\right]^2}{
m_n^2(q^2-m_n^2)}, \label{eq:PiV2}\end{eqnarray}
up to a contact interaction from the 
replacement of the factor of $1/q^2$ in Eq.~(\ref{eq:PiV1}) with
$1/m_n^2$ in Eq.~(\ref{eq:PiV2}).

The analytic solution for 
$V(q,z)$ as the regulator $\epsilon\rightarrow 0$ is \begin{equation}
V(q,z)=\frac{\pi q z}{2J_0(q z_m)} \left(J_1(qz)Y_0(q z_m)-Y_1(qz)J_0(qz_m)
\right).
\end{equation}
so we obtain an analytic solution for the polarization $\Pi_V$ \cite{ArkaniHamed:2000ds,Agashe:2004rs,DaRold:2005zs}, 
\begin{equation}  \label{PiV}
\Pi_V(q^2)=-\frac{\pi R\,\left(J_0(qz_m)Y_0(q\epsilon)-Y_0(qz_m)J_0(q
\epsilon)\right)}{2g_5^2\,J_0(q z_m)}. \end{equation}

\noindent
The rho masses $m_{\rho_n}$ 
are determined by the poles of $\Pi_V(q^2)$, at which 
\begin{equation}
J_0(m_{\rho_n} z_m)=0. \label{eq:rhomasses}\end{equation}
Analytically continuing the bulk-to-boundary propagator $V(q,z)$ in $q$ and 
expanding for large Euclidean momentum, $-q^2\gg 1$, 
near the boundary of AdS$_5$, \begin{equation}
V(q,z)= 1+\frac{(-q^2 z^2)}{4}\log(-q^2 z^2)+\dots, \end{equation}
and \begin{equation}
\Pi_V(q^2)=-\frac{R}{2g_5^2}\log(-q^2\epsilon^2)+\mbox{const.}. 
\label{eq:PiVconf}
\end{equation}
This is the correct logarithmic behavior for the current-current correlator
in the conformal theory.  The identification of Eqs.~(\ref{eq:PiVconf}) and 
(\ref{eq:PiV2}) is reminiscent of a Weinberg sum rule.  The sum over
Kaluza-Klein modes in Eq.~(\ref{eq:PiV2}) can be identified with a sum
over resonances $\rho_{n}^{a\,\mu}$ 
carrying the quantum numbers of the current $J_\mu^a$.  With this
identification, we can read off the decay constants
$F_n$ such that
$\langle 0|J_\mu^a(0)|\rho_n^b\rangle=F_n\,\delta^{ab}\varepsilon_\mu$
for a resonance with transverse polarization $\varepsilon_\mu$:
\begin{equation}
F_n^2=\frac{R}{g_5^2}\left[\psi_n'(\epsilon)/\epsilon\right]^2.
\label{eq:Fn}\end{equation}
If we choose to match this result to the perturbative QCD result 
$\Pi_V(q^2)\approx -\frac{N_c}{24\pi^2} \log(-q^2)$ as in
Ref.~\cite{Erlich:2005qh,DaRold:2005zs}, we then identify \begin{equation}
g_5^2=\frac{12 \pi^2}{N_c}. \label{eq:g5}\end{equation}
If $z_m$ is chosen so that $m_\rho=776$ MeV, then the model predicts the decay constant for the rho to be $F_\rho^{1/2}=329$ MeV~\cite{Erlich:2005qh}, to be compared to the experimental value of $345\pm8$ MeV~\cite{PDG:2014}.  In principle this model predicts properties of an infinite spectrum of radial excitations of the rho, though the model also ignores all resonances other than these, and it is not surprising that  predictions are inaccurate at scales significantly higher than the mass of the lightest rho meson.  

A generic feature of  holographic models based on classical equations of motion is that he Kaluza-Klein resonances are infinitely narrow, as can be seen in this example by the fact that the poles in $\Pi_V$ are at real values of $q^2$.  The classical limit, or supergravity limit, of the AdS/CFT correspondence is a large-$N_C$ limit of the gauge theory.   The width of mesonic resonances in QCD goes to zero as the number of colors goes to infinity, so this feature of holographic models may be considered a remnant of the large-$N_C$ limit, even as $N_C$ is set to 3 in results such as Eq.~(\ref{eq:g5}).  In more elaborate models, couplings of the rho meson to pions can be calculated, from which the width of the rho meson can be inferred.  This begins a  bootstrap approach to holographic model building in which quantum corrections are self-consistently included.  This approach has not been elaborated on in the literature.

\section{A bottom-up model with chiral symmetry breaking \label{abum}}

To reproduce the approximate $SU(2) \times SU(2)$ chiral symmetry of QCD,
the hard-wall model of Refs.~\cite{Erlich:2005qh, DaRold:2005zs} 
includes a higher-dimensional $SU(2) \times SU(2)$
gauge invariance.  Treating isospin as unbroken, we consider the breaking of the chiral symmetry due to the chiral condensate $\sigma=
\langle (\overline{u}_{L}u_R+\overline{u}_{R}u_L) \rangle=\langle (\overline{d}_{L}d_R+\overline{d}_{R}d_L) \rangle$.
Correspondingly, the model includes a set of 5-dimensional fields $X_{Ii}$, $I,i\in(1,2)$, 
which transform as a bifundamental under the $SU(2) \times SU(2)$ gauge 
invariance.  A non-vanishing background for $X_{Ii}$ breaks the
chiral symmetry via the Higgs mechanism.  If we assume the breaking is such
that \begin{equation}
\langle X^{Ii}(x,z)\rangle = X_0(z)\delta^{Ii}, \label{eq:XIi}\end{equation}
then the diagonal subgroup of the gauge group is unbroken by the background.  
This unbroken 
group is identified with isospin, and Kaluza-Klein modes of the corresponding gauge fields are identified with the rho meson and its radial excitations.  The broken sector  describes the axial-vector mesons and pions.  The AdS/CFT correspondence
motivates an identification of quark mass and chiral condensate as 
coefficients in the background $X_0(z)$.  If no other sources of symmetry breaking are introduced in the model, this pattern of symmetry breaking
ensures that approximate relations like the Gell-Mann-Oakes-Renner relation
\cite{GellMann:1968rz},
which are due to the pattern of chiral symmetry breaking, are
satisfied in the model \cite{Erlich:2005qh}.  However, it is important to ensure that the boundary conditions on fluctuations about the background are consistent with the pattern of chiral symmetry breaking \cite{Albrecht:2011xk}.

The symmetry-breaking background solves the equations
of motion for the fields $X^{Ii}$ in the background space-time.  For simplicity
we assume that the scalar fields have a mass $m^2$ but otherwise no potential, though nontrivial potentials have been considered (for example, Ref.~\cite{Gherghetta:2009ac}).  We
further assume that there is little backreaction from the profile
$X_0(z)$ on the space-time geometry.   The
action describing non-gravitational fluctuations in the model is~\cite{Erlich:2005qh, DaRold:2005zs}:
\begin{equation}\label{eq:S5D}
S=\int\!d^5x\,\sqrt{g}\, {\rm Tr}\Bigl\{ |DX|^2  -m^2 |X|^2
- \frac{1}{2g_5^2} (F_L^2 + F_R^2)\Bigr\},
\end{equation}
where $D_\mu X=\partial_\mu X - iA_{L\mu} X + iX A_{R\mu}$,
$A_{L,R}=A_{L,R}^aT^a$, and $F_{L,R}^{\mu\nu}=\partial^\mu A_{L,R}^\nu
-i[A_{L,R}^\mu,A_{L,R}^\nu]$, and indices are contracted with the AdS metric of Eq.~(\ref{eq:AdSmetric}).

The scalar field equations of motion for the background
$X_0(z)$ are:
\begin{equation}
 \frac{d}{dz}\left(\frac{1}{z^3}\frac{d}{dz}X_0\right)+\frac{m^2R^2}{z^5}X_0=0,
\end{equation}
with solutions \begin{equation}
X_0(z)=m_q\,z^{4-\Delta}+\frac{\sigma}{4(\Delta-2)}\,z^\Delta, 
\label{eq:X0}\end{equation}
where we assume $2<\Delta<4$ and we take $m_q$ and $\sigma$ real.

According to the AdS/CFT dictionary, the mass $m$ of a 5-dimensional
$p$-form field depends
on the scaling dimension $\Delta$ of the corresponding operator, according to \req{Deltamup}
 \begin{equation}
m^2=(\Delta-p)(\Delta+p-4). \label{eq:mDelta}\end{equation}
This relation also follows from demanding the correct scaling behavior of correlation
functions deduced by the AdS/CFT correspondence in the conformal
field theory~\cite{Gubser:1998bc, Witten:1998qj}, 
or in the deep Euclidean regime in models of asymptotically free confining theories.
For example, for a conserved current, $\Delta=3$ and $p=1$, 
corresponding to a massless 5-dimensional gauge field as discussed above.  
In AdS/QCD models it is reasonable to leave the effective scaling
dimension $\Delta$ of
most operators as adjustable parameters, because renormalization modifies the
scaling dimension of operators at low energies.
However, conserved currents  
are not renormalized and always correspond to massless 5-dimensional gauge fields in
the dual theory.

For scalar fields like $X_{Ii}$, $p=0$ and the exponent $\Delta$ is related to the  mass $m$ via,
\begin{equation}
m^2=\Delta(\Delta-4). \label{eq:mDelta0}\end{equation}
A QCD operator with 
appropriate quantum numbers to be dual to $X_{Ii}$ is the scalar quark bilinear 
$\langle \overline{q}^{I}_{L}q^{i}_{R}\rangle$ where $(q^1,q^2)=(u,d)$, which in the UV has dimension $\Delta=3$.

In the bottom-up approach we treat the 5-dimensional scalar mass squared, $m^2$, 
as a free parameter\footnote{In the light-front holographic approach the 5-dimensional mass $m$ is not  free parameter, but it is fixed by the holographic mapping to the light-front (See Chapters \ref{ch4} and \ref {ch5}).}.
However, for 
definiteness we can fix $m^2=-3/R^2$, so that $\Delta=3$.
The action Eq.~(\ref{eq:S5D}) with $m^2=-3/R^2$, 
together with the background space-time geometry
Eq.~(\ref{eq:AdSmetric}) between $z=0$ and $z_m$, defines one version of 
the hard-wall AdS/QCD model~\cite{Erlich:2005qh,DaRold:2005zs}.  

The solution $X_0(z)\propto z^{4-\Delta}$ is
not normalizable in the sense defined earlier:  For scalar field 
profiles of the form $X_{Ii}(x,z)=X_0(z)\,X_{Ii}(x)$, the integral over $z$ in
the action is divergent if $X_0(z)\propto z^{4-\Delta}$ if $\Delta>2$.  On the
other hand, the solution
$X_0(z)\propto z^\Delta$ is normalizable.
The coefficients of the two solutions are defined 
in anticipation of their physical interpretation via the AdS/CFT 
correspondence.  The normalizable solution to the equation
of motion corresponds to the state of the system \cite{Balasubramanian:1998de,Klebanov:1999tb}. In this case the vacuum
expectation value of the operator corresponding to the field $X$, which is
proportional to the
chiral condensate $\sigma$  (We assume
that $\sigma$ in Eq.~(\ref{eq:X0}) is real).  
The non-normalizable solution corresponds to
the source for the corresponding operator, in this case the 
(isospin-preserving) quark mass 
\cite{Gubser:1998bc, Witten:1998qj}.
The factor of $1/4(\Delta-2)$ in Eq.~(\ref{eq:X0}) is suggested by the 
AdS/CFT correspondence
\cite{Klebanov:1999tb,Erlich:2008gp}.
As noted by Cherman, Cohen and Werbos \cite{Cherman:2008eh}, the interpretation
of $m_q$ and $\sigma$ with quark mass and chiral condensate is only up to a
rescaling of $m_q$ and $\sigma$.  Matching the holographic prediction of the
two-point correlator of scalar operators $\overline{q}q$ with the QCD 
prediction at large Euclidean momentum, the parameter $m_q$ should be
rescaled by $\sqrt{N_C}/2\pi$ and $\sigma$ by $2\pi/\sqrt{N_C}$ in order for $m_q$
and $\sigma$ to represent the quark mass and chiral condensate, respectively 
\cite{Cherman:2008eh}.
Although the identification of parameters in the scalar field background with
QCD parameters would follow from the AdS/CFT correspondence, 
they may more conservatively be regarded as model parameters allowed to
vary in order to fit experimental data.

We define the vector and axial-vector combinations of the gauge field
$V_M^a=(L_M^a+R_M^a)/\sqrt{2}$ and 
$A_M^a=(L_M^a-R_M^a)/\sqrt{2}$, respectively.  
Under parity the gauge fields transform as
$L_i^a(t,\mathbf{x},z)\leftrightarrow -R_i^a(t,-\mathbf{x},z)$,
$L_0^a(t,\mathbf{x},z)\leftrightarrow R_0^a(t,-\mathbf{x},z)$,
$L_z^a(t,\mathbf{x},z)\leftrightarrow R_z^a(t,-\mathbf{x},z)$.  
The vector combination 
$V_i^a=(L_i^a+R_i^a)/\sqrt{2}$ is odd under parity;  
the axial-vector combination $A_i^a=(L_i^a-R_i^a)/\sqrt{2}$ 
is even.  
Expanding the scalar fields about the background, we may write \begin{equation}
X(x,z)=\left(X_0(z)+\sigma(x,z)\right)\,\exp\left[2i\,T^a\pi^a(
x,z)\right], 
\end{equation}
where $T^a$ are $SU(2)$ generators normalized so that ${\rm Tr}\, T^aT^b=
\delta^{ab}/2$.
In this matrix notation, we also define $L_M=L_M^aT^a$ and $R_M=R_M^aT^a$, and
similarly for $V_M$ and $A_M$.  The Hermitian matrix of fields 
$\sigma(x,z)$ are scalar under Lorentz invariance and
parity, and for the time being we set these fields to zero.  
The equations of motion for $\sigma$, $\pi$ and the gauge fields, with appropriate boundary conditions \cite{Albrecht:2011xk}, determine the spectrum of Kaluza-Klein modes, which are identified with the corresponding meson states.  
The quantum numbers of the mesons are determined by transformations of the
5-dimensional fields 
under symmetries of the 5-dimensional theory.  The gauge fields are odd under charge
conjugation.  Under 5-dimensional parity
the vector $V_\mu^a$ is odd; the axial-vector $A_\mu^a$ is even.  Hence, 
the Kaluza-Klein modes of the transverse part of $V_\mu^a$ are 
identified with the tower of radial excitations of
rho mesons; the Kaluza-Klein modes of the
transverse part of $A_\mu^a$ are identified with the tower of radial 
excitations of $a_1$ mesons. The fields $\pi^a$, which mix with the longitudinal part of $A_\mu^a$, are odd under  charge conjugation and parity. The solutions to the coupled equations of motion for $\pi^a$ and the longitudinal part of $A_\mu^a$, with appropriate boundary conditions, are identified with the pions.

The decay constants for the vector and axial-vector mesons
may be calculated either from the AdS/CFT correspondence as described above, or by examination of the effective action for the Kaluza-Klein modes upon integration of the action over the $z$-coordinate \cite{Sakai:2004cn}.  The model defined above has three free parameters: $m_q$, $\sigma$, and $z_m$.  A global fit  for the lightest pion, rho and $a_1$ masses and decay constants yields agreement with data at better than 10\% \cite{Erlich:2005qh}.  The effective action also determines couplings like $\rho$-$\pi$-$\pi$, which generally do not fare as well.  However, the existence of a non-vanishing $\rho$-$\pi$-$\pi$ coupling in the model predicts a non-vanishing width for the rho meson, which would then shift the corresponding pole of the vector current two-point correlator off the real axis.

\begin{table}[ht]
\centering
\begin{tabular}{|c|c|c|}
\hline
Observable&Measured&Model Fit \\
&(Central Value-MeV)&(MeV)  \\  
\hline  
$m_\pi$&139.6&134 \\
$f_\pi$&92.4&86.6 \\
$m_K$&496&514 \\
$f_K$&113&101 \\
$m_{K_0^*}$&672&697 \\
$f_{K_0^*}$&&36 \\
$m_\rho$&776&789\\
$F_\rho^{1/2}$&345&335 \\
$m_{K^*}$&894&821 \\
$F_{K^*}$& &337 \\
$m_{a_1}$&1230&1270 \\
$F_{a_1}^{1/2}$&433&453 \\
$m_{K_1}$&1272&1402 \\
$F_{K_1^{1/2}}$&&488 \\ [0.8ex]  
\hline
\end{tabular}
\caption{\label{tab:strange} \small Fit of a five-parameter hard-wall model for meson masses and decay constants, from Ref.~\cite{Abidin:2009aj}.}
\end{table}

The addition of the strange quark is straightforward.  The $SU(2) \times SU(2)$
gauge group becomes $SU(3) \times SU(3)$, and to separate the strange quark
mass from the up and down quark masses one can expand about the background
$X_0(z)={\rm diag}(m_q,m_q,m_s)z +\sigma z^3/4$, as in Ref.~\cite{Abidin:2009aj}.
The result of a five-parameter model to fourteen observables is given in Table~\ref{tab:strange} from Ref.~\cite{Abidin:2009aj}.

\section{Top-down models}
\label{sec:top-down}

Top-down AdS/QCD models are based on
configurations of D-branes in string theory are more restrictive than bottom-up models.
Strings ending on D-branes have a
field-theoretic interpretation at low energies, and the AdS/CFT correspondence
provides a dual description of the field theory in terms of a supergravity
or string theory in a fixed background space-time with fluxes of certain fields.
The class of QCD-like theories which emerge from D-brane configurations
with supergravity duals is rather limited, and those theories each
differs from QCD in
a number of important ways.  Top-down models are motivated rather directly by the AdS/CFT correspondence, so this is an appropriate point to briefly review that motivation.

In the top-down approach, a brane configuration in string theory is engineered whose low-energy spectrum
of open-string fluctuations has a known field-theoretic interpretation.  
Via the AdS/CFT correspondence, for some brane constructions 
describing large-$N_C$ gauge theories with large 't Hooft coupling $g^2 N_C$,
a dual description exists in terms of supergravity on
a fixed space-time background supported by fluxes of certain fields \cite{Maldacena:1997re}.   The prototypical example
is ${\cal N}=4$ supersymmetric gauge theory, which has as its dual (in 
the supergravity limit $N_C \rightarrow\infty$ and $g^2 N_C \gg 1$) Type IIB supergravity on
an AdS$_5\times S^5$ space-time background with non-vanishing five-form flux and constant dilaton.

For this prototypical example Maldacena considered the dynamics of a stack of overlapping D3-branes in Type IIB string theory~\cite{Polchinski:1998rr}. The number 3 in ``D3-branes'' indicates that the branes span three spatial dimensions.  Quantizing the open strings which end on the $N_C$ D3-branes, one finds the massless spectrum of ${\cal N}=4$ supersymmetric  $SU(N_C)$ gauge theory in 3+1 dimensions.  The massless spectrum of the closed strings describes Type IIB supergravity, which includes a scalar dilaton field and a spin-2 graviton, in addition to various $p$-form and fermionic fields.  The gauge coupling $g$ in the action for the $SU(N_C)$ gauge theory on the D3-branes is also related to the string coupling $g_s$, by $g_s= g^2$.

If $g_s\rightarrow0$  then perturbative string theory corrections are negligible, and classical supergravity is an appropriate description of the theory.  In addition to a non-vanishing constant dilaton, the solutions to the supergravity equations of motion include a non-vanishing five-form flux (for which the D3-branes act as sources) and a space-time with horizon.   The near-horizon geometry of this space-time is the aforementioned AdS$_5 \times S^5$.  The curvature of both the anti-de Sitter factor and the five-sphere  is given by a distance scale  $R_5 = l_s (g_s N_C)^{1/4}$, where $l_s$ is the string length.  Hence, if $g_s N_C\gg 1$ then the geometries are smooth on scales of the string length, and massive string modes can be neglected so that  fluctuations are described by classical supergravity without higher-derivative operators induced by the massive modes.  In this case the 't Hooft coupling $g^2N$ is large in the gauge theory on the D3-branes.  The supergravity limit, in which classical supergravity describes the fluctuations of the background created by the stack of D-branes, is what we have just described: $g_s\rightarrow0$ with $g_s N_C\gg 1$, or equivalently, $N_C \rightarrow \infty$ with $g^2 N_C \gg1$.

Maldacena's great conceptual leap was to identify the physics of the open strings ({\em i.e.} ${\cal N}=4$ Yang-Mills theory) with the physics of the closed strings in the near-horizon geometry ({\em i.e.} AdS$_5 \times S^5$).
Some evidence for the duality is provided by a matching of symmetries, in particular the $SO(2,4)$ conformal symmetry and the $SO(6)$ $R$-symmetry of the ${\cal N}=4$ Yang-Mills theory, which are identical to the isometries of $AdS_5$ and $S^5$, respectively.  Certain classes of operators in the ${\cal N}=4$ theory  naturally map onto the spherical harmonics on the $S^5$.
The explicit dictionary between the dual theories was proposed independently by Witten \cite{Witten:1998qj} and Gubser, Klebanov and Polyakov \cite{Gubser:1998bc}.   Operators in the field theory correspond to supergravity fields on AdS$_5$. In the supergravity limit $N_C \rightarrow \infty$ with $g^2 N_C \gg 1$, the generating functional for connected correlation functions in the ${\cal N}=4$ theory is identified with the action in the supergravity on the AdS$_5 \times S^5$ background, with the constraint that the supergravity fields approach (3+1)-dimensional configurations on the boundary of AdS$_5$ that are the sources for the corresponding operators in the ${\cal N}=4$ theory.

Karch and Katz \cite{Karch:2002sh} suggested the possibility to add to the basic 
scenario a small number of matter
fields that transform in the fundamental representation of the $SU(N_C)$ gauge
group by adding to the $N$ D3-branes a 
small number of D7-branes.  The light
fluctuations of strings which stretch from the D7-branes to the D3-branes
include scalar fields and fermions which transform in the fundamental representation of the $SU(N_C)$ gauge group, in analogy to the quarks of $SU(3)$ QCD.  The resulting theory is not asymptotically free, but progress is made by treating the D7-branes as probes while ignoring their backreaction on the supergravity background, in which case the difficulties associated with the loss of asymptotic freedom are conveniently evaded.

There now exists a large number of examples of field theories with supergravity
duals, and the basic dictionary has been expanded to include theories with interesting phenomenological features.  Conformal invariance and supersymmetry are not essential.  The field theory can be confining with
chiral symmetry breaking, which in those respects is similar to QCD.  
Some examples of confining theories
with known supergravity 
duals are the ${\cal N}$=1$^*$ theory of Polchinski and Strassler 
\cite{Polchinski:2000uf}, the Klebanov-Strassler cascading gauge theory
\cite{Klebanov:2000hb}, the D4-D6 system of Kruczenski {\em et al.}
\cite{Kruczenski:2003uq}, 
and the D4-D8 system of Sakai and Sugimoto \cite{Sakai:2004cn,Sakai:2005yt}.
The predictions of a top-down model pertain to the specific theory dual to the particular supergravity background defining the model.  The benefit of top-down models is that both sides of the duality can be described, at least in part, independently of the duality.  Top-down AdS/QCD models are engineered to have certain similarities to QCD, but always suffer from a difficulty in separating scales of the desired degrees of freedom from additional degrees of freedom and interactions not present in QCD.

Although the AdS/CFT correspondence is conjectured to extend to smaller $N_C$ by extending the supergravity to the full string theory, for the sake of calculability AdS/QCD models generally ignore the string-theoretic corrections and naively extend the supergravity description of the AdS/CFT correspondence to QCD, with $N_C=3$.  
Certain remnants of the large-$N_C$ approximation remain at first approximation in both top-down and bottom-up AdS/QCD models, such as infinitely narrow resonances.
Additional difficulties in matching aspects of QCD, such as the distribution of jets at high energies, have been discussed by various authors, for example \cite{Csaki:2008dt, Unsal:2008eg}.

\begin{figure}
\centering
\includegraphics[angle=0,width=12.0cm]{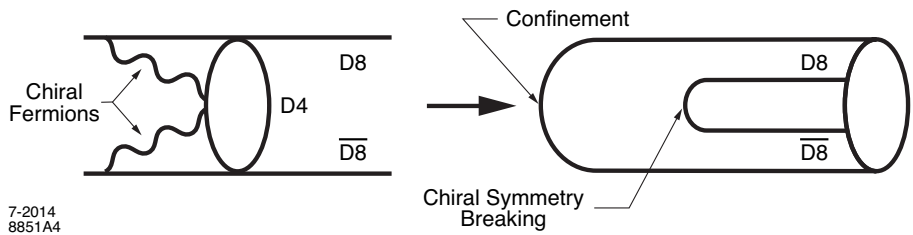}
\caption{\label{fig:1} \small Brane configuration in the Sakai-Sugimoto model.  The D4-branes
create a cigar-like geometry, and the D8 and $\overline{{\rm D8}}$-branes
connect. A mass gap is created by the cigar-like box, and chiral symmetry breaking arises from the joining of the branes responsible for the left and right-handed chiral fermions.  }
\end{figure}

\section{The Sakai-Sugimoto model}

The Sakai-Sugimoto model is so far the top-down application of the AdS/CFT correspondence
most closely related to QCD.  The system includes a stack of $N$ D4-branes in Type IIA string theory~\cite{Polchinski:1998rr} wrapped on a circle on which
fermions satisfy antiperiodic boundary conditions which break the supersymmetry of the theory.
From an effective 3+1 dimensional point of view, the massless spectrum includes
the $SU(N)$ gauge fields, but the fermions and scalars are massive.  At low energies, the theory is described by pure Yang-Mills theory. Without the additional flavor branes introduced by Sakai and Sugimoto, this model was discussed by Witten shortly after the initial AdS/CFT conjecture \cite{Witten:1998zw} in the context of the deconfinement transition in QCD. In that case the circle is the compactified Euclidean time, whose size is inversely related to the temperature of the system.  Kruczenski {\em et al.} \cite{Kruczenski:2003uq} suggested the addition of probe D6 and anti-D6 $(\overline{{\rm D6}})$ branes to the model, which gives rise to fermionic quark fields in the low-energy spectrum, but without the pattern of chiral symmetry breaking observed in nature. 
Instead of the probe D6-branes, the Sakai-Sugimoto model  includes $N_f$ D8-branes and $N_f$ $\overline{{\rm D8}}$-branes
transverse to the circle on which the D4-branes wrap. The D8 and  $\overline{{\rm D8}}$ branes  span all of the nine spatial dimensions of the string theory except the circle on which the D4-branes are wrapped, and they intersect
the D4-branes on 3+1 dimensional manifolds at definite positions along the
circle, as in Fig.~\ref{fig:1}.  The massless fluctuations of open strings
connecting the D4 and D8 or $\overline{{\rm D8}}$-branes at their intersections
describe 3+1 dimensional chiral fermions, with opposite chirality at the D8
and $\overline{{\rm D8}}$-branes.  This is the Sakai-Sugimoto model 
\cite{Sakai:2004cn,Sakai:2005yt}.  There are two  ``ultraviolet'' regions on the D8-branes, for which the geometry is asymptotically different than AdS, and the gauge theory dual to the Sakai-Sugimoto model is not asymptotically free.  In fact, since the geometry in this model is not asymptotically AdS, it cannot account for dimensional power scaling~\cite{Polchinski:2001tt, Brodsky:1973kr, Brodsky:1974vy, Matveev:ra} in QCD for hard scattering. The Sakai-Sugimoto model should be considered a model of QCD only at low energies.

In the supergravity limit the D4-branes generate a space-time with a 
cigar-like topology which effectively cuts off the space-time geometry at the tip of the cigar.
The 9+1 dimensional space-time metric generated by the D4-branes wrapped on a circle is \cite{Witten:1998zw} \begin{equation}
ds^2=\left(\frac{U}{R}\right)^{3/2}\left(\eta_{\mu\nu}dx^\mu dx^\nu+f(U)d\tau^2\right)+\left(\frac{R}{U}\right)^{3/2}\left(\frac{dU}{f(U)}+U^2d\Omega_4^2\right), \end{equation}
where the Greek indices span 3+1 dimensions, $\tau$ is the coordinate on the circle, and $d\Omega_4^2$ is the metric on the unit four-sphere.  The coordinate $U$ plays a role analogous to the radial coordinate $z$ of Anti-de Sitter space in the holographic interpretation of this model.  The function $f(U)$ describes the horizon at some $U=U_{{\rm KK}}$: \begin{equation}
f(U)=1-\frac{U_{{\rm KK}}^3}{U^3}.\end{equation}
As $U$ approaches the horizon at $U_{KK}$ the proper size of the circle in the $\tau$-direction shrinks to zero size.  
For the geometry to be smooth, the location of the
tip of the cigar is correlated with the size of the compact circle on which the 
D4-branes wrap \cite{Witten:1998zw,Kruczenski:2003uq}.  This is one reason that non-QCD states are not decoupled from the hadronic states of interest.
The dilaton $\phi$ also has a profile in the D4-brane background, and plays a role in the action on the D8-branes: \begin{equation}
e^{\phi}=g_s\left(\frac{U}{R}\right)^{3/4}, \end{equation}
where $g_s$ is the string coupling.  The D4-branes also generate a 4-form flux which supports the space-time geometry, but otherwise that flux does not play a role in the present discussion.

In the D4-brane  background, ignoring the backreaction of the D8 and $\overline{{\rm D8}}$-branes on the geometry, the D8 and $\overline{{\rm D8}}$-branes bend as a consequence of their tension so as to minimize the Dirac-Born-Infeld action on the branes.  The D8-brane profile is described by a curve in the $U$-$\tau$ coordinates, which is specified by a function $U=U(\tau)$.  The result is that 
the D8 and $\overline{{\rm D8}}$-branes connect at some minimum value of $U$, which corresponds to the spontaneous breaking of the $SU(N_f) \times SU(N_f)$ chiral symmetry of the theory to a diagonal $SU(N_f)$ isospin symmetry. 

The perturbative massless  spectrum is that of SU($N$)
QCD with $N_f$ flavors of  quarks.  However, Kaluza-Klein modes associated
with the circle direction have masses comparable to the confining scale
in this theory, so the massive spectrum of QCD-like bound states are not 
separated in mass from the spectrum of non-QCD-like Kaluza-Klein modes.
As a result, the  five-dimensional nature of the effective theory on the D4-branes becomes
apparent at the same scale as the hadron masses we are interested in.
This is an important distinction between the Sakai-Sugimoto model and QCD, and typically the unwanted states are ignored when comparing the model to experiment.  

If we ignore the Kaluza-Klein modes around the circle, and integrate out the four dimensions  along the D8-branes but transverse to the D4-branes, then
the fluctuations of the 4+1 dimensional 
$SU(N_f) \times SU(N_f)$ gauge fields on the D8
and $\overline{{\rm D8}}$-branes are identified with vector mesons,
axial-vector mesons, and pions.  The quantum numbers of the corresponding
states can be identified with symmetries of the D-brane system.  The 3+1 
dimensional parity symmetry,
for example, is identified with a parity symmetry in the 
4+1 dimensional theory, which also exchanges the two sets of
$SU(N_f)$ gauge fields.  If we ignore the extra circle direction, then
the effective 3+1 dimensional action on the D8-branes at the intersection with the D4-branes describes the 
effective action for the light mesons, and allows for the comparison
of decay constants ($f_\pi$, $F_\rho$, {\em etc.}) and couplings ({\em e.g.}
$g_{\rho\pi\pi}$), or alternatively some of the chiral low-energy coefficients, with QCD.    
At this stage the setup is similar to that of bottom-up models, except that the D8-branes have two asymptotic UV boundaries (one for each $SU(N_f)$ factor), and
 the AdS$_5$ space-time is replaced by the induced metric on the D8-branes, 
\begin{equation}
ds^2=\left(\frac{U}{R}\right)^{3/2}\eta_{\mu\nu} dx^\mu dx^\nu-\left(\frac{R}{U}\right)^{3/2}\left[\frac{1}{f(U)}+\left(\frac{U}{R}\right)^3\frac{f(U)}{U'(\tau)^2}\right]dU^2  -\left(\frac{R}{U}\right)^{3/2}U^2d\Omega_4^2, \end{equation}
and the four-sphere is integrated over in the effective 4+1-dimensional action, after which it no longer plays an important role.  (This is analogous to the role of the five-sphere in the prototypical AdS$_5 \times S^5$ geometry in the AdS/CFT correspondence.)

Once  the effective 4+1 dimensional action for the gauge fields on the D8-brane is evaluated, the calculation of meson observables proceeds as in bottom-up models.  One  difference of note is that there are two UV boundaries in the Sakai-Sugimoto model, one for each factor of $SU(N_f)$ in the chiral symmetry.  There is one set of $SU(N_f)$ gauge fields in the geometry with two ultraviolet regions rather than two sets of $SU(N_f)$ gauge fields in the hard-wall model, with one ultraviolet region.  The potential to enhance the symmetries of a theory by way of additional ultraviolet regions was exploited in unrelated models in Ref.~\cite{Carone:2008rx}.

Several related descriptions of light baryons in AdS/QCD have been identified.  
The effective 4+1 dimensional gauge theory on the D8-branes, after integrating out the additional 4-sphere in the D8-brane geometry, has small instanton solutions which correspond to D4-branes wrapped on the 4-sphere \cite{Hata:2007mb,Hong:2007kx}.
Also integrating out the remaining extra dimension along the induced geometry on the D8-branes, the Skyrme term in the effective 3+1 dimensional pion Lagrangian emerges directly from the kinetic terms in the action for the $SU(2)_L \times SU(2)_R$ gauge fields \cite{Sakai:2004cn,Sakai:2005yt}.  The baryon number of the Skyrmion is equivalent to the instanton number in the 4+1 dimensional effective theory \cite{Hata:2007mb}, so the Sakai-Sugimoto model provides a connection between various solitonic descriptions of the baryon~\cite{Schechter:1999hg}. A similar discussion in a bottom-up model was given in Ref.~\cite{Pomarol:2008aa}.

Despite the differences between the Sakai-Sugimoto model and QCD, the qualitative, and sometimes quantitative, success of a naive application of the AdS/CFT correspondence  when compared with QCD has helped to further the hope that interesting features of QCD might be better understood by consideration of holographic models.  The geometry on which the D8-branes live is similar to one considered earlier by Son and Stephanov, albeit in latticized form, in the context of extended hidden local symmetry models \cite{Son:2003et}.  The Sakai-Sugimoto model is also  similar to the hard-wall model with chiral symmetry breaking, although the quarks in the Sakai-Sugimoto model are exactly massless. An alternative approach to chiral symmetry breaking with massive quarks in a system with D7 branes in a nontrivial background was suggested in Ref.~\cite{Babington:2003vm}.

\chapter{Summary and Conclusion \label{ch8}}
The AdS/CFT correspondence (or gauge/gravity duality) introduced by  Maldacena~\cite{Maldacena:1997re}  has  given rise to a completely new set of tools for studying the dynamics of strongly coupled quantum field theories such as QCD.  In effect, the strong interactions of quarks and gluons are represented by a simpler classical gravity theory in a higher-dimensional space. Anti-de Sitter space in five dimensions plays a special role in elementary particle physics because it provides an exact geometrical representation of the conformal group.

Although a perfect string theory dual of QCD  is not yet known, bottom-up approaches, starting from a QCD description and searching for a higher-dimensional theory, of which QCD is the boundary theory,   has already provided many new and remarkable insights into QCD.  In particular one has to break by some mechanism the maximal symmetry of  AdS space, because its holographic dual quantum field theory is conformally invariant and cannot incorporate a mass scale like $\Lambda_{\rm QCD}$.

In this report we mainly concentrate  on light-front holographic QCD~\cite{deTeramond:2008ht, Brodsky:2006uqa, deTeramond:2013it}. This approach to strongly coupled quantum field theory exploits the remarkable holographic duality between classical gravity in AdS$_5$ and the semiclassical approximation to light-front quantized QCD.   Light-front Hamiltonian theory, derived from the quantization of the QCD Lagrangian at fixed light-front time $x^+ = x^0+x^3$, provides an ideal  framework for describing bound-states in relativistic theory (Chapter \ref{ch2}). Light-front holography leads to a precise relation between the holographic variable $z$  of the fifth dimension of AdS space,  and the light-front variable $\zeta$, the argument of the boost-invariant light-front wave functions describing the internal structure of hadrons in physical space-time~\cite{Brodsky:2006uqa}.

For massless quarks the classical QCD Lagrangian contains no scale and is conformal invariant. However, the appearance of a scale, or a mass gap, is necessary for confinement. Therefore the origin of a mass scale in an originally conformal theory is a fundamental unsolved problem in the theory of strong interactions.  An indication of the  origin of the QCD mass scale can be drawn from a remarkable paper by  V.~de Alfaro {\it et. al.}~\cite{deAlfaro:1976je} in the context of one-dimensional quantum field theory.  Starting from a conformally invariant action these authors showed how to construct a generalized Hamiltonian from the the three generators of the conformal group in one dimension.   This remarkable result is based in the fact that  the $SO (2,1)$ algebra can be realized in conformal quantum mechanics:  one of the generators of $SO(2,1)$, the rotation in the 2-dimensional space,  is compact and has therefore a discrete spectrum.  Mathematically, this result is based on the isomorphism  of the one-dimensional conformal group ${\it Conf}\!\left(R^1\right)$ with the group $SO(2,1)$. Since the generators of the conformal group $H$, $D$ and $K$ have different dimensions their connection with the generators of $SO(2,1)$ require the introduction of  a scale~\cite{deAlfaro:1976je}, which plays a prominent role in the dAFF procedure.   The evolution parameter corresponding to the dAFF generalized Hamiltonian is proportional to the light front-time and has a finite range~\cite{Brodsky:2013ar}.

The threefold connection of light-front dynamics, classical gravity in a higher-di\-men\-sional space, and a conformal invariant  one-dimensional quantum field theory  provides new  insights into the origin of a fundamental mass scale and the physics underlying confinement dynamics in QCD~\cite{Brodsky:2013ar}. The mapping of the generalized dAFF Hamiltonian to the light front fixes the effective instantaneous light-front potential to a harmonic oscillator form (See Chapter \ref{ch3}). This corresponds to a quadratic dilaton profile in the embedding AdS space. For large separation distances the quadratic effective potential in the front form of dynamics  corresponds to a linear potential in the usual instant form of dynamics~\cite{Trawinski:2014msa}.  The final result is a relativistic light-front  wave equation for arbitrary spin which incorporates essential spectroscopic and dynamical features of hadron physics.

The light-front Hamiltonian equation predicts that the pion has zero mass for massless quarks, and the resulting Regge trajectories have equal constant slope in the radial and orbital quantum numbers as observed experimentally.   There is only one input, the constant $\sqrt \la$ setting the QCD mass scale (See Chapter \ref{ch5}). Our procedure can be extended to non-zero light quarks without modifying to first approximation the transverse dynamics and the universality of the Regge slopes (See Sec. \ref{LQM}). The predicted meson light-front wave function accurately describes diffractive vector meson electroproduction~\cite{Forshaw:2012im} and other observables. The shape of form factors is successfully described (See Chapter \ref{ch6}).

We have derived hadronic bound-state equations for hadrons with arbitrary spin starting from an effective invariant action in the higher-dimensional classical gravitational theory. The mapping of the equations of motion from the gravitational theory to the Hamiltonian equations in light-front quantized QCD has been an important guide to construct the effective actions in the bulk and to separate kinematical and dynamical aspects~\cite{deTeramond:2013it}. For mesons, this separation determines a $J$-dependent constant term in the effective potential. The unmodified AdS geometry reproduces the kinematical aspects of the light-front Hamiltonian, notably the emergence of the light-front angular momentum which is holographically related to the AdS mass in the gravitational theory, and the modification of AdS space in the infrared region encodes the dynamics, including confinement (See Chapter \ref{ch4}).

For baryons the 3-body state is described by an effective two-body light-front Hamiltonian, where the holographic variable is mapped to the invariant separation of one constituent (the active constituent) to the cluster of the rest (the spectators). Therefore, the mapping of AdS equations to the light-front bound-state equations  imply that there is only one relevant angular momentum, the light-front orbital angular momentum $L$ between the active and the spectator cluster, an effective approximation which captures much of the strongly-coupled dynamics. Furthermore, since the action for fermions is linear in the covariant derivatives, no mixing between dynamical and kinematical aspects occurs. Thus, for baryons there is no explicit $J$ dependence in the light-front equations of motion, and consequently the bound-state spectrum of baryons can only depend on $L$ (See Sec. \ref{Baryons}). These remarkable predictions, which are inferred from the geometry of AdS space, are independent of the specific dynamics and account for many  striking similarities and differences observed in the systematics of the meson and baryon spectra~\cite{deTeramond:2013it}. The equality of the slopes of the  linear Regge trajectories and the multiplicity of states for mesons and baryons is explained. We also explain the observed differences in the meson versus the baryon spectra that are due to spin-orbit coupling. For example, the predicted triplet spin-orbit splitting for vector mesons is in striking contrast with the empirical near-degeneracy of baryon states of different total angular momentum $J$; the baryons are classified by the internal orbital angular momentum quantum number $L$ along a given Regge trajecory, not $J$ (See Chapter \ref{ch5}).

The semiclassical approximation described here is not restricted to a specific number of colors. Indeed, in this effective theory the color quantum number does not appear explicitly. However, since the model is an offspring of the original AdS/CFT correspondence~\cite{Maldacena:1997re}, it is reminiscent of an $N_C \to \infty$  theory. This interpretation is also in accordance with the zero width of all states, including the excited ones.

The treatment of the chiral limit in the LF holographic approach to strongly coupled QCD is substantially different from the standard approach. In the standard approach spontaneous chiral  symmetry breaking  plays the crucial role. The massless pion is the Goldstone boson of the broken symmetry and the mass differences between the parity doublets, for example the  $\rho(770)$ -- $a_1(1260)$ and the $ N(939)$ -- $N(1535)$  doublets, is a consequence of the spontaneous chiral symmetry breaking. In light-front holographic QCD discussed here, the vanishing of the pion mass in the chiral limit follows   from the precise cancellation of the light-front kinetic energy and light-front potential energy terms for the quadratic confinement potential (See Sec. \ref{Mesons}). This effective potential  results from the triple correspondence of light-front quantized QCD, gravity in AdS$_5$ space and conformal quantum mechanics~\cite{Brodsky:2013ar}. The mass differences between the parity doublets  also follows from this specific potential. The parity splitting in this framework depends crucially on the light-front orbital quantum numbers. Therefore, in this approach the parity doublets are not degenerate and the trajectories remain parallel as observed experimentally (See Chapter \ref{ch5}).

The mapping of  transition amplitudes in the gravity theory to the light front is also an important aspect of light-front holography (See Chapter \ref{ch6}). In addition to reproducing the essential elements of the transverse dynamics found by the light-front mapping of the Hamiltonian equations, one also obtains new information on the longitudinal dynamics which is relevant, for example, to compute QCD distribution amplitudes and extend the formalism to include light-quark masses (See Sec. \ref{LQM}).

In addition to describing hadronic bound states (normalizable solutions), the gravity theory  allows to extend the confining dynamics to  external currents (non-normalizable solutions).  The ``dressed''  or confined current corresponds to sum an infinite class of Fock states  containing $q \bar q$ pairs in the hadronized current. This leads to the remarkable results that for the soft-wall model  the current is expressed as an infinite sum of poles and the form factor as a product of poles (See Sec. \ref{SWFF}). Notably, the actual number of  poles in form factor is determined by the twist of the hadronic state (the number of constituents). At large space- and time-like $q^2$ the form factor  incorporates the correct power-law fall-off for hard scattering   independent of the specific dynamics and is dictated by the twist~\cite{Polchinski:2001tt}. At low $q^2$ the form factor leads to vector dominance, and therefore there are no divergences in the limit of zero quark masses. Furthermore the analytic expressions obtained for the form factors allows an analytic  continuation into the time-like region (See Sec. \ref{TLFF}).

Finally, in Chapter \ref{ch7} we reviewed bottom-up and top-down holographic models motivated by chiral symmetry breaking in QCD.   We described in this chapter the most relevant aspects of this approach and practical limitations.  Hopefully future studies will help to understand better the connections between both approaches.

\section{Open problems and future applications \label{op}}

The light-front holographic model discussed here gives a satisfactory first-order description of a large bulk of light-hadronic data using essentially a single scale. It is of course desirable to improve the agreement with observations and to better understand some aspects of the framework. We collect the most noticeable points that from our perspective should be addressed. 

\begin{enumerate}

\item
The agreement for the meson trajectories of the $\rho$ and $K^*$ is very satisfactory.  For the  mesons on the $\pi$ and $K$ trajectory the agreement is less satisfactory and the scale $\sqrt{\la}$ has to be increased by 10 \%. The triplet splitting of the $a$ mesons is qualitatively correct, but the predicted splitting is too large.

\item 
An extended computation should also include the isoscalar mesons, the description of which is problematic in most models~\cite{Parganlija:2013xsa}. A proper description should probably include higher twist components.  For example, the $f_0$ could be a superposition of two quarks in a $P$ wave  and four quarks in an $S$ wave.

\item
In the soft-wall model the dilaton profile in the AdS action for fermions does not lead to an effective potential, since it can be absorbed by a field redefinition~\cite{Kirsch:2006he}. Thus confinement must be imposed by introducing an additional term in the action (See Sect. \ref{half}).

\item
The agreement  in the full baryon sector is also very satisfactory. Indeed,  two newly discovered states~\cite{PDG:2014}, the $N(1875)$  and  the $N(1900)$  are well described by the model.  However, the quantum number assignment for the $\nu$ quantum number (See Sec. \ref{Baryons}) is only fixed for the proton trajectory. For the other families the assignment is phenomenological. We expect that a further investigation of the light-front mapping to baryonic states with different quark configurations will explain this successful assignment.

\item
The $\nu$-assignment of the proton mentioned above is based on the fact that it is the ground state. In our approach it corresponds to a bound state of an active quark and the remaining cluster. Therefore the number of effective constituents in the nonperturbative domain is 2, corresponding to an effective twist 2. In the nucleon this corresponds to a quark-diquark cluster decomposition. At short distances all constituents in the proton are resolved and therefore the fall-off of the form factor at high $q^2$ is governed by the number of all constituents, {\it i. e.}, twist 3. It is desirable to understand better the dynamics of the cluster formation in the nucleon and thus get further insight into the transition region of the nonperturbative to the perturbative regime.

\item
The description of electromagnetic form factors in the soft-wall model is very satisfactory if the poles of the confining electromagnetic current ($J=1, L=1$) are shifted to their physical locations (See Sec. \ref{SWFF}),  which corresponds to the predicted bound-state  poles of the $\rho$  ground state and its radial excitations ($J=1, L=0$).

\item
To extend the computation of form-factors in the time-like region  for larger values of $q^2$ one has to include finite decay widths with a correct threshold behavior.  The present analysis only included constant decay widths and Fock states up to twist four (See Sec. \ref{TLFF}).

\item
We have used successfully the  $SU(6)$ spin-flavor symmetry only for the computation of the Dirac form factor. The computation of the Pauli form factor was carried out using the generic expression for a twist-4 hadronic state normalized to the experimental anomalous magnetic moment (See Sec. \ref{FFLFH}). In fact, the value of anomalous magnetic moment and the actual form of the Pauli form factor should result from the theory.

\item
For particles with spin, the AdS wave function is factorized as product of a $z$-independent polarization tensor and a $z$-dependent scalar function (See Chapter \ref{ch4}). This implies that the light-front wave function is the same for all polarizations, which is not a realistic result. This does not affect the spectrum, since the mass eigenvalue is independent of the polarization. But for electroproduction of vector mesons, for instance,  the different polarizations give different contributions, and therefore their precise form is important.  A more elaborate model, for instance starting from a Duffin-Kummer-Petiau equations~\cite{Duffin:1938zz, Kemmer:1939zz, Petiau:1936} in AdS, could solve this problem.

\item
An essential feature for the construction of a confining Hamiltonian from a conformal action is the transition from the original evolution parameter $t$ to a new evolution parameter $\tau$, which is proportional to the light-front lime $x^+$ and  has a finite range (See Chapter \ref{ch3}).  It is natural to identify this $\tau$ as the LF time difference of the confined $q$ and $\bar q$ in the hadron, a quantity which is naturally of finite range and in principle could be measured in double-parton scattering processes.

\item
The emergence of a confining light-front Hamiltonian here was obtained by rather formal arguments. One would like to relate this derivation to more conventional methods used in QCD like the Dyson-Schwinger approach or the summation of H diagrams (two-gluon exchange with gluonic multi-rungs) which are infrared divergent~\cite{Appelquist:1977tw, Appelquist:1977es}.

\end{enumerate}

We have discussed and successfully applied an extension of the light-front mapping to include light-quark masses in the Hamiltonian and the light-front wave functions (LFWFs). The formal procedure could be extended to heavy masses, but then conformal symmetry can no longer be a guiding principle and there is no reason that the harmonic oscillator light-front potential remains valid. Furthermore additional contributions as the one gluon-exchange potential become important. There are many applications in  light-front QCD~\cite{Bakker:2013cea} that  require knowledge of the AdS/light-front wave functions. We list a few of them:

\begin{enumerate}[a)]

\item
The nucleon transition form factor to the first radial excitation using light-front holographic methods has been computed in Ref. \cite{deTeramond:2011qp}. With the 12 GeV upgrade at Jefferson Lab, it will be possible to measure different nucleon form factors to higher excited states at high virtuality~\cite{Aznauryan:2012ba}. The methods of Ref. \cite{deTeramond:2011qp} can be extended to compute these quantities.

\item
The shape of the pion light-front wave function is measured in diffractive dijet reactions $\pi A \to Jet \, Jet \, X$~\cite{Ashery:2005wa}. The data shows a  Gaussian fall-off in $k_\perp$ and a transition at high $k_\perp$ to power-law fall-off. This could be a testing ground of LF holography and the confinement potential.   Using the holographic LFWFs,  we can predict, for example, the slope in $k^2_\perp$ and the change in the shape in $k_\perp$  with $x$.

\item
Using the light-front holographic wave functions obtained here one can in principle compute hadronization at the amplitude level~\cite{Brodsky:2008tk}.

\item
The  holographic light-front wave functions can be used to compute the quark interchange contributions to exclusive hadronic amplitudes~\cite{Gunion:1973ex}.

\item
Since the transverse holographic light-front wave functions form a complete basis, they can be used as a starting point to compute higher order corrections, as for example using the Lippmann-Schwinger equation  or the coupled-cluster method~\cite{Hiller:2014fka}. The light-front wave functions also provide a convenient basis for numerical computations as in the BFLQ approach~\cite{Vary:2009gt}.

\end{enumerate}

There are two conceptually essential points which need further clarification. First, the existence of a weakly-coupled classical gravity with negligible quantum corrections requires that the corresponding dual field theory has a large number of degrees of freedom. In the prototypical AdS/CFT duality ~\cite{Maldacena:1997re} this is realized by taking the limit of a large number of colors $N_C$. In the light front, the bound-state dynamics corresponds to strongly correlated multiple-particle states in the Fock expansion, and the large $N_C$ limit is not a natural concept. The mapping of the AdS bound-state equations to the light-front Hamiltonian is carried out for $N_C = 3$, with remarkable phenomenological success. Following the original holographic ideas~\cite{Bekenstein:1973ur, Hawking:1974sw}  it is thus tempting to conjecture that the required large number of degrees of freedom is provided in this case by the large number of Fock states in light-front dynamics. In fact, in the light-front approach, the effective potential is the result of integrating out all higher Fock states, corresponding to an infinite number of degrees of freedom,  thus absorbing all the quantum effects. The reduction of higher-Fock states to an effective potential is not related to the value of $N_C$. It should also be noted that  the main objective of this report, namely to find a bound state equation in QCD, is bound to incorporate an essential feature of $N_C\to \infty$ QCD:  the feature that all bound states, also the excited ones,  are stable.

There is  another important relation which we have not fully exploited here:  the relation  between the generators of the one-dimensional conformal group and those of the isometries of Anti-de Sitter space in 2 dimensions, AdS$_2$.  The relations between the generators of the triple isomorphism of ${\it Conf}\!\left(R^1\right)$, $SO(2)$, and AdS$_2$ are given in Secs. \ref{group-relation} and \ref{CCMAdS2}. The connection between the  isometries of AdS$_2$ space and the $SO(2,1)$ group of conformal quantum mechanics is the basis of the AdS$_2$/CFT$_1$ correspondence~\cite{Chamon:2011xk}.  In Chapter \ref{ch3} we have shown that the introduction of the scale leading to a confining Hamiltonian corresponds to a ``detuning'' of the relation between the Hamiltonian and the $SO(2,1)$ transformations, see \req{conf}. This detuning, which corresponds to the introduction of a scale and the appearance of a harmonic oscillator potential in the LF Hamiltonian,  has a very simple geometrical interpretation \cite{Dosch:2014wxa} on the AdS$_2$ hyperboloid embedded in a three-dimensional Euclidean space (Fig. \ref{adshyp}).

Recently, the relevance of AdS$_2$ and the emergence of its IR one-dimensional dual quantum field theory~\cite{Faulkner:2009wj} has become manifest through holographic renormalization in the bulk~\cite{Skenderis:2002wp}, the geometric version of the the Wilson renormalization group~\cite{Wilson:1973jj}. In this approach, the holographic flow in the bulk geometry from the boundary theory to the resulting low energy behavior is associated with the holographic coordinate which represents the energy scale~\cite{Peet:1998wn}.  Of particular interest is the formulation of holographic renormalization flows for strongly interacting theories with a cut-off in the bulk, where integration of high energy modes corresponds to integrate the bulk geometry up to some  intermediate scale to extract the low-energy effective theory of the initial boundary theory~\cite{Heemskerk:2010hk, Faulkner:2010jy}. A particularly interesting example, and a line of research worth pursuing,   is the holographic flow of boundary theories to AdS$_2$ geometry in the infrared~\cite{Faulkner:2010jy, Elander:2011vh, Almheiri:2014cka}, since its dual one-dimensional conformal quantum field theory is realized as conformal quantum mechanics (Sect. \ref{CCMAdS2}) with remarkable features which, intertwined  to light-front dynamics and holography  as explained in this report,  captures quite well important properties of the hadronic world.

\chapter*{Acknowledgements}

We would like to thank Fu-Guang Cao, Carl Carlson, Alexandre Deur, Stanislaw Glazek, Diego Hofman, Eberhard Klempt, Volkhard  M\"uller, Joseph Polchinski, Craig Roberts, Robert Shrock, Matthew Strassler,  Peter Tandy and Arkadiusz Trawi\'nski  for  valuable discussions and communications. The work of SJB was supported by the Department of Energy contract DE--AC02--76SF00515. The work of JE was supported by the NSF under Grant PHY--1068008.

\chapter*{Addendum}

After completion of this work, it has become apparent that the striking phenomenological similarities between the baryon and meson spectra, described in this review, reflect underlying supersymmetric relations  responsible for these important features. In fact,  it has been shown recently that the extension of superconformal quantum mechanics to the light front has important applications to hadronic physics~\cite{deTeramond:2014asa, Dosch:2015nwa}.

In particular, it has been shown~\cite{deTeramond:2014asa} that a comparison of the half-integer LF bound-state equations with the Hamiltonian equations of superconformal quantum mechanics~\cite{Fubini:1984hf, Akulov:1984uh}  fixes the form of the LF potential, in full agreement with the phenomenologically deduced form $V(\zeta) = \la_B \, \zeta$. This new development addresses one of the open problems of the approach to strongly-coupled QCD described in this report (See   Sec. \ref{op}). In contrast to conformal quantum mechanics without supersymmetry, which is dual to the bosonic sector of AdS only up to a constant term, which in turn is fixed by embedding the LF wave equations for arbitrary integer spin into AdS. This procedure, originally developed by  Fubini and Rabinovici~\cite{Fubini:1984hf}, is the superconformal extension of the procedure applied by de Alfaro {\it et al.}~\cite{deAlfaro:1976je}.  Following this procedure a new evolution Hamiltonian  is constructed from a generalized supercharge, which is a superposition of the original supercharge together with a spinor operator which occurs only in the superconformal algebra. The resulting  one-dimensional effective theory applied to the fermionic  LF bound-state equations  is equivalent to the semiclassical approximation to strongly coupled dynamics which, as described in this report, follows from the light-front clustering properties of the semiclassical approximation to strongly coupled-QCD dynamics and its holographic embedding in AdS space.

In Ref.~\cite{deTeramond:2014asa}  superconformal quantum mechanics was used to describe baryonic states. In this case, the supercharges  relate the positive and negative chirality components of the baryon wave functions, consistent with parity conservation. Light-front superconformal quantum mechanics can also be used to relate hadronic states with different fermion number~\cite{Dosch:2015nwa}.    In this approach  the nucleon trajectory is the superpartner of the pion trajectory, but the pion, which is massless in the chiral limit, has no supersymmetric partner.  It is important to notice that the quantum-mechanical supersymmetric relations derived in~\cite{deTeramond:2014asa, Dosch:2015nwa} are not a consequence of a supersymmetry of the underlying quark and gluon fields; they are instead  a consequence of the superconformal-confining dynamics of the semiclassical theory described in this Review and the clustering inherent in light-front  holographic QCD.

\appendix

\chapter{Riemannian Geometry and Anti-de Sitter Space \label{RG}}
We  briefly review in this appendix relevant elements of
Riemannian geometry useful in the discussion of Anti-de Sitter
space and applications of the AdS/CFT correspondence.

\section{Basics of non-Euclidean geometry \label{A1}}

The geometric properties of a $D$-dimensional curved space  with coordinates $x^M = \left(x^0, x^1 \cdots x^{D-1} \right)$ are  described by the metric
tensor $g_{MN}(x)$ which defines the space-time metric
\begin{equation}  \label{line}
ds^2 = g_{M N}  dx^M dx^N,
\end{equation}
in each reference frame. In $D$-dimensional
Minkowski space  in Cartesian coordinates, the metric tensor  $\eta_{M N}$ has diagonal components  $(1, -1, \cdots, -1)$.
In non-Euclidean geometry the metric tensor varies from point to
point and its form depends on the coordinate choice.  Since (\ref{line}) is invariant,
the change of the metric
induced by a coordinate transformation $x^M \to {x'}^M(x)$
\begin{equation} \label{differentials}
dx'^M = \frac{\partial x'^M}{\partial x^N} dx^N = \partial_Nx'^M dx^N,
\end{equation}
 is
 \begin{equation} \label{metrics}
 g'_{M N}(x')=  \partial\,'_M x^R \partial\,'_N x^S g_{R S}(x),
 \end{equation}
where we define $\partial_M  \equiv \partial  / \partial x^M$, $\partial\,'_M  \equiv \partial  / \partial x'^M$.
A $D$-components object $V^M$  which  transforms as the differential quantities (\ref{differentials})  is said to be a contravariant vector
 \begin{equation}  \label{contra}
 V^M(x) \to V'^M(x') =  \partial_N  x'^M V^N.
 \end{equation}

The gradient $\partial_M \phi$ of a scalar function   $\phi(x)=\phi'(x')$ transforms according to
 \begin{equation} \label{gradient}
 \partial\,'_M \phi'(x') =  \partial\,'_M x^N  \partial_N \phi(x).
 \end{equation}
 A $D$-components object $V_M$  which  transforms as the gradient  of a scalar function (\ref{gradient}) is said to be a covariant vector
 \begin{equation} \label{co}
V_M(x) \to V'_M(x') =  \partial\,'_M x^N V_N .
\end{equation}
Thus, contravariant vectors are denoted by upper
components, and covariant ones by lower components.
A covariant or contravariant or mixed tensor has several indices
which transform according to (\ref{co}) or (\ref{contra}). For example a
covariant tensor of rank two  transforms as
\begin{equation}
V_{MN} \to V'_{M N} = \partial \,'_M x ^S  \partial \,'_N x^R  V_{RS} \label{co2} .
\end{equation}

The inverse of $g_{MN}$ is the contravariant metric tensor $g^{MN}$;
$ g_{M R} \,g^{R N} = \delta^N_M$,
where $\delta^N_M$ is the mixed unit tensor: $\delta_N^M = 1, N = M$; $\delta_N^M = 0, N \ne M$.
In Minkowski space-time the inverse metric tensor  in
Cartesian coordinates is equal to the original one: $\eta_{M N} = \eta^{M N}$, but this is not the case for a general non-Euclidean
metric. We can transform contravariant vectors in covariant ones and vice
versa. Indices are lowered or raised  by the metric tensor
$V_M = g_{M N} V^N$,  $V^M=  g^{M N} V_N$ .
The scalar product of two vectors
 $A^M = \left(A^0, A^1 \cdots A^{D-1} \right)$ and $B^M = \left(B^0, B^1 \cdots B^{D-1} \right)$
  \begin{equation} \label{product}
 A \cdot B =   A_N B^M =  A^M B_N = g_{M N} A^M B^N = g^{M N} A_M B_N ,
 \end{equation}
 is invariant under coordinate transformations as one can easily verify from (\ref{contra}) and (\ref{co}).

When integrating in a curved space time, the volume element $dV$ should behave as an invariant upon integration over a D-dimensional domain.  In curved space
\begin{equation}
dV =  \sqrt{g} \, dx^0 dx^1 \cdots dx^{D-1},
\end{equation}
where $g$ is  the absolute value of the metric determinant  $g \equiv \vert \rm {det } \, g_{MN}  \vert$.

\subsection{Covariant derivative and parallel transport \label{A11}}

The next step is to define a covariant derivative $D_M$ which transforms covariantly. For example, when $D_M$
acts on a covariant vector $V_N$  the resulting rank-two covariant tensor $D_M V_N$ should transform as
\begin{equation} \label{Dtransf}
D'_M V'_N = \partial\,'_M x^R \partial\,'_N x^S D_R V_S.
\end{equation}
This is not the case for the usual partial derivative
\begin{equation}
\partial\,'_M V'_N = \partial\,'_M x^S \partial\,'_N x^R \partial_S V_R + V_R \, \partial\,'_M  \, \partial\,'_N  x^R ,
\end{equation}
since the second term spoils the general covariance unless the second derivatives vanish: $\partial\,'_M \partial\,'_N x'^R = 0$. This happens only for linear transformations:
$x'^M = \Lambda^M_N x^N \! + a^M$, where $\Lambda^M_N$ and $a^M$ are constants. In non-Euclidean geometry a
vector changes its components under parallel transport when comparing two vectors at the same point (for example when
computing the change in the velocity of a particle). One has to make  a parallel transport of a vector $V^M$ from $x^M$   to an infinitesimally close point $x^M+\epsilon^M$
\begin{equation} V^M(x) \to V^M(x+\epsilon) =  V^M(x) +  \Gamma^M_{K L } V^K(x) \epsilon^L,
\end{equation}
where the $\Gamma^M_{K L}$, so-called  Christoffel symbols for the connection, are functions of the coordinates.  Taking into account  parallel transport, the total change expressed by the covariant derivative is thus
\begin{equation}   \label{Dcontra}
D_M V^N = \partial_M V^N + \Gamma^N_{K M} V^K.
\end{equation}
The expression for the covariant derivative of a covariant vector follows from (\ref{Dco}) and the fact that the scalar product of two vectors is invariant under parallel transport. The result is
\begin{equation} \label{Dco}
D_M V_N = \partial_M V_N - \Gamma^K_{M N} V_K.
\end{equation}

It is possible to choose a local inertial coordinate system $x^M$ at a given point with metric $\eta_{M N}$. Under a coordinate transformation (\ref{metrics})
 \begin{equation} \label{metricsetag}
 g'_{M N} =  \partial\,'_M x^R \partial\,'_N x^S \eta_{R S}.
 \end{equation}
Differentiating (\ref{metricsetag}) with respect to $x'^K$ and permuting indices
\begin{equation}
{\partial \,'}_M {\partial \, '}_N x^R {\partial \,'}_K x^S  \eta_{R S} =
\frac{1}{2}  \left( \partial\,'_M g'_{N K} +\partial\,'_N g'_{M K} - \partial\,'_K g'_{MN} \right).
\end{equation}
From the inverse transformation of  $g_{MN}$ to $\eta_{MN}$ and the relation $\delta_M^N = {\partial \,'}_M x^R   {\partial_R x'} ^N$, we find the useful relation
\begin{equation} \label{Gamma1}
{\partial \,'}_M {\partial \, '}_N x^R = \frac{1}{2} g'^{K L} \left( \partial\,'_M g'_{N L} +\partial\,'_N g'_{M L} - \partial\,'_L g'_{M N} \right)  {\partial \,'}_K x^R .
\end{equation}
Using (\ref{Gamma1}) it is straightforward to prove that the covariant derivative has indeed the right transformation properties under a general coordinate transformation
(\ref{Dtransf}). Furthermore, we find from (\ref{Gamma1}) and  (\ref{Dco}) an expression for the the Christoffel symbols  $\Gamma^K_{MN}$ in terms of the metric tensor
\begin{equation}
\Gamma^K_{M N} =
\frac{1}{2} g^{K L} \left( \partial_M g_{N L} +\partial_N g_{M L} - \partial_L g_{M N} \right) ,
\label{Gamma2}
\end{equation}
where  $\Gamma^K_{M N}= \Gamma^K_{N M}$.
Eq. (\ref{Gamma2}) can also be obtained from the condition that the covariant
derivative of the metric tensor $g_{MN}$ is zero
\begin{equation}
\label{Dg}
 D_K g_{MN} = 0.
 \end{equation}

 It follows from (\ref{Gamma2}) that $\Gamma^R_{M R} = \half g^{R S} \partial_M g_{R S} = \partial_M\ln \sqrt{g}$, where $g$ is the absolute value of the metric determinant
 $g \equiv \vert \rm {det } \, g_{MN}  \vert$. We thus find the expression for the divergence of a vector in curved space-time
 \begin{equation} \label{div}
 D_M V^M = \frac{1}{\sqrt g} \partial_M \left( \sqrt{g} \, V^M \right).
 \end{equation}
It also follows from the  symmetry of the $\Gamma^K_{MN}$   that  the rank-two antisymmetric tensor $D_M V_N - D_N V_M = \partial_M  V_N - \partial_N V_M$.

\subsection{Space-time curvature \label{A12}}

Unlike flat space, the second covariant derivative in curved space depends on the order of derivation:
\begin{equation} \label{DDco}
[D_N, D_K]  V_M =  - R^L_{\, M N K} V_L,
\end{equation}
where the fourth-order  Riemann tensor or curvature tensor $R^L_{\, M N K}$,
\begin{equation}
R^L_{\, M N K}=  \partial_N \Gamma^L_{M K} - \partial_K \Gamma^L_{M N} +
 \Gamma^L_{N R} \Gamma^R_{M K} -  \Gamma^L_{K R} \Gamma^R_{M N} ,
\label{Riemann}
\end{equation}
depends on the coordinate system chosen.
Likewise for a contravariant vector one obtains
\begin{equation} \label{DDcontra}
[D_N, D_K]  V^M =   R^M_{\, L N K} V^L.
\end{equation}
The curvature tensor in antisymmetric in the indices $N$ and $K$;
$R^L_{\, M N K} = - R^L_{\, M K N}$. The tensor is null in Euclidean space  $R^L_{\, MNK}=0$, and conversely, if  $R^L_{\, MNK}=0$ the space is Euclidean.
Is is useful to express the Riemann tensor in covariant form
\begin{equation}
R_{K M L N} = g_{K R} R^R_{\, M L N},
\end{equation}
with the symmetry properties
\begin{eqnarray}
R_{K M L N} &=& - R_{M K L N} \label{R1},\\
R_{K M L N} &=& - R_{K M N L} \label{R2},\\
R_{K M L N} &=& R_{L N K M } \label{R3}.
\end{eqnarray}

The rank two tensor
 \begin{equation}
 R_{MN} = g^{K L} R_{K M L N} = R^L_{\, M L N} ,
 \label{Ricci}
 \end{equation}
 is the Ricci tensor  $R_{M N}  = R_{ N M}$.  Contracting $R_{M N}$ we obtain the scalar curvature $\mathcal{R}$
\begin{equation}
\mathcal{R} = g^{M N} R_{MN} = g^{K L} g^{M N} R_{K M L N},
 \label{scalarcurv}
 \end{equation}
which  encodes  intrinsic properties of space-time. The Einstein tensor given by
\begin{equation}
G_{MN} =  R_{MN} - \frac{1}{2} \mathcal{R}  \, g_{MN} ,
\end{equation}
has the property that $D_N  G^N_M = 0$, known as the Bianchi identity.

To have a better intuition of parallel transport and the Ricci tensor consider the parallel transport of a vector  between two points along different paths in a non Euclidean space. In particular the change $\Delta V_{M}$ of a vector along a closed infinitesimal loop is
\begin{equation}
\Delta V_M = \oint \Gamma_{M N}^L V_L dx^N.
\end{equation}
Using Stokes theorem one obtains
\begin{equation}
\Delta V_M = \frac{1}{2} R^L_{\, M N K} V_L  \, \Delta S^{NK},
\end{equation}
where  $\Delta S^{NK}$ is the infinitesimal area enclosed by the integration contour. The result of parallel transporting a vector is path dependent.

\subsection{Spinors in non-Euclidean geometry \label{A13}}

To describe spin in curved space-time we attach an orthonormal frame, the vielbein, at each point in space-time~\cite{GSW:1987}.
The vielbein at a point $P$  is  a local inertial system.  The spin
connection gives us information on how the vielbein is rotated when
it moves along a curve. It turns out that the vielbein is
essential for treating spinors in non-Euclidean geometry.\footnote{Technically the quantities $\partial  x'^M \! / \partial x^N$ in (\ref{contra})  for transforming  a vector field $V^M$ are the elements of the general linear group $GL(D,R)$, the group of   invertible real $D \times D$ matrices. Thus a vector $V^M$ transforms in the fundamental representation
of $GL(D,R)$. Spinors transform under the special orthogonal group $SO(D)$, a subgroup of $GL(D,R)$, but the spinor representations are not comprised in the representations of $GL(D,R)$. The vielbein formalism allows us to replace the $GL(D,R)$ matrices by a $SO(D)$ matrix representation.}

Let $\xi^A(x)$ be the coordinates of the point $P$ in the local inertial frame. The vielbein  is a set of $D$ orthonormal tangent vectors
 $  e^A_M(x) = \partial_M \xi^A(x) $,
where $x$ are the coordinates of the same point $P$ in a general frame with metric tensor $g_{MN}(x)$, and
 the index  $A$ is a Lorentz index in  tangent space. Lorentz indices are denoted by $A, B \cdots$ and curved space
 indices by $M, N \cdots$. Since the metric tensor in the inertial frame is the Minkowski tensor $\eta_{M N}$,
 the metric tensor in the general
frame is according to (\ref{metrics})   $g_{M N}(x) = \partial_M \xi^A \partial_N \xi^ B  \eta_{A B}$. Thus
 \begin{equation}
 g_{M N}(x)
= \eta_{A B} \, e^A_M(x)  e^B_N(x)  = e^A_M(x) e_{A N}(x),
\label{emetric}
\end{equation}
where $e_{A M} = \eta_{A B} \, e^B_M$, since the tangent indices are raised or lowered by the Minkowski metric $\eta_{AB}$.
The inverse vielbein is denoted $e^M_A$, thus  $e^A_M(x) e_B^M(x) = \delta^A_B$.
The vectors $e^A_M$ form a complete basis at each point $P$.
The vielbein is not uniquely determined since in each point there is an infinity of equivalent inertial coordinate
systems, all related through Lorentz transformations; ${e'}_M^A(x) =  \Lambda^A_B(x) e^B_M(x)$, where $\Lambda^A_B(x)$ is a local Lorentz transformation at the point $P$.

For spinors, the analogue of the Christoffel symbols is the spin connection denoted ${\omega_M}^A_{~B}$. Thus the covariant derivative of a vector with tangent indices $V^A= e^A_M\,V^M$
\begin{equation}
D_M V^A = \partial_M V^A + {\omega_M}^A_{~B} V^B .
\label{eq:DcoV}
\end{equation}
Consistency with (\ref{Dg}) and (\ref{emetric}) implies that the covariant derivative of the vielbein is zero \footnote{The condition (\ref{De}) corresponds to minimal coupling
of the spinors.  Deviations from minimal coupling are measured by the torsion $T^A_{M N}$ defined by $T^A_{M N} = D_M e^A_N - D_N e^A_M$.}
\begin{equation}
D_M e^A_N = 0,
\label{De}
\end{equation}
and thus   $D_M e^A_N = \partial_M e^A_N - \Gamma^L_{MN}e^A_L + {\omega_M}^A_{~B} e^B_N =0$. The spin connection then follows from  (\ref{De})
\begin{equation}
 \omega_M^{\, A B} = e^A_N \partial_M e^{N B} + e^A_L e^{N B} \Gamma^L_{N M}.
\label{eq:spincon}
 \end{equation}

Equations (\ref{Dco}) and (\ref{eq:DcoV}) for the covariant derivative
are equivalent.  To see this we contract (\ref{eq:DcoV}) with $e^N_A$
and use (\ref{eq:spincon}):
\begin{align}
D_MV^N=e^N_A\,D_MV^A&=e^N_A \partial_MV^A+e^N_Ae_K^A(\partial_M e^{K}_B)V^B
+e^N_A e_L^A e^K_B\Gamma^L_{KM}V^B \nonumber \\
&=\partial_M (e^N_A V^A)-(\partial_M e^N_A)V^A+(\partial_M e^N_B)V^B
+\Gamma^N_{KM}(e^K_B V^B) \nonumber \\
&=\partial_M V^N+\Gamma^N_{KM} V^K.\nonumber
\end{align}

We now come to spinors. We consider a spinor field $\psi(x)$ in the spinor representation of the Lorentz group. The generators of the Lorentz
group $\Sigma_{AB}$ in the spinor representation  are
\begin{equation}
\Sigma_{AB}= \frac{i}{4}[ \Gamma_A, \Gamma_B] ,
\label{Sigma}
\end{equation}
where the flat space gamma matrices $\Gamma_A$ are constant matrices obeying the usual
anticommutation relations
\beq \label{gammaflat}
\left\{ \Gamma^A, \Gamma^B \right\} = 2 \eta^{A B}.
\enq
  The covariant derivative of $\psi$ is defined by
\begin{equation}
 D_M \psi = \left(\partial_M - \frac{i}{2} \omega^{AB}_M \Sigma_{AB}\right) \psi .
 \label{Dpsi}
 \end{equation}
 Under a local Lorentz transformation $\psi(x) \to \Lambda(x) \psi(x)$, with $\Lambda = \exp \left( - \frac{i}{2}  \omega_{A B} \Sigma^{A B} \right)$,  the covariant
 derivative $D_M \psi(x)$ also transforms as a spinor, $D_M \psi(x) \to \Lambda(x) D_M\psi(x)$, thus general covariance is maintained.

 The vielbein allows us also to construct space-dependent
$\Gamma$-matrices in non-Euclidean space
\beq \mbox{
$\Gamma$}^M(x) = e^M_A(x) \, \Gamma^A \label{gammane}. \enq
From
(\ref{emetric}) and (\ref{gammaflat}) it follows that the matrices  \mbox{
$\Gamma$}$^M$ obey $\{\mbox{$\Gamma$}^M(x),\mbox{
$\Gamma$}^N(x)\} = 2 g^{MN}(x)$. It can also be proven from (\ref{De})
that the matrices $\mbox{$\Gamma$}^M(x)$  are covariantly constant,
$[D_M, \Gamma^N(x)] =0$. Thus the Dirac operator $ \Dslash$
can be defined as
\beq \Dslash = \Gamma^M(x) \,D_M = e^M_A (x) \Gamma^A
 \,D_M. \label{diracop} \enq

\section{Maximally symmetric spaces \label{A2}}

\subsection{Definition \label{A21}}

A space-time manifold is said to be maximally symmetric if it has the same number of isometries as Euclidean space.
A $D$-dimensional Euclidean space is isotropic and homogeneous. Its metric tensor is invariant under
$D(D - 1)/2$ rotations  (isotropy) and $D$ translations (homogeneity). We have thus altogether
$D(D + 1)/2$ isometries {\it i.e.}, transformations which leave the metric invariant.  This is also the
 maximal number of isometries of a $D$-dimensional Minkowski space, therefore flat space-time
is a maximally symmetric space. This concept of isometry and maximal symmetry can be translated into non-Euclidean
geometry.

If the metric does not change its form under a coordinate transformation $x \to x'$, that is  $g_{MN}(x) \to g'_{MN}(x') = g_{MN}(x')$, then it follows from
 (\ref{metrics}) that
\begin{equation}
 g_{M N}(x')=  \partial\,'_M x^R \partial\,'_N x^S g_{R S}(x).
  \label{isometry}
 \end{equation}
Any change of coordinates that satisfies \ref{isometry} is an isometry. In particular,
 if one looks for local isometries under infinitesimal transformations:
  $x{'}^M = x^M+ \epsilon\,\xi^M(x)$,
  one can show easily that the requirement to be an isometry is given by the covariant structure
    \begin{equation}
    D_M \xi_N + D_N \xi_M = 0,
    \label{killing}
    \end{equation}
  the Killing equation.  Any vector field $\xi_M(x)$ satisfying (\ref{killing}) is said
   to form a Killing vector of the metric $g_{MN}(x)$. The maximum number of independent Killing
   vectors in a $D$-dimensional space is $D(D+1)/2$, thus the Killing vectors are the infinitesimal generators of isometries. A space with a metric that admits this
   maximal number of Killing vectors is  maximally symmetric: it is homogeneous and isotropic for every point.

\subsection{Anti-de Sitter space-time AdS$_{d+1}$ \label{A22}}

Anti-de Sitter space is the maximally symmetric space-time
with negative  scalar curvature. We can construct the metric of a maximally symmetric space by embedding it in a flat space of
one dimension higher. Anti-de Sitter space-time in $d+1$ dimensions,  AdS$_{d+1}$,
can be described as a hyperboloid embedded in a flat $d+2$ dimensional
space-time with an
additional time-like direction, {\em i.e.} the surface

\beq
\label{embedd}
X_{-1}^2+X_0^2 -X_1^2 - \cdots -X_d^2 \equiv\sum_{M=-1,d} X^M\,X_M =R^2,
\enq
with metric induced by the $(d+2)$-dimensional flat metric with signature
(2, $d$)
\begin{equation}
ds^2=dX_{-1}^2+dX_{0}^2-dX_1^2-\dots-dX_d^2. \label{eq:hyp2}
\end{equation}

The isometries of the embedded space are those which leave the hyperboloid invariant; that is, they are isomorphic to the group $SO(2,d)$. This group in turn is isomorphic to the conformal group ${\it Conf}(R^{1,d-1})$. These groups have have $d(d+1)/2$ generators.  The isometries of AdS$_{d+1}$ can be either obtained via the Killing vectors \req{killing}, or, expressed in the embedding coordinates $X$  through the transformations of $SO(2,d)$.

The AdS space time contains closed time-like curves.  For example, for
fixed coordinates $\{X_1,\dots,X_d\}$, any closed path along the circle in the
$(X_0,X_1)$ plane is a closed time-like curve.
Global coordinates ({\em i.e.} coordinates which cover the
entire space time)  are $\rho,\,
  \tau,\,$ and $\Omega_i$ with
  $ 0 \leq \rho < \pi/2,$   $-\pi <\tau \leq \pi$, and  $ -1\leq
  \Omega_i \leq 1$  with   $\sum_{i=1}^{d} \Omega_i^2 =1$. Global coordinates
  are related to the embedding coordinates $X$ by:
  \begin{eqnarray}
 X_{-1} &=& R \frac{ \sin \tau }{\cos \rho}, \\
  X_0&=&R \frac{\cos \tau}{\cos \rho}, \\
   X_i&=& R\,  \Omega_i \tan \rho .
  \end{eqnarray}
In global coordinates the metric is
 \begin{equation}
ds^2=\frac{R^2}{\cos^2\rho}\left(d\tau^2-d\rho^2-\sin^2\rho\,d\Omega^2\right).
\end{equation}

The universal cover of Anti-de Sitter space is obtained by
unwrapping the hyperboloid along the time-like circle
and repeatedly gluing the resulting space-time to itself along the seam
{\em ad infinitum}  in order to eliminate the periodicity,
which also eliminates the closed time like curves.

 Poincar\'e coordinates naturally split the AdS space time into smaller patches.
 The plane $X_{-1}=X_d$ splits the full AdS$_{d+1}$ space in  two patches.
 On these patches we can introduce coordinates by first
  defining the light-cone coordinates
  \begin{equation}  \label{PCi}
   u \equiv \frac{1}{R^2}(X_{-1} - X_d),  \quad v \equiv \frac{1}{R^2}(X_{-1}+X_d),
  \end{equation}
  and by introducing the Poincar\'e coordinates  $x^0=t$, $x^i, \; i=1 \dots d-1$ and
  $z$, which are related to the embedding coordinates by
   \begin{equation}
    x^i= \frac{X_i}{R\,u}, \quad x^0 = \frac{X_{0}}{R \, u}, \quad z = \frac{1}{u}.
  \end{equation}
   In order to obtain  the metric tensor in the Poincar\'e coordinates we
   eliminate $v$ by the embedding condition \req{embedd} and
   obtain
   \begin{eqnarray} \label{PCf}
   X_{-1}&=& \frac{1}{2 z} (z^2 +R^2 + \mathbf{x}^2 - t^2) , \\
   X_{0} &=& \frac{R \,t}{z} , \\
   X_i &=& \frac{R\, x^i}{z} , \\
   X_d &=& \frac{1}{2 z}(z^2 -R^2 + \mathbf{x}^2 - t^2) . \label{PCff}
   \end{eqnarray}

The coordinate $z=R^2/(X_{-1}-X_d)$ is referred to as the holographic  coordinate,
and separates the AdS space-time into two distinct regions: $z>0$ and $z<0$.
Each region is absent of closed time-like curves.
The region $z=0$ belongs to the boundary of the
AdS space time, and we will be interested in the Poincar\'e patch $z>0$.
In Poincar\'e coordinates the AdS metric with coordinates $x^M = \left(x^\mu, z \right)$,
$\mu, \nu  = 0, 1, 2, \cdots, d-1$, takes the form
\begin{equation}
ds^2= g_{MN} dx^M dx^N = \frac{R^2}{z^2}\left(\eta_{\mu \nu} dx^\mu dx^\nu - dz^2\right),
\label{AdSmetric}
\end{equation}
and thus
\begin{equation}
g_{MN} = \frac{R^2}{z^2} \, \eta_{NM}, \quad  g^{MN} =
\frac{z^2}{R^2}  \, \eta^{NM},
 \quad e^A_M = \frac{R}{z} \, \delta^A_M,
  \quad e^M_A = \frac{z}{R} \, \delta^M_A ,
\label{ge}
\end{equation}
where $\eta_{NM}$ has
diagonal components $(1, -1, \cdots, -1)$.
The metric determinant $g = \vert {{\rm det} \,  g_{MN} \vert}$ is $ g = \left(R/z\right)^{2 d + 2}$.
Additional details regarding various coordinates on anti-de Sitter space time
and its Poincar\'e patches can be found in Ref.~\cite{Bayona:2005nq}.

The  Christoffel symbols  \req{Gamma2} for the metric \req{ge} are:
 \beqa \label{gammaads} 
 \Gamma^{L}_{MN} &=& \frac{1}{2} g^{LK}\left(\pa_M g_{KN} +\pa_N g_{KM}- \pa_K  g_{MN}\right)\\
  &=&  \frac{-1}{z}\left(\de^z_M\de^L_N+\de^z_N\de^L_M - \eta^{Lz}\eta_{MN}\right),  \nn 
  \enqa
  and the   covariant derivative \req{Dco} for a vector is
     \beqa
     D_z V_M &=& (\pa_z + \frac{1}{z}) V_M \label{adspar} ,\\
      D_\mu V_z  &=& \pa_\mu V_z  + \frac{1}{z}  V_\mu \nn , \\
   D_\mu V_\nu  &=& \pa_\mu V_\nu  + \frac{1}{z} \et_{\mu\nu} V_z .
   \nn
    \enqa

Every maximally symmetric space has constant curvature $\mathcal{R}$. In fact, for an isotropic and homogeneous space the Ricci
 tensor $R^{L}_{\, MNK}$ can only depend on the metric tensor, and thus by virtue of its symmetry properties \req{R1} - \req{R3} it should have the form
 $R_{K M L N} = \lambda \left(  g_{K L} \,  g_{M N}  -   g_{K N}  \,  g_{M L}  \right)$,
where $\lambda$ is a constant. For AdS$_{d+1}$ the Ricci tensor (\ref{Ricci}) is  $R_{MN} = \lambda \, d \, g_{M N}$ and the scalar curvature (\ref{scalarcurv})
$\mathcal{R} =  d(d +1) \lambda$. A simple but tedious computation using  the expression (\ref{Riemann}) for the Riemann tensor
and \req{gammaads} for the Christoffel symbols gives  $\mathcal{R} = \eta^{zz}\, d(d+1)/R^2$ and thus\footnote{This relation shows that the  signum of the holographic variable in the metric determines the sign of the scalar curvature and hence if the space is a de Sitter or an anti-de Sitter space.}
 \begin{equation}
 R_{K M L N} = - \frac{1}{R^2} \left(  g_{K L}  \,  g_{M N}  -   g_{K N}  \,  g_{M L}  \right) .
 \end{equation}

From the realization of AdS$_{d+1}$ as an hyperboloid in a $d+2$ dimensional space,
(See Eq. \req{embedd}), it  follows that the isometries of AdS$_{d+1}$ are isomorphic to the transformations which leave this hyperboloid invariant, {\it i.e.} the elements of $SO(2,d)$. The Killing vectors of $AdS_{d+1}$ in Poincar\'e coordinates can be constructed from the  $SO(2,d)$
generators 
\beq
L^{M,N}=i\,\left(X^N\frac{\partial}{\partial X_M}-X^M\frac{\partial}{\partial X_N}\right), 
\enq
where  $M,N= -1,0,\dots d$, with the relations \req{PCf} - \req{PCff} between the different coordinates. In the next subsection this is done for the case $d=1$.

\subsection{Relation between  $Conf  \! \left(R^1\right)$, $SO(2,1)$ and the isometries of AdS$_2$ \label{group-relation}}

\begin{figure}[h]
\centering
\includegraphics[angle=0,width=6.8cm]{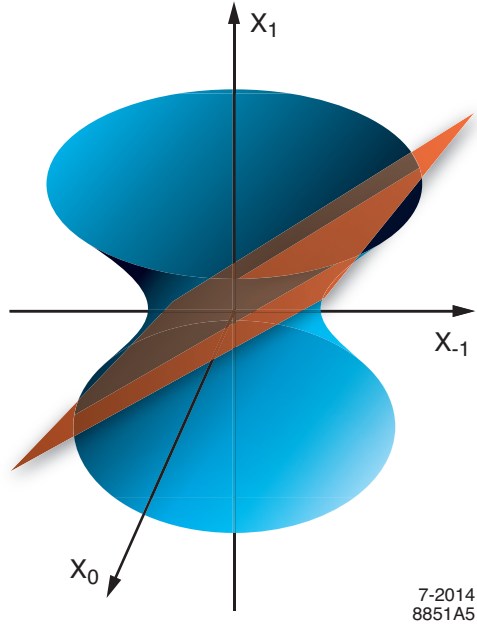} 
\caption{\small \label{adshyp} The space AdS$_2$ embedded as the hyperboloid $X_{-1}^2 + X_0^2 - X_1^2 = R^2$  into a three dimensional space and the plane $X_1-X_{-1}=0$.}
\end{figure}

Since the conformal group plays  an eminent role in our treatment, we give here a  more extensive discussion of the relation between the generators of the conformal group and those of the isometries of  Anti-de Sitter space  in two dimensions, AdS$_2$~\footnote{The relation between the conformal group, AdS$_2$ and  the generators of $SO(2,1)$ is described with great detail in the Senior Thesis of  T. Levine at Brown University~\cite{Levine:2002}. We thank Antal Jevicki for pointing us this work.}.     In Figure \ref{adshyp} we show the embedding of the AdS$_2$, {\it i.e.} $d=1$,  into a three dimensional space. The surface of the hyperboloid is the space AdS$_2$ . The intersection with the plane $X_1-X_{-1}=0$ corresponds to the value $z=\pm \infty$ which separates the AdS$_2$ space into two patches, none of them containing  time-like closed curves. The value $z=0$ correspond to the limit $X_1 \to - \infty$.

The conformal group in one dimension $Conf  \! \left(R^1\right)$ is locally isomorphic  not only  to the group $SO(2,1)$, but also to the isometries of AdS$_2$.  The three generators of the latter, $A^{(i)}$,  can be constructed with the Killing vectors $\xi^{(i)}$ as (See Sec. \ref{A21})
\beq \label{k1}
A^{(i)}= \xi^{(i)\,M} \frac{\pa}{\pa x^M},  \quad  x^1 =t, \, x^2 = z.
\enq
The conditions for the Killing vectors of AdS$_2$ follow from the Killing equation \req{killing}
\beq \label{k2}
 \pa_t \xi^{(i)\,1}-\frac{1}{z} \xi^{(i)\,2}=0 , \quad \pa_z\xi^{(i)\,2}-\frac{1}{z} \xi^{(i)\,1}=0 , \quad  \pa_t\xi^{(i)\,2} -  \pa_z\xi^{(i)\,1}=0,
\enq
with the solutions
\beq \label{k3}
\xi^{(1)} = R\, \Big(1,\; 0\Big), \quad \xi^{(2)} = \Big(t,\; z\Big), \quad \xi^{(3)} = \frac{1}{R} \Big( (t^2+z^2),\;  2 t z\Big).
\enq

The relation between the Killing vectors and the generators of
$SO(2,1)$ is not uniquely fixed by the commutation relations   \req{killing},
therefore it is advantageous to construct them explicitly by going
back to AdS$_2$ as a 2-dimensional hyperboloid embedded in a three
dimensional space (Fig. \ref{adshyp}). The embedding space has two time coordinates
$\{X_{-1},\, X_0\}$ and one space coordinate  $ X_1$ . The space  AdS$_2$ is given by the surface $X_{-1}^2
+X_0^2 - X_1^2 = R^2$ and its  isometries are the transformations
which leave the hyperboloid invariant, that is the rotation
$L^{0,-1}$ in the $(X_0,\, X_{-1})$ plane and the two boosts
$L^{1,0}$ and $L^{1,-1}$. They obey  the commutation relations of the SO(2,1) algebra (Sect. \ref{alg})
\beqa \label{ccl}
\left[ L^{0,-1},\,L^{1,0}\right] &=& i L^{1,-1}, \\
\left[ L^{0,-1},\,L^{1,-1}\right] &=& -i L^{1,0} ,\nn \\
\left[ L^{1,0},\,L^{1,-1}\right] &=&- i L^{0,-1} .\nn
\enqa

The Poincar\'e coordinates are given by (See \req{PCi}--\req{PCff})
\beq \label{pc}
t=\frac{X_0\,R}{X_{-1}-X_1},  \qquad  z= \frac{R^2}{X_{-1} - X_1}.
\enq

The relation between the generators of $SO(2,1)$ and the isometries of AdS$_2$ given by \req{k1},  can now be obtained directly by expressing the rotation and the boost generators in  Poincar\'e coordinates
\beqa
L^{0,-1} = i\,\left(  X_{-1}\frac{\pa}{\pa X_{0}}-X_{0}\frac{\pa}{\pa X_{-1}} \right)
&=& \frac{i}{2}\left\{\left(\frac{t^2+z^2}{R}+R\right)\pa_t + 2\frac{t\,z}{R} \pa_z\right\} , \\
L^{1,0}\;\; =i\,\left( X_{0}\frac{\pa}{\pa X_{1}}+ X_{1}\frac{\pa}{\pa X_{0}} \right) \hspace{0.5cm}
&=& \frac{i}{2}\left\{\left(\frac{t^2+z^2}{R}-R\right)\pa_t + 2\frac{t\,z}{R} \pa_z\right\}\nn ,\\
L^{1,-1}  = i\,\left( X_{-1}\frac{\pa}{\pa X_{1}}+ X_{1}\frac{\pa}{\pa X_{-1}} \right)
&=&i (t\,\pa_t + z\, \pa_z),\nn
\enqa
where $X_0 = X^0$,  $X_{-1} = X^{-1}$ and $X_1 =  - X^1$.  In terms of the generators of the  isometries of AdS$_2$  \req{k1},
\beq
 L^{0,-1} = \frac{i}{2} (A^{(3)}+ A^{(1)}),\quad
 L^{1,0} = \frac{i}{2} (A^{(3)}- A^{(1)}), \quad
 L^{1,-1} = i A^{(2)}.
\enq
Notice that the asymmetry between the representation of the two boosts is due to the choice of the Poincar\'e coordinates in \req{pc}.
In Table \ref{tabiso} the relation between   the generators of the isometries  of AdS$_2$, the generators of $SO(2,1)$ and the generators of the conformal group $H, \, D$, and $K$ (See Chapter \ref{ch3}) are displayed.

\renewcommand{\arraystretch}{1.3}
\begin{table} [h] 
\begin{center}
\begin{tabular}{lll}
 \hline\hline 
{\it Conf}\,($R^1$) & ~~~ $SO(2,1)$ & \hspace{40pt}AdS$_2$   \vspace{0.5pt} \\ 
\hline    \vspace{1.5pt}
$H$ &$ \frac{1}{a} (L^{0,-1} -L^{1,0}) $&$ ~~~~~ \frac{i}{a}A^{(1)}=\frac{i\, R}{a}  \pa_t$ \\
$D$ & $L^{1,-1}$ & $ ~~~~~ i A^{(2)} = i (t \pa_t + z \pa_z)$\\
$K$ & $a  (L^{0,-1} + L^{1,0}) $&$ ~~~~~ i a  A^{(3)}= \frac{i a}{R}\big( (t^2+z^2) \pa_t + 2 z t \pa_z\big)$   \vspace{1.5pt} \\
\hline \hline
\end{tabular}
\end{center}
\caption{\label{tabiso} \small Relation between the generators of the conformal group in one dimension $Conf  \! \left(R^1\right)$, the generators of $SO(2,1)$ and the Killing vectors of AdS$_2$ expressed in Poincar\'e coordinates.}
\end{table}

The generators of the AdS$_2$ isometries depend on the AdS$_2$ Poincar\'e coordinates $t$ and $z$. Using Table \ref{tabiso}, one  can see explicitly the equivalence of the generators of AdS$_2$ isometries at  the AdS$_2$ boundary, $z=0$, with  the representation of the conformal generators $H$, $D$ and $K$ in conformal quantum mechanics given by \req{HDKSev} in Appendix \ref{cfqm}:  $H = i \pa_t$, $D = i t \pa_t$ and  $H = i t^2 \pa_t$, provided that  $a = R$.

\chapter{Light-Front Metric Conventions and Spinors \label{metric}}
The Minkowski metric is written in terms of light-front coordinates 
\begin{equation}
d\sigma^2 = dx^+ dx^- - d \mbf{x}_\perp^2 ,
\end{equation}
with time-like and space-like components $x^+ = x^0 + x^3$ and $ x^- = x^0 - x^3$ respectively.
We write  contravariant four-vectors such as $x^\mu$ as
\begin{equation}
x^\mu = \left(x^+, x^-, x^1, x^2\right) = \left(x^+, x^-, \mbf{x}_\perp\right).
\end{equation}
Scalar products are
\begin{eqnarray} \nonumber
x \cdot p &\! = \!&  x_\mu p^\nu = {\rm g}_{\mu \nu} x^\mu p^\nu \\ \nonumber
&\! = \!& x_+p^+ + x_-p^- + x_1 p^1 + x_2 p^2   \\
& \!= \!& \frac{1}{2} \left(x^+ p^- + x^- p^+\right) - \mbf{x}_\perp \cdot \mbf{p}_\perp,
\end{eqnarray}
with front-form metrics
\begin{equation}
{\rm g}_{\mu \nu} =
  \begin{pmatrix}
  0 &  \frac{1}{2} & 0 & 0 \\
  \tfrac{1}{2} & 0  & 0 & 0 \\
  0 & 0  & -1 & 0 \\
  0 & 0  & 0 & -1 \\
  \end{pmatrix} , ~~~
  {\rm g}^{\mu \nu} =
  \begin{pmatrix}
  0 &  2 & 0 & 0 \\
  2 & 0  & 0 & 0 \\
  0 & 0  & -1 & 0 \\
  0 & 0  & 0 & -1 \\
  \end{pmatrix}.
  \end{equation}
  A covariant vector such as $\partial_\mu$ is
  \begin{equation}
  \partial_\mu = \left(\partial_+, \partial_-, \partial_1, \partial_2\right)=
  \left(\partial_+, \partial_-, \vec \partial_\perp\right).
  \end{equation}
  Thus $\partial^+ = 2 \, \partial_-$ and $\partial ^- = 2 \, \partial_+$.

Useful Dirac matrix elements for light-front helicity spinors  with spin component along the $z$-axis
$(\lambda = \pm 1)$ are~\cite{Lepage:1980fj}
\begin{eqnarray} \label{eq:sn1}
\bar u(p) \gamma^+ u(q) \delta_{\lambda_p, \lambda_q} &=& 2 \sqrt{p^+ q^+} , \\
 \bar u(p) \gamma^+ u(q) \delta_{\lambda_p, - \lambda_q} &=&  0,
\end{eqnarray}
with $ \bar v_\alpha(\ell) \gamma^\mu v_\beta(k) = \bar u_\beta(k)  \gamma^\mu u_\alpha(\ell) $. 
Other useful matrix elements for LF spinors are given in ~\cite{Lepage:1980fj}.

\chapter{Notes on Conformal Quantum Mechanics \label{cfqm}}
 In this appendix we  examine specific features of the framework introduced by de Alfaro {\it et al.}~\cite{deAlfaro:1976je} which are useful for the discussions in Chapter  \ref{ch3}. For the relation between the conformal group and the isometries of AdS$_2$ see Sec. \req{group-relation}. We start with the dAFF action
\beq \label{Aa}  
A = \half \int dt \,  \left(\dot Q^2 - \frac{g}{Q^2}\right) ,
\enq
and the corresponding Lagrangian
\beq \label{La} 
L[Q] = \half \left(\dot Q^2 - \frac{g}{Q^2}\right) ,
\enq 
which is, up to a total derivative, conformally invariant {\it i.e.}, 
\beq \label{r1a} 
L[Q']\,dt'= \left( L[Q] - \frac{d \Om[Q]}{d t} \right)\, dt .
\enq

We use the notation $\dot Q = d Q(t)/dt , \;\; \dot Q' = d Q'(t')/ dt'$.  To proof  \req{r1a} we perform a general conformal transformation
\beq \label{GCa}
t'=\frac{\al t + \be}{\ga t + \de},
\quad ~  Q'(t') = \frac{Q(t)}{\ga t + \de}, \quad ~
\al \de - \be \ga =1, \enq 
with 
\beq
 dt' = \frac{dt}{(\ga t + \de)^2}, \quad \quad  \dot Q'(t') = (\de + \ga t)\dot Q - \ga Q . 
\enq
 One obtains immediately 
\beq \label{TL}  
A = \half \int dt' \left( (\dot Q')^2 - \frac{g}{{Q'}^2} \Big)   =  \half \int dt \Big( \dot Q^2 - \frac{g}{Q^2} -  \frac{d \Om}{d t} \right),
\enq 
with 
\beq  \label{diva} 
\Om = \half \ga  \frac{Q^2}{\de + \ga t} .
\enq
Therefore the action $A$ \req{Aa}  is invariant, up to a surface term,  under the group of conformal transformations in one dimension.

The  constants of motion, which   follow from the invariance of the action \req{Aa} under conformal transformations,  are also the generators of the conformal group expressed in  terms of the field operators $Q$.  The conserved generators can be obtained from  Noether's theorem.  If the Lagrangian is form invariant under infinitesimal transformations {\it i.e.},
\beq
 \label{vn} \quad L'[Q'] = L[Q'] + \frac{ d \Om}{d t}
 \enq
then $J$ obtained from
\beq \label{nt} 
\de J = \left(L- \frac{\pa L}{\pa \dot Q}\dot Q\right) \de t + \frac{\pa L}{\pa \dot Q}\de Q + \Om,
\enq  
is a constant of motion.

The Noether currents of the three independent transformations of
the conformal group in one dimension are:

\begin{enumerate}

\item
For translations $t'-t =  \de t =\ep$ and $\de Q =0$ in \req{GCa}. The group parameters  are \,  $\al= \de = 1, \; \ga=0, \; \be = \ep$.
The Noether current is 
$ J\, \ep = \left(L- \frac{\pa L}{\pa \dot Q}\dot Q\right)  \ep $ and the translation operator in   the variable $t$ 
\beq   \label{Hta}
 H = -J =  \half \left( \dot Q^2 + \frac{g}{Q^2}\right).
 \enq

\item
For dilatations $t'-t =  \de t =\ep t$ \;and\;   $\de Q= Q'-Q = \half \ep \, Q   + O(\ep^2)$ in  \req{GCa}. The group parameters  are  $\al= \sqrt{1+\ep},\; \de = 1/\sqrt{1+\ep} , \;
\ga = \be =0$.
The Noether current is $ J\, \ep = \left(L- \frac{\pa L}{\pa \dot Q}\dot Q\right)  \ep\, t+\frac{\pa L}{\pa \dot Q} \, \frac{\ep \, Q}{2}$ and the dilatation operator
\beq  \label{D} 
D = - J =  \half \left(\dot Q^2 + \frac{g}{Q^2}\right) t -  {\frac{1}{4}} \left( \dot Q \,Q + Q \,\dot Q \right).
\enq

\item
For special conformal transformations  $ t' - t = \de t =\ep \, t^2  + O(\ep^2)$ \; and \; $\de Q = Q'-Q =\ep \, t   Q  + O(\ep^2)$  in  \req{GCa}. From \req{diva} we have
 $\Om = - \half \ep \frac{Q^2}{1-\ep t} $. The group parameters in this case are $\al=  \de = 1, \;\be =0,  \;\ga = -\ep$.
The Noether current is 
$ J\, \ep =  \left(L- \frac{\pa L}{\pa \dot Q}\dot Q\right)  \ep\, t^2+\frac{\pa L}{\pa \dot Q} \ep \, t \, Q - \half \ep \frac{Q^2}{1-\ep t}$ and the generator of the special
conformal transformations
\beq  \label{K} 
K=-J = \half \left( \dot Q^2 + \frac{g}{Q^2} \right) t^2 - \half \left( \dot Q \,Q + Q \, \dot Q \right)\,t + \half Q^2 .
\enq

\end{enumerate}

 Since the  operators must be Hermitean, one has to write the classical product  $\dot
Q \,Q$ as the symmetrized expression $\half ( \dot Q \,Q + Q \, \dot Q )$.
Note that the crucial term in $K$, namely $Q^2$, stems from the
derivative  $d \Om/dt$ in the transformed Lagrangian \req{TL}.

One can check explicitly that the  generators $H_t, D$ and $K$ obey the algebra of the generators of the conformal group 
\beq \label{confal} 
[H,D]= i\,H, \quad [H,K]=2\, i \, D, \quad [K,D]=- i\,  K .
\enq
To proof this, one has to use repeatedly the  commutation relation  $[Q, \dot Q]= i$, {\it e.g.}
 \beqast
[\dot Q^2,\dot Q Q]&=& \dot Q^3 Q - \dot Q \, Q \, \dot Q^2= \dot Q^3 Q - \dot Q (i + \dot Q Q) \dot Q\\
&=&  \dot Q^3 Q -i \dot Q^2 -\dot Q^2 (i + \dot Q Q) = -2 i \dot Q^2. 
\enqast 
A useful  relation is $\dot Q \,Q^{-1} = Q^{-1} \,\dot Q +i Q^{-2}$ which can be proved by multiplying 
both sides by $Q$.

We can now examine the action of the generators of the conformal group on the state-space vectors
\beqa  \label{HDKSev}
   e^{ - i  H\,\ep} |\psi(t)\rangle &=& |\psi(t+\ep)\rangle=|\psi(t)\rangle +  \frac{d}{dt}|\psi(t)\rangle \, \ep + O(\ep^2) ,\\ \nn
   e^{ - i  D \,\ep} |\psi(t)\rangle &=& |\psi(t+t\, \ep)\rangle=|\psi(t)\rangle +  \frac{d}{dt}|\psi(t)\rangle \, \ep \, t + O(\ep^2) , \\ \nn
   e^{ - i  K \,\ep} |\psi(t)\rangle &=& \Big|\psi\Big(\frac{t}{1-\ep\,t}\Big) \Big\rangle=|\psi(t)\rangle +  \frac{d}{dt}|\psi(t)\rangle \,\ep \, t^2  + O(\ep^2). 
\enqa

Using  the canonical commutation relation for the fields,  $[Q, \dot Q]= i$, we can also obtain the evolution generated by the operators $H$, $D$ and $K$ in the
Heisenberg picture
\beqa  \label{HDKHev}
i \left[H, Q(t) \right]  &=& \frac{d Q(t)}{d t}, \\ \nn
 i \left[D, Q(t) \right]  &=& t \,\frac{d Q(t)}{d t} - \half \, Q(t), \\ \nn
  i \left[K, Q(t) \right]  &=&t^2  \frac{d Q(t)}{d t} - t Q(t). 
\enqa

In terms of the new time  variable $\tau$ \req{dtau} and the new field $q(\tau)$ \req{qtau}
\beq 
 d\tau = \frac{dt}{u + v t + w t^2}, \quad \quad  q(\tau) = \frac{Q(t)}{\sqrt{u + v t + w t^2}},
 \enq
one obtains 
 \beqa  \label{AQq}
 A &=& \half \int dt \,  \left(\dot Q^2 - \frac{g}{Q^2}\right) \\  \nn
 &=& \half  \int d\tau \left( \dot q^2 - \frac{g}{q^2} +  \frac{v^2 - 4 u w} {4}  q^2
 + \half \frac{d}{d \tau} \left[ (v+ 2 w \, t(\tau) \,) q^2 \right] \right),
\enqa
Where we have used the identity
\beq  
(v+ 2 w t) q\dot q = \half \pa_\tau[  (v+ 2 w t) q^2] - \half  q^2 \pa_\tau (v + 2w\,  t(\tau)).
\enq 
Thus,   the transformed action differs from the original action only by a surface term which does not
modify  the equations of motion.

\chapter{Useful Formulas for Higher Spin Equations in Anti-de Sitter Space \label{HSWEAdS}}
\section{Arbitrary integer spin \label{app-int}}

\subsection{The action in the local Lorentz frame \label{inertial}}

Using the Christoffel symbols in \ref{A22} one
obtains from Eq.  \req{gammaads}
 \beqa
\label{cov1}
 D_M \Phi_{N_1 \cdots N_J}
&=&  \pa_M \Phi_{\{N\}} + \Omega(z)  \sum_j \Big( \de^z_M
\Phi_{N_j, N_1\cdots N_{j-1}N_{j+1}\cdots N_J}+\\
&&\hspace{2.5cm} \de^z_{N_j} \Phi_ {M N_1\cdots
N_{j-1}N_{j+1}\cdots N_J}+ \eta_{M N_j} \Phi_{z N_1\cdots
N_{j-1}N_{j+1}\cdots N_J}\Big)\nn , \enqa with he warp factor
$\Omega(z) = 1/z$ in AdS space.

The appearance of covariant derivatives \req{cov1} in the action
for higher spin fields  (\ref{action1}, \ref{action2}) leads to
multiple sums and quite complicated expressions. These, however
simplify considerably  if one goes to a local inertial frame with
(Minkowskian) tangent indices. The transformation from general
covariant indices to those with components in the local tangent
space is achieved by the vielbeins, see \req{ge} \beq \label{VB}
\hat \Phi_{A_1 A_2 \cdots A_J}
 = e_{A_1}^{N_1} e_{A_2}^{N_2} \cdots e_{A_J}^{N_J} \,
 \Phi_{N_1 N_2 \cdots N_J},
 \enq
and thus
 \beq \label{inertiala}
  \hat  \Phi_{N_1 \dots N_J}=\left(\frac{z}{R}\right)^J\,
 \,  \Phi_{N_1 \dots N_J}.
\enq Notably,  one can express the covariant derivatives in a
general frame in terms of partial derivatives in a local tangent
frame. One finds for the AdS metric
 \beq \label{dzm}
  D_z \Phi_{N_1 \dots N_J}= \left(\frac{R}{z}\right)^J \pa_z \hat
\Phi_{N_1 \dots N_J},
 \enq
 and
 \begin{multline} \label{dmum}
 \lefteqn{g^{\mu \mu'}  g^{\nu_1 \nu_1'}\dots  g^{\nu_J \nu_J'}
D_\mu \Phi_{\nu_1 \dots \nu_J} \, D_{\mu'} \Phi_{\nu_1' \dots \nu'_J}= } \\
 g^{\mu \mu'}  \eta^{\nu_1 \nu_1'}\dots  \eta^{\nu_J \nu_J'}
 \left( \pa_\mu \hat \Phi_{\nu_1 \dots \nu_J} \,
\pa_{\mu'} \hat \Phi_{\nu_1' \dots \nu'_J} +  g^{zz}   J \,
\Omega^2(z) \, \hat \Phi_{\nu_1 \dots \nu_J}\, \hat \Phi_{\nu_1'
\dots \nu'_J}\right) ,
\end{multline}
where $\Omega(z)= 1/z$ is the  AdS warp factor in the affine
connection.

It is convenient for the application of the Euler-Lagrange
equations (\ref{ELJ}, \ref{ELz}) to split  action ({\ref{action2})
into three terms:

\begin{enumerate}

\item
A term $S^{[0]}_{\it eff}$ which contains only
fields $\Phi_{\nu_1 \dots \nu_J}$ orthogonal to the holographic
direction.
\item
A term $S^{[1]}_{\it eff}$, which is linear in the
fields  $\Phi^*_{z N_2 \cdots N_J}$,   $\Phi^*_{N_1 z \cdots
N_J}$, $\cdots$,   $\Phi^*_{N_1  N_2 \cdots z}$.
\item
The remainder, which   is
quadratic in fields with $z$-components, {\it i.~e.}, it contains
terms such as  $\Phi^*_{z N_2 \dots N_J}\Phi_{z N_2' \dots N_J'}$.
This last term does not contribute to the Euler-Lagrange equations \req{ELz}, since, upon variation of the action, a vanishing term subject to the condition (\ref{no}) is left.

\end{enumerate}

 Using (\ref{action2}), \req{dzm} and \req{dmum}, and making use of the symmetry of the tensor fields,  one finds,
\begin{multline}
 \label{a-wraph}
S^{[0]}_{\it eff} = \int d^{d}x\, dz\,
\left(\frac{R}{z}\right)^{d-1}
 e^{\vp(z)}\,  \eta^{\nu_1  \nu _1'} \cdots  \eta^{\nu_J  \nu _J'}
 \Bigg( - \pa_z \hat \Phi^*_{\nu_1 \dots \nu_J}\,\pa_z \hat \Phi_{\nu_1' \dots \nu_J'}   \\
   + \eta^{\mu \mu'}
\pa _\mu \hat \Phi^*_{\nu_1 \dots \nu_J}\,\pa_{\mu'} \hat
\Phi_{\nu_1' \dots \nu_J'}
 - \left[ \left(\frac{\mu_{\it eff} (z) R}{z}\right)^2  + J \, \Omega^2(z)\right]
\hat \Phi^*_{\nu_ 1 \dots \nu_J}\, \hat \Phi_{\nu_1' \dots
\nu_J'}\Bigg),
\end{multline}
 and
  \begin{multline}
\label{a-subh}
 S_{\it eff}^{[1]} = \int  d^{d}x\, dz\, \left(\frac{R }{z}\right)^{d-1}
  e^{\vp(z)} \,
\Big( - J \, \Omega(z) \,  \eta^{\mu \mu'} \eta^{N_2 \nu \,_2'}
\cdots \eta^{N_J \nu _J'}
  \pa_\mu \hat \Phi^*_{z N_2 \dots N_J} \hat  \Phi_{\mu' \nu _2' \dots \nu_J'}\\
+ J  \, \Omega(z) \,  \eta^{\mu \nu} \eta^{N_2 \nu \,_2'} \cdots
\eta^{N_J \nu_J'}
 \hat \Phi^*_{z N_2 \dots N_J}
 \pa_{\mu} \hat \Phi_{\nu \nu _2' \dots \nu_J'}  \\
 -J(J-1)  \,  \Omega^2(z) \,
\eta^{\mu \nu}
 \eta^{N_3 \nu \,_3'} \cdots \eta^{N_J \nu_J'}
\hat \Phi^*_{z z N_3 \cdots N_J} \hat \Phi_{\mu \nu \nu_3' \cdots
\nu_J'} \Big) .
 \end{multline}
As can be seen from the presence of the affine warp factor
$\Omega(z)$ in (\ref{a-subh}), this  term is only due to the
affine connection  and thus should only contribute to  kinematical
constraints.

From \req{a-wraph}  one  obtains, upon variation with respect to $
\hat \Phi^*_{\nu_1 \dots \nu_J}$ (\ref{ELJ}),
 the equation of motion in the local tangent space
 \beq  \label{PhiJhat}
 \left[  \pa_\mu \pa^\mu
   -  \frac{ z^{d-1}}{e^{\varphi(z)}}   \partial_z \left(\frac{e^{\varphi(z)}}{z^{d-1}} \partial_z   \right)
  +  \frac{(\mu_{\it eff} (z) R)^2 + J }{z^2}  \right]  \hat \Phi_{\nu_1 \dots \nu_J} = 0,
  \enq
where $\pa_\mu \pa^\mu \equiv \eta^{\mu \nu} \pa_\mu \pa_\nu$. Using  \req{a-subh} one finds by variation with respect to
$ \hat \Phi^*_{N_1 \cdots z  \cdots N_J}$  (\ref{ELz}) the kinematical constraints 
\beq \label{scPhihat}
\eta^{\mu \nu} \pa_\mu  \hat \Phi_{\nu \nu_2 \cdots \nu_J}=0, \quad
\eta^{\mu \nu}  \hat  \Phi_{ \mu \nu \nu_3  \cdots \nu_J}=0.
\enq

From \req{PhiJhat} and \req{scPhihat}    one obtains using (\ref{inertiala}) the 
wave equation in a general frame in terms of the original
covariant tensor field $\Phi_{N_1 \cdots N_J}$ given in
(\ref{PhiJ}, \ref{muphi}) and the kinematical constraints  given by \req{scPhi}.

 \subsection{Warped metric \label{warp}}

In the warped metric \req{gw} the vielbein has the form
 \beq
\tilde e^A_M =
 \frac{R}{z} e^{\tilde \vp(z)} \de^A_M ,
\enq The Christoffel symbols for  the warped metric (\ref{gw})
have the same form as (\ref{gammaads}) \beq \Gamma^{L}_{MN} = -
\tilde \Om(z) \left(\de^5_M\de^L_N+\de^5_N\de^L_M -
  \eta^{L5}\eta_{MN}\right) \nn \enq
with the warp factor $\tilde \Omega(z) = 1/z  - \pa_z \tilde
\vp(z)$.

The effective action, invariant with respect to the warped metric
$\tilde g_{MN}$  is \beq \label{action1w}
 \tilde S_{\it eff} = \int d^{d} x \, dz \sqrt{\tilde g} \,  \tilde g^{\{N N'\}} \left(  \tilde g^{M M'}
 D_M \Phi^*_{\{N\}}\, D_{M'} \Phi_{\{N'\}} - \tilde \mu_{\it eff}^2(z) \Phi^*_{\{N\}}\, \Phi_{\{N'\}}\right),
\enq where $\tilde \mu_{\it eff}(z)$ is again an  effective mass.

Again one can go to a local tangent frame
 \beq \label{inert}
 \hat \Phi_{A_1\cdots A_J} = e^{N_1 \cdots N_J}_{A_1\cdots A_J}  \Phi_{N_1 \cdots N_J}=
 \left(\frac{z}{R}\right)^J  \, e^{-J\,\tilde \vp(z)} \, \Phi_{A_1\cdots A_J}
\enq  
and obtain
\begin{multline}
 \label{a-wraphw}
\tilde S_{\it eff}^{[0]}  = \int d^{d} x \, dz \left(\frac{R \,
e^{\tilde \vp(z)}}{z}\right)^{d-1}  \!
 \eta^{\nu_1 \nu_1'} \cdots   \eta^{\nu_J \nu_J'}\Bigg( - \pa_z \hat \Phi^*_{\{\nu\}}\,\pa_z \hat \Phi_{\{\nu'\}}
  +\\ \eta^{\mu \mu'}
\pa _\mu \hat \Phi^*_{\nu_1 \cdot \nu_J}\,\pa_{\mu'} \hat
\Phi_{\nu'_1 \cdot \nu'_J} - \left[   \left(\frac{\mu_{\it eff}(z)
R \, e^{\tilde \vp(z) }}{z} \right)^2 + J  \tilde
\Omega^2(z)\right]
 \hat \Phi^*_{\nu_1 \cdot \nu_J}\, \hat \Phi_{\nu'_1 \cdot \nu'_J}\Bigg).
\end{multline}

Comparing (\ref{a-wraphw}) with the AdS action (\ref{a-wraph}),
one sees that both forms of the action are equivalent provided
that  one sets \beq \label{subw} \tilde \vp(z) ={\textstyle
\frac{1}{d-1}} \vp(z) \quad \mbox{and} \quad (\tilde \mu_{\it
eff}(z) R)^2 e^{2 \tilde \vp}  =(\mu_{\it eff}(z) R)^2 + \tilde
\Omega^2(z) (J -1). \enq

Also  $\tilde S^{[1]}_{\it eff}$ is equivalent  to (\ref{a-subh}),
only $\Om$ is replaced by $\tilde \Om$; since these warp factors
factor out,  their special form is not relevant for the kinematical
conditions derived from \req{ELz}.  Therefore the kinematical
constraints \req{scPhi} follow also from the warped action
\req{action3}.

\section{Arbitrary half integer spin \label{apphalf}}

\subsection{General treatment}

The covariant derivative  of a  Rarita-Schwinger spinor
$\Psi_{\{N\}}$ is given by \beq
 D_M \Psi_{N_1 \cdots N_T} = \pa_M \Psi_{N_1 \cdots N_T}
 - \frac{i}{2} \om_M^{AB} \Si_{AB} \Psi_{N_1 \cdots N_T}  - \, \sum_j \Ga ^L_{M N_j} \Psi_{N_1 N_{j-1} L N_{j=1} \cdots N_T} ,
\enq where $\Sigma_{A B}$ are the generators of the Lorentz group
in the spinor representation \beq \Sigma_{A B} = \frac{i}{4}
\left[\Gamma_A, \Gamma_B\right] , \enq and the tangent space Dirac
matrices obey the usual anti-commutation relation
 \beq \label{Ga} \Ga ^A \, \Ga^B + \Ga^B\, \Ga^A = 2 \, \eta^{AB}. \enq
 The spin connection in AdS is
 \beq  \label{afc}
 w_M^{A B} = \Om(z) \left(\eta^{A z} \delta_M^B - \eta^{B z} \delta^A_M\right),
 \enq
with  $\Om(z) = 1/z$ .

 For  even $d$ one can choose the set of gamma matrices
$\Gamma^A = \left(\Gamma^\mu, \Gamma^z\right)$ with $\Gamma^z =
\Gamma^0 \Gamma^1 \cdots \Gamma^{d-1}$. For $d=4$ one has \beq
\label{gamma4}
 \Ga^\mu = \ga^\mu, \quad \Ga^z =  - \Gamma_z=  -i \, \ga^5 ,
\enq
 where $\ga^\mu$ and  $\ga^5$
are the usual 4-dimensional Dirac matrices with $\gamma^5 \equiv i
\gamma^0 \gamma^1 \gamma^2 \gamma^3$ and $(\ga^5)^2=+1$. The spin
connections are given by
 \beq \label{spinconads}
  \om^{z \al}_\mu = - \om^{\al z}_\mu = \Om(z) \de^\al_\mu ,
  \enq
  all other components $\om^{A B}_M$  are zero.

The covariant derivatives of a Rarita-Schwinger spinor in AdS are
\beqa \label{covF}  D_z \Psi_{N_1 \cdots N_T} &=& \pa_z \Psi_{N_1
\cdots N_T}+ T\, \Om(z) \Psi_{N_1 \cdots N_T}
,\\
  D_\mu \Psi_{N_1 \cdots N_T}&=&
 \pa_\mu \Psi_{N_1 \cdots N_T} + \frac{1}{2 } \Om(z) \Ga_\mu \, \Ga_z  \Psi_{N_1 \cdots N_T}  +\\
&&
 \Om(z) \sum_j \Big( \de^z_{N_j} \Psi_{N_1 N_{j-1},\mu,N_{j+1}, \cdots N_T} +
\eta_{\mu N_j} \Psi_{N_1 N_{j-1},z,N_{j+1}, \cdots N_T}\Big) . \nn
 \enqa
where, as usual the index $z$ denotes the $(d+1)$
holographic direction.

The derivation of the equation of motion  follows the lines outlined in
Secs. \ref{arbitrary} and \ref{app-int}. One introduces fields with
tangent indices using a local Lorentz frame as in Sec.  \ref{inertial}
 \beq
 \hat \Psi_{A_1 \dots A_T}  =e^{N_1}_{A_1} \cdots  e^{N_T}_{A_T} \, \Psi_{N_1\dots N_T}= \left(\frac{z}{R}  \right)^T \Psi_{A_1\dots A_T},
 \enq
and  separate
 the  action into a part $S_{F \it eff}^{[0]}$ containing only spinors
orthogonal to the holographic direction, and a term $S_{F \it
eff}^{[1]}$, containing  terms linear in $\bar \Psi_{z N_2 \dots
N_T}$;  the remainder does not contribute to the Euler-Lagrange
equations \req{ELzf}. Since the fermion action is linear in the
derivatives,  the calculations are considerably simpler than for
the integer spin.  One obtains from \req{covF}
 \begin{multline}\label{afnu}
 S_{F \, \it eff}^{[0]} =   \int d^{d} x \,dz\, \Big(\frac{R}{z}\Big)^{d+1} e^{\vp(z)}  \eta^{\nu_1 \nu_1'}\dots \eta^{\nu_T \nu_T'}
 \bigg( \frac{i}{2} \,e^M_A \, \,\overline  {\hat \Psi}_{\nu_1 \cdots \nu_T} \Ga^A \,\pa_M
{ \hat \Psi}_{\nu'_1 \dots \nu'_T}   \\
   - \frac{i}{2}\, e^M_A  \left(\pa_M \overline {\hat \Psi}_{\nu_1 \dots \nu_T}\right)  \Ga^A\, { \hat \Psi}_{\nu'_1 \cdots \nu'_T }
- \left( \mu + \rho(z) \right)   \overline {\hat \Psi}_{\nu_1
\cdots \nu_T} {\hat \Psi}_{\nu_1' \cdots \nu_T' }
 \bigg),
\end{multline}
and
  \begin{multline} \label{afz}
 S_{F \, \it eff}^{[1]} = -  \int d^d x \,dz\, \Big(\frac{R}{z}\Big)^{d}  e^{ \vp(z)}   \eta^{N_2 N_2'}\cdots  \eta^{N_T  N_T'} \\
 T \,\Om(z) \,
  \Big( \overline{ \hat \Psi}_{z N_2 \cdots N_T}  \,\Gamma^\mu \hat \Psi_{\mu
  N_2' \cdots N_T'}
 + \overline{ \hat \Psi}_{\mu  N_2 \cdots N_T} \,  \Gamma^\mu \hat \Psi_{z
  N'_2 \dots N'_T}\Big),
 \end{multline}
where the factor of the affine connection  $\Om(z) = 1/z$ follows from Eqs. \req{gammaads}
and \req{afc}.  Performing a partial integration, the  action \req{afnu} becomes:
  \begin{multline} \label{afp}
   S_{F \,\it eff}^{[0]}  =  \int d^d x \, dz \, \Big(\frac{R}{z}\Big)^{d} \,e^{\vp(z)}   \,
\eta^{\nu_1 \nu_1'}\cdots \eta^{\nu_T \nu_T'}  \\
 \overline{\hat \Psi}_{\nu_1 \cdots \nu_T}\, \Big( i \et^{NM}  \Ga_M \pa_N +
 \frac{i}{2 z}\Ga_z  \left(d -  z \vp'(z) \right) -  \mu R- \rho(z)   \Big) \hat \Psi_{\nu'_1 \dots \nu'_T} ,
\end{multline}
plus surface terms.

The variation of  \req{afz}  yields indeed
  the Rarita-Schwinger condition in physical space-time \req{RS}
 \beq \label{dirac-SE1a}
 \gamma^\nu \hat \Psi_{\nu  \nu_2 \, \dots \,\nu_T} =0, \enq
and the variation of  \req{afp} provides the AdS Dirac-like wave
equation \beq \label{DEhat} \left[ i \left( z \eta^{M N} \Gamma_M
\partial_N + \frac{d - z \vp'}{2} \Gamma_z \right) - \mu R - R \,
\rho(z)\right]   \hat \Psi_{\nu_1 \dots \nu_T}=0. \enq Going back
to covariant Rarita-Schwinger spinors $ \Psi_{\nu  \nu_2 \, \dots
\,\nu_T} $ one obtains immediately \req{DEz} and \req{dirac-SE1}.

\subsection{Spin-$\frac{3}{2}$ Rarita-Schwinger field in AdS space \label{threehalf}}

The generalization~\cite{Volovich:1998tj, Matlock:1999fy} of the
Rarita-Schwinger action~\cite{Rarita:1941mf}  to AdS$_{d+1}$  is
\beq 
S = \int d^{d} x \,dz\,  \sqrt{g}\; \bar \Psi_N \left( i
\tilde \Ga^{[NMN']}\,D_M -\mu \, \tilde \Ga^{[NN']} \right)\Psi_N, 
\enq where $\tilde \Ga^{[NMN']}$ and $\tilde \Ga^{[NN']}$ are
the antisymmetrical products of three and two Dirac matrices
$\tilde \Ga^M = e^M_A \Ga^A = \frac{z}{R} \, \de^M_A\,\Ga^A$, with
tangent space matrices $\Ga^A$ given by \req{Ga}. From the
variation of this action  one obtains  the generalization of the
Rarita-Schwinger equation \beq \label{aS}
 \left(i\,\tilde \Ga^{[NMN']}\,D_M -\mu\, \tilde  \Ga^{[NN']}\right)\Psi_{N'} = 0 .
 \enq
 The Christoffel symbols in the covariant derivative can  be omitted  due to the  the antisymmetry of the indices in $\tilde \Ga^{[NMN']}$ and only the spin connection must be taken into account. Eq.  \req{aS}  leads to  the Rarita-Schwinger condition~\cite{Volovich:1998tj}
 \beq
\Ga^M\, \Psi_M=0 ,\enq and the generalized Dirac
equation~\cite{Matlock:1999fy} \beq \label{PsihatB} \left[ i
\left( z \eta^{M N} \Gamma_M \partial_N + \frac{d}{2} \Gamma_z
\right) - \mu R \right]   \hat \Psi_A = \Ga_A \hat \Psi_z , \enq
for the spinor with tangent indices $\hat \Psi_A =   \frac{z}{R}
\, \de^M_A\,  \Psi_M$. These equations agree for $T=1,\;
\vp(z)=\rho(z) =0$  and  $\hat \Psi_z=0$ with  Eq. \req{DEhat},
derived from the effective action \req{af}, for $\vp=\rh=0$.

\chapter{Light-Front Holographic Mapping and the Energy-Momentum Tensor \label{EMT}}
\section{Gravitational form factor of composite hadrons}

In Chapter \ref{ch6} we described the matching of the electromagnetic current matrix elements in AdS space  with the corresponding expression derived from light-front quantization in physical space-time, thus establishing a precise relation between wave functions in AdS space and the light-front wave functions describing the internal structure of hadrons. One may ask if the holographic mapping found in~\cite{Brodsky:2006uqa} for the electromagnetic current is specific to the charge distribution within a hadron or a general feature of light-front holographic QCD. In fact, the matrix elements of local operators of hadronic composite systems, such as  currents, angular momentum and the energy-momentum tensor,  have exact Lorentz-invariant representations in the light front in terms of the overlap of light-front wave functions and thus the LF holographic mapping is a general property of LF observables.  In this appendix we show explicitly that one obtains indeed identical holographic mapping using matrix elements of the energy-momentum tensor~\cite{Brodsky:2008pf, Abidin:2008ku}.

\subsection{Meson gravitational form factor in AdS space}

The action for gravity coupled to a scalar field in AdS$_5$  is
\beqa
S  &=&   \frac{1}{\kappa^2} \int \! d^4x \, dz  \sqrt{g} 
\left(\mathcal{R} - 2 \Lambda\right) + S_M  \nn \\
  &=& S_G + S_M,
\label{eq:SAdS}
\enqa
where $g \equiv \vert \rm {det } \, g_{MN}  \vert$ and $\mathcal{R}$ is the scalar curvature (See \ref{A12}), $\kappa$ is the 5-dimensional Newton constant, $g$ is the determinant of the metric tensor, and $\La$ is a bulk `cosmological' constant. The first term in the action $S_G$ describes the  dynamics of the gravitational fields $g_{M N}$ and determines the  background, which is AdS space (See \ref{A22}).   The coordinates of AdS$_5$ are the Minkowski coordinates $x^\mu$ and $z$ labeled $x^M = (x^\mu, z)$,
with $M, N = 0, \cdots 4$. The dynamics of all other fields, the matter fields, is included in $S_M$. 
To simplify the discussion we consider a scalar field in AdS. In this case the matter content is represented by the AdS  action:
\begin{equation}
S_M =  \int \! d^4x \, dz  \sqrt{g } \,
\left( g^{M N} \partial_M \Phi^*\partial_N \Phi 
-  \mu^2 \Phi^* \Phi \right),
\label{eq:SAdSM}
\end{equation}
which describes a meson  in AdS space.    The symmetric and gauge-invariant Hilbert energy-momentum tensor of the mater fields follows from the functional derivative 
\begin{equation} \label{eq:EMTMAdS}
\Theta_{M N}(x^L) = - \frac{2}{\sqrt{g }} \frac{\delta S_M}{\delta g^{M N}(x^L) }  ,
\end{equation}
and is given by
\begin{equation} \label{eq:EMs}
\Theta_{M N} \! = \partial_M \Phi^* \partial_N \Phi + \partial_N \Phi^* \partial_M\Phi 
- g_{M N} \! \left(\partial^L \Phi^* \partial_L \Phi - \! \mu^2 \Phi^* \Phi\right).
\end{equation}

To determine the matrix elements of the energy-momentum tensor for arbitrary momentum transfer, we must first identify the interaction term in the action of the matter fields with an external gravitational source at the AdS boundary~\cite{Abidin:2008ku}. To this end  we  consider a small deformation of the metric about its AdS background: $\bar g_{MN} = g_{MN} + h_{MN}$,  and expand expand $S_M$ to first order in $h_{MN}$ 
\begin{equation}
S_M[h_{M N}] = S_M[0] + \half \int \! d^4 x \, dz \, \sqrt{g } \,  h_{M N} \Theta^{M N} + \mathcal{O}(h^2),
\end{equation}
where we have used the relation $\Theta^{M N} \delta g_{M N} = - \Theta_{M N} \delta g^{M N}$
which follows from $g^{M N} \delta g_{M N} = - g^{M N} \delta g_{M N}$.
Thus, in the weak gravitational approximation the coupling of an external graviton field $h_{M N}$
to matter is given by the interaction term 
\begin{equation} \label{eq:SI}
S_I =  \half \int \! d^4x \, dz \, \sqrt{g} \, h_{M N} \Theta^{M N}.
\end{equation}

Likewise, we can determine the AdS equation of motion of the graviton field $h_{M N}$   by substituting
the modified metric $\bar g_{M N} = g_{M N} + h_{M N}$ into the gravitational action
$S_G$. We find 
\beq \label{eq:SGh}
S_G[h_{M N}] = S_G[0]   + \frac{1}{4 \kappa^2} \!
\int \! d^4 x \, dz   \, \sqrt{g} \, \Big(\partial_N h^{L M} \partial^N h_{L M} 
- \half \partial_L h \, \partial^L h\Big)  + \mathcal{O}(h^2),
\enq
where  the trace $h_L^L$ is denoted by $h$.  In deriving (\ref{eq:SGh}) we have made use of the gauge invariance of the theory $h'_{L M} = h_{L M} +
\partial_L \epsilon_M + \partial_M \epsilon_L$ to impose the harmonic gauge condition $\partial_\ell h^\ell_m = \half \partial_m h$. The action describing the dynamical fields $h_{L M}$ is given by the linearized form
\begin{equation}  \label{eq:Sh}
S_h = \frac{1}{4 \kappa^2} \!
\int \! d^4x \, dz   \sqrt{g} \Big(\partial_N h^{L M} \partial^N h_{L M} 
- \half \partial_L h \, \partial^L h\Big),
\end{equation}
resembling the treatment of an ordinary  gauge field.
The total bulk action describing the coupling of gravity and matter with an external
graviton in the weak field approximation thus has two additional terms: $S = S_G +S_M + S_h + S_I$.

\subsubsection{Hadronic transition matrix elements in AdS and gravitational form factor}

To simplify the discussion, we consider the holographic mapping of matrix elements of the energy-momentum tensor of mesons, where only one gravitational form factor is present, but the results can be extended to other hadrons.  We also describe the bulk AdS geometry with a model with a wall at  a finite distance  $z = 1/\Lambda_{\rm QCD}$ -- the hard wall model of Ref.~\cite{Polchinski:2001tt}. In the higher-dimensional background theory the matrix element of the energy-momentum tensor  for the hadronic transition $P \to P'$, follows from the interaction term  (\ref{eq:SI})  describing the coupling of the meson field  $\Phi_P(x,z)$ with the external graviton field $h_{M N}(x, z)$ propagating in AdS space,
\begin{equation} 
 \int \! d^4x \, dz \sqrt{g}\, h_{M N}  \left(
\partial^{M} \Phi_{P'}^* \partial^{N} \Phi_P+
\partial^{N} \Phi_{P'}^* \partial^{M} \Phi_P \right),
\label{eq:T}
\end{equation} 
where we have dropped the second term in (\ref{eq:EMs}), which vanishes on-shell modulo a surface term.
The hadronic transition matrix element has  the form
\begin{multline}
\label{MGFF}
  \int \! d^4x \, dz  \sqrt{g} \,
  h_{M N}(x,z)  \left(
\partial^{M} \Phi_{P'}^*(x,z) \partial^{N} \Phi_P(x,z) + \partial^{N} \Phi_{P'}^*(x,z) \partial^{M} \Phi_P(x,z) \right) \\
  \sim
  (2 \pi)^4 \delta^4 \left( P'  \! - P - q\right) \epsilon_\nu^\mu  \left( P^\nu P'_\mu + P_\mu  P'^\nu \right)   A(q^2) ,
 \end{multline}
where the meson has initial and final four momentum $P$ and $P'$ respectively and $q$ is the four-momentum transferred to the pion by the graviton with polarization $\epsilon_\nu^\mu$. The expression on the right-hand side of (\ref{MGFF}) represents the space-like gravitational form factor in physical space-time:
\begin{equation}
\left\langle P' \left\vert \Theta_\mu^{\, \nu} \right\vert P \right\rangle 
=  \left( P^\nu P'_\mu + P_\mu  P'^\nu \right) A(q^2).
\end{equation}
It is the  matrix element of the energy-momentum tensor operator in QCD $\Theta_{\mu \nu}$ obtained below in Sec. \ref{MFFLF}, and represents a local coupling to pointlike constituents. Despite  the fact that the expressions for the transition amplitudes are quite different, one can show  that in the semiclassical approximation, discussed in Chapters \ref{ch2} and \ref{ch6}, a precise mapping can be carried out at fixed light-front time for an arbitrary number of partons in the bound state~\cite{Brodsky:2008pf}.

The propagation of the meson in AdS space is described by a normalizable mode $\Phi_P(x^\mu, z) = e^{i P  \cdot x} \Phi(z)$ with invariant  mass $P_\mu P^\mu = M^2$ and plane waves along the physical coordinates $x^\mu$.   The boundary limit of the graviton probe is a plane wave along the Minkowski coordinates with polarization indices  along  physical space-time according to $h_\mu^{\, \nu}(x, z \to 0) = \epsilon_\mu^{\, \nu}(q) \, e^{ i q \cdot x}$.  We thus write
\begin{equation} \label{h}
h_\mu^{\, \nu}(x, z ) = \epsilon_\mu^{\, \nu}(q) \, e^{i q \cdot x} H(q^2, z),
\end{equation}
with
\begin{equation} \label{eq:Hbc}
H(q^2= 0, z) = H(q^2, z = 0) = 1.
\end{equation}
Extracting the overall factor  $(2 \pi)^4 \delta^4 \left( P'  \! - P - q\right)$ from momentum conservation at the vertex, which arises from integration over Minkowski variables in (\ref{MGFF}), we find~\cite{Abidin:2008ku}
\begin{equation} 
A(Q^2)  =   R^3 \!  \int^{1/ \La_{\rm QCD}} \frac{dz}{z^3} \,  H(q^2, z) \Phi^2(z),
\label{eq:AdSA}
\end{equation}
with $A(0) = 1$.
The gravitational form factor in AdS is thus 
represented as the $z$-overlap
of the normalizable modes dual to the incoming
and outgoing hadrons, $\Phi_P$ and $\Phi_{P'}$, with the
non-normalizable mode, $H(q^2, z)$, dual to the external
graviton source~\cite{Abidin:2008ku}; this provides the form of the 
gravitational transition matrix element analogous to the electromagnetic form factor
in AdS~\cite{Polchinski:2002jw}.   It is interesting to notice that in holographic QCD hadrons appear noticeably more compact measured by the gravitational form factor than by the corresponding charge form factor~\cite{Abidin:2008ku, Abidin:2009hr, Abidin:2008hn}.

\subsection{Meson gravitational form factor in light-front QCD \label{MFFLF}}

The symmetric and gauge-invariant expression for the energy-momentum tensor $\Theta_{\mu \nu}$ is obtained by varying  the QCD action with respect to the four-dimensional Minkowski space-time metric ${\rm g}_{\mu \nu}(x)$ 
\begin{equation}
\Theta_{\mu \nu}(x^\rho) = - \frac{2}{\sqrt{\rm g }} \frac{\delta S_{\rm QCD}}{\delta {\rm g}^{\mu \nu}(x^\rho) }  ,
\end{equation}
where $S_{\rm QCD} = \int \! d^4 x \sqrt{\rm g } \, \mathcal{L}_{\rm QCD}$
and ${\rm g} \equiv  \vert {\rm det} \, {\rm  g}_{\mu \nu} \vert $.
From \req{LQCD} we obtain he result:
\beq\label{eq:emt}
\Theta_{\mu \nu} =  \tfrac{1}{2}  
\bar \psi  i \! \left( \gamma_\mu D_\nu + \gamma_\nu D_\mu \right) \psi
- {\rm g}_{\mu \nu} \psi \left( i \Dslash - m\right)\psi  
 - G^a_{\mu \lambda} {G^a_\nu}^{\hspace{0.5pt} \lambda}
 + \tfrac{1}{4} {\rm g}_{\mu\nu}  G^a_{\lambda \sigma} G^{a \hspace{1pt} \lambda \sigma},
\enq
where the first two terms  correspond to the fermionic contribution to the
energy-momentum tensor  and the last two to the gluonic contribution.

In the front form, the gravitational form factor is conveniently computed  from the matrix elements of the plus-plus component of the energy momentum tensor at LF time $x^+=0$:
\beq
 \left\langle P' \left\vert \Theta^{++}(0) \right\vert P \right\rangle = 2 \left(P^+\right)^2  A\left(q^2\right),
\enq 
where $P'= P + q$.  In the LF gauge $A^+ = 0$ the fermionic component of the operator $\Theta^{++}$ is 
\begin{equation} \label{eq:LFemt}
\Theta^{++}(x) =  \frac{i}{2} \sum_f \bar \psi_f(x) \gamma^+ 
\overleftrightarrow\partial^{\!+} \psi_f(x),
\end{equation}
where an integration by parts is carried out to write $\Theta^{++}$ in its hermitian operator 
form. The sum 
in (\ref{eq:LFemt}) extends over all the types of quarks $f$ present in the hadron \footnote{The plus-plus component of the energy-momentum  does not connect Fock states with different numbers of constituents in the $q^+=0$ frame~\cite{Drell:1969km}. In the semiclassical AdS/CFT correspondence there are no quantum effects, and 
only the valence Fock state contributes to the hadronic wave function. In this approximation we need to consider only the quark contribution to the energy 
momentum tensor in \req{eq:LFemt}.  Notice also that the second term of the energy-momentum tensor (\ref{eq:emt}) does
not appear in the expression for $\Theta^{++}$ since the metric component ${\rm g}^{++}$ is zero in the light-front (Appendix \ref{metric}).}.
The expression for the operator $\Theta^{++}(0)$ in the particle number representation follows from the momentum expansion of the Dirac field $\psi(x)$ in terms of creation
and annihilation operators given by (\ref{eq:psiop}).  The matrix element the operator $\Theta^{++}$ is then computed by expanding the initial and final meson states $\vert \psi_M(P^+ \! , \mbf{P}_\perp)\rangle$  in terms of its Fock components \req{eq:LFWFexp}. Using the normalization condition \req{eq:normFC} for each individual constituent,  and after integration over the intermediate variables in the $q^+ = 0$ frame  we find the expression  for the gravitational form factor of a meson~\cite{Brodsky:2000ii, Brodsky:2008pf}
\beq \label{eq:MGFF}
A_M(q^2) = \sum_n \int \big[d x_i\big] \left[d^2 \mbf{k}_{\perp i}\right]   
\sum_{f=1}^n x_f  \,
 \psi^*_{n/M} (x_i, \mbf{k}'_{\perp i},\lambda_i)
\psi_{n/M} (x_i, \mbf{k}_{\perp i},\lambda_i),
\enq
where the sum is over all the partons in each Fock state $n$. 
The variables of the light-cone Fock components in the
final state are given by $\mbf{k}'_{\perp i} = \mbf{k}_{\perp i} 
+ (1 - x_i)\, \mbf{q}_\perp $ for a struck  constituent quark and 
$\mbf{k}'_{\perp i} = \mbf{k}_{\perp i} - x_i \, \mbf{q}_\perp$ for each
spectator. Notice that each type of parton contributes to the gravitational form factor with struck constituent light-cone momentum fractions $x_f$, 
instead of the electromagnetic constituent charge $e_f$ which appears in the electromagnetic form factor.
Since the longitudinal momentum fractions of the constituents add to one, $\sum_f x_f = 1$,
the momentum sum rule is satisfied at $q =0$: $A(0) = 1$. The formula is exact if the sum is over all Fock states $n$.

In order to compare with AdS results it is  convenient to express the LF
expressions in the transverse impact representation since the
bilinear forms may be expressed in terms of the product of light-front wave
functions  with identical variables. We substitute (\ref{eq:LFWFb}) in the formula (\ref{eq:MGFF}). 
Integration over $k_\perp$ phase space gives us $n - 1$ delta
functions to integrate over the $n - 1$ intermediate transverse variables with the result ~\cite{Brodsky:2008pf}
\beq \label{eq:MGFFbb} 
A_M(q^2) =  \sum_n \prod_{j=1}^{n-1}\int d x_j d^2 \mbf{b}_{\perp j}   \sum_{f=1}^n x_f
\exp \! {\Bigl(i \mbf{q}_\perp \! \cdot \sum_{k=1}^{n-1} x_k \mbf{b}_{\perp k}\Bigr)} 
\left\vert  \psi_{n/M}(x_j, \mbf{b}_{\perp j}, \lambda_j)\right\vert^2,
\enq
corresponding to a change of transverse momentum $x_j \mbf{q}_\perp$ for each
of the $n-1$ spectators.

For a baryon,  the spin-conserving form factor $A(q^2)$ is the analog of the Dirac form factor $F_1(q^2)$. It allows one to measure the momentum  fraction carried by each constituent. There is also a spin-flip form factor $B(q^2)$,  the analog of the Pauli form factor $F_2(Q^2)$ of a nucleon, which provides a  measure of the orbital angular momentum carried by each quark and gluon constituent of a hadron at $q^2=0$. An important constraint is $B(0) = 0$, the vanishing of the anomalous gravitomagnetic moment of fermions~\cite{Kobzarev:1962wt, Teryaev:1999su}. For a composite bound state this means that the anomalous gravitomagnetic moment of a hadron vanishes when summed over all the constituents. The explicit verification of these relations, Fock state by Fock state, can be obtained in the light-front quantization of QCD in  light-cone gauge~\cite{Brodsky:2000ii}.  Physically $B(0) =0$ corresponds to the fact that the sum of the $n$ orbital angular momenta $L$ in an $n$-parton Fock state must vanish since there are only $n-1$ independent orbital angular momenta.

\subsection{Light-front holographic mapping}

The mapping of AdS transition amplitudes to light-front QCD transition matrix elements is much simplified for two-parton  hadronic states. The light-front expression for a meson  form factor in impact space is given by (\ref{eq:MGFFbb}) and includes the contribution of each struck parton with longitudinal momentum fraction $x_f$. For $n=2$, there are two terms which contribute to the $f$-sum in   (\ref{eq:MGFFbb}). 
Exchanging $x \leftrightarrow 1-x$ in the second term and integrating  over angles we find
\beq \label{eq:PiGFFb} 
A_\pi(Q^2) =  4 \pi \int_0^1 \frac{dx}{(1-x)}  \int \zeta d \zeta 
J_0 \! \left(\! \zeta q \sqrt{\frac{1-x}{x}}\right)  \vert \psi_{q \bar q / M}\! (x,\zeta)\vert^2,
\enq
where $\zeta^2 =  x(1-x) \mathbf{b}_\perp^2$ and $A_M(0) = 1$.

We now consider the expression for the hadronic gravitational form factor in 
AdS space (\ref{eq:AdSA}).  Since the energy momentum tensor $\Theta^{M N}$ is gauge invariant, we may impose a more
restricted gauge condition in order to simplify the calculations and use the general covariance of the theory to obtain the final result. We choose the harmonic-traceless gauge
 $\partial_L h^L_M = \half \partial_M h = 0$ and
we consider the propagation inside AdS space of a graviton probe  $h_{M N}$ with vanishing metric components
along the $z$-coordinate  $h_{zz} = h_{z \mu} = 0$.  
The set of linearized Einstein equations from (\ref{eq:Sh}) reduce  for  $H(Q^2,z)$ in \req{h} to the wave equation~\cite{Abidin:2008ku}
\beq
\left[ \frac{d^2}{dz^2} -  \frac{3}{z}  \, \frac{d}{dz} - Q^2 \right]   H(Q^2, z)  = 0,
\enq
which describes the propagation of the external graviton inside AdS space. Its solution subject to the boundary conditions (\ref{eq:Hbc}) is  $(Q^2 = - q^2 > 0)$
\begin{equation} \label{eq:H}
H(Q^2, z) = \half  Q^2 z^2  K_2(z Q),
\end{equation}
the result obtained by Abidin and Carlson~\cite{Abidin:2008ku}.  Using the integral representation of $H(Q^2,z)$ from \req{JKint}
\begin{equation} 
H(Q^2, z) =  2 \! \int_0^1\!  x \, dx \, J_0\!\left(\!z Q\sqrt{\frac{1-x}{x}}\right) ,
\end{equation}
the AdS gravitational form factor \req{eq:AdSA} can be expressed as
\begin{equation} 
A(Q^2)  =  2  R^3 \! \int_0^1 \! x \, dx  \! \int \! \frac{dz}{z^3} \, 
J_0\!\left(\!z Q\sqrt{\frac{1-x}{x}}\right) \left \vert\Phi_\pi(z) \right\vert^2 .
\end{equation}

We now compare with  the light-front QCD  gravitational form factor \req{eq:PiGFFb} using the expression of the  light-front wave function  (\ref{eq:psiphi})
\begin{equation} \label{psiphi}   
\psi(x,\zeta, \varphi) = e^{i L \varphi} X(x) \frac{\phi(\zeta)}{\sqrt{2 \pi \zeta}} , 
\end{equation}
 which we use to  factor out the longitudinal  and transverse modes $\phi(\zeta)$ and $X(x)$ from the LFWF  in \req{eq:PiGFFb}. Both expressions for the gravitational form factor are identical for arbitrary values of $Q$ provided that~\cite{Brodsky:2008pf}
\beq 
\phi(\zeta) = \left(\frac{R}{\zeta}\right)^{-3/2} \Phi(\zeta)  \quad \quad {\rm and}  \quad\quad  X(x) = \sqrt{x(1-x)}.
\enq
This comparison allows us to identify the transverse impact LF variable $\zeta$ with the holographic variable $z$, $z \to \zeta = \sqrt{x(1-x)} \vert \mbf b_\perp \vert$~\footnote{Extension of the results to arbitrary $n$ follows from the $x$-weighted definition of the transverse impact variable of the $n-1$ spectator system given by Eq. (\ref{zetan}).}.  The results are identical to those obtained from the mapping of the electromagnetic form factor in Sec. \ref{EMLFHM}.

\chapter{ Propagators in the Limiting Theory of AdS$_5$ \label{generating}}
\section{AdS boundary conditions and gauge/gravity correspondence}

The formal statement of the duality between a gravity theory on a $(d+1)$-dimensional Anti-de Sitter $AdS_{d+1}$ space and the strong coupling limit of a conformal field theory (CFT)
on the $d$-dimensional  flat space boundary 
at $z=0$, is expressed in terms of the $d+1$ partition function for
a field $\Phi(x ,z)$ propagating in the bulk
\begin{equation}
Z_{\rm grav}[\Phi] =
\int \mathcal{D}[\Phi] e^{ i S_{\rm grav}[\Phi]},
\label{eq:Zgrav}
\end{equation}
and the $d$-dimensional generating functional of correlation functions
of the boundary conformal theory in presence of an external source $j(x)$:
\begin{equation}
  Z_{CFT}[j]  = \left< \exp\left(i \int d^dx \, j(x) \mathcal{O}(x)\right) \right>,
\label{eq:Z}
\end{equation}
where $\mathcal{O}$ is a local interpolating operator.  The  interpolating
operators $\mathcal{O}$ of the boundary quantum field theory are constructed
from local products of  fields and
their covariant derivatives, taken at the same point in
four-dimensional space-time.

The correlation function $\left< \mathcal{O}(x_1) \cdots  \mathcal{O}(x_n)\right>$ follows from the functional derivatives
of the generating functional of the connected Green functions $\log Z_{CFT}[j]$:
\beq
\left< \mathcal{O}(x_1) \cdots  \mathcal{O}(x_n)\right> =  (-i)^n   \frac{\delta}{\delta j(x_1)} \cdots   \frac{\delta}{\delta j(x_n)} \log Z_{CFT}[j].
\enq

As a specific example consider a scalar field in AdS.
In the boundary limit $z \to 0$, the independent solutions behave as
\begin{equation} \label{eq:Phiz0}
\Phi(x , z) \to z^\De \,\Phi_+(x) + z^{d - \De} \,\Phi_-(x),
\end{equation}
where $\De$ is the scaling dimension.
The non-normalizable solution $\Phi_-$ has the leading boundary behavior
and is the boundary value of the bulk
field $\Phi$ which couples to a QCD interpolating operator:
\beq
\lim_{z \to 0} z ^{\De - d} \Phi(x,z) = j(x),
\enq
where $j(x) = \Phi_-(x)$, a finite quantity.  The normalizable solution $\Phi_+$ is the response
function and corresponds to the physical
states~\cite{Balasubramanian:1998sn}.

The precise relation of the gravity theory on AdS space
to the conformal field theory at its boundary 
is~\cite{Gubser:1998bc, Witten:1998qj}
\begin{equation}
  Z_{CFT}\left[j\right] = Z_{\rm grav}\left[\Phi_{[z = \ep]} \to j \right],
 \label{eq:grav-CFT}
\end{equation}
where the partition function (\ref{eq:Zgrav}) on $AdS_{d+1}$ is integrated 
over all possible configurations
$\Phi$ in the bulk which approach its boundary value $j$.
In the classical limit  we neglect the contributions from quantum fluctuations 
to the gravity partition function, then the generating functional of the  four-dimensional gauge theory $\log Z[j]$
(\ref{eq:Z}) is precisely equal to the classical (on-shell) gravity action $ S_{\rm grav}^{\rm on-shell} \big[\Phi^{\rm cl}_{[z =\ep]} \to j \big]$:
\beq \label{ZS}
\log Z[j] = i \, S_{\rm grav}^{\rm on-shell}\left[\Phi^{\rm cl}_{[z =\ep]} \to j \right] ,
\enq
evaluated in terms of the classical solution  $\Phi^{\rm cl}$ to the bulk equation of motion.
This defines  the semiclassical approximation to the conformal field theory.  In the bottom-up phenomenological approach, the effective action
 in the bulk is  usually modified for large values of $z$ to incorporate confinement and is truncated at the quadratic level.

Consider the AdS action for the scalar field $\Phi$ \footnote{For a description of the correlators of spinor and vector fields see for example Ref.   \cite{{Muck:1998iz}}.}
\beq \label{SS}
S_{\rm grav} = \half \int \! d^d x \, dz  \,\sqrt{g} 
  \left( g^{M N} \partial_M \Phi \partial_N \Phi -  \mu^2   \Phi^2 \right)  ,
\enq
where  $g= \left(\frac{R}{z}\right)^{2d+2}$.
Integrating by parts, and using the equation of motion, the bulk
contribution to the action vanishes, and one is left with a non-vanishing surface
term in the  boundary
\begin{equation} 
S^{\rm on-shell}_{\rm grav} =
- \half \int d^d x \big(\sqrt{g}  \Phi g^{zz} \partial_z \Phi \big)_{z = \ep}.
\label{eq:SUV}
\end{equation}

We can compute the expectation value of $\mathcal O$
\beq
\left \langle{\mathcal O}(x)  \right\rangle_{j} = 
-i   \frac{\delta}{\delta j(x)} S^{\rm on-shell}_{\rm grav}  = - i \lim_{z \to 0} z^{d - \De}  \, \frac{\delta}{\delta \Phi(x,z)} S^{\rm on-shell}_{\rm grav},
\enq
and thus
\beq
\left \langle{\mathcal O}(x)  \right\rangle_{j}  \sim \Phi_+(x).
\enq
One finds that 
$\Phi_+(x)$ is related to the expectation values of $\mathcal O$
in the presence of the source $j$~\cite{Balasubramanian:1998sn}.
The exact relation depends on the normalization of the fields 
chosen~\cite{Klebanov:1999tb}. The field $\Phi_+$ thus acts as a
classical field, and it is the boundary limit of the normalizable string
solution which propagates in the bulk.

In the bottom-up phenomenological approach, the effective action
 in the bulk is usually modified for large values of $z$ to incorporate confinement and is usually truncated at the quadratic level.

\section{Two-point functions for arbitrary spin and Migdal procedure}

We start from the action \req{a-wraph}.  Performing a partial
integration in the variable $z$ and using the equation of motion \req{PhiJhat} one obtains,
after a Fourier transform in $x$, the on-shell gravity action
\beq
 S_{\rm grav}^{\rm on-shell}[\hat \Phi^{\rm cl}] = \int d^4q
\left(\frac{R}{z}\right)^{d-1} e^{\vp(z)}\;
 \ep^{\nu_1 \dots\nu_J} (-q) \,\hat  \Ph_J(-q,z) \;
\ep_{\nu_1 \dots\nu_J} (q) \pa_z \hat  \Ph_J(q,z) ,
\enq where
$\ep_{\nu_1 \dots\nu_J}(q) $ is the polarization vector with
 \beq \label{spincond} 
q^{\nu_1}  \ep_{\nu_1 \dots\nu_J}= 0 , \quad \eta^{\nu_1\nu_2}  \ep_{\nu_1 \nu_2 \dots\nu_J}=0,
 \enq
and $\hat \Phi_J(q,z)$ is the solution of the equation of motion in the inertial frame (See \req{PhiJhat}):
 \beq \left[ -q^2 - \frac{z^{d-1}}{e^\vp}
\pa_z\left( \frac{e^\vp}{z^{d-1}} \pa_z\right) + \frac{(\mu_{\it
eff}(z)R)^2 + J}{z^2}\right]\hat \Phi(q,z) = 0.
\enq

We go back to covariant tensor  with $\Phi_J = \left(\frac{R}{z}\right)^{d-1}\hat \Phi$  (See  \req{inertiala}),
 \begin{multline}
 S_{\rm grav}^{\rm on-shell}[\Phi^{\rm cl}] =  \int d^4q
\left(\frac{R}{z}\right)^{d-1-2J} e^{\vp(z)}\;
\Big ( \ep^{\nu_1 \dots\nu_J} (-q) \, \Ph_J(-q,z) \\
 \times  \ep_{\nu_1 \dots\nu_J} (q) \pa_z \Ph_J(q,z) +\frac{J}{z}
\ep^{\nu_1 \dots\nu_J} (-q) \, \Ph_J(-q,z) \; \ep_{\nu_1
\dots\nu_J} (q)  \Ph_J(q,z) \Big).\nn 
\end{multline}

In order to satisfy the condition ${\Phi^{\rm cl}[q,z\to 0]= j(q)}$ we put:
\beq \Phi_{J }^{ \rm cl} (q,z) \,\ep_{\nu_1 \dots\nu_J}=
j_{\nu_1 \dots\nu_J}(q) \, \lim_{\ep\to 0} \,
\frac{\Phi_J(q,z)}{\Phi_J(q,\ep)}.
 \enq 
 We then obtain
 \begin{multline}
 S_{\rm grav}^{\rm on-shell}[\Phi^{\rm cl}]  =  \int d^4q
\left(\frac{R}{z}\right)^{d-1-2J} e^{\vp(z)}
 \Big(\frac{\pa_z \Phi(q,z)}{\Phi(q,\ep) }+ \frac{J}{z}\Big) \\
 \times \si_{\nu_1\nu_1'\dots\nu_J \nu'_J}(q) \, j^{\nu_1
\dots\nu_J}(-q) \, j^{\nu'_1 \dots\nu'_J}(q), \nn
 \end{multline}
 where the
spin tensor $\si_{\nu_1\nu_1'\dots\nu_J \nu'_J}(q)$ reflects the
conditions \req{spincond}. The term $J/z$ in the action is
independent of $q$ and therefore gives only rise to a contact
term; it will, like all  contact terms, be  discarded in the
following.

The propagator of the quantum field $\phi$ is obtained from $\log Z[j]$ by
functional differentiation using the equality \req{ZS}
\beqa \langle \phi_{\nu_1
\dots\nu_J}(q)   \phi_{\nu'_1 \dots\nu'_J}(-q)\rangle &=&
\frac{\de}{\de  j^{\nu_1 \dots\nu_J}(-q) }
 \frac{\de}{\de j^{\nu'_1 \dots\nu'_J}(q)} \log Z[j] \\ &=&
2 i \, \si_{\nu_1\nu_1'\dots\nu_J \nu'_J}(q)\;
\left(\frac{R}{z}\right)^{3-2J} e^{\vp(z)}
\frac{\pa_z \Phi(q,z)}{\Phi(q,\ep) } \nn\\
&\equiv&  i \, \si_{\nu_1\nu_1'\dots\nu_J \nu'_J}(q)\;  \Si(q^2).
\enqa

For the conserved vector current ($\mu = 0;\, J=L=1$) one  starts
with the AdS action \beq S_{\rm AdS}= \frac{1}{2} \int d^4 x\, dz
\left(\frac{R}{z}\right)^5 F^{MN} \,F_{MN} .\enq
In the hard wall model with  Dirichlet boundary conditions at $z_0$  one obtains~\cite{Erlich:2006hq} 
\beq \label{hw}
 \frac{1}{R}\Sigma(q^2) =  2 q^2 \, \lim_{z\to 0} \left(\log(q\, z)
- \frac{\pi \, Y_1(q z_0)}{2\, J_1(q z_0)}  \right).
 \enq

For the soft wall model the solutions are
   \beq  \label {PhiU}
   \Phi(z) = z^{2+L-J}\,  e^{-(\la + |\la|)z^2/2}\,U(a,L+1,|\la|\,z^2),
  \enq
with $a=\frac{1}{4}\left(\frac {-q^2}{|\la|} + 2 L +2
-\frac{\la}{|\la|}(2- 2J )\right)$.
The function $U(a,b,z)$ in \req{PhiU}, is the solution of Kummer's equation which vanishes for $z\to \infty$  \cite{AS64}.
One  obtains for the propagator of the conserved current in the soft wall model \cite{Cata:2006ak, Jugeau:2013zza}
\beq \label{digamma}
\frac{1}{R} \Sigma(q^2) =-q^2 \psi\left(-\frac{q^2}{4
\la}+1\right) -q^2 \log(|\la| z^2),
\enq 
where $\psi(z)$ is the meromorphic
Digamma function, $\psi(z) = \pa_z\log(\Ga(z)) $ which has poles at
$z=0, -1, -2 \cdots$.  The
term which is infinite in the limit $z\to 0$ is a contact term,
which will be discarded.

As has been shown in \cite{Erlich:2006hq} the
propagator of the conserved vector current obtained in
the hard-wall model of AdS$_5$, Eq. \req{hw}, is identical to the result obtained
by Migdal \cite{Migdal:1977nu, Migdal:1977ut} in its hadronization
scheme of perturbative QCD. In the  $N_C \to \infty $ limit of QCD
the hadronic cuts should vanish and only the hadron poles survive.
Furthermore  conformal invariance makes it possible to find explicit
perturbative expressions up to order $\al_s^4$ which are diagonal
in the rank of the hadron interpolating operators~\cite{Dosch:1977qh, Dosch:1977cu}. Migdal has
constructed propagators  which asymptotically reproduce
the perturbative result, but have only poles on the positive real
axis of $q^2$. As a prescription to obtain this answer, he used the
Pad\'e approximation where the logarithm of the perturbative
result is approximated by a quotient of power series. As stressed
in \cite{Erlich:2006hq} the hard-wall model of AdS$_5$ is a
convenient framework  to achieve this ``meromorphization''. The logarithmic
cuts occurring in the Bessel function $Y_1$ are cancelled by the
logarithmic term in \req{hw}. It was considered as a deficiency that
the Migdal regularization scheme did not lead to linear Regge
trajectories and induced Migdal not to pursue this investigations
further~\footnote{M. Shifman, private communication}. 

Insisting on equally spaced poles on the real axis (Regge daughters) an obvious
choice for a meromorphization is the digamma function $\psi(z)$,
see \req{digamma}. It has poles at
$z=0, \, -1, \, -2 \cdots $  and behaves in the deep Euclidean
region $ z \gg 1 $ like $\log z^2$. Therefore, one can
also see the soft-wall model of AdS$_5$ as a correspondence of the
{\it meromorphization program} of perturbative QCD. Since the
asymptotic expansion of the digamma function is
  \beq \label{digas}
\psi(z) \sim \log z -\frac{1}{2 z} - \sum_{n=1}^\infty
\frac{B_{2n}}{2 n z^{2n}}=\log z -\frac{1}{2 z} -\frac{1}{12 z^2}
+\dots, \quad z\to \infty, \; |\arg z| < \pi,
 \enq
the soft wall  model  is closer to the QCD sum rule method of Shifman, Vainshtein and
Zakharov~\cite{Shifman:1978bx} than to the Migdal model~\cite{Cata:2006ak, Jugeau:2013zza}. Also in the sum rule approach,
 corrections with inverse powers of $Q^2$  are added to the perturbative
expression  in order to improve the perturbative result in the not so deep Euclidean region. Note that there is also a power correction proportional to $1/Q^2$ which does not correspond to a classical vacuum expectation value (See \cite{Narison:2001ix}). Finite-energy sum rules also lead under model assumptions to a linear rise in $M^2$ for the radial excitations: $M^2 \sim n$~\cite{Kataev:1982xu, Gorishnii:1983zi}.

Next we discuss the fields with $L=0$ \footnote{For a treatment in the soft wall model with negative
dilaton profile see  \cite{Colangelo:2008us,Ratsimbarison:2010wr}.} which play an important
role in the phenomenology of LF holographic QCD, where $\la >0$.
For the vector current which interpolates the $\rho$  in light-front AdS/QCD, {\it i.e.} the
field with quantum numbers $J=1, L=0$, ($(\mu R)^2=-1$) one obtains in the soft wall model (up to finite and diverging
contact terms)
\beq\label{twist-softwall}
 \frac{1}{R} \Sigma(q^2) = \frac{1}{(z \log (z))^2 }\left[
\psi\left(-\frac{q^2}{4 \la}+\half\right) + O\left(\frac{1}{\log
z}\right) \right]+ O(1).
 \enq

The leading term in the expression for the
generating functional $\log Z[j]$, proportional to the square bracket
in \req{twist-softwall} is a meromorphic function with the poles
as predicted by the equation of motion. The
gauge gravity  prescription~\cite{Gubser:1998bc,Witten:1998qj} given by  \req{ZS} is equivalent to
 \beq  
 \label{GPWp} 
 \log Z[j]= \lim_{z\to 0} \, (z \log z)^2\;  i \,S_{\rm grav}^{\rm on-shell}\left[\Phi^{\rm cl}_{[z =\ep]} \to j \right].
 \enq
 The factor  $z^2 \, \log ^2 z$ cancels the infinity  in \req{twist-softwall} as $z \to \infty$
and the $ O(1) $ contributions vanish. Note that in this case the polarization tensor
is given by $g_{\mu\nu} - q_\mu \,q_\nu/q^2$.

For a general tensor field with $L>0$ we obtain:
\beq\label{twist-softwall-L}
 \Sigma(q^2) = C_L R^{3-2 J} z^{2(J+L)-4} \; \la^L\;
\frac{\Ga\left(\frac{-q^2 +2(J+L)\la}{4
\la}\right)}{\Ga\left(\frac{-q^2 +2(J-L)\la}{4 \la}\right)}\;
\psi\left(\frac{-q^2 +2(J+L)\la}{4 \la}\right),
 \enq 
where $C_L$ is an $L$-dependent number. The ratio of the Gamma functions is a
polynomial in $q^2$ of degree $L$. For $L>0$ the relation \req{ZS} becomes
\beq
\label{GPWp} 
\log Z[j]= \lim_{z\to 0} (z\, )^{4-2(J+L)}\; i \, S_{\rm grav}^{\rm on-shell}\left[\Phi^{\rm cl}_{[z =\ep]} \to j \right] .
 \enq 
In the hard-wall model we obtain similar results as
in the soft-wall model, but the meromorphic digamma function is
replaced by a meromorphic combination of $J$ and $Y$ Bessel functions.
Specifically for the case $L>0$  we obtain:
\beq\label{twist-softwall-L}
  \Sigma(q^2) = C'_L R^{3-2 J} z^{2(J+L)-4} \; q^{2 L}\;
\frac{\pi Y_L(q z_0) -2  \log(q z_0) \, J_L(q z_0)}{J_L(q z_0)}.
\enq

To summarize: Both the soft- and hard-wall models in LF holographic
QCD can be seen as a \emph{meromorphization} of the perturbative QCD
expression. The philosophy of the two procedures is, however, quite
different. The Migdal procedure, corresponding to the hard-wall model,  stays as close as possible to perturbative QCD. The
Pad\'e approximation is chosen in such a way as to optimize the agreement with the perturbative expression in the deep Euclidean
region. In the soft-wall model, where the two-point function is
proportional to the digamma function,  there are  power
corrections to the (perturbative) logarithm, see \req{digas}. Therefore, the soft-wall model is closer in spirit to the QCD sum rule
method~\cite{Shifman:1978bx}  where also power corrections are
added to the perturbative result~\cite{Cata:2006ak, Jugeau:2013zza}. The essential input is  the
equal spacing of the Regge daughters as  in the Veneziano
model~\cite{Veneziano:1968yb}. The slope of the trajectory fixes
the genuine non-perturbative quantity, the scale $\la$.  Finally,
let us remark that both the hard- and the soft-wall models share
essential features of QCD in the $N_C\to \infty$ limit: The
propagators are meromorphic functions and the higher $n$-point
functions  vanish.

\chapter{Some Useful Formul\ae   \label{formulae}}
\section{Solutions of the equations of motion in AdS space \label{solutions}}

The equation of motion for arbitrary spin has the generic form given by \req{PhiJM}
\beq  \label{PhiJMa}
-\frac{z^k}{e^{\vp(z)}} \, \pa_z \left(\frac{e^{\vp(z)}}{z^k} \, \pa_z \Phi(z)\right)
 +\left(\frac{(m  R)^2}{z^2} - M^2\right) \Phi(z) = 0.
 \enq
 where $k= d-2 J -1$. This equation can be rewritten as
 \beq  \label{mastereq} 
 \left( -z ^2 \,  \pa_z^2 +\left(k - z \vp'(z)\right)  z \pa_z + (m R)^2 - z^2 M^2\right) \Phi(z) = 0 .
\enq 
 By rescaling the field  according to  $\Phi(z) = z^{k/2} e^{-\vp(z)/2}\, \phi(z)$ this equation  can be brought into a Schr\"odinger form
\beq \label{schroedinger}
\left( - \frac{d^2}{d z^2} + \frac{ 2 k + k^2 + 4 (m R)^2 }{4 z^2} + \frac{ z^2 \,\vp'^2 + 2 z^2 \vp'' - 2 k z \vp'}{4 z^2} \right) \phi(z) = M^2 \phi(z),
\enq
which shows the structure of the light-front Hamiltonian. 

For the hard-wall model  one has  $\vp(z) = 0$ and  can, for  $M^2>0$ bring \req{mastereq} into the form of a Bessel equation~\cite{AS64} by rescaling
$\Phi(z) = z^{(k+1)/2}\,f(y), \quad y=M z$:
\beq
 y^2 f''(y) + y f'(y) +(y^2-\nu^2)f(y) =0,
\enq with $\nu^2 =(mR)^2 + \frac{1}{4}(k+1)^2$.
Solutions  are the Bessel functions of the first and second kind, $J_\nu(y)$ and $Y_\nu(y)$. Only the Bessel functions of the first kind are regular at $z=0$
\beq  
J_\nu(y) = \frac{(y/2)^\nu }{\Ga(\nu +1)}+O(y^{2(\nu+1)}),   \quad \nu \neq -1, -2, \cdots. 
\enq
For normalization one can use the integral formula
   \beq
   \int_0^{y_0} dy \, y\, J^2_L(y) =  \frac{1}{2} y_0^2 \Big((J_L^2(y_0)
   - J_{L-1}(y_0) \,J_{L+1}(y_0) \Big).
  \label{normhard}
  \enq
In the hard wall model $J_\nu(M z_0)=0$ and the zeroes of the Bessel functions  $J_\nu(y)$   determine  the hadron spectrum: Each Bessel function has an infinite set of zeroes.  For small values of  $\nu$ we can use the approximation~\cite{AS64}
\beq \be_{\nu,r} \approx \left(k + \frac{\nu}{2} -
    \frac{1}{4}\right) \pi - \frac{4 \nu^2 -1}{8 \pi\left(r + \frac{\nu}{2}
    - \frac{1}{4} \right)}.
\enq

For the case $M^2=-Q^2 <0$ we obtain  by rescaling  $\Phi(z) = z^{(k+1)/2}\,f(y), \, y=Q \, z$  the modified Bessel equation
\beq
 y^2 f''(y) + y f'(y) -(y^2+\nu^2)f(y) = 0,
\enq 
with $\nu^2 =(mR)^2 - \frac{1}{4}(k+1)^2$.
Its  two independent solutions are the modified Bessel functions $I_\nu(y)$ and $K_\nu(y)$~\cite{AS64}. The function $I_\nu(y)$ increases asymptotically like $e^{y}$ for $y \to \infty$, $K_\nu(y)$ is singular at $y=0,\;  K_0(y) $  diverges logarithmically at $y=0$.

In the case of the soft-wall model one has the dilaton profile $\vp(z) = \la \, z^2$ and \req{mastereq} can be brought into the following  form by rescaling
$\Phi(z) = z^{(k+1)/2} \, e^{- \la z^2/2} f(z)$:
\beq \label{osc}
\left(-  \frac{d^2}{d z^2} - \frac{1}{z} \,\frac{d}{dz} +\frac{L^2}{z^2} + \la^2 z^2\right) f(z) =  \left(M^2  +(k-1) \la\right) {\color{darkgreen} f(z),}
\enq
with $L^2=(m R)^2 +\frac{(k+1)^2}{4}$. 

The operator of the left hand side  of \req{osc} is the Hamiltonian of an harmonic oscillator in two dimensions with angular momentum $L$.
The  normalized eigenfunctions   of the harmonic oscillator are 
\beq
f_{n L}(z)=\sqrt{\frac{2 n!}{(n+L)!}}\, z^L\, e^{-|\la| z^2/2} \,  L^L_n(|\la| z^2),
\enq
and  its eigenvalues are
$E_{n L}= (2n + L +1)|\la|$. The spectrum of eigenvalues for $M^2$ is thus given by
\beq \label{specsw} 
M^2_{n L}= (4 n + 2 L +2)|\la| -(k-1)|\la| .
\enq

If one rescales 
$\Phi(z)= z^{(k+1)/2 + L} g(y) $ with $y= |\la|  \,z^2$ and $L^2= (mR)^2 + \frac{(k+1)^2}{4}$  one obtains the equation
\beq \label{kummer}
y \, g''(y) +(b + \frac{\la}{|\la|} y) g'(y) - a\,g(y)=0,
\enq
with $a= -\frac{1}{4}\frac{\la}{|\la|} \left(2 L + k + 1\right) - \frac{M^2}{4 |\la|}, \; \; b= L+1$.

For $\la <0$ this is Kummer's equation~\cite{AS64} with the solutions $M(a,b,y)$ and $U(a,b,y)$. $M(a,b,z)$ increases exponentially for
$z \to + \infty$ and leads to a divergent solution; thus, only the hypergeometric function $U$ is of interest for us. For $a=-n$, the confluent hypergeometric function $U$ is regular at $z=0$  and is given by~\cite{AS64}
\beq
U(-n,L+1,y)= \frac{n!}{(L+1)(L+2) \cdots (L+n)}  L^L_n(y).\enq
The condition $a=-n$  yields the spectrum \req{specsw} for $\la <0$; one  thus recovers the result obtained above. Equation \req{kummer} is  valid for arbitrary values of $a$ and hence also for negative values of $M^2= -Q^2$.

For $\la >0 $ on can transform \req{kummer} into Kummer's equation  by additional rescaling by the factor $e^{-\la z^2}$,  that is  $\Phi(z)= z^{(k+1)/2 + L} e^{-\la z^2} g(y), \; y = |\la| \, z^2$. Then one obtains from \req{mastereq} Kummer's equation \req{kummer},  but with
 $a= \frac{1}{4} \left(2 L +3  - k \right) - \frac{M^2}{4 \la }$.

\begin{table}[h]
\begin{center}
\begin{tabular}{l|l}
\hline \hline
\multicolumn{2}{c}{$\vp \equiv 0$}\\
\hline
$M^2>0$ & $ \Phi(z)=z^{d/2-J}\left(A\,J_L(Mz)+B\,Y_L(Mz) \right)$  \\ \vspace{3pt}
&$\phi(z) = \frac{1}{N} z^{1/2} J_L(Mz)$\\ \vspace{3pt}
&$ N^2=z_0^2\left( J_L(Mz_0)^2 - J_{L-1}(Mz_0)\,J_{L+1}(M z_0)\right)$\\ \vspace{3pt}
&$ M_{n L} = \beta_{L k}/z_0$\\ \vspace{3pt}
$Q^2=-M^2 >0$& $\Phi(z)= z^{d/2-J}\left(A\,K_L(Qz)+B\,I_L(Qz) \right)$\\ \vspace{3pt}
$d=4, L=1,J=1$& $\Ph_1(z) = Q z\, K_1(Qz)$\\ \vspace{3pt}
& $\Ph_1(0)=1$\\ 
\hline
\multicolumn{2}{c}{$\vp = \la z^2 $}\\ 
\hline
$M^2>0$ & $ \Phi_{n,L}(z) =  z^{d/2-J+L}\,L^{L}_n( |\la|\,z^2) e^{-(|\la|+ \la) z^2/2}$\\ \vspace{3pt}
&$\phi_{n,L}(z) = \frac{1}{N}  z^{L+1/2}\,L^{L}_n( |\la|\,z^2) e^{-|\la| z^2/2}$\\ \vspace{3pt}
&$N=\sqrt{\frac{(n+L)!}{2\, n!}} |\la|^{-(L+1)/2}$\\ \vspace{3pt}
&$M^2_{n,L} = (4n + 2L +2)|\la| +2 \la(J-1)$\\ \vspace{3pt}
$Q^2=-M^2>0$&$\Phi(z) =z^{d/2+L-J}\,  e^{-(\la + |\la|)z^2/2}\,U(a,L+1,|\la|\,z^2)$\\ \vspace{3pt}
&$a=\frac{1}{4}\left(\frac {Q^2}{|\la|} + 2 L +2 -\frac{\la}{|\la|}(d - 2J -2)\right)$\\ \vspace{3pt}
 $d=4,L=1, J=1$&$\Phi_1(z)= 
{\Gamma \left(1 + \frac{Q^2}{4 |\la|}\right)} \, e^{-(\la + |\la|) z^2/2} \,U\left(\frac{Q^2}{4 |\la|}, 0,|\la| z^2\right) $\\ \vspace{3pt}
& $\Ph_1(0)=1$\\ 
\hline  \hline
\end{tabular}
\end{center}
\caption{\label{sol} \small  General form of the solutions of the AdS wave equations for integer spin and their spectra. The AdS field $\Phi$ denotes the solution  of   Eq. \req{PhiJM}. The solution $\phi$  is rescaled as $\phi = z^{(2 J+1-d)/2} e^{\vp(z)/2} \Phi$ in order to satisfy an equation of the Schr\"odinger type \req{schroedinger}. The normalization factor $N$ is determined from 
$\int dz \, \phi^2(z) = 1$ and is regular at $z=0$.}
\end{table}

For the electromagnetic current ($d=4$)  we have $k=1, \; L=1$ and the  the solution of \req{kummer} is~\cite{AS64}
\beq
g(y) = U\left(1+ \frac{Q^2}{4 |\la|}, 2,y\right) =  \frac{1}{y} \,U\left(\frac{Q^2}{4 |\la|}, 0,y\right) .
 \enq
The solution $\Phi$, normalized to  $\Phi(0)=1$ can be written as
\beq
\Phi(z) = {\Gamma \left(1 + \frac{Q^2}{4 |\la|}\right)} \, e^{-(\la + |\la|) z^2/2} \,U\left(\frac{Q^2}{4 |\la|}, 0,|\la| z^2\right) .
\enq
The solutions of the differential equations relevant for the soft- and hard-wall models are summarized in Table \ref{sol}.

\subsection{A useful integral}

The integral relation between the Bessel function of the first kind $J_\alpha(x)$ and the modified Bessel function $K_\alpha(x)$ is usually given by the Hankel-Nicholson integral \cite{AS64}
\begin{equation} \label{HN}
\int_0^\infty \frac{t^{\nu+1} J_\nu(z t)}
 {\left(t^2 + a^2\right)^{\mu+1}} \, dt  =
\frac{z^\mu a^{\nu - \mu}}{2^\mu \Gamma(\mu+1)} K_{\nu-\mu}(a z).
\end{equation}
Changing the variable $t$ according to $x = \frac{a^2}{t^2 + a^2}$ we can recast the integral \req{HN} as
\beq \label{JKint}
\int_0^1 dx \, x^{\mu  - 1} \left(\frac{1- x}{x}\right)^{\nu/2}  \! J_\nu\negthinspace\left( \!  a z \sqrt{\frac{1-x}{x}}\right)  = \frac{ (a z)^\mu}{2^{\mu - 1} \Gamma(\mu + 1)} K_{\nu - \mu}(a z).
\enq

\chapter{Integrability and  Light-Front Effective Hamiltonians \label{LFint}}
Integrability of a physical system is related to its symmetries. In holographic QCD the conformal symmetry often means integrability: the solution to the differential equations describing 
the system can be  expressed is terms of analytical functions.  In this appendix we shall follow the remarkable integrability methods introduced by L. Infeld in a 1941 paper~\cite{Infeld:1941, Infeld:1951mw}. The key observation in Infeld's paper is the realization that integrability follows immediately if the equation describing a physical model can be factorized in terms of linear operators. These operators, the ladder operators, generate all the eigenfunctions once the basic eigenfunction is know.  In the following we will describe how to construct effective light-front Hamiltonians corresponding to the hard and soft-wall models discussed in this report  from the algebra of bosonic  or fermionic  linear operators~\cite{Brodsky:2008pg, Teramond:2007conf}.  In particular, we describe here a different approach to the soft-wall model which results from a minimal   extension of the conformal algebraic structures. This method is particularly useful in the fermionic sector where the corresponding linear wave equations become exactly solvable~\cite{Brodsky:2008pg, Teramond:2007conf}.

\section{Light-front effective  bosonic Hamiltonians \label{LFEBH}}

\subsection{Light-front hard-wall model \label{LFHW}}

To illustrate the algebraic construction procedure consider first, as a simple example, the light-front Hamiltonian form \req{LFWEbis} in the conformal limit:
\begin{equation} \label{eq:LFH}
H_{LF}^\nu = - \frac{d^2}{d \zeta^2} - \frac{1 -4 \nu^2}{4 \zeta^2},
\end{equation}
with hadronic mass eigenvalues and eigenstates determined by the
eigenvalue equation
\begin{equation} \label{eq:EE}
H_{LF}^\nu \,  \phi_\nu(\zeta) = M^2_\nu \, \phi_\nu(\zeta).
\end{equation}

If $\nu > 0$ the  Hamiltonian  (\ref{eq:LFH}) can be
expressed as a bilinear form
\begin{equation}
H_{LF}^\nu =  b_\nu b^\dagger_\nu  ,
\end{equation}
where
\begin{equation}  \label{b}
b = \frac{d}{d \zeta} + \frac{\nu + \half}{\zeta} ,
\end{equation}
and its adjoint 
\begin{equation} \label{bdag}
b^\dagger  = - \frac{d}{d \zeta} + \frac{\nu + \half}{\zeta}  ,
\end{equation}
with $\left(\frac{d}{d \zeta} \right)^\dagger = - \frac{d}{d \zeta}$.

Since the Hamiltonian is a bilinear form, its eigenvalues are positive definite
\beq \nonumber
\langle \phi \left\vert H_{LF}^\nu \right\vert \phi \rangle
=  \int d\zeta \vert b^\dagger_\nu \phi(z) \vert^2 \ge 0.
\enq
Consequently $M^2 \ge 0$ if  $ \nu^2 \ge 0$.  The critical value  $\nu = 0$  corresponds to the lowest possible stable solution. If $\nu^2 <0$ the system is not bounded from below.

From the eigenvalue equation (\ref{eq:EE}) we obtain the wave equation
\begin{equation} \label{eq:LFWEx}
\left(- \frac{d^2}{d x^2} - \frac{1 - 4 \nu^2}{4 x^2} \right)
\phi_\nu(x) = \phi_\nu(x) ,
\end{equation}
where $x = \zeta M$.
In terms of the
operators $b_\nu$ and $b_\nu^\dagger$ (\ref{eq:LFWEx}) is equivalent to
\begin{equation}
b_\nu b_\nu^\dagger \vert \nu \rangle = \vert \nu \rangle .
\end{equation}
Multiplying both sides on the left by $b_\nu^\dagger$,
\begin{equation}
b_\nu^\dagger b_\nu \{b_\nu^\dagger \vert \nu \rangle\} 
= \{b_\nu^\dagger \vert \nu \rangle \}.
\end{equation}
It is simple to verify that
\begin{equation} \label{eq:bdb}
b_\nu^\dagger b_\nu = b_{\nu+1} b_{\nu+1}^\dagger,
\end{equation}
and thus
\begin{equation}
 b_{\nu+1} b_{\nu+1}^\dagger \{b_\nu^\dagger \vert \nu \rangle\} 
= \{b_\nu^\dagger \vert \nu \rangle \}.
\end{equation}
Consequently
\begin{equation}
b_\nu^\dagger \vert \nu \rangle = c_\nu \vert \nu + 1 \rangle,
\end{equation}
or
\begin{equation} \label{eq:bdx}
\left(- \frac{d}{dx} + \frac{\nu + \half}{x}\right) \phi_\nu(x) =
c_\nu \phi_{\nu+1}(x),
\end{equation}
with $c_\nu$ a constant.  Thus $b_\nu^\dagger$ is the raising operator. Likewise, one can show that $b_\nu$ is the lowering operator,
\begin{equation}
b_\nu \vert \nu + 1 \rangle = c_\nu \vert \nu  \rangle,
\end{equation}
or
\begin{equation} \label{eq:bx}
\left(\frac{d}{dx} + \frac{\nu + \half}{x}\right) \phi_{\nu+1}(x) =
c_\nu \phi_{\nu}(x),
\end{equation}
with $c_\nu$ a constant.

Writing
\begin{equation}
\phi_\nu(x) = C \sqrt{x} F_\nu(x),
\end{equation}
and substituting in (\ref{eq:bdx}) we get
\begin{equation}
\frac{\nu}{x} F_\nu(x) - F'_\nu(x) \sim  F_{\nu+1}(x),
\end{equation}
a relation which defines a Bessel function $Z_{\nu + 1}(x)$  of rank $\nu + 1$ in terms of
a Bessel of rank $\nu$, $Z_\nu(z)$,~\cite{AS64}
\beq
\frac{\nu}{x} Z_\nu(x) - Z'_\nu(x) =  Z_{\nu+1}(x).
\enq
Thus the normalizable solution to the eigenvalue equation  \req{eq:EE} is
\beq
\phi_\nu(\zeta) = C_\nu \sqrt{z}  J_\nu(\zeta M)
\enq
with $C_\nu$ a constant. The eigenvalues are obtained from the boundary conditions and are given in terms of the roots of the Bessel functions.

\subsection{Light-front soft-wall model \label{LFSWB}}

We can introduce a scale by modifying the operators \ref{b}  and \ref{bdag} while keeping an integrable system. Let us consider the extended operator
\begin{equation} \label{bla}
b_\nu = \frac{d}{d \zeta} + \frac{\nu + \half}{\zeta} + \la \zeta,
\end{equation}
and its adjoint 
\begin{equation} \label{bladag}
b^\dagger_\nu = -\frac{d}{d \zeta} + \frac{\nu + \half}{\zeta} + \la \zeta.
\end{equation}

Since we have introduced a scale $\la$ in the problem, the effective Hamiltonian has the general form
\beq
H_{LF}^\nu = b_\nu b^\dagger_\nu  + C(\la),
\enq
where the constant term $C(\la)$ depends on the spin representations.  Since the Hamiltonian is a bilinear form, modulo a constant,  its eigenvalues are positive definite $M^2 \ge 0$ provided that  $\nu^2 \ge 0$ and $C(\la) \ge - 4 \la$.

Let us consider the case where the ground state is massless (the pion). In this case $C(\la) = - 4 \la$ and the LF effective Hamiltonian is given by  
\begin{equation} \label{LFHla}
H_{LF}^\nu(\zeta) = - \frac{d^2}{d \zeta^2} - \frac{1 - 4 \nu^2}{4 \zeta^2}
+ \la^2 \zeta^2 + 2 \la (\nu - 1),
\end{equation}
which is identical to the LF Hamiltonian from \req{LFWEbis} with effective potential \req{UJ},  
\beq
U(\zeta) =  \la^2 \zeta^2 + 2 \la (\nu - 1),
\enq
for $\nu = J = L$.

Following the analysis of  Sec. \ref{LFHW} it is simple to show that the operator $b_\nu^\dagger$ acts as the creation operator,
\begin{equation}
b_\nu^\dagger \vert \nu \rangle = c_\nu \vert \nu + 1 \rangle,
\end{equation}
or
\begin{equation} \label{eq:bdkx}
\left(- \frac{d}{d \zeta} + \frac{\nu + \half}{\zeta}
+ \la \zeta \right) \phi_\nu(\zeta) =
c_\nu \phi_{\nu+1}(\zeta).
\end{equation}
with $c_\nu$ a constant.

We also consider the operator
\begin{equation}
a_\nu = - \frac{d}{d \zeta} + \frac{\nu + \half}{\zeta} - \la \zeta,
\end{equation}
and its adjoint
\begin{equation}
a^\dagger_\nu = \frac{d}{d \zeta} + \frac{\nu + \half}{\zeta} - \la \zeta.
\end{equation}
It is also simple to verify that the operator $a_\nu$
lowers the radial quantum number $n$ by one unit and raises $\nu$ by one unit
\begin{equation}
a_\nu \vert n, \nu \rangle \sim \vert n - 1, \nu + 1 \rangle.
\end{equation}
Notice that the state $\vert n - 1, \nu + 1 \rangle$ obtained by 
application of the operator $a_\nu$ is degenerate with the state 
$\vert n , \nu \rangle$. For a given $\nu$ the lowest possible state 
corresponds to $n = 0$. Consequently the state $\vert n=0, \nu \rangle$
is annihilated by the action of the operator $a_\nu$
\begin{equation}
a_\nu \vert n=0, \nu\rangle = 0,
\end{equation}
or equivalently
\begin{equation}
\left(\frac{d}{d \zeta} - \frac{\nu + \half}{\zeta} 
+ \la \zeta\right) \phi^{n=0}_\nu(\zeta) =0,
\end{equation}
with solution
\begin{equation}
\phi^{n=0}_\nu(\zeta) = C_\nu \zeta^{1/2 + \nu} e^{-\la \zeta^2/2},
\end{equation}
where $C_\nu$ is a constant. Writing
\begin{equation}
\phi_\nu(\zeta) = C_\nu \zeta^{1/2 + \nu} e^{-\la \zeta^2/2} G_\nu(\zeta),
\end{equation}
and substituting in (\ref{eq:bdkx}) we get
\begin{equation}
2 x G_\nu(x) - G'(x) \sim x G_{\nu + 1}(x),
\end{equation}
with $x = \sqrt{\la} \zeta$, 
a relation which defines the confluent hypergeometric function $U(n, \nu + 1, x)$ in terms of $U(n, \nu, x)$~\cite{AS64}
\beq
  U(n, \nu, x)  - U'(n, \nu, x) =  U(n, \nu + 1, x) ,
\enq
or equivalently
\beq
  2 x \, U(n, \nu, x^2)  - \frac{d U(n, \nu, x^2)}{dx}  =  2 x \, U(n, \nu + 1, x^2) .
\enq
Thus the normalizable solution of the eigenvalue equation  $b_\nu b_\nu^\dagger \, \phi= M^2 \phi$ is
\begin{equation}
\phi_{n, \nu}(\zeta) = C_\nu \, \zeta^{1/2 + \nu} e^{-\la \zeta^2/2}  L_n^\nu(\la \zeta^2),
\end{equation}
with $C_\nu$ a constant.
The solution also follows from the iterative application of the ladder operators, the Rodriguez formula for the Laguerre polynomials  (See~\cite{Arik:2008}). We find
\beq
\phi(\zeta)_{n, \nu}  \sim  \zeta^{1/2 - \nu} e^{\la \zeta^2/2} \left(\frac{1}{\zeta} \frac{d}{d \zeta} \right)^n \zeta^{2(n + \nu)} e^{- \la \zeta^2},
\enq
with eigenvalues
\beq
M^2 = 4 \la (n + \nu + 1).
\enq

\section{Light-front effective fermionic Hamiltonians}

	In this section we extend the algebraic procedure described in Sec. \ref{LFEBH} to construct light-front effective Hamiltonians for LF baryonic modes. We will describe first the conformal case where we have an exact prescription from the mapping of AdS wave equations  (See Secs. \ref{half} and \ref{half-integer-map}). Then, as for for the case of LF bosonic modes described above, we extend the conformal limiting case to include a scale while maintaining integrability of the Hamiltonian eigenvalue equations~\cite{ Brodsky:2008pg, Teramond:2007conf}. This procedure turns out to be particularly useful since in AdS the confining  dilaton background is absorbed by a rescaling of the Dirac field (Sec. \ref{half}), and thus we have little guidance in this case from the gravity side. However, as we shall show below, a consistent solution can be found by imposing the correct transformation properties for half-integer spin.

\subsection{Light-front hard-wall model}

We consider an effective light-front Dirac-type equation to describe a
baryonic state in holographic QCD. In the conformal limit
\begin{equation} \label{eq:DEzeta}
\left( D_{LF} - M \right) \psi(\zeta), 
\end{equation}
where  the  Dirac operator  $D_{LF}$  is the hermitian operator
\beq \label{DEzeta}
D_{LF} =  -i \al   \left(-\frac{d}{d \zeta} + \frac{\nu + \half}{\zeta} \ga \right),
 \enq
and $\al$ and $\ga$ are matrices to be determined  latter. 

It is useful to define the matrix-valued (non-Hermitian)  operator
\begin{equation}  \label{bf}
\mbf{b} = \frac{d}{d \zeta} + \frac{\nu + \half}{\zeta} \ga ,
\end{equation}
and its adjoint 
\begin{equation} \label{bfdag}
\mbf{b}^\dagger  = - \frac{d}{d \zeta} + \frac{\nu + \half}{\zeta} \ga .
\end{equation}
Premultiplying the linear Dirac wave equation (\ref{DEzeta}) by the operator $D_{LF} + M$ we should recover the LF Hamiltonian eigenvalue equation:
\begin{equation} \label{DLFH}
H_{LF}  \,  \psi = D_{LF}^2 \psi = M^2  \psi,
\end{equation}
which imply  that
\begin{eqnarray} \label{eq:DM1}
\alpha^\dagger = \alpha,~~~ \alpha^2 = 1,\\ \label{eq:DM2}
\gamma^\dagger = \gamma,~~ \gamma^2 = 1,\\ \label{eq:DM3}
\{\alpha, \gamma\} = 0.
\end{eqnarray}
Consequently the matrices $\alpha$ and $\gamma$ are four-dimensional Dirac
matrices.

The effective light-front Hamiltonian $H_{LF} = D_{LF}^2 =  \mbf{b}_\nu \mbf{b}^\dagger_\nu$ is thus given by
\begin{equation}  \label{eq:D2LFH}
H_{LF}^\nu 
= - \frac{d^2}{d \zeta^2} 
+ \frac{\left(\nu + \half\right)^2}{\zeta^2} - \frac{\nu + \half}{\zeta^2} \gamma.
\end{equation}
The positivity of the product of operators   imply that $\langle \psi \vert H^\nu_{LF} \vert \psi \rangle \ge 0$, and thus
$M^2 \ge 0$ if  $\nu^2 \ge 0$, identical to the stability bound for the scalar case.

To satisfy  the wave equation (\ref{eq:DEzeta}) for each component $\psi_\al$ we require that the matrix $\gamma$ satisfies the equation
\begin{equation}
\gamma \, u_\pm = \pm u_\pm,
\end{equation} 
where $u_\pm$ are four-component chiral spinors.  Consequently the matrix $\gamma$ is the four dimensional chirality operator $\gamma_5$. 
The LF equation  (\ref{eq:D2LFH}) thus leads to the eigenvalue equation
\begin{equation}
H_{LF} \psi_\pm = M^2 \psi_\pm,
\end{equation}
where
\begin{equation} \label{eq:LFWEp}
\left(- \frac{d^2}{d \zeta^2} - \frac{1- 4 \nu^2}{4 \zeta^2} \right)\psi_+(\zeta) =   M^2 \psi_+(\zeta)
\end{equation}
and
\begin{equation} \label{eq:LFWEm}
\left( - \frac{d^2}{d \zeta^2} - \frac{1 - 4 (\nu+1)^2}{4 \zeta^2}   \right) \psi_-(\zeta) = M^2   \psi_-(\zeta). 
\end{equation}
These are two uncoupled equations for the upper and lower spinor components,
$\psi_+$ and $\psi_-$, with the solution
\begin{equation}
\psi_+(\zeta) \sim \sqrt{\zeta} J_\nu(\zeta M), ~~~~~
\psi_-(\zeta) \sim \sqrt{\zeta} J_{\nu+1}(\zeta M).
\end{equation}
The plus and minus  components are not independent since they
must also obey the first order Dirac equation (\ref{eq:DEzeta}). In the Weyl representation where the chirality operator $\gamma$ is diagonal  ($i \alpha = \gamma \beta$) we have
\begin{equation} 
 i \alpha =
  \begin{pmatrix}
  0& I\\
- I& 0
  \end{pmatrix}, ~~~~~
 \beta =
  \begin{pmatrix}
  0& I\\
  I& 0
  \end{pmatrix},~~~~~
\gamma =  
  \begin{pmatrix}
  I&   0\\
  0&  -I
  \end{pmatrix},
\end{equation}
where $I$ a two-dimensional unit matrix.  The linear equation (\ref{eq:DEzeta}) is equivalent to the system of coupled  equations
\begin{eqnarray} \label{eq:D1l}
- \frac{d}{d\zeta} \psi_- -\frac{\nu+\half}{\zeta}\psi_- &=&
M \psi_+, \\ \label{eq:D2l}
  \frac{d}{d\zeta} \psi_+ -\frac{\nu+\half}{\zeta}\psi_+ &=&
M \psi_-,
\end{eqnarray}
a result which is identical with the results wich follow from the Dirac AdS wave  equation in the conformal limit~\cite{Brodsky:2008pg, Teramond:2007conf} (See  Sec. \ref{half-integer-map}). Solving the coupled equations (\ref{eq:D1l} - \ref{eq:D2l}) and making use of the relation between Bessel
functions
\begin{equation} 
J_{\nu+1}(x) = \frac{\nu}{x} J_\nu(x) - J'_\nu(x),
\end{equation}
we obtain the solution
\begin{equation} \label{eq:DEs}
\psi(\zeta) = C \sqrt{\zeta} \left[ J_\nu(\zeta M) u_+
   +   J_{\nu+1}(\zeta M) u_- \right],
\end{equation}
with normalization
\begin{equation} \label{eq:spnorm}
\int d\zeta \,  \psi^2 _+(\zeta)   =  \int d \zeta  \, \psi^2_-(\zeta) .
\end{equation}
identical for the plus and minus components.

\subsection{Light-front soft-wall model}

We write the Dirac equation  
\beq \label{DEzetala}
\left(D_{LF} - M \right)\psi(\zeta) = 0,
\enq
and construct an extended LF Dirac operator $D_{LF}$ following the same procedure used for the bosonic case in Sec.  \ref{LFSWB},  where we introduced a scale by making the substitution
$\frac{\nu + 1/2}{\zeta} \to \frac{\nu + 1/2}{\zeta} + \la \zeta$.   Thus the extended Dirac operator
\beq \label{DEzeta}
D_{LF} =  -i \al   \left(-\frac{d}{d \zeta} + \frac{\nu + \half}{\zeta} \ga +  \la \zeta \gamma \right),
 \enq
and the extended matrix-valued non-Hermitian  operators
\beqa  \label{bfla}
\mbf{b} &=& \frac{d}{d \zeta} + \frac{\nu + \half}{\zeta} \ga  +  \la \zeta \gamma, \\
\mbf{b}^\dagger & =&  - \frac{d}{d \zeta} + \frac{\nu + \half}{\zeta} \ga +  \la \zeta \gamma .
\enqa

The  effective  light-front Hamiltonian $H_{LF} = D^2_{LF} = \mbf{b} \mbf{b}^\dagger$ is given by
\begin{equation} \label{eq:LFHs}
H_{LF}= - \frac{d^2}{d \zeta^2} 
+ \frac{\left(\nu + \half\right)^2}{\zeta^2} - \frac{\nu +
  \half}{\zeta^2} \gamma + \la^2 \zeta^2 +
\la (2 \nu + 1) + \la \gamma, 
\end{equation}
with $\gamma$ the chirality matrix $\gamma u_\pm = \pm u$.

The eigenvalue  equation $H_{LF} \psi_\pm = M^2 \psi_\pm$ leads to the uncoupled light-front wave equations 
\begin{eqnarray} \label{eq:LFDEpk}
\left(- \frac{d^2}{d \zeta^2} - \frac{1- 4 \nu^2}{4 \zeta^2}  + \la^2 \zeta^2 + 2 (\nu+1) \la \right) \psi_+(\zeta) &=&  M^2  \psi_+(\zeta), \\ 
\label{eq:LFDEmk}
\left(- \frac{d^2}{d \zeta^2} - \frac{1 - 4 (\nu+1)^2}{4 \zeta^2}  + \la^2 \zeta^2 + 2 \nu \la \right) \psi_-(\zeta) &=&  M^2 \psi_-(\zeta) , 
\end{eqnarray}
with solutions
\begin{eqnarray}
\psi_+(\zeta) &\sim& z^{\frac{1}{2} + \nu} e^{-\la \zeta^2/2}
  L_n^\nu(\la \zeta^2) ,\\
\psi_-(\zeta) &\sim&  z^{\frac{3}{2} + \nu} e^{-\la \zeta^2/2}
 L_n^{\nu+1}(\la \zeta^2), 
\end{eqnarray}
and  eigenvalues
\begin{equation}
M^2 = 4 \la (n + \nu + 1),
\end{equation}
identical for both plus and minus eigenfunctions.

Using the $2 \times 2$
representation of the Dirac matrices given in the previous section we find
 the system of coupled linear equations
\begin{eqnarray} \label{eq:cD1k}
- \frac{d}{d\zeta} \psi_- -\frac{\nu+\half}{\zeta}\psi_- 
- \la \zeta \psi_-&=&
M \psi_+, \\ \label{eq:cD2k}
  \frac{d}{d\zeta} \psi_+ -\frac{\nu+\half}{\zeta}\psi_+ 
- \la \zeta \psi_+ &=&
M \psi_-. 
\end{eqnarray}
This result is identical with the results from the AdS wave equation in presence of a potential $V(z) = \la z$~\cite{ Brodsky:2008pg, Teramond:2007conf} (See  Sec. \ref{half-integer-map}).

Solving the coupled equations (\ref{eq:cD1k} - \ref{eq:cD2k})  making use of  the relation between associated Laguerre functions
\begin{equation}
L_{n-1}^{\nu+1}(x) + L_n^\nu(x) = L_n^{\nu+1}(x),
\end{equation}
 we find
\begin{equation}
\psi(\zeta) = C  z^{\frac{1}{2} + \nu} e^{-\la \zeta^2/2}
\left[ L_n^\nu\left(\la \zeta^2\right) u_+
   + \frac{\sqrt{\la} \zeta}{\sqrt{n +\nu + 1}}  
L_n^{\nu+1}\left(\la \zeta^2\right) u_- \right],
\end{equation}
with normalization
\begin{equation} \label{eq:spnorm}
\int d\zeta \,  \psi^2 _+(\zeta)   =  \int d \zeta  \, \psi^2_-(\zeta)  .
\end{equation}

It is important to notice that, in contrast to the bosonic case,  one cannot add a constant term to the light-front effective Hamiltonian with dependence on the spin representations for baryons.    This constraint  follows from the fact that the plus and minus components are not independent and obey the first order linear equation  \ref{DEzetala}. This additional requirement has the notable consequence that, in contrast to bosons, there is no spin-orbit coupling in the light-front holographic model for baryons as discussed in Chapters \ref{ch4} and \ref{ch5}.

\chapter{Equations of Motion for $\bf p$-Form Fields in AdS \label{pform}}
In this appendix we describe the properties of massive $p$-form
fields propagating in AdS space. As we discussed in Chapter \ref{ch4}, obtaining the general form of the
equations of motion for higher-spin fields in AdS space may become
quite complex. It is thus useful to study the simplified structure
of the differential equations of $p$-form fields in AdS, which for
$p = 0$ and $p = 1$ represent spin 0 and spin 1 respectively (Sec. \ref{scal}). The
compact formalism of differential forms is particularly convenient
to describe the solutions of higher $p$-form actions in
AdS$_{d+1}$ space. In this notation a $p$-form field $\bf A$ is
\begin{equation}
{\bf A}=    \mathcal{A}_{M_1 M_2 \cdots M_p} \,
dx^{M_1}\wedge \cdots \wedge dx^{M_p},
\end{equation}
in the dual basis $dx^M$. The tensor field $\mathcal{A}_{M_1
M_2 \cdots M_p}$ is a totally antisymmetric tensor of rank
$p$, and the sum is over $M_1 < \cdots < M_p$.  AdS$_{d+1}$
coordinates are the Minkowski coordinates $x^\mu$ and $z$ which we
label $x^M$, with $M, N = 0, \cdots , d$.

The field strength ${\bf F}$ of the $p$ form {\bf A}  is the $p +
1$ form given by the exterior derivative
\begin{equation}
{\bf F} =  d {\bf A} = \partial_M \mathcal{A}_{M_1 M_2
\cdots M_p} \, dx^{M_1} \wedge \cdots \wedge dx^{M_p}
\wedge dx^{M} ,
\end{equation}
with sum over $M$ and $M_1 < \cdots < M_p$. The 
invariant Lagrangian density must be a $d+1$ form. This leads to
the action in geometrical units
\begin{equation} \label{eq:Skform}
S = \frac{1}{2} \int_{AdS_{d+1}}  \! \left( {\bf F} \wedge ^* \! {\bf F}
- \mu^2 {\bf A} \wedge ^* \! \! {\bf A}\right),
\end{equation}
where $\mu$ is the AdS  mass.

The \emph{wedge product} $\wedge$  of a $p$-form
 $\bf  A$ and a $q$-form $\bf B$ is the $p + q$ form $\bf A \wedge
 B$.  In tensor notation:
 \beq
 {\bf A} \,\wedge {\bf B} = \frac{(p+q)!}{p!\,q!} A_{[M_1 \dots
 M_p} \,B_{M_{p+1} \dots M_{p+q}]}dx^{M_1}\wedge \dots
 \wedge ^dx^{M_{p+q}}.
\enq

The \emph{hodge star} operator
establishes a correspondence between $p$-forms to $d+1 -p$ forms
on Ad$S_{d+1}$. In tensor notation the Hodge dual of $\bf A$ is
obtained by contracting the indices of $\bf A$ with the
$d+1$-dimensional completely antisymmetric Levi-Civita tensor
\begin{equation}
(^*\negthinspace {\bf A})_{M_1 M_2 \cdots M_{d+1-p} }=
 \frac{1}{p!} ({\bf A})^{N_1 N_2 \cdots N_p} \eta_{N_1 N_2 \cdots N_p,
M_1  \cdots M_{d+1-p}},
\end{equation}
where the Levi-Civita tensor is
\beq
 \eta_{M_1 \cdots M_p} =\sqrt{g}  \epsilon_{M_1 \cdots M_p},
\enq 
with $g \equiv \vert \rm {det } \, g_{MN}  \vert$ and $\epsilon_{M_1 \cdots M_p}$ the totally
antisymmetric tensor density with entries $\pm 1$.

The classical equations of motion
for $\bf A$ follow from the variation of  the action
(\ref{eq:Skform}). Making use of the Stokes theorem we obtain:
 \beq
 \int_{AdS_{d+1}} \! d\left( {\bf A} \wedge ^* \!d{\bf A}\right)=0,
  \enq
 from which follows
  \beq
 \int_{AdS_{d+1}}  \! \left( d{\bf A} \wedge ^* \!d {\bf A}\right)=
 (-1)^p\, \int_{AdS_{d+1}}  \! \left( {\bf A} \wedge d^* \!d {\bf
 A}\right),
 \enq
 yielding
\begin{equation} \label{eq:eomPhiS}
(-)^p d\, ^* \! d {\bf A} + \mu^2 \, ^* \! {\bf A} = 0.
\end{equation}
Since $dd \,\bf{A} =0$, this equation implies for $\mu \neq 0$
\begin{equation}
d\, ^* \! {\bf A} = 0.
\end{equation}

In tensor notation the equations of motion for $\bf A$ are expressed as the set of $p+1$
coupled differential equations~\cite{l'Yi:1998eu}
\begin{eqnarray} \label{eq:PhiStz} \label{eq:eomPhi1}
\left[ z^2 \partial_z^2 - (d + 1 - 2 p) z \, \partial_z - z^2 \partial_\rho \partial^\rho
      - (\mu R)^2 +  d + 1 - 2 p \right] \! \mathcal{A}_{z  \alpha_2 \cdots \alpha_p}
      \! &=& \! 0,
      \\  \label{eq:eomPhi2}
 \left[ z^2 \partial_z^2 - (d + 1 - 2 p) z \, \partial_z - z^2 \partial_\rho \partial^\rho
      - (\mu R)^2 + d + 1 - 2 p \right] \! \mathcal{A}_{\alpha_1  z \cdots \alpha_p}
      \! &=& \! 0, \\
      \nonumber  \cdots  \\ \label{eq:eomPhi3}  \nonumber
 \left[ z^2 \partial_z^2 - (d - 1 -2 p) z \, \partial_z - z^2 \partial_\rho \partial^\rho
      - (\mu R)^2  \right]  \mathcal{A}_{\alpha_1  \alpha_2 \cdots \alpha_p}
     \! &=& \! 2z \bigl( \partial_{\mu_1}  \mathcal{A}_{z  \alpha_2 \cdots \alpha_p}  \\
      + \, \, \partial_{\mu_2} \mathcal{A}_{\alpha_1  z \cdots \alpha_p}
      + \cdots\bigr),
\end{eqnarray}
where $\al, \rho = 0,1,2, \cdots, d - \!1$ and
the notation $z  \alpha_2 \cdots \alpha_p$ means $M_1=z$, $M_2= \alpha_2$,  $\cdots$,
$M_p = \alpha_p$, etc.

We introduce fields with  Lorentz tangent indices  $A, B = 0, \cdots , d$,
\begin{equation} \label{eq:ti}
 \mathcal{\hat A}_{A_1 A_2 \cdots A_p}
 = e_{A_1}^{M_1} e_{A_2}^{M_2} \cdots e_{A_p}^{M_p} \,
 \mathcal{A}_{M_1 M_2 \cdots M_p}
 =  \left(\frac{z}{R}\right)^p  \negthinspace \mathcal{A}_{A_1 A_2 \cdots A_p},
 \end{equation}
where $e_A^M$ is the vielbein  (See Sect. \ref{A13}). In terms of $\mathcal{\hat A}$
we obtain from  (\ref{eq:eomPhi1}-\ref{eq:eomPhi3})  the set of $p+1$ differential equations~\cite{l'Yi:1998eu}
\begin{eqnarray} \label{eq:eomPhiT1}
\left[ z^2 \partial_z^2 - (d - 1) z \, \partial_z - z^2 \partial_\rho \partial^\rho
      - (\mu R)^2 + p(d - p) \right]
      \Big(\frac{\mathcal{\hat A}_{z  \alpha_2 \cdots \alpha_p}}{z}\Big)  &=& 0, \\
 \left[ z^2 \partial_z^2 - (d - 1) z \, \partial_z - z^2 \partial_\rho \partial^\rho
      - (\mu R)^2 + p(d - p) \right]
      \Big(\frac{\mathcal{ \hat A}_{\alpha_1  z \cdots \alpha_p}}{z}\Big)  &=& 0, \\    \nonumber
     \cdots \\    \label{eq:PhipT} \nonumber
\left[ z^2 \partial_z^2 - (d - 1) z \, \partial_z - z^2 \partial_\rho \partial^\rho
      - (\mu R)^2 + p(d - p) \right] \mathcal{\hat A}_{\alpha_1  \alpha_2 \cdots \alpha_p}
      &=& 2z \bigl( \partial_{\mu_1} \mathcal{\hat A}_{z  \alpha_2 \cdots \alpha_p}  \\ \label{eq:eomPhiT3}
      + \,\, \partial_{\mu_2}  \mathcal{\hat A}_{\alpha_1  z \cdots \alpha_p}
      + \cdots\bigr).
      \end{eqnarray}
      
 Consider the plane-wave solution $\mathcal{A}_P(x^\mu, z)_{\alpha_1 \cdots \alpha_p}  \! = \! e^{ i P \cdot x} \,
\mathcal{A}(z)_{\alpha_1  \cdots \alpha_p}$, with 4-momentum}
$P_\mu$, invariant hadronic mass $P_\mu P^\mu = M^2$ and
spin indices $\alpha$ along the space-time coordinates, that is
$\mathcal{A}_{z  \alpha_2 \cdots \alpha_p} = \mathcal{A}_{\alpha_1
z \cdots \alpha_p} = \cdots = 0$. In this case the system of
coupled differential  equations
(\ref{eq:eomPhi1}-\ref{eq:eomPhi3}) reduce to the  homogeneous
wave equation
\begin{equation} \label{eq:eomPhipz}
\left[ z^2 \partial_z^2 - (d\! -\! 1 \!- \!2 p) z \, \partial_z + z^2 M^2
\!  -  (\mu R)^2 \right] \!  \mathcal{A}_{\alpha_1 \alpha_2 \cdots \alpha_p}  = 0.
\end{equation}

In tangent space the coupled differential equations  (\ref{eq:eomPhiT1}-\ref{eq:eomPhiT3}) for all polarization indices along the Poincar\'e coordinates reduces to
\begin{equation}
\big[ z^2 \partial_z^2 - (d - 1) z \, \partial_z + z^2 M^2
      - (\mu R)^2 + p(d - p) \big] \mathcal{\hat A}_{\alpha_1  \alpha_2 \cdots \alpha_p} = 0.
\end{equation}
Its solution is
 \begin{equation}
\mathcal{\tilde A}(z)_{\alpha_1 \alpha_2 \cdots \alpha_p}  =  C   z^\frac{d}{2}
 J_{\Delta-\frac{d}{2}} \!
 \left(z M  \right) \epsilon_{\alpha_1 \alpha_2 \cdots \alpha_p},
 \end{equation}
with conformal dimension~\cite{l'Yi:1998eu}
\begin{equation} \label{eq:DeltaS}
\Delta = \frac{1}{2} \bigl( d \pm \sqrt{ (d - 2 p)^2 + 4 \mu^2 R^2} \bigr).
\end{equation}
Thus the relation  
\beq \label{Deltamup}
(\mu R)^2 = (\Delta-p)(\Delta-d+p),
\enq
 for a $p$-form field with dimension $\Delta$.   The  relation \req{eq:DeltaS} agrees with the conventions in Refs.  \cite{Aharony:1999ti, Freedman:1999gp}. For a spinor field in AdS the mass-dimension relation is~\cite{Henningson:1998cd}
\beq
\Delta = \half \left(d + 2  \vert \mu R \vert  \right) .
\enq
The relation for spin-$\threehalf$ is unchanged~\cite{Volovich:1998tj, Koshelev:1998tu}.


\end{document}